\documentclass[a4paper,11pt]{article}
\usepackage{jheppub} 
\usepackage{lineno}

\usepackage{cleveref}
\usepackage{diffcoeff}
\usepackage{physics2}
\usephysicsmodule{ab}
\usephysicsmodule{ab.legacy}
\usephysicsmodule{braket}
\usephysicsmodule{diagmat}
\usepackage{mathtools}
\usepackage{lscape}
\usepackage{array}
\usepackage{tabularray}
\usepackage{comment}
\usepackage{afterpage}
\usepackage{nicematrix}

\allowdisplaybreaks

\NewDocumentCommand{\comm}{s m m}{\IfBooleanTF{#1}{\bab*{#2, #3}}{\bab{#2, #3}}}
\NewDocumentCommand{\acomm}{s m m}{\IfBooleanTF{#1}{\Bab*{#2, #3}}{\Bab{#2, #3}}}
\NewDocumentCommand{\Tr}{}{\operatorname{Tr}}
\NewDocumentCommand{\expval}{s O{} m}{\IfBooleanTF{#1}{\braket*[1,#2]{#3}}{\braket[1,#2]{#3}}}
\NewDocumentCommand{\dn}{s o m}{\IfNoValueTF{#2}{\mathrm{d}#3}{\mathrm{d}^{#2}#3}\IfBooleanTF{#1}{}{\,}}
\NewDocumentCommand{\txtA}{}{\text{A}}
\NewDocumentCommand{\txtB}{}{\text{B}}
\NewDocumentCommand{\txtI}{}{\text{I}}
\NewDocumentCommand{\txtR}{}{\text{R}}

\NewDocumentCommand{\omegabar}{}{\overline{\omega}}
\NewDocumentCommand{\supi}{}{{(i)}}
\NewDocumentCommand{\supT}{}{{(T)}}
\NewDocumentCommand{\bvec}{m}{{\boldsymbol{#1}}} 
\NewDocumentCommand{\adj}{}{\operatorname{adj}} 
\NewDocumentCommand{\pv}{}{\operatorname{p.v.}} 
\NewDocumentCommand{\T}{}{{\operatorname{T}}} 
\NewDocumentCommand{\residue}{}{{\operatorname{Res}}}

\difdef { f, fp, s, sp } { | }
{
outer-Ldelim = \left . ,
outer-Rdelim = \right |,
sub-nudge = 0 mu
}

\title{\boldmath Field mixing in a thermal medium: A quantum master equation approach}

\author{Shuyang Cao}
\affiliation{Department of Physics \& Astronomy, University of Pittsburgh,\\
100 Allen Hall
3941 O'Hara St
Pittsburgh, PA 15260, USA}

\emailAdd{shuyang.cao@pitt.edu}

\abstract{ 
We studied the nonequilibrium dynamics of the indirect mixing of two (pseudo-)scalar fields induced by their couplings to common decay channels in a medium. The effective non-Markovian quantum master equation (QME) for the two fields' reduced density matrix is derived to leading order in the couplings of the two fields with the medium, but to all orders of the couplings among degrees of freedom in the medium. The self-energy and noise-kernel in the QME satisfy a fluctuation-dissipation relation. The solutions show that an initial expectation value (condensate) of one field induces a condensate of the other field through the indirect mixing and that the populations and coherence of the two fields thermalize and approach to non-vanishing values asymptotically. The nearly-degenerate field masses and coupling strengths resonantly enhance the quantum beats and asymptotic coherence, and induce a prominent dynamics of the vacuum after the switch-on of the couplings. We argue that a time-dependent definitions of particles due to the changing vacuum must be introduced so as to obtain results consistent with the calculations of equilibrium states in the asymptotic limit. A coupling strength hierarchy breaks down the resonant enhancement in the nearly-degenerate case but leads to different power countings of the coupling strengths in the magnitudes of the observables and time-scales in the evolution, suggesting the possibility of detecting extremely long-lived particles using prepared short-lived particles within a practical experimental period.
}

\begin{document}
\maketitle
\flushbottom

\section{Introduction}

The mixing of two fields induced by their couplings to common decay channels is of broad fundamental interest within the context of CP violation and/or baryogenesis. It may also serve as the mixing mechanism for particles in the Standard Model (SM) and/or hypothetical dark sectors beyond the Standard Model (BSM), via the exchange of (hypothetical) mediators in the ``portal'' connecting different sectors, namely, common intermediate states with which the fields on different sides of the portals couple.

A paradigmatic example in vacuum is the mixing of flavored meson-antimesons as a consequence of the common intermediate states in the weak interaction box diagrams, which has been heavily studied in colliders \cite{Bigi_Sanda_2009, 10.1093/oso/9780198503996.001.0001, Nierste2009-vy, DDbarmixing, CPQuark, PILAFTSIS199761, KITTEL2012682}. In the early Universe, such mixing phenomena occur in a bath where thermal corrections become important.

More general field mixing beyond the particle-antiparticle mixing may be ubiquitous in cosmology. One instance is axion-like particles (ALPs). The axion is a CP-odd pseudoscalar particle proposed in BSM extensions as a possible solution to the strong CP problem in Quantum Chromodynamics (QCD) \cite{PhysRevLett.38.1440,PhysRevD.16.1791,PhysRevLett.40.223,PhysRevLett.40.279}. Various BSM extensions also propose other hypothetical particles with properties similar to that of the QCD Axion, which are collectively referred to as ALPs \cite{BANKS1996173,RINGWALD2012116,MARSH20161,Marsh:2017hbv,Diez-Tejedor2017-qy,PhysRevD.93.025027,RevModPhys.93.015004,Sikivie2008}. Both of the QCD Axion and ALPs are compelling cold dark matter candidates \cite{PRESKILL1983127,ABBOTT1983133,DINE1983137,doi:10.1126/sciadv.abj3618}. Like the QCD axion, ALPs couple to gauge fields via Chern-Simons terms as a consequence of the chiral anomaly, such as $\vec{E}\cdot\vec{B}$ in the case of photons and $\tilde{G}^{\mu\nu;b} \tilde{G}_{\mu\nu;b}$ in the case of gluons. Their common couplings with gauge fields herald possible mixing of varions ALPs and/or the QCD axion via common intermediate states of the gauge fields. This mixing may connect different BSM dark sectors, yielding interesting dynamics among them worth studying.

The mixing between the SM and dark sectors is also possible, which may leave signatures in current observations. For example, at the QCD phase transition, the lightest quarks $u,d$ bind into the lightest meson, pion. This pseudoscalar isotriplet are copiously produced at $T_{QCD} \simeq 150 \,\text{--}\, 170 \text{MeV} \geq m_\pi \simeq 140 \text{MeV}$. Out of the three pion states $\pi^0$ and $\pi^\pm$, the neutral pion $\pi^0$ couples to photons via the $U(1)$ triangle anomaly in the same way as ALPs, resulting in a $\simeq 99\%$ branching ratio for its decay into two photons \cite{10.1093/ptep/ptac097}. The QCD axion and/or ALPs may be produced non-thermally in the early Universe, e.g., by a misalignment mechanism \cite{MARSH20161,RINGWALD2012116,PhysRevD.84.103501,PhysRevLett.124.251802,PhysRevD.102.015003}. Their coherent states may seed a small pion condensate after the QCD phase transition through the mixing between $\pi^0$ and the QCD Axion or ALPs \cite{PhysRevD.108.025012}.

Furthermore, it has been realized that axions may play a role in condensed matter physics, as emergent quasiparticles in topological insulators where magnetic fluctuations couple to electromagnetism just like axions \cite{PhysRevB.78.195424,Li2010-oz,Nenno2020-la,10.1063/5.0038804}, as axionic charge density waves in Weyl semimetals \cite{Gooth2019-mj,PhysRevB.104.174406}, as an emergent axion response in multilayered metamaterials with tunable couplings \cite{PhysRevB.108.115101} or in multiferroics \cite{PhysRevResearch.3.033236,Balatsky2023-wh}. These ``synthetic'' axions may mix with the cosmic axion in the same way as pions or ALPs in the early Universe. This possibility suggests the study of their mixing via a photon bath (i.e., cosmic microwave background), which may yield an alternative probe for cosmic ALPs in condensed matter experiments.

The theory of flavoured meson-antimeson mixing in vacuum via weak interaction intermediate states stems from the pioneering study of CP violation by Lee, Oehme and Yang \cite{PhysRev.106.340} which is based on the theory of atomic linewidth developed by Weisskopf and Wigner \cite{Weisskopf1930-ml,Weisskopf1930-ao,SACHS1963239,CPpuzzle,PhysRevD.42.3712}. It has been the cornerstone of the analysis of the mixing dynamics of flavored mesons and CP violation \cite{Bigi_Sanda_2009,10.1093/oso/9780198503996.001.0001,Nierste2009-vy,DDbarmixing,CPQuark}. However, using an effective Schr\"odinger equation, this theory is a single-particle description of equal-mass flavoured meson-antimeson mixing. It cannot be readily extended to study field mixing in a more general situation where masses are not necessarily equal, the initial state may be a condensate and the background may be a thermal bath, which requires a multi-particle description.

Motivated by the ubiquity and broad relevance of field mixing in particle physics, cosmology and condensed matter physics, we developed in this article a more general method using quantum master equation (QME) \cite{10.1093/acprof:oso/9780199213900.001.0001,PhysRevD.107.063518,PhysRevD.96.034021,PhysRevD.97.074009,PhysRevD.108.L011502,PhysRevD.91.056002,PhysRevD.101.034011} to study the indirect mixing of generic (pseudo-)scalar fields due to their couplings to common decay channels. In this method the mixing fields constitute an open quantum subsystem interacting with external environments made up of the common decay channels. We distinguish the indirect mixing from the direct mixing that we define as a result of explicit mixing terms in the Lagrangian, such as off-diagonal elements in mass matrices or kinetic mixing terms. In this article we focus on the indirect mixing. The QME method provides an alternative framework and an independent confirmation for the results in \cite{PhysRevD.108.025012,PhysRevD.109.036038} where the correlation functions in field mixing are studied using the Keldysh–Schwinger formalism. A distinct advantage of QME is that it yields the equations of motion for populations and coherence more readily and exhibits the thermalization and decoherence more clearly. The strategy adopted and conclusions drawn for populations and coherence may be extended to the study of open quantum systems in quantum information as QME is also a ubiquitous tool thereof \cite{Nielsen_Chuang_2010}.

Our objectives in this study are two-fold: \textbf{(I)} Derive an effective non-Markovian QME up to leading-order for the reduced density matrix of the subsystem. Show that the equation features a fluctuation-dissipation relation, which is rarely uncovered in the context of QME albeit its frequent appearances in Langevin-type equations of quantum Brownian motion using the Keldysh–Schwinger formalism \cite{FEYNMAN1963118,CALDEIRA1983587,Boyanovsky_2015,PhysRevD.109.036038,PhysRevD.108.025012}. Solve the effective equation without resorting to Markov, rotating-wave and/or secular approximations, which typically imply neglecting memory effects and off-resonance contributions. \textbf{(II)} Give the explicit leading order solutions with generic initial conditions (e.g., coherent states caused by misalignment mechanism) and a finite-temperature background. Discuss decoherence, thermalization and quantum beats in field amplitudes, populations and coherence in the cases of both non-degenerate and nearly-degenerate particle masses. None of these discussions can be easily accommodated in the traditional theory of flavored meson-antimeson mixing.

The article is organized as follows: In \cref{sec:effective-quantum-master-equation-of-field-mixing} we derive the effective non-Markovian QME for the reduced density matrix of the mixing fields up to leading order perturbations, show the generalized fluctuation-dissipation relation in the equation, and discuss the trace-preserving and hermicity-preserving properties of the equation. In \cref{sec:evolutions-of-field-amplitudes-and-their-mixing} we solve the evolution of field amplitudes and discuss the influence of the magnitude of the mass degeneracy on the field mixing effect. In \cref{sec:populations-and-coherence} we solve the evolution of populations and coherence, show the Kronecker-product relation between their quasi-normal modes and that of the field amplitudes, and discuss the thermalization, quantum beats and asymptotic medium-induced coherence in the cases of non-degenerate and nearly-degenerate masses. In \cref{sec:effects-of-a-coupling-strength-hierarchy}, we show the effects of a coupling strength hierarchy on the phenomena presented in \cref{sec:evolutions-of-field-amplitudes-and-their-mixing,sec:populations-and-coherence} that are originally enhanced by the nearly-degenerate masses. In \cref{sec:discussion}, we discuss the possible improvement on the asymptotic results of the evolution, show that the vacuum background is time-dependent in the nearly-degenerate case and the consequent new definitions of particles guarantee the non-negativity of populations in the framework we proposed, compare our results with that of the rotating-wave approximation and connect the proposed framework to the Lindblad form and the Weisskopf-Wigner theory. In \cref{sec:summary-of-results-and-conclusions} we summarize this article and draw conclusions. The relations and symmetries used in this article are presented in \cref{app:Lehmann-representation-of-correlation-functions,app:symmetries-of-self-energy,app:symmetries-of-noise-kernel,app:generalized-fluctuation-dissipation-relation}. We give a calculation of single species in \cref{app:single-species} to help clarify our approach to solve the equation. Explicit expressions of the results in this article are listed in \cref{app:explicit-results-for-one-point-correlation-functions} for one-point correlation functions and in \cref{app:explicit-results-in-evolutions-of-number-densities-and-coherence} for populations and coherence.

Throughout this article we use the mostly minus signature of the Minkowski metric $\eta_{\mu\nu} = \operatorname{diag}\pab*{1,-1,-1,-1}$. Bold letters are used for spatial Euclidean vectors. The contraction of compatible vectors, either four-vectors or Euclidean vectors, is written as dot product in which metrics are implicitly chosen, e.g, $k\cdot x = k^0 x_0 - \bvec{k}\cdot \bvec{x} = k^0 x_0 - \sum_{i=1}^3 k^i x_i$. For matrix symbols, column vectors are indicated by letters with arrows on the top and matrices are represented by bold letters. $F(\bvec{k},t)$ is the spatial Fourier transform of a general function $F(\bvec{x},t)$ in a $3+1$D space and $F(\bvec{k},k_0)$ (or $F(k)$) is its $3+1$D Fourier transform. We denote the Laplace transform of $F(\bvec{k},t)$ as $F(\bvec{k},s)$. It is distinguished from the $3+1$D Fourier transform via its arguments and the context. $k_0$ in the Fourier transform is always a real number and $s$ in the Laplace transform is a complex number on the $s$-plane.

\section{The effective quantum master equation of field mixing}
\label{sec:effective-quantum-master-equation-of-field-mixing}

Consider two (pseudo-)scalar fields $\phi_1$ and $\phi_2$ which couple to common decay channels in a medium. We denote the degrees of freedom in this medium generically as $\chi$. The total Hamiltonian is
\begin{equation}
    H = H_{\phi_1} + H_{\phi_2} + H_\chi + H_\txtI \,.
    \label{eqn:total-Hamiltonian}
\end{equation}

$H_{\phi_1}$ and $H_{\phi_2}$ are the free-field Hamiltonians of the two scalar fields.
\begin{equation}
    H_{\phi_i} = \int \dn[3]{x} \Bab{ \frac{1}{2} \pi_{\phi_i}^2 + \frac{1}{2} \pab{\nabla\phi_i}^2 + \frac{1}{2} m_i^2 \phi_i^2 },  \quad i = 1,2 \, .
\end{equation}
Without loss of generality, we assume $m_1 \geq m_2$. In what follows, we refer to the set of fields $\phi_1,\phi_2$ collectively as $\phi \equiv \Bab{\phi_1,\phi_2}$ to simplify notations. They constitute the ``subsystem'' in this open-system problem. The goal in this section is to obatin an effective QME for the reduced density matrix of this subsystem after tracing out the $\chi$ degrees of freedom.

$H_\chi$ is the Hamiltonian of the medium in the absence of the subsystem. It includes not only the free-field Hamiltonian of the generic field(s) $\chi$, but also all interactions among them inside the medium. For instance, if the medium is a plasma of photons and electrons, $H_\chi$ also includes the electron-photon coupling $e \bar{\psi} \gamma^\mu A_\mu \psi$ besides the free-field terms of electrons and photons.

$H_\txtI$ is the interaction between the subsystem and the medium. We assume it to be
\begin{equation}
    H_\txtI = \int \dn[3]{x} \Bab{ J_1[\phi_1] \mathcal{O}_1[\chi] + J_2[\phi_2] \mathcal{O}_2[\chi] } \,.
\end{equation}
where $J_{1,2}$ and $\mathcal{O}_{1,2}$ are Hermitian operator functions of their arguments. Examples of such a form include axion-photon coupling $g_{a\gamma\gamma} a\, \bvec{E}\cdot\bvec{B}$, axion-gluon coupling $g_\text{aGG}\, a\, G^{\mu\nu;b} G_{\mu\nu;b}$, or neutral Kaon decay $K, \overline{K} \to 2\pi, 3\pi$ in a low-energy effective field theory. The small coupling constant $g_1$ (or $g_2$) is included in $\mathcal{O}_1[\chi]$ (or $\mathcal{O}_2[\chi]$) for notational convenience. In this article, we will assume $g_1 \sim g_2$ and refer to them collectively as ``$g$'' when counting orders of perturbations, unless a coupling strength hierarchy is explicitly introduced.

In the Schr\"odinger picture (denoted by a superscript ``$(\text{S})$''), the von Neumann equation describes the evolution of the total density matrix of the whole system.
\begin{equation}
    \dot{\hat{\rho}}_{\text{tot}}^{(\text{S})} = -i \comm{H^{(\text{S})}}{\hat{\rho}_{\text{tot}}^{(\text{S})}} \,.
\end{equation}
We use $e^{-i H_0 t}$ with $H_0 = H_{\phi_1} + H_{\phi_2} + H_\chi$ to define the interaction picture (denoted by a superscript ``$(\text{int})$''). In the interaction picture, operators evolve as
\begin{equation}
    \hat{\rho}_{\text{tot}}^{(\text{int})}
    \coloneqq e^{i H_0 t} \, \hat{\rho}_{\text{tot}}^{(\text{S})}(t) \, e^{-i H_0 t}
    \,,
    \qquad
    \hat{\mathcal{A}}_{\text{tot}}^{(\text{int})}
    \coloneqq e^{i H_0 t} \, \hat{\mathcal{A}}^{(\text{S})} \, e^{-i H_0 t}
    \,.
    \label{eqn:interaction-picture}
\end{equation}
where $\hat{\mathcal{A}}$ is a general operator. The von Neumann equation in the interaction picture is
\begin{equation}
    \dot{\hat{\rho}}_{\text{tot}}^{\text{(int)}} = -i \comm{H_\txtI^{\text{(int)}}(t)}{\hat{\rho}_{\text{tot}}^{\text{(int)}}(t)} \,.
    \label{eqn:rho-exact-equation-in-interaction-picture}
\end{equation}
From now on, we will always work in the interaction picture and drop the superscript ``(int)''. Integrating \cref{eqn:rho-exact-equation-in-interaction-picture} and inserting the result back to the right hand side (RHS) of \cref{eqn:rho-exact-equation-in-interaction-picture} yields
\begin{equation}
    \dot{\hat{\rho}}_{\text{tot}} (t) = - i \comm{H_\txtI(t)}{\hat{\rho}_{\text{tot}}(0)} - \int_0^t \comm{H_\txtI(t)}{\comm{H_\txtI(t')}{\hat{\rho}_{\text{tot}}(t')}} \dl{t'} \,.
    \label{eqn:rho-eqn-exact-one-insertion}
\end{equation}
Repeating the integration and insertion results in a formal solution of $\rho_{\text{tot}}(t)$.
\begin{equation}
    \rho_{\text{tot}}(t) = \sum_{n=0}^\infty 
    \underbrace{ \int_0^t \dn*{t_n} \cdots \int_0^{t_2} \dn{t_1} }_{\text{$n$ integrals}} 
    \underbrace{ [ -i H_\txtI(t_n), \cdots, \comm{-i H_\txtI(t_2)}{\comm{-iH_\txtI(t_1)}{\hat{\rho}_{\text{tot}}(0)}} \cdots] \vphantom{\int_0^{t_n}} }_{\text{$n$ commutators}}
\end{equation}
The $n$-th term of this series is of the order $\order{H_\txtI^n} \sim \order{g^n}$. To find the equation of motion up to the leading order perturbation, we stop at \cref{eqn:rho-eqn-exact-one-insertion} after integrating one time. It is worth noting that no approximation has been done yet and \cref{eqn:rho-eqn-exact-one-insertion} is still exact.

The QME of the reduced density matrix of the subsystem is obtained after taking a partial trace over the $\chi$ degrees of freedom of the medium in \cref{eqn:rho-eqn-exact-one-insertion}.
\begin{equation}
    \dot{\hat{\rho}} = -i \Tr_\chi \Bab{ \comm{H_\txtI(t)}{\hat{\rho}_{\text{tot}}(0)} } - \int_0^t \Tr_\chi \Bab{ \comm{H_\txtI(t)}{\comm{H_\txtI(t')}{\hat{\rho}_{\text{tot}}(t')}} } \dl{t'} \,.
    \label{eqn:trace-over-chi}
\end{equation}
where $\hat{\rho} = \Tr_\chi \Bab{ \hat{\rho}_{\text{tot}}(t) }$ is the reduced density matrix. However, \cref{eqn:trace-over-chi} is not closed because its RHS still depends on the total density matrix $\hat{\rho}_{\text{tot}}$ instead of the reduced density matrix $\hat{\rho}$. To proceed, we follow refs.  \cite{10.1093/acprof:oso/9780199213900.001.0001,Gardiner2010-js,10.1063/1.522979}.

First, we assume that at $t=0$ the system is in a Kronecker product state of the subsystem and the medium
\begin{equation}
    \hat{\rho}_{\text{tot}}(0) = \hat{\rho}_\phi(0) \otimes \hat{\rho}_\chi(0) \,.
    \label{eqn:tensor-product-initial-state}
\end{equation}
So the first term of the RHS of \cref{eqn:trace-over-chi} becomes
\begin{equation}
    -i \Tr_\chi \Bab{ \comm{H_\txtI(t)}{\hat{\rho}_{\text{tot}}(0)} }
    \quad\longrightarrow\quad
    - i \sum_{a=1,2} \int \dn[3]{x}  \, \comm{J_a(t)}{\hat{\rho}_{\phi}(0)} \expval{\mathcal{O}_a}_\chi \,.
    \label{eqn:qme-first-term}
\end{equation}
where we defined $\expval{\cdots}_\chi \coloneqq \Tr_\chi\Bab{\cdots \hat{\rho}_\chi(0)}$. This means that we prepare the states of the subsystem and the medium separately first, then put them in contact and switch on the interaction $H_\txtI$ at $t=0$. Such a setup belongs to the quench dynamics in the sense of a sudden change of the Hamiltonian \cite{doi:10.1146/annurev-conmatphys-031016-025451,Castro-Alvaredo2019-hj}, in which a growth of correlations between the subsystem and the medium is expected. It is possible that $\expval{\mathcal{O}_a}_\chi$ vanished identically due to symmetry. For example, the parity of $\mathcal{O}_a = \bvec{E}\cdot\bvec{B}$ in axion-photon coupling is odd and $\expval*{\bvec{E}\cdot\bvec{B}}_\chi = 0$ if we assume $\hat{\rho}_\chi(0)$ is in a thermal state whose parity is even, as discussed in \cite{PhysRevD.106.123503,PhysRevD.107.063518,PhysRevD.107.083531}. Even if $\expval{\mathcal{O}_a}_\chi$ doesn't vanish, we can always remove \cref{eqn:qme-first-term} from \cref{eqn:trace-over-chi} without affecting the system's dynamics by adding an energy shift in the system's Hamiltonian \cite{10.1063/1.5115323}. This is equivalent to normal-order $\mathcal{O}_a$ upon the medium's initial state $\hat{\rho}_\chi(0)$ \cite{Boyanovsky_2015}. Therefore, we will not include this term in our following calculations. It has been pointed out that besides the usual divergences to be absorbed by the renormalization of masses, wave functions and coupling strengths, extra divergences may arise in non-equilibrium evolutions when $t \to 0$ (i.e., at the initial time) and invalidate the initial state assumed, which we refer to as initial time singularities. They may be fixed via Bogoliubov transformation \cite{PhysRevD.63.045023,PhysRevD.57.6398} or via an adiabatic switch-on of the interactions \cite{PhysRevD.81.065030,PhysRevD.83.085011}. As will be shown in \cref{sec:evolutions-of-field-amplitudes-and-their-mixing,sec:populations-and-coherence,sec:effects-of-a-coupling-strength-hierarchy}, in our calculation, the time-dependence of the solutions are of the form of exponentially decaying oscillations. Therefore, the solutions are free of initial time singularities when we set $t\to 0$ and the divergences in the factors of the solutions cause the renormalization of masses, wave functions and coupling strengths.

Second, to get rid of $\hat{\rho}_{\text{tot}}(t')$ in \cref{eqn:trace-over-chi}, we invoke the Born approximation \cite{10.1063/1.5115323,10.1093/acprof:oso/9780199213900.001.0001} where $H_\txtI$ is assumed to be perturbatively small and the medium is considered as a large reservoir such that the influence of the subsystem on the reservoir is negligible. Then, the total density matrix is always factorizable and the state of the medium always remains unchanged, i.e.,
\begin{equation}
    \hat{\rho}_{\text{tot}}(t) \approx \hat{\rho}_\phi(t) \otimes \hat{\rho}_\chi(0) \,.
    \label{eqn:Born-approximation}
\end{equation}
This factorization allows us to identify the reduced density matrix $\hat{\rho}$ with the density matrix of the subsystem $\hat{\rho}_\phi$.
\begin{equation}
    \hat{\rho}(t) \approx \Tr_\chi\Bab{\hat{\rho}_\phi(t) \otimes \hat{\rho}_\chi(0)} = \hat{\rho}_\phi(t) \,.
\end{equation}
Consequently, we can rewrite the partial traces of the four terms in the nested commutators of \cref{eqn:trace-over-chi} in the following way.
\begin{subequations}
\begin{align}
    \Tr_\chi\Bab{H_\txtI(t) H_\txtI(t') \hat{\rho}_{\text{tot}}(t')}
    & \mapsto \sum_{a,b=1,2} \int \dn[3]x \dn[3]{x'} J_a(x) J_b (x') \hat{\rho}(t') G_{ab}^>(x,x') \,,
    \\
    \Tr_\chi\Bab{\hat{\rho}_{\text{tot}}(t') H_\txtI(t') H_\txtI(t)}
    & \mapsto \sum_{a,b=1,2} \int \dn[3]x \dn[3]x' \hat{\rho}(t') J_b(x') J_a(x) G_{ab}^<(x,x') \,,
    \\
    \Tr_\chi\Bab{H_\txtI(t) \hat{\rho}_{\text{tot}}(t') H_\txtI(t')}
    & \mapsto \sum_{a,b=1,2} \int \dn[3]x \dn[3]{x'} J_a(x) \hat{\rho}(t') J_b(x') G_{ab}^<(x,x') \,,
    \\
    \Tr_\chi\Bab{H_\txtI(t') \hat{\rho}_{\text{tot}}(t') H_\txtI(t)}
    & \mapsto \sum_{a,b=1,2} \int \dn[3]x \dn[3]{x'} J_b(x') \hat{\rho}(t') J_a(x) G_{ab}^>(x,x') \,.
\end{align}
\end{subequations}
where $x\coloneqq(t,\bvec{x})$ and $x'\coloneqq(t',\bvec{x}')$. Although $a,b$ are dummy variables, we always use $J_a(x)$ and $J_b(x')$ for consistency. $G_{ab}^{\lessgtr}(x,x')$ are correlation functions of the medium with opposite orders of time, which also appear in the Keldysh-Schwinger formalism \cite{PhysRevD.108.025012,PhysRevD.109.036038}.
\begin{equation}
    G_{ab}^>(x,x') \coloneqq \expval{\mathcal{O}_a(x) \mathcal{O}_b(x')}_\chi \,,
    \qquad
    G_{ab}^<(x,x') \coloneqq \expval{\mathcal{O}_b(x') \mathcal{O}_a(x)}_\chi \,.
    \label{eqn:G<>-definition}
\end{equation}
Since $H_\chi$ includes all interactions among the fields of the medium and is used to define the interaction picture, the correlation functions $G_{ab}^{\lessgtr}(x,x')$ are exact to all orders in the couplings of the fields in the medium except their couplings to the subsystem.

Gathering the results from the above two approximations, the QME for the reduced density matrix $\hat{\rho}$ becomes
\begin{equation}
    \dot{\hat{\rho}}(t) = - \sum_{a,b=1,2} \int_0^t \dn*{t'} \int \dn[3]x \dn[3]{x'} \comm{J_a(x)}{ \pab[Big]{J_b(x') \hat{\rho}(t') G_{ab}^>(x,x') - \hat{\rho}(t') J_b(x') G_{ab}^<(x,x')} } \,.
    \label{eqn:master-eqn-G-<>}
\end{equation}
The trace of the commutator in \cref{eqn:master-eqn-G-<>} vanishes because of the cyclic symmetry of a trace, which implies
\begin{equation}
    \diff*{\Tr\Bab{\hat{\rho}}}{t} = 0 \,.
    \label{eqn:trace-preserving}
\end{equation}
\cref{eqn:trace-preserving} suggests that albeit the approximations made above, \cref{eqn:master-eqn-G-<>} still preserves the total probability of the subsystem. Notice that $[G_{ab}^>(x,x')]^\dagger = G_{ab}^<(x,x')$ given $\mathcal{O}_a$ and $\mathcal{O}_b$ are Hermitian. The RHS of \cref{eqn:master-eqn-G-<>} is Hermitian if we assume $\hat{\rho}(t')$ and $J_a(x),J_b(x')$ are Hermitian. Consequently, $\dot{\hat{\rho}}(t)$ is Hermitian. This indicates that \cref{eqn:master-eqn-G-<>} preserves the hermicity of $\hat{\rho}(t)$ in spite of the above approximations. To make it more explicit, we decompose $G_{ab}^\lessgtr(x,x')$ into the Hermitian and anti-Hermitian parts as below,
\begin{align}
    \mathcal{N}_{ab}(x,x')
    & \coloneqq \frac{1}{2} \bab{G_{ab}^>(x,x') + G_{ab}^<(x,x')} \equiv \frac{1}{2} \expval{\acomm{\mathcal{O}_a(x)}{\mathcal{O}_b(x')}}_\chi \,.
    \label{eqn:noise-kernel-definition}
    \\
    i \Sigma_{ab}(x,x')
    & \coloneqq G_{ab}^>(x,x') - G_{ab}^<(x,x') \equiv \expval{\comm{\mathcal{O}_a(x)}{\mathcal{O}_b(x')}}_\chi \,.
    \label{eqn:self-energy-definition}
\end{align}
Note that both $\Sigma_{ab}(x,x')$ and $\mathcal{N}_{ab}(x,x')$ are Hermitian. Inserting \cref{eqn:noise-kernel-definition,eqn:self-energy-definition} into \cref{eqn:master-eqn-G-<>} yields
\begin{multline}
    \dot{\hat{\rho}}(t) = \sum_{a,b=1,2} \int \dn[3]{x} \dn*[3]{x'} \int_0^t \dn{t'} \bigg\{ -\frac{i}{2}    \comm{J_a(x)}{ \acomm{J_b(x')}{\hat{\rho}(t')} } \Sigma_{ab}(x,x') 
    \\
    -  \comm{J_a(x)}{\comm{J_b(x')}{\hat{\rho}(t')}} \mathcal{N}_{ab}(x,x')
    \bigg\} \,.
    \label{eqn:qme}
\end{multline}

Suppose $\mathcal{O}(\Bab{\bvec{x}_i}, t)$ is an arbitrary operator acting on the Hilbert space of the subsystem, depending on zero or more spatial coordinates $\Bab{\bvec{x}_i} \coloneqq \bvec{x}_1,\bvec{x}_2,\dots$ and time $t$. There are two contributions in the time derivative of its expectation value $\expval{\mathcal{O}} \coloneqq \Tr\Bab{\mathcal{O}(\Bab{\bvec{x}_i},t) \hat{\rho}(t)}$
\begin{equation}
    \diff**{t}{\expval{\mathcal{O}}} = \expval{\dot{\mathcal{O}}} + \Tr\Bab{\mathcal{O} \dot{\hat{\rho}}} \,.
    \label{eqn:formal-d<O>/dt}
\end{equation}
Inserting \cref{eqn:qme} yields the equation of motion for $\expval{\mathcal{O}}$ up to leading order perturbations.
\begin{multline}
    \diff**{t}{\expval{\mathcal{O}}} = \expval{\dot{\mathcal{O}}} + \sum_{a,b=1,2} \int \dn[3]{y} \dn*[3]{y'} \int_0^t \dn*{t'} \bigg\{
        - \frac{i}{2}  \Tr\Bab[Big]{ \acomm{ \comm{\mathcal{O}(\Bab{\bvec{x}_i},t)}{ J_a(y) } }{J_b(y')} \hat{\rho}(t') } \Sigma_{ab}(y,y')
    \\
        - \Tr\Bab[Big]{ \comm{ \comm{\mathcal{O}(\Bab{\bvec{x}_i},t)}{ J_a(y) } }{J_b(y')} \hat{\rho}(t') } \mathcal{N}_{ab}(y,y')
    \bigg\} \,.
    \label{eqn:expval-O-dt}
\end{multline}
where we defined $y=(\bvec{y},t), y'=(\bvec{y}',t')$ and transformed the nested (anti-)commutators using the cyclic symmetry of the trace.
\begin{equation}
\begin{aligned}
    \Tr\Bab[Big]{ \mathcal{O} \comm{J_a}{ \acomm{J_b}{\hat{\rho}} } }
    & \;\rightarrow\;
    \Tr\Bab[Big]{ \comm{\mathcal{O}}{J_a} \acomm{J_b}{\hat{\rho}} }
    \;\rightarrow\;
    \Tr\Bab[Big]{ \acomm{\comm{\mathcal{O}}{J_a}}{J_b} \hat{\rho} } \,,
    \\
    \Tr\Bab[Big]{ \mathcal{O} \comm{J_a}{ \comm{J_b}{\hat{\rho}} } }
    & \;\rightarrow\;
    \Tr\Bab[Big]{ \comm{\mathcal{O}}{J_a} \comm{J_b}{\hat{\rho}} }
    \;\rightarrow\;
    \Tr\Bab[Big]{ \comm{\comm{\mathcal{O}}{J_a}}{J_b} \hat{\rho} } \,.
\end{aligned}
\label{eqn:structure-transformation}
\end{equation}
In \cref{eqn:structure-transformation}, we dropped the operators' arguments temporarily to highlight the transformation. The new structure helps us make use of commutation relations in future calculations.

In this article, we are mostly interested in the scenario where the subsystem is linearly coupled to a thermal medium, which is motivated by various phenomena in early universe, e.g., ALPs' coupling with CMB photons, evolutions of kaon/pion after the QCD phase transition. Therefore, throughout the rest of this article,  we will assume $J_a[\phi_a] \equiv \phi_a$ and $\hat{\rho}_\chi$ to be a thermal ensemble at temperature $T=1/\beta$.
\begin{equation}
    \hat{\rho}_\chi(0) = \frac{e^{-\beta H_\chi }}{Z_\chi} \,,
    \qquad
    Z_\chi = \Tr_\chi\Bab{ e^{-\beta H_\chi} }
    \,. 
    \label{eqn:thermal-chi}
\end{equation}
where $Z_\chi$ is the partition function. A thermal medium is both space-time translationally and rotationally invariant. It results in the space-time translational symmetry and the rotational symmetry of $G_{ab}^\lessgtr(x-x')$, $\Sigma_{ab}(x-x')$ and $\mathcal{N}_{ab}(x-x')$ (see \cref{app:Lehmann-representation-of-correlation-functions,app:symmetries-of-self-energy,app:symmetries-of-noise-kernel}). Note that we have written their arguments ``$(x,x')$'' as ``$(x-x')$''. It will be shown in \cref{sec:evolutions-of-field-amplitudes-and-their-mixing,sec:populations-and-coherence} that $\Sigma_{ab}(x-x')$ describes the dissipation in the evolution and $\mathcal{N}_{ab}(x-x')$ describes the fluctuation due to the medium. Therefore, we call $\Sigma_{ab}(x-x')$ self-energy and $\mathcal{N}_{ab}(x-x')$ noise-kernel from now on. Both of them are due to the same interaction $H_\txtI$ hence are connected by a generalized fluctuation-dissipation relation (see \cref{app:generalized-fluctuation-dissipation-relation}). The same relation was previously shown for a Langevin equation of mixing fields which was obtained in the Keldysh-Schwinger formalism \cite{PhysRevD.108.025012,PhysRevD.109.036038}.

These properties lead to the starting point of our study.
\begin{multline}
    \diff**{t}{\expval{\mathcal{O}}} = \expval{\dot{\mathcal{O}}} + \sum_{a,b=1,2} \int \dn[3]{y} \dn*[3]{y'} \int_0^t \dn*{t'} \bigg\{
        - \frac{i}{2}  \Tr\Bab[Big]{ \acomm{ \comm{\mathcal{O}(\Bab{\bvec{x}_i},t)}{ \phi_a(y) } }{\phi_b(y')} \hat{\rho}(t') } \Sigma_{ab}(y-y')
    \\
        - \Tr\Bab[Big]{ \comm{ \comm{\mathcal{O}(\Bab{\bvec{x}_i},t)}{ \phi_a(y) } }{\phi_b(y')} \hat{\rho}(t') } \mathcal{N}_{ab}(y-y')
    \bigg\} \,.
    \label{eqn:expectation-value-evolution}
\end{multline}

\section{Field amplitudes and their mixing}
\label{sec:evolutions-of-field-amplitudes-and-their-mixing}

\subsection{Equations of motion and solutions of field amplitudes}
\label{subsec:equations-of-motion-and-solutions-of-field-amplitues}

The fields in the subsystem may feature nonzero amplitudes along the evolution. For example, ALPs may be produced non-thermally as a condensate due to the misalignment mechanism, which gives ALPs an initially non-zero amplitude. To study the evolution of the field amplitudes $\expval{\phi_c}, c=1,2$, we set $\mathcal{O} = \phi_c$ in \cref{eqn:expectation-value-evolution}.

The term $\expval*{\dot{\mathcal{O}}}$ in \cref{eqn:expectation-value-evolution} requires us to find the time-dependence of the field operators $\phi_c$ first, which is given by the interaction picture defined in \cref{eqn:interaction-picture}. We substitute $\phi_c$ and $\pi_c$ into the ``$\hat{\mathcal{A}}$'' in \cref{eqn:interaction-picture} and take the time-derivative, which yileds
\begin{align}
    \dot{\phi}_c
    & = i \comm{H_{\phi_1} + H_{\phi_2} + H_{\chi}}{\phi_c}
      = \pi_c \,,
      \label{eqn:phi-dot}
    \\
    \dot{\pi}_c
    & = i \comm{H_{\phi_1} + H_{\phi_2} + H_{\chi}}{\pi_c}
      = \nabla^2 \phi_c - m_c^2 \phi_c \,.
      \label{eqn:pi-dot}
\end{align}
where the repeated indices in \cref{eqn:pi-dot} doesn't imply a summation and we also used
\begin{equation}
\begin{aligned}
    \comm{\nabla_{\bvec{y}}\, \phi_a(\bvec{y},t)}{\pi_b(\bvec{x},t)} 
    & = i \nabla _{\bvec{y}} \pab{\delta(\bvec{y}-\bvec{x})} \,,
    \\
    \int \dn[3]y F(\bvec{y}) \nabla_{\bvec{y}} \pab{\delta(\bvec{y}-\bvec{x})} 
    & = - \nabla_{\bvec{y}} F(\bvec{y}) \Big|_{\bvec{y}=\bvec{x}} \,,
    \quad \text{$F(\bvec{y})$ is a general function.}
\end{aligned}
\end{equation}
The equation for $\pi_c$ is obtained to close the equations.

We also have to solve the commutators inside the integral of \cref{eqn:expectation-value-evolution}. Since the equal-time commutator of two field operators is zero, i.e., $\comm{\phi_c(\bvec{x},t)}{\phi_a(\bvec{y},t)} = 0$, both traces inside the integral vanish, resulting in $\Tr\Bab*{\phi_c \dot{\hat{\rho}}}=0$. The equation of motion for field amplitudes becomes
\begin{equation}
    \diff**{t}{\expval{\phi_c}} = \expval{\pi_c} \,.
    \label{eqn:phi-amp-dt-leading-order}
\end{equation}
where we used \cref{eqn:phi-dot,eqn:expectation-value-evolution}. We want to stress that the vanishing of the quantum correction $\Tr\Bab*{\phi_c \dot{\hat{\rho}}}$ in \cref{eqn:phi-amp-dt-leading-order} is a coincidence and the reason is twofold. First, the effective equation \cref{eqn:expectation-value-evolution} is up to the leading order perturbation, therefore only contains two structures $\Tr\Bab{\acomm{\comm{\cdot}{\cdot}}{\hat{\rho}}}$ and $\Tr\Bab{\comm{\comm{\cdot}{\cdot}}{\hat{\rho}}}$. Should we include higher order perturbations in \cref{eqn:expectation-value-evolution}, new structures would appear and in general would not lead to a vanishing $\Tr\Bab*{\phi_c \dot{\hat{\rho}}}$. Second, we are calculating field amplitudes, hence, encounter a term $\comm{\phi_c(\bvec{x},t)}{\phi_a(\bvec{y},t)}$ that vanishes identically. If we compute $\expval{\pi_c}$ as in next paragraph, or the populations $\expval*{\widehat{N}_c}$ as in \cref{subsec:equations-of-motion-and-formal-solutions}, the commutator is not zero and the correction will persist in the equation of motion.

The $\expval{\pi_c}$ in \cref{eqn:phi-amp-dt-leading-order} means an equation for $\expval{\pi_c}$ is needed to close the equations of motion, which is obtained by setting $\mathcal{O} = \pi_c$ and inserting \cref{eqn:pi-dot} into \cref{eqn:expectation-value-evolution}. The result is
\begin{equation}
    \diff**{t}{\expval{\pi_c}} = \nabla^2 \expval{\phi_c} - m_c^2 \expval{\phi_c} - \sum_{b=1,2} \int \dn*[3]{y'} \int_0^t \dn{t'} \Sigma_{cb}(\bvec{x}-\bvec{y}',t-t') \expval{\phi_b}(\bvec{y}',t')
    \label{eqn:pi-amp-dt-leading-order}
\end{equation}
where we used $\comm{\pi_c(\bvec{x},t)}{\phi_a(\bvec{y},t)}=-i \delta(\bvec{x}-\bvec{y}) \delta_{c,a}$ for the commutators in \cref{eqn:expectation-value-evolution} and noticed that $\Tr\Bab{\acomm{\phi_b(\bvec{y}',t')}{\hat{\rho}(t')}} = 2 \expval{\phi_b}(\bvec{y}',t')$ and $\Tr\Bab{\comm{\phi_b(\bvec{y}',t')}{\hat{\rho}(t')}} =0$ due to the cyclic symmetry of a trace.

Lastly, a closed equation for $\expval{\phi_c}$ is obtained by taking another time derivative of \cref{eqn:phi-amp-dt-leading-order} and plugging in \cref{eqn:pi-amp-dt-leading-order}.
\begin{equation}
    \diff**[2]{t}{\expval{\phi_c}} - \nabla^2 \expval{\phi_c} + m_c^2 \expval{\phi_c} + \sum_{b=1,2} \int \dn*[3]{y'} \int_0^t \dn{t'} \Sigma_{cb}(x-y') \expval{\phi_c}(y') = 0 \,.
    \label{eqn:phi-amp-evolution}
\end{equation}
where $x=(\bvec{x},t), y'=(\bvec{y}',t')$. This equation is exactly the same equation for field amplitudes as the one analyzed in \cite{PhysRevD.108.025012,PhysRevD.109.036038}, which is obtained at the second order perturbation in the Keldysh-Schwinger formalism. Therefore, the solution of \cref{eqn:phi-amp-evolution} is a resummation of the second order contributions.

It is simpler to solve \cref{eqn:phi-amp-evolution} in the spatial momentum space.
\begin{equation}
    \diff**[2]{t}{\expval{\phi_{c,\bvec{k}}}} + \omega_{c,\bvec{k}}^2 \expval{\phi_c} + \sum_{b=1,2}  \int_0^t \dn{t'} \Sigma_{cb}(\bvec{k},t-t') \expval{\phi_{c,\bvec{k}}}(t') = 0 \,.
    \label{eqn:phi-k-dt2}
\end{equation}
where we defined $\omega_{c,\bvec{k}} = \sqrt{m_c^2 + \abs*{\bvec{k}}^2}$ and used the mode expansion
\begin{equation}
    \phi_c(\bvec{x},t)
    = \frac{1}{\sqrt{V}} \sum_{\bvec{q}} \phi_{c,\bvec{q}}(t) e^{i\bvec{q}\cdot\bvec{x}} \,,
    \qquad
    \Sigma_{cb}(\bvec{x},t)
    = \frac{1}{\sqrt{V}} \sum_{\bvec{q}} \Sigma_{cb}(\bvec{q},t) e^{i\bvec{q}\cdot\bvec{x}} \,.
\end{equation}
The solution of \cref{eqn:phi-k-dt2} is obtained by a Laplace transform (denoted as $\mathcal{L}[\cdot]$) as befits an initial-value problem.
\begin{equation}
    s^2 \expval{\phi_c}(s) - s \eval{\expval{\phi_c}}_{t=0^-} - \eval{\expval{\pi_c}}_{t=0^-} + \omega_c^2 \expval{\phi}_c + \sum_{b=1,2} \Sigma_{cb}(s) \expval{\phi_c}(s) = 0 \,.
    \label{eqn:amplitude-equation-element-s-domain}
\end{equation}
where $s$ is the Laplace variable and \cref{eqn:phi-amp-dt-leading-order} was used for $\eval{\diff*{\expval{\phi_c}}{t}}_{t=0^-}$. Since there is only one momentum label $\bvec{k}$ in the equation, we dropped it for simpler notations.

To solve \cref{eqn:amplitude-equation-element-s-domain}, we write it in a matrix form after defining $\expval{\boldsymbol{\phi}} = \bab{\expval{\phi_1}, \expval{\phi_2}}^\T$ and $\expval{\boldsymbol{\pi}} = \bab{\expval{\pi_1}, \expval{\pi_2}}^\T$.
\begin{equation}
    \mathbf{G}_\phi^{-1}(s) \cdot \expval{\boldsymbol{\phi}} = s \expval{\boldsymbol{\phi}} \eval{}_{t=0^-} + \eval{\expval{\boldsymbol{\pi}}}_{t=0^-} \,.
    \label{eqn:phi-equation-matrix-form}
\end{equation}
where the matrix $\mathbf{G}_\phi^{-1}(s)$ was read off from \cref{eqn:amplitude-equation-element-s-domain} directly.
\begin{equation}
    \mathbf{G}_\phi^{-1}(s) = 
    \begin{pmatrix}
        s^2 + \omega_1^2 + \Sigma_{11}(s) & \Sigma_{12}(s)
        \\
        \Sigma_{21}(s) & s^2 + \omega_2^2 + \Sigma_{22}(s)
    \end{pmatrix}
    \,.
    \label{eqn:G-A-inverse-s}
\end{equation}
Inverting the matrix $\mathbf{G}_\phi^{-1}(s)$ gives the solution in the $s$-domain.
\begin{equation}
    \expval{\boldsymbol{\phi}}(s) = \mathbf{G}_\phi(s) \cdot \pab[Big]{ s \expval{\boldsymbol{\phi}} \eval{}_{t=0^-} + \eval{\expval{\boldsymbol{\pi}}}_{t=0^-} } \,.
    \label{eqn:phi-solution-s-domain}
\end{equation}
The solution of $\expval{\boldsymbol{\phi}}$ in the time domain is obtained by taking the inverse Laplace transform.
\begin{equation}
    \expval{\boldsymbol{\phi}}(t) = \dot{\mathbf{G}}_\phi(t) \cdot \expval{\boldsymbol{\phi}}(0^-) + \mathbf{G}_\phi(t) \cdot \expval{\boldsymbol{\pi}}(0^-) \,.
    \label{eqn:phi-t-solution}
\end{equation}
Inserting $\expval{\boldsymbol{\phi}}(t)$ into \cref{eqn:phi-amp-dt-leading-order} yields the solution of the momentum $\expval{\boldsymbol{\pi}}$.
\begin{equation}
    \expval{\boldsymbol{\pi}}(t) = \dot{\mathbf{G}}_\phi(t) \cdot \expval{\boldsymbol{\pi}}(0^-) + \ddot{\mathbf{G}}_\phi(t) \cdot \expval{\boldsymbol{\phi}}(0^-) \,.
    \label{eqn:pi-t-solution}
\end{equation}
The inverse Laplace transform of the Green's function $\mathbf{G}_\phi(s)$ is defined as,
\begin{equation}
    \mathbf{G}_\phi(t) = \frac{1}{2\pi i} \int_\mathcal{C} \dn{s} e^{s t} \mathbf{G}_\phi(s) \,.
    \label{eqn:G-A-t}
\end{equation}
in which $\mathcal{C}$ denotes the Bromwich contour parallel to the imaginary axis. The contour $\mathcal{C}$ has to reside in the region $\Re\Bab{s}>0$ such that $\mathbf{G}_\phi(s)$, or equivalently, $\mathbf{G}_\phi^{-1}(s)$ and $\Sigma_{ab}(s)$ converge (see \cref{app:symmetries-of-self-energy}). Therefore, we choose the contour to be $s=i\omega + \epsilon, -\infty < \omega < \infty, \epsilon \to 0^+$. The limit $\epsilon \to 0^+$ results in a decomposition of the self-energy $\Sigma_{ab} (i\omega+\epsilon)$ (see \cref{app:symmetries-of-self-energy}).
\begin{equation}
    \lim_{\epsilon \to 0^+} \Sigma_{cd}(i\omega + \epsilon) = \Sigma_{\txtR,cd}(i\omega) + i \Sigma_{\txtI,cd}(i\omega) \,.
    \label{eqn:Sigma-ab-decomposition-epsilon->0}
\end{equation}
Throughout the rest of this article, such a decomposition is always implied after the inverse Laplace transform. \cref{eqn:G-A-t} is then evaluated using the residue theorem where we close the contour along a large simicircle at infinity with $\Re\Bab{s} < 0$ after noticing $t>0$. The contributions to the integral come from both the poles and multi-particle branch cuts inside the closed contour. To find the poles, we invert $\mathbf{G}_\phi^{-1}(s)$ using
\begin{equation}
    \mathbf{G}_\phi(s) = \frac{ \adj\Bab[big]{ \mathbf{G}_\phi^{-1}(s) } }{ \det\Bab[big]{ \mathbf{G}_\phi^{-1}(s) } } \,.
    \label{eqn:G-A-s-adj-det}
\end{equation}
where $\adj\Bab{\cdot}$ is the adjugate matrix and $\det\Bab{\cdot}$ is the determinant. Note that both $\det\Bab*{ \mathbf{G}_\phi^{-1}(s)}$ and the elements of $\adj\Bab*{\mathbf{G}_\phi^{-1}}(s)$ are polynomials of the elements of $\mathbf{G}_\phi^{-1}(s)$. Given the elements of $\mathbf{G}_\phi^{-1}(s)$ don't include fractions, there is no fraction in $\det\Bab*{ \mathbf{G}_\phi^{-1}(s)}$ or $\adj\Bab*{\mathbf{G}_\phi^{-1}}(s)$. As a consequence, all poles of $\mathbf{G}_\phi(s)$ come from
\begin{equation}
    \det\Bab[big]{\mathbf{G}_\phi^{-1}(s)} = 0 \,.
    \label{eqn:det-G-A^-1-s=0}
\end{equation}
The zeroth order (bare) poles are obtained by setting $g_1=g_2=0$ in \cref{eqn:det-G-A^-1-s=0}.
\begin{equation}
    s_{a_1}^{(0)} = -i\omega_1 \,, \qquad
    s_{a_1^\dagger}^{(0)} = i\omega_1 \,, \qquad
    s_{a_2}^{(0)} = -i\omega_2 \,, \qquad
    s_{a_2^\dagger}^{(0)} = i\omega_2 \,.
    \label{eqn:zero-order-poles-A}
\end{equation}
where the subscripts indicate to which operator these poles are associated. Explicit expressions of the results up to $\order{g^2}$ are listed in \cref{subapp:pole-and-residues-in-the-Green's-function-of-field-amplitudes} where the subscripts are kept consistent with those in \cref{eqn:zero-order-poles-A}. It should be pointed out that all four leading order poles (see \cref{eqn:A-poles-s}) reside in the closed contour, i.e., $\Re\Bab{s_i} < 0$, hence contribute to the integral in \cref{eqn:G-A-t}. This is a consequence of the assumption that the medium consists of common decay channels of the two (pseudo-)scalar fields. Due to the nonlinearity of the self-energy $\Sigma_{ab}(s)$, there may be more solutions to \cref{eqn:det-G-A^-1-s=0} than the four leading-order poles. By counting the powers of the coupling strength $g$, we can divide all solutions to \cref{eqn:det-G-A^-1-s=0} into two groups, a group where all solutions are of order $s \sim \order{g^n}, n \geq 0$ and a group where all solutions are of order $s \sim \order{1/g^n}, n > 0$. For solutions scaling as $s \lesssim \order{1}$, we treat $\Sigma_{ab}(s)$ as perturbatively small quantities and solve \cref{eqn:det-G-A^-1-s=0} perturbatively order by order up to $\order{g^2}$. These steps yield the four leading poles mentioned earlier. For solutions scaling as $s\sim \order{1/g^n}, n > 0$ (if exist), such an order means they generate dynamics that are much faster than any physical process we are considering and their effects fastly average out. Therefore, we only consider the four leading-order poles in the following calculation.


We will invoke the Breit-Wigner approximation to ignore the contribution from multi-particle branch cuts, for which one has to assume the distances between poles and cuts are of order $\order{1}$. Under this assumption, the difference of $\mathbf{G}_\phi(s)$ on the two sides of a cut is of order $\order{g^2}$ since it is caused by the $\order{g^2}$ self-energy. This implies that the contribution from all multi-particle branch cuts is of order $\order{g^2}$, for which we assume implicitly that the number of cuts is of order $\order{1}$.

In summary, after using the residue theorem in \cref{eqn:G-A-t} and igoring contributions from the multi-particle branch cuts and the poles of order $\order{1/g^n}$, we find
\begin{equation}
    \mathbf{G}_\phi(t) = \sum_{i=a_1,a_2,a_1^\dagger,a_2^\dagger} \mathbb{G}_{\phi,i} e^{s_{i} t} \,.
    \label{eqn:G-A-Breit-Wigner}
\end{equation}
where $\mathbb{G}_{\phi,i}$ are residues of $\mathbf{G}_\phi(s)$ at the corresponding poles. $\mathbb{G}_{\phi,i}$ are obtained up to leading order and their explicit expressions are listed in \cref{subapp:pole-and-residues-in-the-Green's-function-of-field-amplitudes}. Two properties of the poles and residues are worth noting. Since both $\omega_{c,\bvec{k}}$ and $\Sigma_{cd}(\bvec{k},i\omega)$ are rotationally symmetric (see \cref{app:symmetries-of-self-energy}), so are the poles and residues, i.e.,
\begin{equation}
    s_{\mathcal{A}_{\bvec{k}}} = s_{\mathcal{A}_{\abs{\bvec{k}}}} \,,
    \qquad
    \mathbb{G}_{\phi,\mathcal{A}_\bvec{k}} = \mathbb{G}_{\phi,\mathcal{A}_{\abs{\bvec{k}}}} \,,
    \qquad\text{for}\;\;
    \mathcal{A} = a_1, a_2, a_1^\dagger, a_2^\dagger \,.
    \label{eqn:amplitude-pole-|k|-dependence}
\end{equation}
where we restored the momentum label $\bvec{k}$ dropped previously. $\omega_{a,\bvec{k}}^2$ and $\Sigma_{cd}(\bvec{k},i\omega)$ also possess a conjugate symmetry under $\omega \to -\omega$ (see \cref{app:symmetries-of-self-energy}). The poles and residues acquire the same symmetry via their expressions in \cref{eqn:A-poles-s,eqn:residue-G-phi}.
\begin{equation}
    s_{\mathcal{A}}^* = s_{\mathcal{A}^\dagger} \,,
    \qquad
    \mathbb{G}_{\phi,\mathcal{A}}^* = \mathbb{G}_{\phi,\mathcal{A}^\dagger} \,,
    \qquad
    \text{for}\;\;
    \mathcal{A} = a_1, a_2
    \label{eqn:conjugate-symmetry-of-amplitude-poles-residues}
\end{equation}
It immediately follows that the Green's function $\mathbf{G}_\phi(t)$ is a real function, which ensures that the solutions in \cref{eqn:phi-t-solution,eqn:pi-t-solution} are real hence physical for the real fields $\phi_c$.

We refer to each $\mathbb{G}_{\phi,i} e^{s_i t}$ (denoted as $\mathbf{G}_{\phi,i}(t)$) as a quasi-normal mode (QNM) of the field amplitudes. Since $\Re\Bab{s_i} < 0$, all QNMs feature exponential decaying oscillations. Inserting them into \cref{eqn:phi-t-solution,eqn:pi-t-solution}, we find that the evolutions of the field amplitudes and momenta are a set of exponentially decaying oscillations starting with $\expval*{\phi_c}(0^-)$ and $\expval*{\pi_c}(0^-)$ and the off-diagonal elements of the QNMs due to the non-diagonal self-energy indicate a field mixing in these evolutions. These non-diagonal QNMs imply that even if initially the condensate might only exist in one of the two fields, a condensate in the other field would still be induced at a later time and yield a non-zero amplitude.

In addition, a conjugate symmetry of the four QNMs follows \cref{eqn:conjugate-symmetry-of-amplitude-poles-residues}, which suggests that we can group the four QNMs into two pairs that correspond to the two field degrees of freedom modified by the interaction with the medium. From them two pairs of renormalized frequencies and decay rates can be extracted,
\begin{equation}
    \Omega_i = \abs[big]{\Im\Bab[big]{s_{a_i}}} = \abs[big]{\Im\Bab[big]{s_{a_i^\dagger}}} \,,
    \qquad
    \Gamma_i = - \Re\Bab[big]{s_{a_i}} = - \Re\Bab[big]{s_{a_i^\dagger}} \,,
    \qquad
    i = 1,2 \,.
    \label{eqn:amplitudes-frequencies-decay-rates}
\end{equation}
where $\Re\Bab{\cdot}$ and $\Im\Bab{\cdot}$ mean the real and imaginary part of a complex number. Note that $\Gamma_{i}$ defined here are the decay rates for amplitudes. They are half of the particle number decay rates in the usual sense, as will be shown when we calculate populations in \cref{sec:populations-and-coherence}.

\subsection{Amplitude mixing in the non-degenerate and nearly-degenerate cases}
\label{subsec:amplitude-mixing-in-the-non-degenerate-and-nearly-degenerate-cases}

We have discovered in \cref{subsec:equations-of-motion-and-solutions-of-field-amplitues} that the field amplitudes mix via the non-diagonal QNMs in the evolution. It turns out that the mixing strength depends on the magnitude of the field masses' degeneracy, i.e., $m_1^2 - m_2^2$ (or equivalently, $\omega_1^2 - \omega_2^2$), which was not specified when we set up the system. Fixing its power in terms of $g$ naturally leads to the non-degenerate and nearly-degenerate cases. In what follows, we will only discuss $\mathbb{G}_{\phi,i} e^{s_{i} t}, i=a_1^\dagger,a_2^\dagger$ given the conjugate symmetry of the QNMs of the field amplitudes (see \cref{eqn:conjugate-symmetry-of-amplitude-poles-residues}).

\subsubsection*{The non-degenerate case}

The subsystem is called non-degenerate if the mass difference is much larger than self-energy corrections, i.e.,  $m_1^2 - m_2^2 \sim \order*{1} \gg \Sigma_{ab} \sim \order*{g^2}$. The poles (see \cref{eqn:A-poles-s}) can be further simplified in this limit as below.
\begin{equation}
    s_{a_1^\dagger} = i \pab{ \omega_1 + \frac{\Sigma_{\txtR,11}(i\omega_1)}{2\omega_1} } - \frac{\Sigma_{\txtI,11}(i\omega_1)}{2\omega_1} \,,
    \quad
    s_{a_2^\dagger} = i \pab{ \omega_2 + \frac{\Sigma_{\txtR,22}(i\omega_2)}{2\omega_2} } - \frac{\Sigma_{\txtI,22}(i\omega_2)}{2\omega_2} \,.
\end{equation}
where we used the decomposition of $\Sigma_{cd}(i\omega)$ in \cref{eqn:Sigma-ab-decomposition-epsilon->0} to better show the frequency corrections and decay rates. It is easy to notice that these poles only exhibit perturbative corrections from the diagonal terms of self-energy, in other words, corrections from only a field's own interaction with the medium, as if the two fields evolved in the thermal medium independently. Thus, in the non-degenerate case, the mixing effect is only manifested in the off-diagonal elements of the residues $\mathbb{G}_{\phi,i}$, which are
\begin{equation}
    \mathbb{G}_{\phi,a_1^\dagger} = \frac{1}{2 s_{a_1^\dagger}}
        \eval{
        \begin{pmatrix}
            1  & \frac{\Sigma_{12}(s)}{\omega_1^2 - \omega_2^2}
            \\
            \frac{\Sigma_{21}(s)}{\omega_1^2 - \omega_2^2} & 0
        \end{pmatrix}
        }_{s=i\omega_1}
    \,,
    \qquad
    \mathbb{G}_{\phi,a_2^\dagger} = \frac{1}{2s_{a_2^\dagger}}
        \eval{
        \begin{pmatrix}
            0 & \frac{\Sigma_{12}(s)}{\omega_2^2 - \omega_1^2}
            \\
            \frac{\Sigma_{21}(s)}{\omega_2^2 - \omega_1^2} & 1
        \end{pmatrix}
        }_{s=i\omega_2}
    \,.
    \label{eqn:G-phi-a^dagger-non-degenerate}
\end{equation}
In the non-degenerate case, both poles and $\omega_1^2 - \omega_2^2$ are of order $\order{1}$. Thus, the off-diagonal terms in \cref{eqn:G-phi-a^dagger-non-degenerate} are of order $\order{g^2}$, indicating a perturvatively small mixing effect.

\subsubsection*{The nearly-degenerate case}

The subsystem is called nearly-degenerate if the mass difference is of the same order as self-energy corrections, i.e., $m_1^2 - m_2^2 \sim \Sigma_{ab} \sim \order*{g^2} $. In this limit, we introduce $\omegabar = (\omega_1 + \omega_2)/2$ and $\delta = \omega_1 - \omega_2$ and find that
\begin{equation}
    \delta = \frac{\omega_1^2 - \omega_2^2}{\omega_1 + \omega_2} \sim \order{g^2} \,,
    \qquad\quad
    \Delta^2(s) \sim \order*{g^2} \,,
    \qquad\quad
    D(s) \sim \order*{g^2} \,.
    \label{eqn:delta-power-counting-nearly-degenerate}
\end{equation}
where $\Delta^2(s)$ and $D(s)$ are defined in the poles' explicit expressions (see \cref{eqn:Omega^2-D-Delta^2-definitions}). Up to leading order, the poles in the nearly-degenerate case are
\begin{equation}
    s_{a_1^\dagger} = i\omegabar + i \frac{D(i\omegabar) + \Sigma_{11}(i\omegabar) + \Sigma_{22}(i\omegabar)}{4\omegabar} \,,
    \qquad
    s_{a_2^\dagger} = i\omegabar + i \frac{-D(i\omegabar) + \Sigma_{11}(i\omegabar) + \Sigma_{22}(i\omegabar)}{4\omegabar} \,.
    \label{eqn:amplitude-pole-nearly-degenerate}%
\end{equation}
In contrast to the non-degenerate case, the poles in \cref{eqn:amplitude-pole-nearly-degenerate} receive corrections from all elements of the self-energy, in other words, corrections due to both fields' interactions with the medium. This manifests a strong mixing effects in the poles.

The same strong mixing effect is manifested in the residue matrices, too, which are
\begin{equation}
    \mathbb{G}_{\phi,a_1^\dagger} = \frac{1}{2i\omegabar}
    \eval{
    \begin{pmatrix}
        \frac{1}{2} + \frac{\Delta^2(s)}{2D(s)} & 
        \frac{\Sigma_{12}(s)}{D(s)}
        \\[0.5em]
        \frac{\Sigma_{21}(s)}{D(s)} & 
        \frac{1}{2} - \frac{\Delta^2(s)}{2D(s)}
    \end{pmatrix} }_{s=i\omegabar}
    ,
    \qquad
    \mathbb{G}_{\phi,a_2^\dagger} = \frac{1}{2i\omegabar}
    \eval{
    \begin{pmatrix}
        \frac{1}{2} - \frac{\Delta^2(s)}{2D(s)} &
        - \frac{\Sigma_{12}(s)}{D(s)}
        \\[0.5em]
        - \frac{\Sigma_{21}(s)}{D(s)} &
        \frac{1}{2} + \frac{\Delta^2(s)}{2D(s)}
    \end{pmatrix} }_{s=i\omegabar}
    .
    \label{eqn:phi-residue-nearly-degenerate}
\end{equation}
All elements in the residues are of order $\order{1}$ given the power counting in \cref{eqn:delta-power-counting-nearly-degenerate}, which indicates a strong mixing effect in the QNMs. The enhancement due to the $\order{g^2}$ term $D(s)$ in the denominators explains why a leading order perturbation effect is manifested by $\order{1}$ terms. This enhancement also implies the complete $\order{g^2}$ contributions in \cref{eqn:phi-residue-nearly-degenerate} will include terms like $ \frac{1}{D(s_{a_1})} (s_{a_1}-s_{a_1}^{(0)})\diff**{s}{\Sigma_{ab}(s_{a_1}^{(0)})}$, which are actually next-leading order perturbation effects at the level of the equation of motion and are inconsistent with \cref{eqn:phi-amp-evolution}. Therefore, we only keep $\order{1}$ terms in \cref{eqn:phi-residue-nearly-degenerate} for consistency.

\section{Populations and coherence}
\label{sec:populations-and-coherence}

\subsection{Equations of motion and formal solutions}
\label{subsec:equations-of-motion-and-formal-solutions}

To study multi-time correlation functions, such as $\expval*{\phi_c(x_1)\phi_d(x_2)}$ or $\expval*{\phi_c(x_1)\pi_d(x_2)}$ with $x_i=(\bvec{x}_i,t_i), i=1,2$, the quantum regression theorem is usually used in the context of quantum master equation. However, this theorem is proved for Markovian QMEs \cite{10.1093/acprof:oso/9780199213900.001.0001} but not generally valid for non-Markovian QMEs \cite{PhysRevA.90.022110,PhysRevLett.94.200403,PhysRevA.106.022214}. Therefore, in this section, we primarily focus on the equal-time correlation functions, in particular, the bilinear form of creation/annihilation operators at equal time, e.g. $a_{c,\bvec{k}}^\dagger(t) a_{d,\bvec{q}}(t)$ and $a_{c,\bvec{k}}(t) a_{d,\bvec{q}}(t)$, after noticing that the solutions to \cref{eqn:phi-dot,eqn:pi-dot} for the field and momentum operators merit a Fourier expansion \cite{Greiner1997-xn}.
\begin{subequations}
\begin{align}
    \phi_c(\bvec{x},t)
    & = \frac{1}{\sqrt{V}} \sum_{\bvec{q}} \phi_{c,\bvec{q}} e^{i\bvec{q}\cdot\bvec{x}} \,,
    && \phi_{c,\bvec{q}} = \frac{1}{\sqrt{2\omega_{c,\bvec{q}}}} \bab{ a_{c,\bvec{q}}(t) + a_{c,-\bvec{q}}^\dagger(t) } \,.
    \\
    \pi_c(\bvec{x},t) 
    & = \frac{1}{\sqrt{V}} \sum_{\bvec{q}} \pi_{c,\bvec{q}} e^{i\bvec{q}\cdot\bvec{x}} \,,
    && \pi_{c,\bvec{q}} = -i \sqrt{\frac{\omega_{c,\bvec{q}}}{2}} \bab{ a_{c,\bvec{q}}(t) - a_{c,-\bvec{q}}^\dagger(t) } \,.
\end{align}
\label{eqn:mode-expansion}%
\end{subequations}
where the creation/annihilation operators in the interaction picture are given by \cref{eqn:interaction-picture}
\begin{equation}
    a_{c,\bvec{q}}(t) = a_{c,\bvec{q}}^{(\text{S})} e^{-i\omega_{c,\bvec{q}}t} \,,
    \qquad\quad
    a_{c,\bvec{q}}^\dagger(t) = a_{c,\bvec{q}}^{(\text{S})\dagger} e^{i \omega_{c,\bvec{q}}t} \,.
\end{equation}
In what follows, the argument ``$(t)$'' will be dropped for convenience but are always implied.

The zero spatial dependence of the bilinear forms of creation and annihilation operators makes it possible to rewrite \cref{eqn:expectation-value-evolution} in the momentum space to take advantage of the commutation relations,
\begin{multline}
    \diff**{t}{\expval{\mathcal{O}}} = \expval{\dot{\mathcal{O}}} + \sum_{a,b=1,2} \sum_{\bvec{p}} \int_0^t \dn{t'} \bigg\{
        - \frac{i}{2} \Tr\Bab[Big]{ \acomm{ \comm{\mathcal{O}(t)}{\phi_{a,-\bvec{p}}(t)} }{ \phi_{b,\bvec{p}}(t') } \hat{\rho}(t') } \Sigma_{ab}(\bvec{p},t-t')
    \\
        - \Tr\Bab[Big]{ \comm{ \comm{\mathcal{O}(t)}{\phi_{a,-\bvec{p}}(t)} }{ \phi_{b,\bvec{p}}(t') } \hat{\rho}(t') } \mathcal{N}_{ab}(\bvec{p},t-t')
    \bigg\} \,.
    \label{eqn:expval-O-evolution-momentum-space}
\end{multline}
where we inserted \cref{eqn:mode-expansion} into \cref{eqn:expectation-value-evolution}. In general, new operators in addition to $\mathcal{O}$ itself will appear in the RHS of \cref{eqn:expval-O-evolution-momentum-space}, like the $\pi_c$ in \cref{eqn:phi-amp-dt-leading-order}. It requires us to assign a set of operators to $\mathcal{O}$ to obtain a closed system of equations, like the set $\Bab{ \phi_c(x), \pi_c(x) ; c=1,2 }$ for the field amplitudes in \cref{sec:evolutions-of-field-amplitudes-and-their-mixing}. For the bilinear forms of creation/annihilation operators, the momentum-preserving operators is one possible set fulfilling the requirement.
\begin{equation}
    \Bab{a_{c,\bvec{p}}^\dagger a_{d,\bvec{p}}, \;
        a_{c,\bvec{p}} a_{d,\bvec{p}}^\dagger , \;
        a_{c,\bvec{p}}^\dagger a_{d,-\bvec{p}}^\dagger , \;
        a_{c,\bvec{p}} a_{d,-\bvec{p}} , \;
        ; \; c,d=1,2, \; \bvec{p} = \pm \bvec{k}}\,.
    \label{eqn:all-quadratic-terms}
\end{equation}
which is found by letting $\mathcal{O} = a_{1,\bvec{k}}^\dagger a_{1,\bvec{k}}$ first then keeping setting $\mathcal{O}$ to the new operators that appear on the RHS of \cref{eqn:expval-O-evolution-momentum-space} until these equations of expectation values are closed. It is worth noting that this procedure terminates after finite steps because we assumed linear couplings $J_a[\phi_a] = \phi_a$ such that the output of the nested (anti-)commutators are classical numbers or quadratic terms of creation/annihilation operators. If $J_a[\phi_a]$ is not linear in $\phi_a$, new operators in the output will be of higher degrees and the procedure will not stop unless further approximations are imposed. We want to point out that the operators in \cref{eqn:all-quadratic-terms} describe the populations and coherence of the subsystem. They also yield the Fourier transform of equal-time same-position two-point correlation functions once we notice that $\int \dn[3]x \mathcal{A}_c(\bvec{x},t) \mathcal{B}_d(\bvec{x},t) = \sum_{\bvec{k}} \mathcal{A}_{\bvec{k}}(t) \mathcal{B}_{-\bvec{k}}(t)$ and use \cref{eqn:mode-expansion}. So we will refer to the expectation values of the operators in \cref{eqn:all-quadratic-terms} as two-point correlation functions, too.

The nested (anti-)commutators in the first trace of \cref{eqn:expval-O-evolution-momentum-space} yield new quadratic terms in the form of an anti-commutator, $\Tr\Bab{ \acomm{\mathcal{A}}{\mathcal{B}} \hat{\rho}(\tau) }$, where $\mathcal{A}$ are $\mathcal{B}$ are creation or annihilation operators. We can rewrite this form as $2\expval{\mathcal{A}\mathcal{B}} * f(\tau)$ if $\mathcal{A}$ and $\mathcal{B}$ commute or $2 \expval{\mathcal{A}\mathcal{B}} * f(\tau) + h(\tau)$ if $\mathcal{A}$ and $\mathcal{B}$ don't commute, where $f(\tau),h(\tau)$ are time-dependent classical numbers. The extra term $h(\tau)$ prevents us from writing all equations in a uniform way. To avoid this inconvenience, we choose to solve the evolution for $\expval{\acomm{\mathcal{A}}{\mathcal{B}}}$ instead of $\expval{\mathcal{A}\mathcal{B}}$ so that only one form of the unknowns, namely, $\expval{\acomm{\mathcal{A}}{\mathcal{B}}}$, appear in the equations. To proceed, we introduce the following two special equal-time correlation functions.
\begin{equation}
    A_{cd, \bvec{k}}(t) \coloneqq \expval*{ \acomm*{a_{c,\bvec{k}}^\dagger(t)}{a_{d,\bvec{k}}(t)} } \,,
    \qquad\quad
    B_{cd,\bvec{k}} \coloneqq \expval{ \acomm{ a_{c,\bvec{k}}(t) }{ a_{d,-\bvec{k}}(t) } } \,.
    \label{eqn:A-cd-k-B-cd-k-definitions}
\end{equation}
Their equations of motion are obtained after assigning them to $\expval{\mathcal{O}}$ in \cref{eqn:expval-O-evolution-momentum-space}.
\begin{align}
    \diff**{t}{A_{cd,\bvec{k}}}(t) =
    & i \pab{\omega_c - \omega_d} A_{cd,\bvec{k}}(t)
    - i \sum_{a,b=1,2} \int_0^t \Big[ -\delta_{ca} \pab{ B_{db,\bvec{k}}(t') + A_{db,\bvec{k}}^*(t') } e^{-i\omega_d(t-t')} 
    \nonumber \\*
    & \quad + \delta_{da} \pab{ A_{cb,\bvec{k}}(t') 
    + B_{cb,\bvec{k}}^*(t') } e^{i\omega_c(t-t')} \Big] \widetilde{\Sigma}_{ab}(\abs{\bvec{k}},t-t') \,\dn{t'}
    + \mathcal{I}_{A_\bvec{k},cd}(t) \,.
    \label{eqn:A-cd-k-dt}
    \\
    \diff**{t}{B_{cd,\bvec{k}}(t)} =
    & -i(\omega_c + \omega_d) B_{cd,\bvec{k}}(t)
    -i\sum_{a,b=1,2} \int_0^t \Big[ \delta_{ca} \pab{B_{db,-\bvec{k}}(t') + A_{db,-\bvec{k}}^*(t')} e^{-i\omega_d(t-t')}
    \nonumber \\*
    & \quad + \delta_{da} \pab{ B_{cb,\bvec{k}}(t') + A_{cb,\bvec{k}}^*(t') } e^{-i\omega_c(t-t')} \Big] \widetilde{\Sigma}_{ab}(\abs{\bvec{k}},t-t') \, \dn{t'} 
    + \mathcal{I}_{B_\bvec{k},cd}(t) \,.
    \label{eqn:B-cd-k-dt}
\end{align}
where the inhomogeneous terms are obtained after the substitution $t' \to t-t'$,
\begin{align}
    \mathcal{I}_{A_\bvec{k},cd} (t)
    & = + \sum_{a,b=1,2} \int_0^t \dn{t'} 2 \bab{ \delta_{ca} \delta_{db} e^{-i\omega_b t'} + \delta_{da} \delta_{cb} e^{i\omega_b t'} } \widetilde{\mathcal{N}}_{ab}(\abs{\bvec{k}},t') \,.
    \\
    \mathcal{I}_{B_\bvec{k},cd} (t)
    & = - \sum_{a,b=1,2} \int_0^t \dn{t'} 2 \bab{\delta_{da} \delta_{cb} + \delta_{ca} \delta_{db}} e^{-i\omega_b t'} \widetilde{\mathcal{N}}_{ab}(\abs{\bvec{k}},t') \,.
\end{align}
and for notational convenience, we defined
\begin{equation}
    \widetilde{\Sigma}_{ab}(\bvec{k}, t) \coloneqq \frac{\Sigma_{ab}(\bvec{k},t)}{\sqrt{2 \omega_{a,\bvec{k}} 2 \omega_{b,\bvec{k}}}}  \,,
    \qquad\quad
    \widetilde{\mathcal{N}}_{ab}(\bvec{k},t) \coloneqq \frac{\mathcal{N}_{ab}(\bvec{k},t)}{\sqrt{2 \omega_{a,\bvec{k}} 2 \omega_{b,\bvec{k}}}} \,.
    \label{eqn:tilde-Sigma-N}
\end{equation}
The rotational symmetry of the self-energy and noise kernel (see \cref{app:symmetries-of-self-energy,app:symmetries-of-noise-kernel}) is also used such that $\Sigma_{ab}(\pm\bvec{k},t) = \Sigma_{ab}(\abs{\bvec{k}},t)$ and $\mathcal{N}_{ab}(\pm\bvec{k},t) = \mathcal{N}_{ab}(\abs{\bvec{k}},t)$.

By taking the complex conjugates of $A_{cd,\bvec{k}}$ and $B_{cd,\bvec{k}}$ and/or relabelling $\bvec{k} \to -\bvec{k}$, we are able to generate all anti-commutators (listed below) made up of terms in \cref{eqn:all-quadratic-terms}.
\begin{equation}
    A_{cd,\bvec{k}} \,, \quad A_{cd,-\bvec{k}} \,, \quad
    A_{cd,\bvec{k}}^* \,, \quad A_{cd,-\bvec{k}}^* \,, \quad
    B_{cd,\bvec{k}} \,, \quad B_{cd,-\bvec{k}} \,, \quad
    B_{cd,\bvec{k}}^* \,, \quad B_{cd,-\bvec{k}}^* \,.
    \label{eqn:all-A-k-B-k}
\end{equation}
Applying the same operations to \cref{eqn:A-cd-k-dt,eqn:B-cd-k-dt} yields their equations of motion so that the system of equations is closed. The definitions in \cref{eqn:A-cd-k-B-cd-k-definitions} implies that
\begin{equation}
    A_{cd,\bvec{k}}^*(t) = A_{dc,\bvec{k}}(t) \,,
    \qquad\quad
    B_{cd,\bvec{k}}(t) = B_{dc,-\bvec{k}}(t) \,.
    \label{eqn:A-k-B-k-symmetries}
\end{equation}
which reduce the number of independent equations from eight to four. In what follows, we keep the equations of $A_{cd,\bvec{k}}, B_{cd,\bvec{k}}, A_{cd,-\bvec{k}}$ and $B_{cd,\bvec{k}}^*$. The equations of $A_{cd,\bvec{k}}$ and $B_{cd,\bvec{k}}$ have been given in \cref{eqn:A-cd-k-dt,eqn:B-cd-k-dt} and the other two equations are obtained via relabelling $\bvec{k} \to -\bvec{k}$ and/or taking complex conjugate in \cref{eqn:A-cd-k-dt,eqn:B-cd-k-dt}. We present the detailed steps and resulting equations in \cref{subapp:equations-of-motion-two-point-correlations}. These equations take the same form as \cref{eqn:A-cd-k-dt,eqn:B-cd-k-dt}, namely, a convolutional-differential equation with an inhomogeneous term. This observation motivates us to use the same techniques as in \cref{sec:evolutions-of-field-amplitudes-and-their-mixing} to solve these equations.

The strategy is as follows. Write the 16 unknown correlation functions, which are currently grouped in four $2\times 2$ matrices, as a $16\times 1$ column vector, denoted as $\overrightarrow{\mathcal{D}}(t)$. Then, the 16 coupled equations can be written formally as
\begin{equation}
    \diff**{t}{\overrightarrow{\mathcal{D}}} = i \mathbf{\Omega} \cdot \overrightarrow{\mathcal{D}}(t) -  \int_0^t \dn{t'} \mathbf{K}(t-t') \cdot \overrightarrow{\mathcal{D}}(t') + \overrightarrow{\mathcal{I}}(t) \,.
    \label{eqn:16-16-matrix-equation}
\end{equation}
where $\mathbf{\Omega}, \mathbf{K}$ are $16\times 16$ matrices and $\overrightarrow{\mathcal{I}}$ is the corresponding $16\times 1$ column of inhomogeneous terms. Their elements can be directly read off from the equations of motion in \cref{subapp:equations-of-motion-two-point-correlations} in a manner similar to how \cref{eqn:G-A-inverse-s} was obtained. The solution of \cref{eqn:16-16-matrix-equation} is obtained by a Laplace transform (denoted as $\mathcal{L}[\cdot]$) as befits an initial-value problem.
\begin{equation}
    s \overrightarrow{\mathcal{D}}(s) - \overrightarrow{\mathcal{D}}(0^-) = i \mathbf{\Omega} \cdot \overrightarrow{\mathcal{D}}(s) - \mathbf{K}(s) \cdot \overrightarrow{\mathcal{D}}(s) + \overrightarrow{\mathcal{I}}(s) \,.
    \label{eqn:16-equation-s-domain}
\end{equation}
where $\overrightarrow{\mathcal{D}}(0^-) \coloneqq \overrightarrow{\mathcal{D}}(t=0^-)$ and the solutions in the $s$-domain is
\begin{equation}
    \overrightarrow{\mathcal{D}}(s) = \mathbf{G}_\mathcal{D}(s) \cdot \bab{ \overrightarrow{\mathcal{D}}(0^-) + \overrightarrow{\mathcal{I}}(s) } \,,
    \qquad
    \mathbf{G}_\mathcal{D}^{-1}(s) \coloneqq s - i \mathbf{\Omega} + \mathbf{K}(s) \,.
    \label{eqn:formal-solution-s-domain}
\end{equation}
The solution in the time domain is obtained by taking the inverse Laplace transform $\mathcal{L}^{-1}[\cdot]$.
\begin{equation}
    \overrightarrow{\mathcal{D}}(t) = \mathbf{G}_\mathcal{D}(t) \cdot \overrightarrow{\mathcal{D}}(0^-) + \int_0^t \dn{t'} \mathbf{G}_\mathcal{D}(t-t') \cdot \overrightarrow{\mathcal{D}}(t') \,,
    \qquad
    \mathbf{G}_\mathcal{D}(t) \coloneqq \mathcal{L}^{-1}\bab{ \mathbf{G}_\mathcal{D}(s) } \,.
    \label{eqn:formal-soluiton-time-domain}
\end{equation}
Lastly, repeating the same arguments as those for \cref{eqn:G-A-s-adj-det,eqn:G-A-Breit-Wigner}, we can write the Green's function in the $s$-domain and the time-domain as
\begin{equation}
    \mathbf{G}_\mathcal{D}(s) = \frac{\adj\Bab{\mathbf{G}_\mathcal{D}^{-1}(s)}}{\det\Bab{\mathbf{G}_\mathcal{D}^{-1}(s)}} \,,
    \qquad
    \mathbf{G}_\mathcal{D}(t) = \sum_{i=1}^{16} \mathbb{G}_{\mathcal{D},i}\, e^{s_{\mathcal{D},i} t} \,.
    \label{eqn:G-D}
\end{equation}
where $s_i$ are the 16 physical poles given by $\det\Bab*{\mathbf{G}_\mathcal{D}^{-1}(s)} = 0$ and $\mathbb{G}_{\mathcal{D},i}$ are the corresponding residues.

\subsection{Quasi-normal modes of populations and coherence}
\label{subsec:quasi-normal-modes-of-populations-and-coherence}

Finding the quasi-normal modes (QNM) $\mathbb{G}_{\mathcal{D},i}\, e^{s_{\mathcal{D},i} t}$ of $\mathbf{G}_\mathcal{D}(t)$ in \cref{eqn:G-D} analytically is an extremely difficult task since the determinant $\det\Bab*{\mathbf{G}_\mathcal{D}^{-1}(s)}$ includes approximately $16! \approx 2\times 10^{13}$ terms. However, it turns out that after keeping only the leading order perturbations in the QNMs so as to be consistent with the order of perturbation in \cref{eqn:expectation-value-evolution}, we can invert $\mathbf{G}_\mathcal{D}^{-1}(s)$ block-wisely and divide the task of finding the QNMs into smaller ones.

To proceed, we first specify the order of the two-point correlation functions in $\overrightarrow{\mathcal{D}}(t)$.
\begin{equation}
    \overrightarrow{\mathcal{D}}^\T = \pab{\overrightarrow{A}^\T,\; \overrightarrow{B}^\T} \,,
    \qquad
    \overrightarrow{A}^\T = \pab{ \overrightarrow{A}_\bvec{k}^\T,\; \overrightarrow{A}_{-\bvec{k}}^\T } \,,
    \qquad
    \overrightarrow{B}^\T = \pab{ \overrightarrow{B}_\bvec{k}^\T,\; \overrightarrow{B}^*_{\bvec{k}}{}^\T }
    \,.
    \label{eqn:A-B-separation}
\end{equation}
where $\overrightarrow{F}_{\bvec{p}}^\T = \pab*{ F_{11,\bvec{p}}, F_{12,\bvec{p}}, F_{21,\bvec{p}}, F_{22,\bvec{p}} }$ for $\overrightarrow{F}_\bvec{p} = \overrightarrow{A}_\bvec{k}, \overrightarrow{A}_{-\bvec{k}}, \overrightarrow{B}_{\bvec{k}}, \overrightarrow{B}_{\bvec{k}}^*$. Note that this order naturally separates the two-point correlation functions into a slowly-oscillating and a fast-oscillating part since the bare frequencies of part A are $\pab*{\omega_{c,\bvec{k}} - \omega_{d,\bvec{k}}}$ and the bare frequencies of part B are $\pm\pab*{\omega_{c,\bvec{k}} + \omega_{d,\bvec{k}}}$. This separation results in a block-row form of \cref{eqn:16-equation-s-domain}.
\begin{equation}
    \begin{pmatrix}
        \mathbf{G}_\txtA^{-1} & \mathbf{K}_{\txtA\txtB} \\
        \mathbf{K}_{\txtB\txtA} & \mathbf{G}_\txtB^{-1}
    \end{pmatrix}
    \cdot
    \begin{pmatrix}
        \overrightarrow{A} \\
        \overrightarrow{B}
    \end{pmatrix}
    =
    \begin{pmatrix}
        \overrightarrow{A}(0^-) \\
        \overrightarrow{B}(0^-)
    \end{pmatrix}
    +
    \begin{pmatrix}
        \overrightarrow{\mathcal{I}}_\txtA \\
        \overrightarrow{\mathcal{I}}_\txtB
    \end{pmatrix}
    \,.
    \label{eqn:decomposed-G-D-equation}
\end{equation}
where the four blocks are related to $\mathbf{G}_\mathcal{D}(s)$ in \cref{eqn:formal-solution-s-domain} as
\begin{equation}
    \mathbf{G}_\mathcal{D}^{-1}(s)
    \eqqcolon
    \begin{pmatrix}
        s - i \mathbf{\Omega}_\txtA + \mathbf{K}_{\text{AA}} &  \mathbf{K}_{\text{AB}} \\
        \mathbf{K}_{\text{BA}} & s - i \mathbf{\Omega}_\txtB + \mathbf{K}_{\text{BB}}
    \end{pmatrix}
    \eqqcolon
    \begin{pmatrix}
        \mathbf{G}_\txtA^{-1} & \mathbf{K}_{\txtA\txtB} \\
        \mathbf{K}_{\txtB\txtA} & \mathbf{G}_\txtB^{-1}
    \end{pmatrix}
    \,.
\end{equation}
with
\begin{equation}
    \mathbf{\Omega} \eqqcolon
    \begin{pmatrix}
        \mathbf{\Omega}_\txtA & 0 \\
        0 & \mathbf{\Omega}_\txtB
    \end{pmatrix}
    \,,
    \qquad\quad
    \mathbf{K} \eqqcolon
    \begin{pmatrix}
        \mathbf{K}_{\text{AA}} & \mathbf{K}_{\text{AB}} \\
        \mathbf{K}_{\text{BA}} & \mathbf{K}_{\text{BB}}
    \end{pmatrix}
    \,.
\end{equation}
in which we have used the property that $\mathbf{\Omega}$ is diagonal. The explicit expressions of these matrices are given in \cref{subapp:Greens-function-in-s-domain} and $\overrightarrow{\mathcal{I}}_\txtA,\overrightarrow{\mathcal{I}}_\txtB$ are given explicitly in \cref{subapp:inhomogeneous-terms-in-s-domain}. Using Gaussian elimination in \cref{eqn:decomposed-G-D-equation} yields
\begin{equation}
    \overrightarrow{A}(s) = \mathcal{G}_\txtA \cdot \pab{ \overrightarrow{A}(0^-) + \overrightarrow{\mathcal{I}}_\txtA } - \mathcal{G}_\txtA \cdot \mathbf{K}_{\txtA\txtB} \cdot \mathbf{G}_\txtB \cdot \pab{ \overrightarrow{B}(0^-) + \overrightarrow{\mathcal{I}}_\txtB } 
    \,.
    \label{eqn:A-s-G-A}
\end{equation}
where
\begin{equation}
    \mathcal{G}_\txtA = \pab{ 1 - \mathbf{G}_\txtA \cdot \mathbf{K}_{\txtA\txtB} \cdot \mathbf{G}_\txtB \cdot \mathbf{K}_{\txtB\txtA} }^{-1} \cdot \mathbf{G}_\txtA \,.
    \label{eqn:G-A}
\end{equation}
We want to emphasize that because the elements of $\mathcal{G}_\txtA(s)$ contain fractions, the arguments for finding poles of $\mathbf{G}_\phi(s)$ or $\mathbf{G}_\mathcal{D}(s)$ do not apply to $\mathcal{G}_\txtA(s)$. $\mathcal{G}_\txtA(s)$ still has the same 16 poles as $\mathbf{G}_\mathcal{D}(s)$ since  no truncation has been done yet although $\mathcal{G}_A(s)$ is a $8\times8$ matrix. This is better seen in the Taylor expansion of $\mathcal{G}_\txtA(s)$.
\begin{equation}
    \mathcal{G}_\txtA = \sum_{n=0}^\infty \bab{ \mathbf{G}_\txtA \cdot \mathbf{K}_{\txtA\txtB} \cdot \mathbf{G}_\txtB \cdot \mathbf{K}_{\txtB\txtA} }^n \cdot \mathbf{G}_\txtA \,.
    \label{eqn:G-A-Taylor-expansion}
\end{equation}
Since the nonzero elements of $\mathbf{K}_{\txtA\txtB}$ and $\mathbf{K}_{\txtB\txtA}$ are self-energies, they are of order $\order{g^2}$ and do not have poles. Therefore, $\mathbf{G}_\txtA(s)$ and $\mathbf{G}_\txtB(s)$ each contribute 8 poles to $\mathcal{G}_\txtA(s)$, which we collectively refer to as $\Bab*{s_\text{slow}}$ and $\Bab*{s_\text{fast}}$ because they feature low and high frequencies as mentioned below \cref{eqn:A-B-separation}. We can use \cref{eqn:G-A-Taylor-expansion} to estimate the residues of $\mathcal{G}_\txtA(s)$.
\begin{equation}
    \residue\pab{\mathcal{G}_\txtA, s} = 
    \begin{cases}
        \residue\pab{\mathbf{G}_\txtA , s} \sim \order{1} \,, & \forall s \in \Bab*{s_\text{slow}}
        \\
        \mathbf{G}_\txtA(s) \cdot \mathbf{K}_{\txtA\txtB}(s) \cdot \residue\pab{ \mathbf{G}_\txtB , s } \cdot \mathbf{K}_{\txtB\txtA}(s) \cdot \mathbf{G}_\txtA(s) \sim \order{g^4} \,, & \forall s \in \Bab*{s_{\text{fast}}}
    \end{cases}
    \,.
    \label{eqn:residue-G-A-at-sA-sB}
\end{equation}
Anticipating that $\mathcal{G}_\txtA(t)$ will take the same form as $\mathbf{G}_\phi(t)$ in \cref{eqn:G-A-Breit-Wigner} or $\mathbf{G}_\mathcal{D}(t)$ in \cref{eqn:G-D}, we conclude that only $\residue\pab{\mathcal{G}_\txtA, s}\eval{}_{s\in\Bab{s_\text{slow}}}$ contribute to $\mathcal{G}_\txtA(t)$ up to the leading order perturbation. In the following calculation, we will use 
\begin{equation}
    \mathcal{G}_\txtA(s) \approx \mathbf{G}_\txtA(s) \,.
    \label{eqn:G-A-approx-G-A-0}
\end{equation}
But this doesn't mean $\overrightarrow{A}(t)$ only has slowly oscillating components. Inserting \cref{eqn:G-A-approx-G-A-0} into \cref{eqn:A-s-G-A} yields
\begin{equation}
    \overrightarrow{A}(s) = \mathbf{G}_\txtA \cdot \pab{ \overrightarrow{A}(0^-) + \overrightarrow{\mathcal{I}}_\txtA } - \mathbf{G}_\txtA \cdot \mathbf{K}_{\txtA\txtB} \cdot \mathbf{G}_\txtB \cdot \pab{ \overrightarrow{B}(0^-) + \overrightarrow{\mathcal{I}}_\txtB }  \,.
    \label{eqn:A-s-elimination}
\end{equation}
The sub-leading (second) term contains all 16 poles coming from $\mathbf{G}_\txtA$ and $\mathbf{G}_\txtB$ and it contributes at the leading order perturbation since
\begin{equation}
\begin{aligned}
    \residue\pab{\mathbf{G}_\txtA,s} \cdot \mathbf{K}_{\txtA\txtB}(s) \cdot \mathbf{G}_\txtB(s)
    & \sim \order{g^2} \,, \quad \forall s\in\Bab*{s_{\text{slow}}}
    \,,
    \\
    \mathbf{G}_\txtA(s) \cdot \mathbf{K}_{\txtA\txtB}(s) \cdot \residue\pab{ \mathbf{G}_\txtB, s }
    & \sim \order{g^2} \,, \quad \forall s\in\Bab*{s_{\text{fast}}}
    \,.
\end{aligned}
\end{equation}
Similar arguments also apply to part B, following which we find that
\begin{equation}
    \mathcal{G}_\txtB
    = \pab{ 1 - \mathbf{G}_\txtB \cdot \mathbf{K}_{\txtB\txtA} \cdot \mathbf{G}_\txtA \cdot \mathbf{K}_{\txtA\txtB} }^{-1} \cdot \mathbf{G}_\txtB
    \approx \mathbf{G}_\txtB \,.
    \label{eqn:G-B-approx-G-B-0}
\end{equation}
and
\begin{equation}
    \overrightarrow{B}(s) = \mathbf{G}_\txtB \cdot \pab{ \overrightarrow{B}(0^-) + \overrightarrow{\mathcal{I}}_\txtB } - \mathbf{G}_\txtB \cdot \mathbf{K}_{\text{BA}} \cdot \mathbf{G}_\txtA \cdot \pab{ \overrightarrow{A}(0^-) + \overrightarrow{\mathcal{I}}_\txtA }  \,.
    \label{eqn:B-s-elimination}
\end{equation}
In \cref{subapp:equations-of-motion-and-Green's-functions}, we apply the same procedure to a subsystem of only one species for a more straightforward clarification of the above arguments, in which the sizes of matrices are smaller so that we can write out all results explicitly.

So far, we have divided the task of finding the QNMs of $\mathbf{G}_\mathcal{D}$ into finding the QNMs of $\mathbf{G}_\txtA$ and $\mathbf{G}_\txtB$. The equations of motion listed in \cref{subapp:equations-of-motion-two-point-correlations} show that $A_{\bvec{k},cd}$ (or $A_{-\bvec{k},cd}$) only couples with $B_{\bvec{k},cd}$ and $B_{\bvec{k},cd}^*$ and vice versa. It implies that $\mathbf{G}_\txtA^{-1}$ and $\mathbf{G}_\txtB^{-1}$ are block-diagonalized and so are $\mathbf{G}_\txtA$ and $\mathbf{G}_\txtB$ (see the explicit expressions in \cref{subapp:Greens-function-in-s-domain}),
\begin{equation}
    \mathbf{G}_\txtA^{-1} = 
    \begin{pmatrix}
        \mathbf{G}_{A_\bvec{k}}^{-1} & 0 \\
        0 & \mathbf{G}_{A_{-\bvec{k}}}^{-1}
    \end{pmatrix}
    \,,
    \qquad\quad
    \mathbf{G}_\txtB^{-1} =
    \begin{pmatrix}
        \mathbf{G}_{B_\bvec{k}}^{-1} & 0 \\
        0 & \mathbf{G}_{B_\bvec{k}^*}^{-1}
    \end{pmatrix}
    \,.
    \label{eqn:G-A-0-G-B-0-block-diagonalized}
\end{equation}
which suggests a further division of the task of finding the QNMs. Particularly, we found that $\mathbf{G}_{A_\bvec{k}}^{-1} = \mathbf{G}_{A_{-\bvec{k}}}^{-1}$, which is a consequence of assuming the state of the medium is rotationally invariant (see \cref{app:Lehmann-representation-of-correlation-functions,app:symmetries-of-self-energy}) that results in rotationally invariant self-energy corrections. In the end, finding 16 QNMs of a $16\times 16$ matrix $\mathbf{G}_\mathcal{D}$ boils down to finding 4 QNMs for each of the four $4\times4$ matrices $\mathbf{G}_{A_\bvec{k}}, \mathbf{G}_{A_{-\bvec{k}}}, \mathbf{G}_{B_\bvec{k}}, \mathbf{G}_{B_\bvec{k}^*}$. These QNMs are obtained using the same techniques as in \cref{subsec:equations-of-motion-and-solutions-of-field-amplitues} and the explicit results are listed in \cref{subapp:poles-of-Green's-functions,subapp:residues-at-poles-of-equal-time-two-point-correlation-functions}, where we label the poles and residues so that when the couplings are switched off, they become the bare frequencies and non-interacting residues of the operators indicated in the subscripts , e.g, $s_{\acomm*{a_{1,\bvec{k}}}{a_{2,\bvec{k}}}}^{(0)} = -i(\omega_1 + \omega_2)$. Same as in \cref{subsec:amplitude-mixing-in-the-non-degenerate-and-nearly-degenerate-cases}, the non-degenerate and nearly-degenerate cases need to be discussed separately.

Comparing the poles of one-point correlations (denoted as $s_{\mathcal{A}}$, see \cref{subsec:amplitude-mixing-in-the-non-degenerate-and-nearly-degenerate-cases}) and the poles of two-point correlations (denoted as $s_{\acomm*{\mathcal{A}}{\mathcal{B}}}$, see \cref{table:poles-of-two-point-correlation-functions}), it is straightforward to notice that in both the non-degenerate and nearly-degenerate cases,
\begin{equation}
\label{eqn:pole-relations}%
    s_{\acomm{\mathcal{A}}{\mathcal{B}}} = s_{\mathcal{A}} + s_{\mathcal{B}} \,.
\end{equation}
This is a part of the Kronecker-product relation between the QNMs of the equal-time two-point correlation functions (denoted as $\mathbf{G}_{\acomm*{\mathcal{A}}{\mathcal{B}}}(t) \coloneqq \mathbb{G}_{\acomm*{\mathcal{A}}{\mathcal{B}}}\, e^{s_{\acomm*{\mathcal{A}}{\mathcal{B}}} t}$) and the QNMs of the dimensionless one-point correlations functions $\expval*{a_{c,\bvec{k}}}, \expval*{a_{c,\bvec{k}}^\dagger}, c=1,2$ (denoted as $\mathbf{G}_\mathcal{A}(t)$). To show this relation, we first obtain the solution of $\expval*{a_{c,\bvec{k}}}, \expval*{a_{c,\bvec{k}}^\dagger}$ by noticing that
\begin{equation}
    \expval{\boldsymbol{\phi}_\bvec{k}}(t) = U_\phi \cdot \pab[big]{ \expval{\boldsymbol{a}_\bvec{k}}(t) + \expval*{\boldsymbol{a}_\bvec{k}^\dagger}(t) } \,,
    \qquad
    \expval{\boldsymbol{\pi}_\bvec{k}}(t) = U_\pi \cdot \pab[big]{\expval{\boldsymbol{a}_\bvec{k}}(t) - \expval*{\boldsymbol{a}_\bvec{k}^\dagger}(t) } \,.
    \label{eqn:phi-pi-a-relation}
\end{equation}
where $\expval{\boldsymbol{\phi}_\bvec{k}}$ and $\expval{\boldsymbol{\pi}_\bvec{k}}$ were defined in \cref{eqn:phi-equation-matrix-form} and for notational convenience, we also defined $\expval*{\boldsymbol{a}_{\bvec{k}}} = \bab{ \expval{a_{1,\bvec{k}}}, \expval{a_{2,\bvec{k}}}}^\T$, $\expval*{\boldsymbol{a}_{\bvec{k}}^\dagger} = \bab*{ \expval*{a_{1,\bvec{k}}^\dagger}, \expval*{a_{2,\bvec{k}}^\dagger}}^\T$ and
\begin{equation}
    U_\phi = 
    \begin{pmatrix}
        \frac{1}{\sqrt{2\omega_1}} & 0 \\
        0 & \frac{1}{\sqrt{2\omega_2}}
    \end{pmatrix}
    \,,\qquad
    U_\pi = 
    \begin{pmatrix}
        -i \sqrt{\frac{\omega_1}{2}} & 0 \\
        0 & -i \sqrt{\frac{\omega_2}{2}}
    \end{pmatrix}
    \,.
\end{equation}
Inverting \cref{eqn:phi-pi-a-relation} and inserting \cref{eqn:phi-t-solution,eqn:pi-t-solution} yield $\expval*{\boldsymbol{a}_\bvec{k}}$ and $\expval*{\boldsymbol{a}^\dagger_\bvec{k}}$ as below.
\begin{subequations}
\begin{align}
    \expval{\boldsymbol{a}_\bvec{k}}(t)
    & = F_{a,a}[\mathbf{G}_\phi(t)] \cdot \expval{\boldsymbol{a}_{\bvec{k}}}(0^-)
      + F_{a,a^\dagger}[\mathbf{G}_\phi(t)] \cdot \expval*{\boldsymbol{a}_{\bvec{k}}^\dagger}(0^-) \,,
    \\
    \expval*{\boldsymbol{a}_\bvec{k}^\dagger}(t)
    & = F_{a^\dagger,a^\dagger}[\mathbf{G}_\phi(t)] \cdot \expval*{\boldsymbol{a}_{\bvec{k}}^\dagger}(0^-)
      + F_{a^\dagger,a}[\mathbf{G}_\phi(t)] \cdot \expval*{\boldsymbol{a}_{\bvec{k}}}(0^-) \,.
\end{align}
\end{subequations}
The explicit expressions of $F_{\mathcal{A},\mathcal{B}}[\mathbf{G}_\phi(t)], \mathcal{A},\mathcal{B} = a,a^\dagger$ are given in \cref{subapp:Green's-functions-of-dimensionless-one-point-correlation-functions}, where we find that regardless of the magnitude of degeneracy, $F_{a,a}[\mathbf{G}_\phi(t)] \sim \order{1}$ and $F_{a,a^\dagger}[\mathbf{G}_\phi(t)] \sim \order{g^2}$, suggesting that $F_{a,a}[\mathbf{G}_\phi(t)]$ is the Green's function of $\expval*{\boldsymbol{a}_\bvec{k}}$. Furthermore, $F_{a,a}[\cdot]$ is a linear map. Substituting the QNMs of $\mathbf{G}_\phi(t)$ (see \cref{subsec:equations-of-motion-and-solutions-of-field-amplitues}) yields $F_{a,a}[\mathbf{G}_\phi(t)] = \sum_{i} F_{a,a}[\mathbf{G}_{\phi,i}(t)]$ with $i=a_1,a_2,a_1^\dagger,a_2^\dagger$, whose explicit results (see \cref{subapp:Green's-functions-of-dimensionless-one-point-correlation-functions}) show that
\begin{equation}
    F_{a,a}[\mathbf{G}_{\phi,a_1}(t)] \sim F_{a,a}[\mathbf{G}_{\phi,a_2}(t)] \sim \order{1} \,,
    \quad
    F_{a,a}[\mathbf{G}_{\phi,a_1^\dagger}(t)] \sim F_{a,a}[\mathbf{G}_{\phi,a_2^\dagger}(t)] \sim \order{g^4}
    \,.
\end{equation}
Therefore, up to the leading-order perturbation, we define the Green's function of $\expval*{\boldsymbol{a}_\bvec{k}}(t)$ as $\mathbf{G}_{\boldsymbol{a}_\bvec{k}}(t) \coloneqq \sum_{i=1,2} \mathbf{G}_{a_{i,\bvec{k}}}(t)$, whose QNMs are 
\begin{equation}
    \mathbf{G}_{a_{1,\bvec{k}}}(t) 
    \coloneqq F_{a,a}[\mathbf{G}_{\phi,a_1}(t)] 
    = \mathbb{G}_{a_{1,\bvec{k}}} e^{s_{a_{1,\bvec{k}}} t} \,,
    \;\;
    \mathbf{G}_{a_{2,\bvec{k}}}(t) 
    \coloneqq F_{a,a}[\mathbf{G}_{\phi,a_2}(t)] 
    = \mathbb{G}_{a_{2,\bvec{k}}} e^{s_{a_{2,\bvec{k}}} t}
    \,.
    \label{eqn:quasi-normal-modes-of-a-k}
\end{equation}
Similar arguments result in $\mathbf{G}_{\boldsymbol{a}_\bvec{k}^\dagger}(t) \coloneqq \sum_{i=1,2} \mathbf{G}_{a_{i,\bvec{k}}^\dagger}(t)$ as the Green's function of $\expval*{\boldsymbol{a}_\bvec{k}^\dagger}$ up to the leading-order perturbation, whose QNMs are
\begin{equation}
    \mathbf{G}_{a_{1,\bvec{k}}^\dagger}(t) 
    \coloneqq F_{a^\dagger,a^\dagger}[\mathbf{G}_{\phi,a_1^\dagger}(t)]
    \,,
    \qquad
    \mathbf{G}_{a_{2,\bvec{k}}^\dagger}(t)
    \coloneqq F_{a^\dagger,a^\dagger}[\mathbf{G}_{\phi,a_2^\dagger}(t)]
    \,.
    \label{eqn:quasi-normal-modes-of-a-k^dagger}
\end{equation}
$\mathbf{G}_{\boldsymbol{a}_\bvec{k}}(t)$ and $\mathbf{G}_{\boldsymbol{a}_\bvec{k}^\dagger}(t)$ are rotationally symmetric because of the rotational invariance of $U_\phi, U_\pi$ and $\mathbf{G}_\phi(t)$ (see \cref{eqn:amplitude-pole-|k|-dependence}). Using the results in \cref{subapp:Green's-functions-of-dimensionless-one-point-correlation-functions}, we can verify the conjugate symmetry of these QNMs.
\begin{equation}
    \bab{\mathbf{G}_{a_{c,\bvec{k}}}(t)}^* = \mathbf{G}_{a_{c,\bvec{k}}^\dagger}(t), \qquad c=1,2 \,.
    \label{eqn:conjugate-symmetry-dimensionless-one-point-correlation-functions}
\end{equation}

Comparing the explicit results of the QNMs of the two-point correlation functions (denoted as $\mathbf{G}_{\acomm{\mathcal{A}}{\mathcal{B}}}(t)$, see \cref{subapp:poles-of-Green's-functions,subapp:residues-at-poles-of-equal-time-two-point-correlation-functions}) and that of the dimensionless one-point correlation functions (denoted as $\mathbf{G}_{\mathcal{A}}(t)$, see \cref{subapp:Green's-functions-of-dimensionless-one-point-correlation-functions}), one will find that
\begin{equation}
    \mathbf{G}_{\acomm{\mathcal{A}}{\mathcal{B}}}(t) = \mathbf{G}_{\mathcal{A}}(t) \otimes \mathbf{G}_{\mathcal{B}}(t)
    \qquad\text{where}\qquad
    \mathcal{A},\mathcal{B} = a_{1,\pm\bvec{k}}, a_{2,\pm\bvec{k}}^\dagger, a_{1,\pm\bvec{k}}, a_{2,\pm\bvec{k}}^\dagger
    \,.
    \label{eqn:modes-relations}
\end{equation}
It follows that
\begin{subequations}
\begin{alignat}{2}
    \mathbf{G}_{A_\bvec{k}}(t)
        & \equiv \sum_{c,d=1,2} \mathbf{G}_{\acomm*{a_{c,\bvec{k}}^\dagger}{a_{d,\bvec{k}}}}(t)
        && = \mathbf{G}_{\boldsymbol{a}_{\bvec{k}}^\dagger}(t) \otimes \mathbf{G}_{\boldsymbol{a}_{\bvec{k}}}(t)
        \,,
    \label{eqn:G-A-k-relations}
    \\
    \mathbf{G}_{A_{-\bvec{k}}}(t)
        & \equiv \sum_{c,d=1,2} \mathbf{G}_{\acomm*{a_{c,-\bvec{k}}^\dagger}{a_{d,-\bvec{k}}}}(t)
        && = \mathbf{G}_{\boldsymbol{a}_{-\bvec{k}}^\dagger}(t) \otimes \mathbf{G}_{\boldsymbol{a}_{-\bvec{k}}}(t)
        \,,
    \\
    \mathbf{G}_{B_\bvec{k}}(t)
        & \equiv \sum_{c,d=1,2} \mathbf{G}_{\acomm*{a_{c,\bvec{k}}}{a_{d,-\bvec{k}}}}(t)
        && = \mathbf{G}_{\boldsymbol{a}_{\bvec{k}}}(t) \otimes \mathbf{G}_{\boldsymbol{a}_{-\bvec{k}}}(t)
        \,,
    \\
    \mathbf{G}_{B_{\bvec{k}}^*}(t)
        & \equiv \sum_{c,d=1,2} \mathbf{G}_{\acomm*{a_{c,\bvec{k}}^\dagger}{a_{d,-\bvec{k}}^\dagger}}(t)
        && = \mathbf{G}_{\boldsymbol{a}_{\bvec{k}}^\dagger}(t) \otimes \mathbf{G}_{\boldsymbol{a}_{-\bvec{k}}^\dagger}(t)
        \,.
\end{alignat}
\label{eqn:relations-of-quasi-normal-modes}%
\end{subequations}
Note that these Kronecker-product relations are indicated by the subscripts of the form $\acomm{\mathcal{A}}{\mathcal{B}}$ in the second column, which are in turn indicated by the definitions of the subscripts in the first column (see \cref{eqn:A-cd-k-B-cd-k-definitions}).

\subsection{Power counting and secular contributions of solutions}
\label{subsec:power-counting-and-secular-contributions-of-solutions}

The solutions of $\overrightarrow{A}$ and $\overrightarrow{B}$ in the $s$-domain has been shown in \cref{eqn:A-s-elimination,eqn:B-s-elimination}. In this subsection, we present the strategy to obtain their solutions in the time domain and count the orders of perturbations in the solutions. We will discuss the homogeneous parts accounting for the initial conditions (denoted by a superscript ``$i$'') and the inhomogeneous parts accounting for the fluctuations of the medium (denoted by a superscript ``$T$'') separately. In the $s$-domain, they are defined as
\begin{equation}
    \overrightarrow{A}(s) = \overrightarrow{A}^\supi(s) + \overrightarrow{A}^\supT(s) \,,
    \qquad\quad
    \overrightarrow{B}(s) = \overrightarrow{B}^\supi(s) + \overrightarrow{B}^\supT(s) \,.
    \label{eqn:separate-initial-thermal-contributions}
\end{equation}
with
\begin{subequations}
\begin{align}
    \overrightarrow{A}^\supi(s)
    & \coloneqq \underbrace{ \mathbf{G}_{\txtA}(s) }_{\sim \order{1}} \cdot \overrightarrow{A}(0^-)
        - \underbrace{ \mathbf{G}_{\txtA}(s) \cdot \mathbf{K}_{\text{AB}}(s) \cdot \mathbf{G}_{\txtB}(s) }_{ \sim \order{g^2} } \cdot \overrightarrow{B}(0^-) \,,
    \\
    \overrightarrow{A}^\supT(s)
    & \coloneqq \underbrace{ \mathbf{G}_{\txtA}(s) \cdot \overrightarrow{\mathcal{I}}_A(s) }_{\text{naively $\sim \order{g^2}$}} - \underbrace{ \mathbf{G}_{\txtA}(s) \cdot \mathbf{K}_{\text{AB}}(s) \cdot \mathbf{G}_{\txtB}(s) \cdot \overrightarrow{\mathcal{I}}_{B}(s) }_{\text{naively $\sim \order{g^4}$}} \,,
    \\
    \overrightarrow{B}^\supi(s)
    & \coloneqq \underbrace{ \mathbf{G}_{\txtB}(s) }_{\sim \order{1}} \cdot \overrightarrow{B}(0^-) - \underbrace{ \mathbf{G}_{\txtB}(s) \cdot \mathbf{K}_{\text{BA}}(s) \cdot \mathbf{G}_{\txtB}(s) }_{\sim \order{g^2}} \cdot \overrightarrow{A}(0^-) \,,
    \\
    \overrightarrow{B}^\supT(s)
    & \coloneqq \underbrace{ \mathbf{G}_{\txtB}(s) \cdot \overrightarrow{\mathcal{I}}_B(s) }_{\sim \order{g^2}} - \underbrace{ \mathbf{G}_{\txtB}(s) \cdot \mathbf{K}_{\text{BA}}(s) \cdot \mathbf{G}_{\txtA}(s) \cdot \overrightarrow{\mathcal{I}}_A(s) }_{\text{naively $\sim \order{g^4}$}} \,.
\end{align}
\label{eqn:A-B-i-T-power-counting}%
\end{subequations}
where we counted the orders of perturbations for each term using $\mathbf{K} \sim \overrightarrow{\mathcal{I}} \sim \order{g^2}$ (see \cref{subapp:Greens-function-in-s-domain,subapp:inhomogeneous-terms-in-s-domain}). This estimation is based on counting orders of perturbation effects at the level of the equation of motion which we refer to as ``naive power counting''. However, it is not necessarily correct in the long time limit as will be shown that secular contributions emerge in the solutions. We use $\overrightarrow{A}^\supT(s)$ as an example to clarify the mechanism.

In \cref{subsec:quasi-normal-modes-of-populations-and-coherence} we showed that up to leading order $\mathbf{G}_{\txtA}(t) = \sum_{s_i\in \Bab*{s_{\text{slow}}}} \mathbb{G}_{A,i} e^{s_i t}$ where $\Bab*{s_{\text{slow}}}$ are the poles of $\mathbf{G}_{\txtA}(s)$ defined above \cref{eqn:residue-G-A-at-sA-sB}. Taking the Laplace transform of this leading-order result yields
\begin{equation}
    \mathbf{G}_{\txtA}(s) = \mathcal{L}[\mathbf{G}_{\txtA}(t)] = \sum_{s_i\in\Bab*{s_{\text{slow}}}} \mathbb{G}_{A,i} \frac{1}{s - s_i} \,.
\end{equation}
Therefore, the leading term of $\overrightarrow{A}^\supT(s)$ can be written as
\begin{equation}
    \mathbf{G}_{\txtA}(s) \cdot \overrightarrow{\mathcal{I}}_A(s)
    = \sum_{s_i\in \Bab*{s_{\text{slow}}}} \mathbb{G}_{A,i} \cdot \frac{  \overrightarrow{N}_A(s) }{-s_i} \pab{\frac{1}{s} - \frac{1}{s - s_i}} \,.
    \label{eqn:G-A-I-A-separation-s-domain}
\end{equation}
where we used that $\overrightarrow{\mathcal{I}}_{A}(s)$ is of the form $\frac{1}{s} \overrightarrow{N}_A(s)$ with $\overrightarrow{N}_A(s) \sim \order{g^2}$ (see \cref{subapp:inhomogeneous-terms-in-s-domain}). Note that in the non-degenerate case, four poles in $\Bab*{s_{\text{slow}}}$ are of order $\order{g^2}$ and the other four poles are of order $\order{1}$, in the nearly-degenerate case all poles in $\Bab*{s_{\text{slow}}}$ are of order $\order{g^2}$ (see \cref{table:poles-of-two-point-correlation-functions}). Hence, an appropriate power counting for $\mathbf{G}_{\txtA}(s) \cdot \overrightarrow{\mathcal{I}}_A(s)$ is
\begin{equation}
    \mathbf{G}_{\txtA}(s) \cdot \overrightarrow{\mathcal{I}}_A(s) \sim
    \begin{cases}
        \order{1} + \order{g^2} \;, & \text{non-degenerate case} \\
        \order{1} \;, & \text{nearly-degenerate case}
    \end{cases}
    \,.
    \label{eqn:G-A-I-A-power-counting}
\end{equation}
To show this is a secular effect, we take the inverse Laplace transform of \cref{eqn:G-A-I-A-separation-s-domain} where the contour is the same as that in \cref{sec:evolutions-of-field-amplitudes-and-their-mixing} in which both $s=0$ and $s=s_i$ reside.
\begin{equation}
    \mathcal{L}^{-1}[\mathbf{G}_{\txtA}(s) \cdot \overrightarrow{\mathcal{I}}_A(s)] = \sum_{s_i\in\Bab*{s_{\text{slow}}}} \mathbb{G}_{A,i} \cdot \pab{ \frac{\overrightarrow{N}_{A}(0)}{-s_i} - \frac{\overrightarrow{N}_A(s_i)}{-s_i} e^{s_i t} } \,.
    \label{eqn:inverse-Laplce-transform-G-A-I-A}
\end{equation}
Since $\Re\Bab{s_i} < 0$, this result approaches to a $\order{1}$ fixed point exponentially starting with a perturbatively small value. Thus, this is a $\order{1}$ secular contribution in the long time limit although it is a $\order{g^2}$-perturbation effect at the level of the equation of motion.

A similar argument holds for the sub-leading term of $\overrightarrow{A}^\supT(s)$. Using the same technique in \cref{subsec:equations-of-motion-and-solutions-of-field-amplitues,subsec:equations-of-motion-and-formal-solutions}, we write $\mathcal{L}^{-1}\bab*{\mathbf{G}_{\txtA}(s) \cdot \mathbf{K}_{\text{AB}}(s) \cdot \mathbf{G}_{\txtB}(s)}(t)$ as a sum of QNMs.
\begin{equation}
    \mathcal{L}^{-1}\bab*{\mathbf{G}_{\txtA}(s) \cdot \mathbf{K}_{\text{AB}}(s) \cdot \mathbf{G}_{\txtB}(s)}(t) = \sum_{s_i \in \Bab*{s_{\text{slow}}} \cup \Bab*{s_{\text{fast}}}} \mathbb{G}_{\text{AB},i}\, e^{s_i t} \,.
    \label{eqn:inverse-Laplace-transform-GA-KAB-GB}
\end{equation}
where
\begin{equation}
    \mathbb{G}_{\text{AB},i} = 
    \begin{cases}
        \residue\pab{\mathbf{G}_{\txtA}(s), s_i} \cdot \mathbf{K}_{\text{AB}}(s_i) \cdot \mathbf{G}_{\txtB}(s_i) \;, & s_i \in \Bab*{s_{\text{slow}}}
        \\
       \mathbf{G}_{\txtA}(s_i) \cdot \mathbf{K}_{\text{AB}}(s_i) \cdot  \residue\pab{\mathbf{G}_{\txtB}(s),s_i} \;, & s_i \in \Bab*{s_{\text{fast}}}
    \end{cases}
    \,.
\end{equation}
$\Bab*{s_{\text{fast}}}$ are the poles of $\mathbf{G}_{\txtB}(s)$ defined above \cref{eqn:residue-G-A-at-sA-sB}. Note that $\mathbb{G}_{\text{AB},i} \sim \order{g^2}$. The Laplace transform of \cref{eqn:inverse-Laplace-transform-GA-KAB-GB} is
\begin{equation}
     \mathbf{G}_{\txtA}(s) \cdot \mathbf{K}_{\text{AB}}(s) \cdot \mathbf{G}_{\txtB}(s) = \sum_{s_i \in \Bab*{s_{\text{slow}}} \cup \Bab*{s_{\text{fast}}}} \mathbb{G}_{\text{AB},i} \frac{1}{s - s_i} \,.
\end{equation}
Therefore, the sub-leading term of $\overrightarrow{A}^\supT(s)$ can be written as
\begin{equation}
    \mathbf{G}_{\txtA}(s) \cdot \mathbf{K}_{\text{AB}}(s) \cdot \mathbf{G}_{\txtB}(s) \cdot \overrightarrow{\mathcal{I}}_{B}(s) = \sum_{s_i \in \Bab*{s_{\text{slow}}} \cup \Bab*{s_{\text{fast}}}} \mathbb{G}_{\text{AB},i} \cdot \frac{\overrightarrow{N}_B(s)}{-s_i} \pab{ \frac{1}{s} - \frac{1}{s - s_i} } \,.
\end{equation}
The $\order{g^2}$ poles in $\Bab*{s_{\text{slow}}}$ induce a secular effect in the same way as in \cref{eqn:G-A-I-A-separation-s-domain,eqn:G-A-I-A-power-counting} but the poles in $\Bab*{s_{\text{fast}}}$ do not induce any secular effect because they are all of order $\order{1}$ (see \cref{table:poles-of-two-point-correlation-functions}). Hence, the appropriate power counting for $ \mathbf{G}_{\txtA}(s) \cdot \mathbf{K}_{\text{AB}}(s) \cdot \mathbf{G}_{\txtB}(s) \cdot \overrightarrow{\mathcal{I}}_{B}(s)$ is
\begin{equation}
    \mathbf{G}_{\txtA}(s) \cdot \mathbf{K}_{\text{AB}}(s) \cdot \mathbf{G}_{\txtB}(s) \cdot \overrightarrow{\mathcal{I}}_{B}(s) \sim \order{g^2} + \order{g^4} \,.
\end{equation}
Applying the same arguments to $\overrightarrow{B}^\supT(s)$, one can find that
\begin{equation}
    \mathbf{G}_{\txtB}(s) \cdot \overrightarrow{\mathcal{I}}_B(s) \sim \order{g^2} \,, 
    \qquad
    \mathbf{G}_{\txtB}(s) \cdot \mathbf{K}_{\text{BA}}(s) \cdot \mathbf{G}_{\txtA}(s) \cdot \overrightarrow{\mathcal{I}}_A(s) \sim \order{g^2} + \order{g^4} \,.
\end{equation}

We can summarize the behaviours of $\overrightarrow{A}(t)$ and $\overrightarrow{B}(t)$ as follows. $\mathbf{G}_{\txtA}(t), \mathbf{G}_{\txtB}(t)$ and $\mathcal{L}^{-1}\bab*{\mathbf{G}_{\txtA} \cdot \mathbf{K}_{\text{AB}} \cdot \mathbf{G}_{\txtB}}(t), \mathcal{L}^{-1}\bab*{\mathbf{G}_{\txtB} \cdot \mathbf{K}_{\text{BA}} \cdot \mathbf{G}_{\txtA}}(t)$ are all sums of QNMs in the form of exponential decays with/without oscillations (see \cref{subapp:poles-of-Green's-functions}). So the contributions from the initial conditions ($\overrightarrow{A}^\supi, \overrightarrow{B}^\supi$) decay exponentially with oscillations. The contributions from fluctuations of the medium ($\overrightarrow{A}^\supT$, $\overrightarrow{B}^\supT$) approach to a non-zero fixed point exponentially with oscillations in the evolution. Secular contributions emerge such that $\overrightarrow{A}^\supT \sim \order{1}$ and $\overrightarrow{B}^\supT(t) \sim \order{g^2}$ asymptotically.

The secular effect induced by the $\order{g^2}$ poles may enhance the magnitudes of $\order{g^4}$-perturbation effects in ``naive power counting'' to be of order $\order{g^2}$. This causes ambiguities when one claims to calculate solutions up to the second order of perturbations. Because $\order{g^2}$ contributions may come from $\order{g^4}$-perturbation effects in the equation of motion due to secular effects, which are not covered by the leading-order perturbed equation of motion (\cref{eqn:expectation-value-evolution}) used throughout this article. It should also be stressed that another source of missing $\order{g^2}$ contributions is the Breit-Wigner approximation we used to find the QNMs of Green's functions. The previously neglected contributions from multiple-particle cuts are of order $\order{g^2}$, too. They yield $\order{g^2}$ contributions in $\mathbf{G}_{\txtA}(t) \cdot \overrightarrow{A}(0^-)$ of $\overrightarrow{A}^\supi(t)$ and $\mathbf{G}_{\txtB}(t) \cdot \overrightarrow{B}(0^-)$ of $\overrightarrow{B}^\supi(t)$. The poles of order $\order{1/g^n}(n>0)$ neglected in the same approximation (see \cref{subsec:equations-of-motion-and-solutions-of-field-amplitues}) yield $\order{g^{2+n}}$ contributions in $\overrightarrow{A}^\supT(t)$ and $\overrightarrow{B}^\supT(t)$ following the arguments in this subsection, which confirms that they can be neglected in a calculation up to $\order{g^2}$.

Therefore, to assure a complete and consistent solution, we only compute the solutions up to $\order{1}$ contributions in terms of their sizes and leave the task of find complete solutions up to $\order{g^2}$ contributions to future work. At the leading order, the only contribution in part $B$ is $\overrightarrow{B}^\supi(t) = \mathbf{G}_{\txtB}(t) \cdot \overrightarrow{B}(0^-)$. The leading-order contributions in part $A$ come from both the initial condition and the medium. Notice that $\mathbf{G}_{\txtA}(s)$ is block-diagonal. At the leading order, it is sufficient to compute $\overrightarrow{A}_\bvec{k}(t)$.
\begin{equation*}
    \overrightarrow{A}(s) = \mathbf{G}_{\txtA}(s) \cdot \pab{\overrightarrow{A}(0^-) + \overrightarrow{\mathcal{I}}_A(s)}
    \;\;\Leftrightarrow\;\;
    \overrightarrow{A}_\bvec{p}(s) = \mathbf{G}_{A_\bvec{p}}(s) \cdot \pab{\overrightarrow{A}_{\bvec{p}}(0^-) + \overrightarrow{\mathcal{I}}_{A_\bvec{p}}(s)}\,,
    \;\; \bvec{p} = \pm \bvec{k}
    \,.
\end{equation*}
It is clear that the homogeneous part $\overrightarrow{A}_\bvec{k}^\supi(t) = \mathbf{G}_{A_\bvec{k}}(t) \cdot \overrightarrow{A}_\bvec{k}(0^-)$ is simply a sum of the QNMs of $\mathbf{G}_{A_\bvec{k}}(t)$ weighted by the intial condition. So we will focus on the inhomogeneous part $\overrightarrow{A}_\bvec{k}^\supT(s) = \mathbf{G}_{A_\bvec{k}}(s) \cdot \overrightarrow{\mathcal{I}}_{A_\bvec{k}}(s)$ in the next two subsections, for which we introduce the following definitions to keep notations simple.
\begin{equation}
    A_1 \coloneqq A_{11,\bvec{k}} \,, \quad
    A_2 \coloneqq A_{12,\bvec{k}} \,, \quad
    A_3 \coloneqq A_{21,\bvec{k}} \,, \quad
    A_4 \coloneqq A_{22,\bvec{k}} \,.
    \label{eqn:A-cd-abbreviations}
\end{equation}
Correspondingly, we abbreviate the labels of the poles $s$, residues $\mathbb{G}$, inhomogeneous terms $\overrightarrow{\mathcal{I}}_{A_\bvec{k}}$ or $\overrightarrow{N}_{A_\bvec{k}}$ as in \cref{table:label-abbreviations}.
\begin{table}[htbp!]
    \centering
    \begin{tabular}{c c c c c}
        \hline
         label 
         & $\acomm*{a_{1,\bvec{k}}^\dagger}{a_{1,\bvec{k}}}$ 
         & $\acomm*{a_{1,\bvec{k}}^\dagger}{a_{2,\bvec{k}}}$
         & $\acomm*{a_{2,\bvec{k}}^\dagger}{a_{1,\bvec{k}}}$
         & $\acomm*{a_{2,\bvec{k}}^\dagger}{a_{2,\bvec{k}}}$
         \\
         \hline
         $i$ 
         & 1 & 2 & 3 &4
         \\
         \hline
    \end{tabular}
    \caption{Abbreviations of labels}
    \label{table:label-abbreviations}
\end{table}

\noindent
For example, $
    s_1 \coloneqq s_{\acomm*{a_{1,\bvec{k}}^\dagger}{a_{1,\bvec{k}}}} \,,
    s_2 \coloneqq s_{\acomm*{a_{1,\bvec{k}}^\dagger}{a_{2,\bvec{k}}}} \,,
    s_3 \coloneqq s_{\acomm*{a_{2,\bvec{k}}^\dagger}{a_{1,\bvec{k}}}} \,,
    s_4 \coloneqq s_{\acomm*{a_{2,\bvec{k}}^\dagger}{a_{2,\bvec{k}}}}
$.

When computing the Green's functions, we used $\Sigma_{ab}(\bvec{k},s) \approx \Sigma_{ab}(\bvec{k},s^{(0)})$ to keep only the leading-order perturbation.  Since $\overrightarrow{\mathcal{I}}_{A_{\bvec{k}}} \sim \overrightarrow{N}_{A_\bvec{k}}(s) \sim \widetilde{\mathcal{N}}_{ab}(s+i\omega_b) \sim \order{g^2}$, it seems that one can also use $\overrightarrow{N}_{A_\bvec{k}}(s) \approx \overrightarrow{N}_{A_{\bvec{k}}}(s^{(0)})$ to calculate the leading order contributions in $\overrightarrow{A}_{\bvec{k}}^\supT(s)$. We would like to emphasize that this is only valid when the temperature is high enough, i.e., $1/\beta \gg \order{g^2}$. To show it, we notice that in the inverse Laplace transform $\mathcal{N}_{ab}(i\omega)$ should be understood as $\lim_{\epsilon\to0^+}\mathcal{N}_{ab}(i\omega+\epsilon) = \mathcal{N}_{\txtR,ab}(i\omega) + i \mathcal{N}_{\txtI,ab}(i\omega)$ and
\begin{equation}
    \mathcal{N}_{\txtR,ab}(i\omega)
    = \pab{n(\omega) + \frac{1}{2}} \Sigma_{\txtI,ab}(i\omega) \,,
    \quad
    \mathcal{N}_{\txtI,ab}(i\omega)
    = - \pv \int \frac{\dn{k_0}}{2\pi} \frac{\pab{n(k_0) + \frac{1}{2}} \rho_{ab}(\bvec{k},k_0)}{\omega + k_0} \,.
\label{eqn:Bose-distribution-in-noise-kernel}
\end{equation}
where we quoted results from \cref{app:symmetries-of-noise-kernel} and recognized that $\coth\pab*{\beta \omega /2} = 2n(\omega) + 1$. Assuming frequency shift $\delta$ is of order $\order{g^2}$, we find that
\begin{equation}
    1 - \frac{n(\omega+\delta)}{n(\omega)} \approx  (n(\omega )+1) \beta \delta  \,,
    \qquad\text{in the limit}\quad
    \beta \to 0, \delta \to 0 \,.
    \label{eqn:Bose-delta}
\end{equation}
The discrepancy between $n(\omega)$ and $n(\omega+\delta)$ is $\gtrsim \order{1}$ if $\beta \gtrsim \order{1/g^2}$, i.e., at low temperature. It indicates that when the interaction energy, which causes corrections to bare frequencies, is comparable with or dominant over the thermal energy indicated by temperature $1/\beta$, it is no longer valid to use the free-field kinetic energy $\omega_1$ or $\omega_2$ to compute the Bose-Einstein distribution. This situation violates the assumption made for the Born approximation in \cref{sec:effective-quantum-master-equation-of-field-mixing} that the disturbance to the medium is negligible. Therefore, our calculation throughout this article is valid when the temperature is high enough, i.e., $1/\beta \gg \order{g^2}$. Under this assumption, for the rest of this article, we use $\mathcal{N}_{ab}(s) \approx \mathcal{N}_{ab}(s^{(0)})$.

\subsection{Thermalization and long-time medium-induced coherence}
\label{subsec:thermalization-and-long-time-coherence}

We first discuss the non-degenerate case. Summarizing the results in \cref{subapp:poles-of-Green's-functions,subapp:residues-at-poles-of-equal-time-two-point-correlation-functions}, one can find that $\mathbf{G}_{A_\bvec{k}}(t)$ is diagonal up to the leading order contributions in this case, $\mathbf{G}_{A_\bvec{k}}(t) = \operatorname{diag} \pab{ e^{s_1 t}, e^{s_2 t}, e^{s_3 t}, e^{s_4 t} }$. Inserting it into \cref{eqn:inverse-Laplce-transform-G-A-I-A} yields
\begin{equation}
    A_i^\supT(t) = \frac{N_i(s=0)}{-s_i} - \frac{N_i(s=s_i)}{-s_i} e^{s_it} \,, \qquad i = 1,2,3,4 \,.
    \label{eqn:A-T-t-non-degenerate}
\end{equation}
where the definitions in \cref{eqn:A-cd-abbreviations,table:label-abbreviations} are used. We dropped the ``$\bvec{k}$'' for simpler notations in \cref{eqn:A-T-t-non-degenerate} since only one momentum is involved. For $i=1$, we find that
\begin{equation}
    \frac{N_1(s=0)}{-s_1}
    = -\frac{2 i \left(\mathcal{N}_{11}\left(-i \omega _1\right)+\mathcal{N}_{11}\left(i \omega _1\right)\right)}{\Sigma _{11}\left(-i \omega _1\right)-\Sigma _{11}\left(i \omega _1\right)}
    = \frac{2 \mathcal{N}_{\text{R},11}\left(i \omega _1\right)}{\Sigma _{\text{I},11}\left(i \omega _1\right)}
    = 2 n\left(\omega _1\right)+1
    \,.
\end{equation}
where at the first ``='' we used the results in \cref{subapp:inhomogeneous-terms-in-s-domain,subapp:poles-of-Green's-functions}, at the second ``='' we split the self-energies and noise-kernels into ``R'' parts and ``I'' parts using \cref{eqn:Sigma-epsilon->0,eqn:noise-kernel-epsilon->0} and made use of their symmetries (see \cref{eqn:Sigma-R-I-symmetry,eqn:noise-kernel-R-I-symmetry}), at the third ``='' we used \cref{eqn:fluctuatio-dissipation-relation-s-domain} and recognized that $\coth\pab{\beta \omega_1 / 2} = 2n(\omega_1) + 1$ with $n(\omega_1)$ being the Bose-Einstein distribution at temperature $1/\beta$. Similar steps for the second term in \cref{eqn:A-T-t-non-degenerate} yield $N_1(s=s_1)/(-s_1) = 2n(\omega_1) + 1$ where we used $\mathcal{N}_{ab}(s) \approx \mathcal{N}_{ab}(s^{(0)})$ at the leading order with caveats discussed in \cref{eqn:Bose-delta}, and $s_1 = - \Sigma_{\txtI,11}(i\omega_1)/\omega_1$ which we will refer to as $-\Gamma_{s_1}$ below. Putting the two terms of $A_1^\supT(t)$ together results in
\begin{equation}
    A_1^\supT(t) = (2n(\omega_1) + 1) \pab{1 - e^{-\Gamma_{s_1} t}} \,.
\end{equation}
We combine the result with the contribution of the initial condition $A_1^\supi(t) = e^{s_1 t} A_1(0^-)$.
\begin{equation}
    A_1(t) = A_1(0^-) e^{-\Gamma_{s_1} t} + (2n(\omega_1) + 1) \pab{1 - e^{-\Gamma_{s_1} t}} \,.
    \label{eqn:A-1-t-non-degenerate-case}
\end{equation}
Notice that by definition, $A_1 \equiv A_{11,\bvec{k}} = \expval*{\acomm*{a_{1,\bvec{k}}^\dagger}{a_{1,\bvec{k}}}} = 2 \expval*{\widehat{N}_{1,\bvec{k}}} + 1$ where $\widehat{N}_{1,\bvec{k}} = a_{1,\bvec{k}}^\dagger a_{1,\bvec{k}}$ is the number operator of field 1 at momentum $\bvec{k}$. Introducing this relation to $A_1(t)$, we find that the population of field $\phi_1$  approaches to a thermal distribution at temperature $1/\beta$ while the initial number density distribution fades away.
\begin{equation}
    \expval*{\widehat{N}_{1,\bvec{k}}}(t) = \expval*{\widehat{N}_{1,\bvec{k}}}(0^-) e^{-\Gamma_{s_1} t} + n(\omega_1) (1 - e^{-\Gamma_{s_1} t}) \,.
    \label{eqn:solution-number-density-field-1}
\end{equation}
A similar result for field $\phi_2$ can be obtained by calculating $A_4(t)$ and recognizing that $A_4(t) = 2\expval*{\widehat{N}_{2,\bvec{k}}} + 1$ with $\widehat{N}_{2,\bvec{k}} = a_{2,\bvec{k}}^\dagger a_{2,\bvec{k}}$.
\begin{equation}
    \expval*{\widehat{N}_{2,\bvec{k}}}(t) = \expval*{\widehat{N}_{2,\bvec{k}}}(0^-) e^{-\Gamma_{s_4} t} + n(\omega_2) (1 - e^{-\Gamma_{s_4} t}) \,.
    \label{eqn:solution-number-density-field-2}
\end{equation}
where we defined $\Gamma_{s_4} \coloneqq -s_4 = \Sigma_{\txtI,22}(i\omega_2)/\omega_2$ in the same manner as we defined $\Gamma_{s_1} = - s_1$.

Unfortunately, such a clean result doesn't exist for $A_2(t)$. A similar argument yields
\begin{equation}
    A_2(t) = 
    A_2(0^-) e^{(i\Omega_{s_2} - \Gamma_{s_2}) t} 
    + \frac{\mathcal{N}_{12}\left(i \omega _2\right)+\mathcal{N}_{21}\left(-i \omega _1\right)}{(-s_2)\sqrt{\omega _1 \omega _2} } 
    - \frac{\mathcal{N}_{12}\left(i \omega _1\right)+\mathcal{N}_{21}\left(-i \omega _2\right)}{(-s_2)\sqrt{\omega _1 \omega _2}} e^{(i\Omega_{s_2}-\Gamma_{s_2}) t}
    \,.
    \label{eqn:A-2-t-solution}
\end{equation}
where we defined $s_2 = i\Omega_{s_2} - \Gamma_{s_2}$ as below using \cref{eqn:Sigma-epsilon->0}.
\begin{equation}
    \Omega_{s_2} = \pab{ \omega_1 + \frac{\Sigma _{\text{R},11}\left(i \omega _1\right)}{2 \omega _1} } - \pab{\omega_2 + \frac{\Sigma _{\text{R},22}\left(i \omega _2\right)}{2 \omega _2} } \,,
    \quad
    \Gamma_{s_2} = \frac{\Sigma _{\text{I},11}\left(i \omega _1\right)}{2 \omega _1} + \frac{\Sigma _{\text{I},22}\left(i \omega _2\right)}{2 \omega _2} \,.
\end{equation}
Note that $s_2 \sim \order{1}$. So, $A_2(t) = A_2(0^-) e^{(i\Omega_{s_2} - \Gamma_{s_2}) t}$ up to the leading order contribution.

By definition, $A_2(t) = A_{12,\bvec{k}}(t) = \expval*{\acomm*{a_{1,\bvec{k}}^\dagger}{a_{2,\bvec{k}}}} = 2 \expval*{a_{1,\bvec{k}}^\dagger a_{2,\bvec{k}}}$. Therefore, in analogy with the Stokes parameters in quantum optics \cite{Scully_Zubairy_1997}, $A_2(t)$ is a measure of inter-field coherence. \cref{eqn:A-2-t-solution} shows that $A_2(t)$ approaches to a $\order{g^2}$ fixed point via an exponentially decaying oscillation at relaxing rate $\Gamma_{s_2}$ and frequency $\Omega_{s_2}$. Noticing the Bose-Einstein distribution included in the noise-kernel (see \cref{eqn:Bose-distribution-in-noise-kernel}), we argue that this asymptotic value is the inter-field coherence in the subsystem of a thermal state at temperature $1/\beta$, which we refer to as medium-induced coherence because it is induced by the effective coupling mediated by the medium. We want to re-iterate that \cref{eqn:A-2-t-solution} doesn't include all $\order{g^2}$ contributions because of the ambiguity discussed in \cref{subsec:power-counting-and-secular-contributions-of-solutions} .

The solution of $A_3(t)$ can be found by taking a complex conjugate of $A_2(t)$ since $A_3(t) \equiv A_{21,\bvec{k}}(t) = A_{12,\bvec{k}}^*(t) \equiv A_2^*(t)$. With the solution of $\overrightarrow{A}_{\bvec{k}}(t)$ at hand, we can make the analogy to quantum optics more explicit. Let $\widehat{S}_{j,\bvec{k}}, j=0,1,2,3$ be the four Stokes parameters \cite{Scully_Zubairy_1997}. They are related to $\overrightarrow{A}_\bvec{k}(t)$ in the following way.
\begin{subequations}
\begin{align}
    \expval[big]{\widehat{S}_{0,\bvec{k}}} 
    & \equiv \expval[big]{a_{1,\bvec{k}}^\dagger a_{1,\bvec{k}}} + \expval[big]{a_{2,\bvec{k}}^\dagger a_{2,\bvec{k}}} 
      =  \frac{A_{11,\bvec{k}} + A_{22,\bvec{k}}}{2} - 1 \,,
    \\
    \expval[big]{\widehat{S}_{1,\bvec{k}}} 
    & \equiv \expval[big]{a_{1,\bvec{k}}^\dagger a_{1,\bvec{k}}} - \expval[big]{a_{2,\bvec{k}}^\dagger a_{2,\bvec{k}}} 
      = \frac{A_{11,\bvec{k}} - A_{22,\bvec{k}}}{2} \,,
    \\
    \expval[big]{\widehat{S}_{2,\bvec{k}}} 
    & \equiv \expval[big]{a_{1,\bvec{k}}^\dagger a_{2,\bvec{k}}} + \expval[big]{a_{2,\bvec{k}}^\dagger a_{1,\bvec{k}}} 
      = \frac{A_{12,\bvec{k}} + A_{21,\bvec{k}}}{2} \,,
    \\
    \expval[big]{\widehat{S}_{3,\bvec{k}}} 
    & \equiv (-i) \pab{ \expval[big]{a_{1,\bvec{k}}^\dagger a_{2,\bvec{k}}} - \expval[big]{a_{2,\bvec{k}}^\dagger a_{1,\bvec{k}}} }
      = \frac{A_{12,\bvec{k}} - A_{21,\bvec{k}}}{2i} \,.
\end{align}
\label{eqn:Stokes-parameters-non-degenerate}%
\end{subequations}
Therefore, we successfully connect the correlation functions in our calculation to observables in the subsystem.

\subsection{Quantum beats and enhanced long-time medium-induced coherence}
\label{subsec:quantum-beats-and-enhanced-long-time-coherence}

In the nearly-degenerate case, all poles of $\overrightarrow{A}_\bvec{k}(t)$ are of order $\order{g^2}$. It leads to $\overrightarrow{N}_{A_\bvec{k}}(s_i) \approx \overrightarrow{N}_{A_\bvec{k}}(0)$ at the leading order with the caveats discussed in \cref{eqn:Bose-delta}. Inserting the result into \cref{eqn:inverse-Laplce-transform-G-A-I-A,eqn:separate-initial-thermal-contributions} yields the following solution of $A_{\bvec{k}}(t)$.
\begin{equation}
    \overrightarrow{A}_{\bvec{k}}(t)
    = \sum_{i=1}^4 \mathbb{G}_{i} e^{s_i t} \cdot \overrightarrow{A}_{\bvec{k}}(0^-) + \sum_{i=1}^4 \frac{\mathbb{G}_{i} \cdot \overrightarrow{N}_{A_{\bvec{k}}}(s=0)}{-s_i} \pab{ 1 - e^{s_i t} }
    \,.
    \label{eqn:A-k-solution-nearly-degenerate-case}
\end{equation}
The explicit expression of $\overrightarrow{A}_\bvec{k}(t)$ is obtained after inserting the results in \cref{app:explicit-results-in-evolutions-of-number-densities-and-coherence} into \cref{eqn:A-k-solution-nearly-degenerate-case}. But doing so doesn't yield further simplifications or more illuminating results as \cref{eqn:solution-number-density-field-1}. Thus, we keep \cref{eqn:A-k-solution-nearly-degenerate-case} as it is, which is enough for us to clarify the conclusions. All poles $s_i$ in \cref{eqn:A-k-solution-nearly-degenerate-case} being $\sim\order{g^2}$ also implies that $\overrightarrow{N}_{A_\bvec{k}}(0)/(-s_i)\sim\order{1}, \forall\, i=1,2,3,4$. Given that all elements of $\mathbb{G}_{i}$ are of order $\order{1}$ in the nearly-degenerate case (see \cref{subapp:residues-at-poles-of-equal-time-two-point-correlation-functions}), we reach the following estimation for all four components of $\overrightarrow{A}_\bvec{k}(t)$.
\begin{equation}
    A_i(t) \sim \sum_{j=1}^4 \order{1} A_i(0^-) e^{s_i t} + \sum_{j=1}^4 \order{1} \pab{ 1 - e^{s_i t} } \,.
    \label{eqn:A-i-t-power-counting-nearly-degenerate-case}
\end{equation}
in which ``$\order{1}$'' represents terms of order $\order{1}$ that are not necessarily the same. Using the relations in \cref{eqn:pole-relations,eqn:conjugate-symmetry-of-amplitude-poles-residues}, it is not difficult to find that $s_1 = -2\Gamma_1, s_4 = -2\Gamma_2$ and $s_2 = s_3^* = i(\Omega_1 - \Omega_2) - (\Gamma_1+\Gamma_2)$ where $\Omega_i,\Gamma_i,i=1,2$ are defined in \cref{eqn:amplitudes-frequencies-decay-rates}, which means two QNMs of $\overrightarrow{A}_\bvec{k}(t)$ are exponential decays and the other two are exponentially decaying oscillations. Therefore, relating the solutions to observables of the subsystem as in \cref{subsec:thermalization-and-long-time-coherence}, we find strong oscillations in both the populations (seen in $A_1(t)$ and $A_4(t)$) and inter-field coherence (seen in $A_2(t)$ and $A_3(t)$), whose magnitudes are of order $\order{1}$ initially and decay exponentially. We refer to the strong oscillations as quantum beats.

Replace $\omega_1$ and $\omega_2$ in $\overrightarrow{N}_{A_\bvec{k}}(s=0)$ with $\omegabar \pm \delta$ as in \cref{eqn:delta-power-counting-nearly-degenerate} and truncate the result to the leading order with caveats discussed in \cref{eqn:Bose-delta}. We find that
\begin{equation}
    \overrightarrow{N}_{A_\bvec{k}}(s=0)
    = \frac{2n(\omegabar) + 1}{2\omegabar} \overrightarrow{\Sigma}_{A_\bvec{k}} \,,
    \qquad\text{with}\quad
    \overrightarrow{\Sigma}_{A_\bvec{k}} \coloneqq
    \begin{pmatrix}
        2 \Sigma _{\text{I},11}\left(i \omegabar\right) \\
        \Sigma _{\text{I},12}\left(i \omegabar\right)+\Sigma _{\text{I},21}\left(i \omegabar\right) \\
        \Sigma _{\text{I},12}\left(i \omegabar\right)+\Sigma _{\text{I},21}\left(i \omegabar\right) \\
        2 \Sigma _{\text{I},22}\left(i \omegabar\right) \\
    \end{pmatrix}
   \,.
   \label{eqn:nearly-degenerate-noise-kernel}
\end{equation}
where we first split the $\mathcal{N}_{ab}(s)$ into ``R'' part and ``I'' parts using \cref{eqn:noise-kernel-epsilon->0} and extracted the Bose-Einstein distribution by using \cref{eqn:fluctuatio-dissipation-relation-s-domain} and recognizing $\coth{\pab{\beta \omega / 2}} = 2n(\omega) + 1$. Hence, we claim that both the populations and coherence thermalize and approach to asymptotic values that belong to a thermal subsystem at temperature $1/\beta$. These asymptotic values are of order $\order{1}$ as shown in \cref{eqn:A-i-t-power-counting-nearly-degenerate-case}, which means the medium-induced coherence is enhanced. As a comparison, in the non-degenerate case (see \cref{subsec:thermalization-and-long-time-coherence}), although the populations and inter-field coherence also thermalize, oscillations didn't show up in the populations and the asymptotic inter-field coherence was perturbatively small.

\section{Effects of a coupling strength hierarchy}
\label{sec:effects-of-a-coupling-strength-hierarchy}

We showed in previous sections that there are strong mixing effects, quantum beats, and enhanced medium-induced coherence when the field masses are nearly-degenerate. It should be pointed out that the assumption made in \cref{sec:effective-quantum-master-equation-of-field-mixing}, that the coupling strengths are nearly degenerate, i.e., $g_1\sim g_2$, is another essential ingredient for these phenomena. This assumption is not generally true, for example, in the case of cosmic ALP and synthetic axion mixing. In this section, we introduce a hierarchy in the coupling strengths and discuss its effect on these phenomena.

More precisely, we assume $1 \gg g_1 \gg g_2$ without loss of generality and $m_1^2 - m_2^2 \lesssim \order{g_1^2}$. The coupling strength order is chosen arbitrarily and is irrelevant to the mass order $m_1 \geq m_2$ assumed in \cref{sec:effective-quantum-master-equation-of-field-mixing}. We will reach the same conclusion if assuming $1 \gg g_2 \gg g_1$. With the hierarchy, the power counting of the self-energy is
\begin{equation}
    \Sigma_{11}(s) \sim \order{g_1^2}
    \;\gg\;
    \Sigma_{12}(s), \Sigma_{21}(s) \sim \order{g_1 g_2}
    \;\gg\;
    \Sigma_{22}(s) \sim \order{g_2^2}
    \,.
    \label{eqn:self-energy-hierarchy}
\end{equation}
If only field $\phi_1$ or $\phi_2$ evolves in the medium, \cref{eqn:self-energy-hierarchy} means that their decay rates satisfy
\begin{equation}
    \Gamma_{\phi_1} \sim \order{g_1^2}
    \;\gg\;
    \Gamma_{\phi_2} \sim \order{g_2^2}
    \,.
\end{equation}
Thus, field $\phi_1$ is a short-lived particle and field $\phi_2$ is a long-lived particle.

\subsection{Field amplitudes}
\label{subsec:effecits-of-the-hierarchy-in-coupling-strengths}

we further approximate \cref{eqn:amplitude-pole-nearly-degenerate} up to the order $\order{(g_2/g_1)^2}$. The results depend on the relative size of the mass difference $m_1^2-m_2^2$ to the self-energy. Because of the conjugate symmetry among the amplitudes' poles, we only show the results of $s_{a_1^\dagger}$ and $s_{a_2^\dagger}$.
\begin{enumerate}
    \item When $\omega_1^2 - \omega_2^2 \sim \order{g_1^2}$,
    \begin{subequations}
    \label{eqn:amplitude-poles-g1^2}%
    \begin{align}
        s_{a_1^\dagger} & = 
            \underbrace{\vphantom{\bigg|} i\omega_1 }_{\sim \order{1}} 
            + \underbrace{\vphantom{\bigg|} \frac{i \Sigma_{11}(s)}{4\omegabar} }_{\sim \order{g_1^2}}
            + \underbrace{\vphantom{\bigg|} \frac{i}{2\omegabar} \frac{\Sigma_{12}(s) \Sigma_{21}(s)}{\omega_1^2 - \omega_2^2 + \Sigma_{11}(s)} }_{\sim \order{g_2^2}}
            \bigg|_{s=i\omegabar}
            \,.
        \\
        s_{a_2^\dagger} & = 
            \underbrace{\vphantom{\bigg|} i\omega_2 }_{\sim\order{1}}
            + \underbrace{\vphantom{\bigg|} \frac{i}{2\omegabar} \pab{ \Sigma_{22}(s) - \frac{\Sigma_{12}(s) \Sigma_{21}(s)}{\omega_1^2 - \omega_2^2 + \Sigma_{11}(s)} } }_{\sim\order{g_2^2}}
            \bigg|_{s=i\omegabar}
            \,.
    \end{align}
    \end{subequations}
    
    \item When $\omega_1^2 - \omega_2^2 \sim \order{g_1 g_2}$ or $\omega_1^2 - \omega_2^2 \sim \order{g_2^2}$ or $\omega_1^2 - \omega_2^2 \ll \order{g_2^2}$,
    \begin{subequations}
    \label{eqn:amplitude-poles-g1g2}%
    \begin{align}
        s_{a_1^\dagger} & =
            \underbrace{\vphantom{\bigg|} i \omega_1 }_{\sim\order{1}}
            + \underbrace{\vphantom{\bigg|} \frac{i \Sigma_{11}(s)}{2\omegabar} }_{\sim\order{g_1^2}}
            + \underbrace{\vphantom{\bigg|} \frac{i}{2\omegabar} \frac{\Sigma_{12}(s) \Sigma_{21}(s)}{\Sigma_{11}(s)} }_{\sim\order{g_2^2}}
            \bigg|_{s=i\omegabar}
            \,.
        \\
        s_{a_2^\dagger} & =
            \underbrace{\vphantom{\bigg|} i\omega_2 }_{\sim\order{1}}
            + \underbrace{\vphantom{\bigg|} \frac{i}{2\omegabar} \pab{ \Sigma_{22}(s) - \frac{\Sigma_{12}(s) \Sigma_{21}(s)}{\Sigma_{11}(s)}  } }_{\sim\order{g_2^2}}
            \bigg|_{s=i\omegabar}
            \,.
    \end{align}
    \end{subequations}
\end{enumerate}
In the above results, we have used the following relations which are valid regardless of the size of the mass difference.
\begin{equation}
    \omegabar + \frac{\omega_1^2 - \omega_2^2}{4\omegabar} = \omega_1 \,,
    \qquad\qquad
    \omegabar - \frac{\omega_1^2 - \omega_2^2}{4\omegabar} = \omega_2 \,.
\end{equation}
Since $\Sigma_{ab}(i\omegabar)$ are complex in general, it is clear in \cref{eqn:amplitude-poles-g1^2,eqn:amplitude-poles-g1g2} that regardless of the size of the mass difference, the leading-order frequency corrections and decay rates are
\begin{equation}
    \Omega_1 - \omega_1 \sim \Gamma_1 \sim \order{g_1^2} \,,
    \qquad
    \Omega_2 - \omega_2 \sim \Gamma_2 \sim \order{g_2^2} \,.
\end{equation}
where $\Omega_i,\Gamma_i,i=1,2$ are defined in \cref{eqn:amplitudes-frequencies-decay-rates}. In summary, although both fields receive influence from the other one via the self-energy, the sizes of the corrections to the poles remain similar to the case where they evolve alone in the medium. In other words, when the masses are nearly degenerate but the coupling strength are not, the mixing doesn't change the lifetimes of the two fields by orders of magnitude. Field $\phi_1$ remains short-lived and field $\phi_2$ remains long-lived.

The coupling strength hierarchy also affects the residue matrices of the QNMs of the field amplitudes. Combining \cref{eqn:Omega^2-D-Delta^2-definitions,eqn:self-energy-hierarchy} yields that when $m_1^2 - m_2^2 \lesssim \order{g_1^2}$,
\begin{equation}
    \Omega^2(s) \sim \order{1} \,,
    \qquad\quad
    \Delta^2(s) \sim \order{g_1^2} \,,
    \qquad\quad
    D(s) \sim \order{g_1^2} \,.
\end{equation}
which implies
\begin{equation}
    D(s) + \Delta^2(s) \sim \order{g_1^2} \,,
    \qquad\quad
    D(s) - \Delta^2(s) = \frac{4 \Sigma_{12}(s) \Sigma_{21}(s)}{D(s) + \Delta^2(s)} \sim \order{g_2^2} \,.
\end{equation}
Inserting these estimations into \cref{eqn:phi-residue-nearly-degenerate} yields
\begin{equation}
    \mathbb{G}_{\phi,a_1} \sim \mathbb{G}_{\phi,a_1^\dagger} \sim
        \begin{pmatrix}
            \order{1} & \order[big]{\frac{g_2}{g_1}}
            \\[0.5em]
            \order[big]{\frac{g_2}{g_1}} & \order[big]{\pab[big]{\frac{g_2}{g_1}}^2}
        \end{pmatrix}
        \,,
    \quad
    \mathbb{G}_{\phi,a_2} \sim \mathbb{G}_{\phi,a_2^\dagger} \sim
        \begin{pmatrix}
            \order[big]{\pab[big]{\frac{g_2}{g_1}}^2} & \order[big]{\frac{g_2}{g_1}}
            \\[0.5em]
            \order[big]{\frac{g_2}{g_1}} & \order{1}
        \end{pmatrix}
        \,.
    \label{eqn:power-counting-G-phi}
\end{equation}
The $\order{g_2/g_1}$ off-diagonal elements indicate that the mixing is weak when there is a coupling strength hierarchy. Obviously, when the coupling strengths are nearly degenerate, i.e., $g_1\sim g_2$, the strong mixing effect is recovered. Therefore, we conclude that a strong mixing happens only when both the masses and coupling strengths of the two fields are nearly degenerate. It should be pointed out that this conclusion is solely based on counting the powers of $g_1$ and $g_2$, which doesn't exclude the possibility that in a particular system, the mass difference and all elements of the self-energy are about the same order of magnitude by coincidence in spite of a hierarchy in the coupling strengths. A strong mixing will emerge in this special scenario, too.

\subsection{Populations and coherence}

We follow the conclusion in \cref{subsec:power-counting-and-secular-contributions-of-solutions} and discuss only the leading order results of the populations and coherence. Applying the arguments in last subsection to the results in \cref{subapp:Green's-functions-of-dimensionless-one-point-correlation-functions}, we find that the power counting of the QNMs of the dimensionless one-point correlation functions is the same as that of the field amplitudes.
\begin{equation}
    \mathbf{G}_{a_{1,\bvec{k}}}(t) \sim \mathbf{G}_{a_{1,\bvec{k}}^\dagger}(t) \sim
        \begin{pmatrix}
            \order{1} & \order[big]{\frac{g_2}{g_1}}
            \\[0.5em]
            \order[big]{\frac{g_2}{g_1}} & \order[big]{\pab[big]{\frac{g_2}{g_1}}^2}
        \end{pmatrix}
        \,,
    \;
    \mathbf{G}_{a_{2,\bvec{k}}}(t) \sim \mathbf{G}_{a_{2,\bvec{k}}^\dagger}(t) \sim
        \begin{pmatrix}
            \order[big]{\pab[big]{\frac{g_2}{g_1}}^2} & \order[big]{\frac{g_2}{g_1}}
            \\[0.5em]
            \order[big]{\frac{g_2}{g_1}} & \order{1}
        \end{pmatrix}
        \,.
\end{equation}
Then, the Kronecker-product relation in \cref{eqn:A-cd-abbreviations,eqn:G-A-k-relations} implies that $\mathbf{G}_{A_\bvec{k}}(t)$ becomes diagonal at the leading order in the nearly-degenerate case if $g_1 \gg g_2$, which results in
\begin{equation}
    A_i^\supT(t) = \frac{N_i(s=0)}{-s_i} - \frac{N_i(s=s_i)}{-s_i} e^{s_it}\,, \qquad i=1,2,3,4\,.
    \label{eqn:A-T-t-hierarchy}
\end{equation}
where the label abbreviations (see \cref{table:label-abbreviations}) are used. This result takes the same form as that of the non-degenerate case in \cref{subsec:thermalization-and-long-time-coherence}, in which the quantum beats disappear in the populations up to the leading order. However, the time-scales and magnitudes in \cref{eqn:A-T-t-hierarchy} are different from that in \cref{eqn:A-T-t-non-degenerate}. To estimate $A_i^\supT(t)$, we first count the powers of $g_1$ and $g_2$ in the poles of $\mathbf{G}_{A_\bvec{k}}(t)$, which is obtained by inserting the results in \cref{eqn:amplitude-poles-g1^2,eqn:amplitude-poles-g1g2} into the relation in \cref{eqn:pole-relations}.
\begin{subequations}
\begin{align}
    s_1 & \equiv s_{\acomm*{a_{1,\bvec{k}}^\dagger}{a_{1,\bvec{k}}}} = s_{a_{1,\bvec{k}}^\dagger} + s_{a_{1,\bvec{k}}} \sim \order{g_1^2} + \order{g_2^2} \,,
    \\
    s_2 & \equiv s_{\acomm*{a_{1,\bvec{k}}^\dagger}{a_{2,\bvec{k}}}} = s_{a_{1,\bvec{k}}^\dagger} + s_{a_{2,\bvec{k}}} \sim i(\omega_1 - \omega_2) + \order{g_1^2} + \order{g_2^2} \,,
    \\
    s_3 & \equiv s_{\acomm*{a_{2,\bvec{k}}^\dagger}{a_{1,\bvec{k}}}} = s_{a_{2,\bvec{k}}^\dagger} + s_{a_{1,\bvec{k}}} \sim -i(\omega_1 - \omega_2) + \order{g_1^2} + \order{g_2^2} \,,
    \\
    s_4 & \equiv s_{\acomm*{a_{2,\bvec{k}}^\dagger}{a_{2,\bvec{k}}}} = s_{a_{2,\bvec{k}}^\dagger} + s_{a_{2,\bvec{k}}} \sim \order{g_2^2} \,.
\end{align}
\label{eqn:population-coherence-poles-power-counting-hierarchy}%
\end{subequations}
Note that $\omega_1 - \omega_2 = \frac{\omega_1^2 - \omega_2^2}{\omega_1 + \omega_2}$ and $\omega_1 + \omega_2 \sim \order{1}$ imply that $\omega_1 - \omega_2 \sim \omega_1^2 - \omega_2^2 \lesssim \order{g_1}^2$. Similar to \cref{eqn:self-energy-hierarchy}, a hierarchy in the noise-kernel (see \cref{eqn:noise-kernel-definition}) emerges due to $g_1 \gg g_2$.
\begin{equation}
    \mathcal{N}_{11}(s) \sim \order{g_1^2}
    \;\gg\;
    \mathcal{N}_{12}(s), \mathcal{N}_{21}(s) \sim \order{g_1 g_2}
    \;\gg\;
    \mathcal{N}_{22}(s) \sim \order{g_2^2}
    \,.
\end{equation}
which leads to the following power counting in \cref{eqn:A-T-t-hierarchy}.
\begin{equation}
    N_1(s) \sim \order{g_1^2}
    \;\gg\;
    N_2(s), N_3(s) \sim \order{g_1 g_2}
    \;\gg\;
    N_4(s) \sim \order{g_2^2}
    \,.
\end{equation}
Aggregating all estimations yields
\begin{subequations}
\begin{align}
    A_1^\supT(t) \sim & \order{1} - \order{1} \exp\Bab{ - \order{g_1^2} t } \,,
    \\
    A_2^\supT(t) \sim A_3^\supT(t) \sim & \order{\frac{g_2}{g_1}} - \order{\frac{g_2}{g_1}} \exp\Bab{ i \order{g_1^2} t - \order{g_1^2} t } \,,
    \\
    A_4^\supT(t) \sim & \order{1} - \order{1} \exp\Bab{ - \order{g_2^2} t } \,.
\end{align}
\label{eqn:A-T-t-power-counting-hierarchy}%
\end{subequations}
in which $\order{1}, \order{g_1^2}, \dots$, etc. represent (not necessarily the same) terms of the indicated orders. \cref{eqn:population-coherence-poles-power-counting-hierarchy} shows that when the masses are nearly-degenerate but the coupling strengths are not, although the relaxing rate of the populations of the short-lived particle (long-live particle) receives corrections from the long-lived particle (short-lived particle) via the self-energy, its size doesn't change by orders of magnitude and still remains $\sim\order{g_1^2}$ ($\sim \order{g_2^2}$). The asymptotic values of the inter-field coherence are of order $\order{g_2/g_1}$ instead of $\order{1}$ as in \cref{subsec:quantum-beats-and-enhanced-long-time-coherence}. An interesting phenomenon is that the relaxing rates of the inter-field coherence are not decreased by the coupling strength hierarchy and are always of order $\order{g_1^2}$. This implies that albeit the populations of the long-lived and short-lived particles evolve at very different time scales, the inter-field coherence is established approximately within the lifetime of the short-live particle, which suggests the possibility of detecting extremely long-lived particles within a practical period. We want to alert readers that the power counting $\order{g_2/g_1}$ does not imply that we can decrease the coupling strength between the short-lived particle and the medium to enhance the medium-induced coherence. Instead, it is just another way to say decreasing the mass difference since the nearly-degenerate case is defined as $m_1^2 - m_2^2 \lesssim \order{g_1^2}$.

A similar phenomenon also exists in the non-degenerate case if $g_1 \gg g_2$. That is to say, the inter-field coherence is established within the lifetime of the short-lived particle although the timescales of the populations' evolution are vastly different, i.e., $\order{1/g_1^2}$ and $\order{1/g_2^2}$. The difference in this situation is that the relaxing rates of the populations of the two fields don't receive corrections from the other field and the asymptotic inter-field coherence is of order $\order{g_1 g_2}$ instead $\order{g_2/g_1}$ as shown in \cref{eqn:A-2-t-solution}.

\section{Discussion}
\label{sec:discussion}

\subsection{Asymptotic stationary states of the evolution}

It was shown in \cref{subsec:thermalization-and-long-time-coherence,subsec:quantum-beats-and-enhanced-long-time-coherence,subsec:equations-of-motion-and-solutions-of-field-amplitues} that the field amplitudes, populations and coherence all approach to fixed points asymptotically. Given the limits exist, if we are only interested in the asymptotic results, we are able to skip all the complications discussed above and get better results using the final-value theorem of Laplace transform \cite{Fleisch_2022}, i.e.,
\begin{equation}
    \lim_{t\to\infty} f(t) = \lim_{s\to0} s F(s)\,, \qquad\text{if both limits exist.}
\end{equation}
in which $F(s) = \mathcal{L}\bab{f(t)}$. Applying the final-value theorem to \cref{eqn:phi-solution-s-domain} yields
\begin{equation}
    \expval{\boldsymbol{\phi}}(t=\infty) 
    = \lim_{s\to0} s \mathbf{G}_\phi(s) \cdot \pab[big]{ \eval{ s \expval{\boldsymbol{\phi}} }_{t=0^-} + \eval{ \expval{\boldsymbol{\pi}} }_{t=0^-} } \,.
    \label{eqn:phi-final-value}
\end{equation}
We know that all poles of $\mathbf{G}_\phi(s)$ are off the imaginary axis (see \cref{subsec:equations-of-motion-and-solutions-of-field-amplitues}). Thus, $\expval{\boldsymbol{\phi}}(\infty) = 0$, which is consistent with the results in \cref{sec:evolutions-of-field-amplitudes-and-their-mixing}.

Applying the theorem to \cref{eqn:formal-solution-s-domain} (populations and coherence) yields
\begin{equation}
    \overrightarrow{\mathcal{D}}(t=\infty) 
    = \lim_{s\to0} s \, \mathbf{G}_\mathcal{D}(s)\ \cdot \overrightarrow{\mathcal{D}}(0^-) + \lim_{s\to0} \mathbf{G}_\mathcal{D}(s) \cdot \overrightarrow{N}(s)
    \,.
    \label{eqn:D-final-value}
\end{equation}
where we used $\overrightarrow{I}(s) = \frac{1}{s} \overrightarrow{N}(s)$ (see \cref{subapp:inhomogeneous-terms-in-s-domain}). We know that all poles of $\mathbf{G}_{\mathcal{D}}(s)$ are off the imaginary axis (see \cref{subsec:quasi-normal-modes-of-populations-and-coherence,subapp:poles-of-Green's-functions}). Thus, the limit associated with $\overrightarrow{\mathcal{D}}(0^-)$ vanishes, which explains why we find in \cref{subsec:thermalization-and-long-time-coherence,subsec:quantum-beats-and-enhanced-long-time-coherence} that the initial conditions fade away and don't contribute to the final states. Since both $\mathbf{G}_\mathcal{D}(s=0)$ and $\overrightarrow{N}(s=0)$ exist, the asymptotic value of $\overrightarrow{\mathcal{D}}(t)$ is $\overrightarrow{\mathcal{D}}(t=\infty) = \mathbf{G}_\mathcal{D}(s=0) \cdot \overrightarrow{N}(s=0)$.

There are more advantages than the obvious simplicity brought by the final-value theorem. The first one is that we don't need to approximate the Green's functions as a sum of QNMs, which are obtained at the cost of ignoring contributions from multi-particle cuts and $\order{1/g^n}$ poles. Thus, the results in \cref{eqn:phi-final-value,eqn:D-final-value} are exact up to the second-order perturbation at the level of the equation of motion. Note that $\order{g^4}$-perturbation effects with $\order{g^2}$ magnitudes due to secular effects are still not included in \cref{eqn:phi-final-value,eqn:D-final-value} because they are not provided in \cref{eqn:expectation-value-evolution}. The second advantage is that we only need to compute $\overrightarrow{N}(s=0)$ so the caveats of perturbing Bose-Einstein distribution (see \cref{eqn:Bose-delta}) are completely avoided. It is worth noting the Laplace transforms $\mathbf{G}_\mathcal{D}(s)$ and $\overrightarrow{N}(s)$ only exist on the right half $s$-plane (see \cref{subsec:equations-of-motion-and-solutions-of-field-amplitues,subsec:equations-of-motion-and-formal-solutions,app:symmetries-of-noise-kernel,app:symmetries-of-self-energy}), which means we can only set $s\to 0^+ + 0i$ when taking the limit. Therefore, $\Sigma_{ab}(i \omega)$ and $\mathcal{N}_{ab}(i \omega)$ should be still understood as $\Sigma_{\txtR,ab}(i\omega) + i \Sigma_{\txtI,ab}(i\omega)$ and $\mathcal{N}_{\txtR,ab}(i\omega) + i \mathcal{N}_{\txtI,ab}(i\omega)$, which confirms the consistency between the results in \cref{eqn:phi-final-value,eqn:D-final-value} and the results in \cref{sec:evolutions-of-field-amplitudes-and-their-mixing,sec:populations-and-coherence}. The third advantage is that we can rewrite the asymptotic values of $\overrightarrow{\mathcal{D}}(t)$ in the following way.
\begin{equation}
    \mathbf{G}_\mathcal{D}^{-1}(s=0) \cdot \overrightarrow{\mathcal{D}}(t=\infty) = \overrightarrow{N}(s=0) \,.
    \label{eqn:linear-equations-final-value}
\end{equation}
Symbolically, this doesn't make much difference. However, the elements of $\mathbf{G}_\mathcal{D}^{-1}(s)$ are directly read off from the equation of motion in the $s$-domain, which means they are obtained earlier than $\mathbf{G}_{\mathcal{D}}(s)$. Solving linear equations is usually easier than matrix inversion in particular when $\mathbf{G}_\mathcal{D}^{-1}(s)$ contains many zeros. Furthermore, in this procedure, we do not need to distinguish the non-degenerate and nearly-degenerate cases. Lastly, it should be pointed out that an oscillation in the long time limit doesn't count as a well-defined limit for the final-value theorem and will not be captured by this method. However, we have previously confirmed $\expval*{\boldsymbol{\phi}}(t)$ and $\overrightarrow{\mathcal{D}}(t)$ are asymptotically stationary at the leading order.

\subsection{Dynamics of the vacuum background in the nearly-degenerate case}
\label{subsec:dynamics-of-the-vacuum-background-in-the-nearly-degenerate-case}

We know from the definitions in \cref{eqn:A-cd-k-B-cd-k-definitions} that the correlation functions $A_{cd,\bvec{k}}(t)$ (or equivalently, $\overrightarrow{A}_\bvec{k}(t)$) include contributions from the vacuum background because of the canonical commutation relations. This contribution is also hinted by ``$2n(\omega_1) + 1$'' in \cref{eqn:A-1-t-non-degenerate-case} and ``$2n(\omegabar) + 1$'' in \cref{eqn:nearly-degenerate-noise-kernel}. Therefore, we refer to the ``zero-particle'' part in $\overrightarrow{A}_{\bvec{k}}(t)$ as the vacuum contribution and denote it with a superscript ``(vac)''. Correspondingly, we define the excitation part in $\overrightarrow{A}_{\bvec{k}}(t)$ as $\overrightarrow{A}_{\bvec{k}}^{\text{(exc)}}(t) = \overrightarrow{A}_\bvec{k}(t) - \overrightarrow{A}_{\bvec{k}}^{\text{(vac)}}(t)$. It is easy to find in \cref{subsec:thermalization-and-long-time-coherence} that in the non-degenerate case $\overrightarrow{A}_\bvec{k}^{\text{(vac)}} = \pab{1,0,0,1}^\T$ is constant. But in the nearly-degenerate case (see \cref{subsec:quantum-beats-and-enhanced-long-time-coherence}), the vacuum contribution becomes time-dependent.
\begin{equation}
    \overrightarrow{A}_\bvec{k}^{\text{(vac)}}(t)
    = \sum_{i=1}^4 \mathbb{G}_{i} e^{s_i t} \cdot \pab[big]{1,0,0,1}^\T
    + \sum_{i=1}^4 \frac{\mathbb{G}_{i}}{-s_i} \pab{ 1 - e^{s_i t} } \cdot
    \frac{1}{2\omegabar} \cdot \overrightarrow{\Sigma}_{A_\bvec{k}} \,.
    \label{eqn:A-k-t-vacuum}
\end{equation}
where the label abbreviations is \cref{table:label-abbreviations} were used for convenience and $\overrightarrow{\Sigma}_{A_\bvec{k}}$ was given in \cref{eqn:nearly-degenerate-noise-kernel}. It must be emphasized that this is not the transient dynamics that leads to the renormalization of fields because it shares the same frequencies and $\order{g^2}$ relaxation rates as that of particles in the subsystem.

The definitions of $A_{11,\bvec{k}}(t)$ and $A_{22,\bvec{k}}(t)$ (or $A_1(t)$ and $A_4(t)$ using notations in \cref{eqn:A-cd-abbreviations}) indicate that the energy density of the two fields are $\mathcal{E}_{\phi_1}(t) = \frac{1}{2} \omegabar A_1(t)$ and $\mathcal{E}_{\phi_2}(t) = \frac{1}{2} \omegabar A_4(t)$ up to the leading order. Correspondingly, the zero-point energies of the subsystem are $\mathcal{E}_{\phi_1}^{\text{(vac)}}(t) = \frac{1}{2} \omegabar A_1^{\text{(vac)}}(t)$ and $\mathcal{E}_{\phi_2}^{\text{(vac)}}(t) = \frac{1}{2} \omegabar A_4^{\text{(vac)}}(t)$, which is the reason why we call \cref{eqn:A-k-t-vacuum} the vacuum contribution. Therefore, the effective coupling between the two fields induced by their interactions with the medium causes corrections not only to particles but also to the vacuum of the subsystem. The originally unperturbed vacuum of the subsystem (at $t<0$) is not the true vacuum of the subsystem after the couplings are switched on (at $t>0$). \cref{eqn:A-k-t-vacuum} shows the dynamics of the transition from the unperturbed vacuum to the new vacuum after the interactions are switched on. As discussed in \cref{sec:populations-and-coherence}, there are strong quantum beats and asymptotic medium-induced coherence of the vacuum in this transition, which is a consequence of the nearly-degenerate masses and coupling strengths.

Particles are defined as excitations out of the vacuum of the subsystem. Thus, the vacuum's evolution leads to time-dependent definitions of particles in the subsystem. This explains why $A_1(t) = 2\expval*{\widehat{N}_1}(t) + 1$ at $t=0$ but its vacuum contribution deviates from the unity at a later time. In particular, $A_1^{\text{(vac)}}(t)-1\sim\order{1}$. However, we have shown that the subsystem approaches to thermal equilibrium asymptotically. If one is ignorant of the previous dynamics and carries out calculations directly based on the asymptotic equilibrium state, the convention would be to quantize the subsystem in a way such that $\comm*{\tilde{a}_{c,\bvec{k}}}{\tilde{a}^\dagger_{d,\bvec{k}}} = \delta_{cd}$ where $\tilde{a}_{c,\bvec{k}}$ and $\tilde{a}^\dagger_{d,\bvec{k}}$ are creation and annihilation operators of the quanta. Notice that in this case the Hamiltonian is $H_{c,\bvec{k}} = \omegabar\pab[big]{\tilde{a}_{c,\bvec{k}}^\dagger \tilde{a}_{c,\bvec{k}} + \frac{1}{2}}$ where the zero-point energy is half of the energy quantum. Therefore, we introduce normalized populations to account for the time-dependence of particles' definitions, where normalized operators are indicated by a tilde on the top.
\begin{equation}
    \expval[Big]{\widehat{\widetilde{N}}_1}(t) \coloneqq \frac{\mathcal{E}_{\phi_1}(t)}{2\mathcal{E}_{\phi_1}^{\text{(vac)}}(t)} - \frac{1}{2} = \frac{ A_1^{\text{(exc)}}(t)}{2 A_1^\text{(vac)}(t)} \,,
    \quad
    \expval[Big]{\widehat{\widetilde{N}}_2}(t) \coloneqq \frac{\mathcal{E}_{\phi_2}(t)}{2\mathcal{E}_{\phi_2}^{\text{(vac)}}(t)} - \frac{1}{2} = \frac{ A_4^{\text{(exc)}}(t)}{2 A_4^\text{(vac)}(t)} \,.
    \label{eqn:normalized-number-density}
\end{equation}
One can easily check that asymptotically $\expval*{\widehat{\widetilde{N}}_c}(\infty) = n(\omegabar), c=1,2$, which is the canonical leading-order result for the populations of a thermal subsystem. Such a result cannot be obtained if we identify the vacuum contribution as $\Tr\Bab*{1 \cdot \rho(t)}$ using $A_{cd,\bvec{k}} = \Tr\Bab*{ \pab*{ 2 a_{c,\bvec{k}}^\dagger a_{d,\bvec{k}} + \delta_{cd}} \rho(t) }$, which is 1 all the time. Because \cref{eqn:normalized-number-density} is the populations of particles instantly defined on the instant vacuum of the subsystem while $\expval*{a_{c,\bvec{k}}^\dagger a_{c,\bvec{k}}},c=1,2$ are the populations of particles defined on the unperturbed vacuum of the system, which does not coincide with the true vacuum at a later time. Going to the Heisenberg picture, the $\tilde{a}_{c,\bvec{k}}(t)$ in the normalized number operator $\widehat{\widetilde{N}}_c$ is not connected to $a_{c,\bvec{k}}(t=0)$ by the unitary operator $\exp\Bab*{-i H t}$ because the vacuum on which $\tilde{a}_{c,\bvec{k}}(t)$ is defined is evolving. Only at $t=0$ do we have $\tilde{a}_{c,\bvec{k}} = a_{c,\bvec{k}}$ since the vacuums of $\tilde{a}_{1,\bvec{k}}$ and $a_{1,\bvec{k}}$ are the same one initially. Following the re-definitions of populations, the Stokes parameters in \cref{eqn:Stokes-parameters-non-degenerate} need to be extended as below.
\begin{equation}
\begin{aligned}
    \expval[Big]{\widehat{\widetilde{S}}_{0,\bvec{k}}} 
    & = \frac{1}{2} \pab{A_{11,\bvec{k}}^{\text{(exc)}} + A_{22,\bvec{k}}^{\text{(exc)}}} \,,
    &\quad
    \expval[Big]{\widehat{\widetilde{S}}_{1,\bvec{k}}} 
    & = \frac{1}{2} \pab{A_{11,\bvec{k}}^{\text{(exc)}} - A_{22,\bvec{k}}^{\text{(exc)}}} \,,
    \\
    \expval[Big]{\widehat{\widetilde{S}}_{2,\bvec{k}}} 
    & = \frac{1}{2} \pab{A_{12,\bvec{k}}^{\text{(exc)}} + A_{21,\bvec{k}}^{\text{(exc)}}} \,,
    &\quad
    \expval[Big]{\widehat{\widetilde{S}}_{3,\bvec{k}}} 
    & = \frac{1}{2i} \pab{A_{12,\bvec{k}}^{\text{(exc)}} - A_{21,\bvec{k}}^{\text{(exc)}}} \,.
\end{aligned}
\end{equation}

\subsection{Non-negativity of populations}

Let $\rho(\mathcal{H})$ be the space of all density matrices in the Hilbert space $\mathcal{H}$. The evolution of a system in $\mathcal{H}$ can be viewed as a map of the space onto itself, $\mathcal{V}: \rho(\mathcal{H}) \to \rho(\mathcal{H})$. To ensure that the output of the map is a density matrix, the map should satisfy two properties: 1) trace preserving, i.e., $\Tr\Bab*{\mathcal{V}\, \hat{\rho}} = \Tr\Bab*{\hat{\rho}}, \forall \hat{\rho} \in \rho(\mathcal{H})$, which was proven in \cref{eqn:trace-preserving} for the equation of motion we used in this article, 2) completely positive \cite{10.1063/1.5115323}. The exact reduced dynamics of a subsystem after tracing out other degrees of freedom, i.e., $\hat{\rho}(t) = \Tr_\chi \Bab*{ e^{-iHt} \, \hat{\rho}_{\text{tot}}(0)\, e^{iHt} }$, is a completely positive map if the total system begins with a product state between the subsystem and the medium $\hat{\rho}_{\text{tot}}(0) = \hat{\rho}_\phi(0) \otimes \hat{\rho}_\chi(0)$ \cite{PhysRevLett.73.1060,Alicki2007-nc}. After approximations, this property is not generally preserved by a QME with memory effects \cite{PhysRevA.70.010304}. Breaking the complete positivity of a map may cause unphysical effects, for example, the negative population in the Redfield equation \cite{De_la_Pradilla2024-tb,10.1063/1.463831}. While it is beyond the scope of this article to rigorously discuss whether the QME (\cref{eqn:qme}) is a generator of a completely positive map, we take a step back and show that at least the populations given in the solutions remain non-negative all the time.

The non-negativity of populations in the non-degenerate case is obvious in \cref{eqn:solution-number-density-field-1,eqn:solution-number-density-field-2}. The nearly-degenerate case takes a bit more algebra. We first notice that $\mathbf{G}_{A_\bvec{k}}(t)$ is a sum of exponential functions and rewrite \cref{eqn:A-k-solution-nearly-degenerate-case} as below using \cref{eqn:nearly-degenerate-noise-kernel}.
\begin{equation}
    \overrightarrow{A}_\bvec{k}(t) = \mathbf{G}_{A_\bvec{k}}(t) \cdot \overrightarrow{A}_\bvec{k}(0^-) + \int_0^t \dn{\tau} \mathbf{G}_{A_\bvec{k}}(\tau) \cdot \frac{2n(\omegabar) + 1}{2\omegabar} \overrightarrow{\Sigma}_{A_\bvec{k}} \,.
\end{equation}
Only $A_1(t)$ (or $A_4(t)$) is needed to clarify the non-negativity of populations. So we introduce the symbol $\bab{\cdot}_{\text{row 1}}$ to denote the first row of a matrix. Let the first row of $\mathbf{G}_{\boldsymbol{a}_\bvec{k}}(t)$ be $\bab{ f_1(t), f_2(t) }$. \cref{eqn:conjugate-symmetry-dimensionless-one-point-correlation-functions,eqn:G-A-k-relations} imply that $\bab*{ \mathbf{G}_{A_\bvec{k}}(t) }_{\text{row 1}} = \bab*{ f_1^*(t), f_2^*(t) } \otimes \bab*{ f_1(t), f_2(t) }$. Inserting the result into the contribution from the initial condition $A_1^{(i)}(t)$ yields
\begin{equation}
    A_1^{\text{($i$,vac)}}(t) = \abs{f_1(t)}^2 + \abs{f_2(t)}^2 \,, \quad
    A_1^{\text{($i$,exc)}}(t) = \Tr\Bab{ \abs{f_1(t) a_{1,\bvec{k}}(0) + f_2(t) a_{2,\bvec{k}}(0)}^2 \rho(0) } \,.
\end{equation}
Obviously both $A_1^{\text{($i$,vac)}}(t)$ and $A_1^{\text{($i$,exc)}}(t)$ are non-negative. The non-negativity of the inhomogeneous part is proved by noticing that the integrand is non-negative. To see it, we use \cref{eqn:Sigma-R-I-definitions,eqn:rho-time-inversion-symmetry,eqn:rho-<>-rho} to rewrite $\overrightarrow{\Sigma}_{A_\bvec{k}}$ as
\begin{equation}
    \overrightarrow{\Sigma}_{A_\bvec{k}} = \frac{1}{1 + n(\omegabar)} \overrightarrow{\rho}^> + \frac{1}{n(\omegabar)} \overrightarrow{\rho}^< \,.
\end{equation}
where $\overrightarrow{\rho}^> = \pab{
        \rho_{11}^>(\omegabar) , \rho_{12}^>(\omegabar) , \rho_{21}^>(\omegabar) , \rho_{22}^>(\omegabar) }$ and $
    \overrightarrow{\rho}^< = \pab{
        \rho_{11}^<(\omegabar) , \rho_{21}^<(\omegabar) , \rho_{12}^<(\omegabar) , \rho_{22}^<(\omegabar) }$.
The momentum label $\bvec{k}$ in $\rho^\lessgtr(k)$ is dropped to simplify notations. It can be shown using the spectral representations (see \cref{eqn:rho-ab->,eqn:rho-ab-<}) that
\begin{align*}
    \bab{\mathbf{G}_{A_\bvec{k}}(\tau)}_{\text{row 1}} \cdot \overrightarrow{\rho}^>
    & = \frac{(2\pi)^4}{Z_\chi} \sum_{n,m} e^{-\beta E_n} \abs[Big]{ \braket*[3]{m}{\sum_{a=1}^2 f_a(\tau) \mathcal{O}_a(0)}{n} }^2  \delta^{(4)}(k - (P_m-P_n)) \,.
    \\
    \bab{\mathbf{G}_{A_\bvec{k}}(\tau)}_{\text{row 1}} \cdot \overrightarrow{\rho}^<
    & = \frac{(2\pi)^4}{Z_\chi} \sum_{n,m} e^{-\beta E_n} \abs[Big]{ \braket*[3]{m}{\sum_{a=1}^2 f_a(\tau) \mathcal{O}_a(0)}{n} }^2 \delta^{(4)}(k - (P_n-P_m)) \,.
\end{align*}
Obviously both expressions are non-negative, implying the non-negativity of the vacuum part $A_1^{\text{($T$,vac)}}(t)$ and the excitation part $A_1^{\text{($T$,exc)}}(t)$ of the inhomogeneous term $A_1^{(T)}(t)$. Therefore, we conclude that $A_1^{\text{(vac)}}(t) \geq 0$ and $A_1^{\text{(exc)}}(t) \geq 0$. Inserting them into \cref{eqn:normalized-number-density} shows the non-negativity of the populations of field $\phi_1$. One can show $\expval*{\widehat{\widetilde{N}}_2} \geq 0$ after using similar arguments in the fourth row of $\mathbf{G}_{A_\bvec{k}}(t)$.

\subsection{Comparison with rotating-wave approximation in equations of motion}
\label{subsec:comparison-with-rotating-wave-approximation-in-equations-of-motion}

We mentioned that $\overrightarrow{A}(t)$ and $\overrightarrow{B}(t)$ constitute the slowly-oscillating part (at frequencies $\sim 0, \pm (\omega_1 - \omega_2)$) and the fast-oscillating part (at frequencies $\sim \pm 2\omega_1, \pm2\omega_2, \pm(\omega_1 + \omega_2)$) when defining them in \cref{eqn:A-B-separation}. A common tool in this situation is the rotating-wave approximation (RWA) in which one argues that fast oscillating terms in an equation yield only negligible contributions. This approximation simplifies the equation of $\overrightarrow{A}(t)$ as below.
\begin{multline}
    \diff**{t}{A_{cd,\bvec{k}}}(t) =
    i \pab{\omega_{c,\bvec{k}} - \omega_{d,\bvec{k}}} A_{cd,\bvec{k}}(t)
    - i \sum_{a,b=1,2} \int_0^t \Big[ -\delta_{ca} \, A_{bd,\bvec{k}}(t') e^{-i\omega_{d,\bvec{k}}(t-t')} 
    \\*
    + \delta_{da} A_{cb,\bvec{k}}(t')  e^{i\omega_{c,\bvec{k}}(t-t')} \Big] \widetilde{\Sigma}_{ab}(\abs{\bvec{k}},t-t') \, \dn{t'}
    + \mathcal{I}_{A_\bvec{k},cd}(t) \,.
\end{multline}
where we used $\overrightarrow{A} = \bab*{ \overrightarrow{A}_\bvec{k}, \overrightarrow{A}_{-\bvec{k}} }^\T$ and \cref{eqn:A-k-equation}. The equation of $A_{cd,-\bvec{k}}$ is obtained by relabeling $\bvec{k}\to-\bvec{k}$. There is no simplification due to RWA in the equations of $B_{cd,\bvec{k}}$ and $B_{cd,\bvec{k}}^*$. However, since $A_{cd,\pm\bvec{k}}$ are now independent of part B, they should be treated as known inhomogeneous terms in the equation of part B.

In this article, we have kept avoiding using RWA in our calculation. This allows us to compare our results with the results of RWA and scrutinize its validity. The solutions after RWA can be obtained using the same techniques as in \cref{sec:populations-and-coherence}. In the $s$-domain, they are
\begin{subequations}
\begin{align}
    \overrightarrow{A}(s)
    & = \mathbf{G}_{\txtA}(s) \cdot \pab{ \overrightarrow{A}(0^-) + \overrightarrow{\mathcal{I}}_{A}(s) } \,,
    \\
    \overrightarrow{B}(s) 
    & = \mathbf{G}_{\txtB}(s) \cdot \pab{ \overrightarrow{B}(0^-) + \overrightarrow{\mathcal{I}}_B(s) } - \mathbf{G}_{\txtB}(s) \cdot \mathbf{K}_{\text{BA}}(s) \cdot \overrightarrow{A}(s) \,.
\end{align}
\end{subequations}
where we use the same notations as in \cref{sec:populations-and-coherence}. Comparing these results with those in \cref{subsec:power-counting-and-secular-contributions-of-solutions}, one would immediately notice that the missing contribution caused by RWA is $-\mathbf{G}_{\txtA}(s) \cdot \mathbf{K}_{\text{AB}}(s) \cdot \mathbf{G}_{\txtB}(s) \cdot \pab*{\overrightarrow{B}(0^-) + \overrightarrow{\mathcal{I}}_A(s)}$, which is of order $\order{g^2}$ as shown in \cref{subsec:power-counting-and-secular-contributions-of-solutions}. Thus, the result of RWA is only exact up to the leading order contribution. One may argue that we may begin with an incoherent state where $\overrightarrow{B}(0^-)$ is zero exactly. But we have shown that secular contributions emerge in terms associated with $\overrightarrow{\mathcal{I}}_A(s)$, causing $-\mathbf{G}_{\txtA}(s) \cdot \mathbf{K}_{\text{AB}}(s) \cdot \mathbf{G}_{\txtB}(s) \cdot \overrightarrow{\mathcal{I}}_A(s)$ to be of order $\order{g^2}$. This contribution persists regardless of the initial condition. In addition, RWA always completely ignores the contributions from $\overrightarrow{B}(s)$ to $\overrightarrow{A}(s)$ no matter how many orders of perturbations we keep in the equation of motion. This means that RWA can only produce results exact up to the leading order disregarding the efforts one may take to keep more perturbations in the equation of motion.

A similar conclusion holds for the poles calculated using RWA. In \cref{subsec:quasi-normal-modes-of-populations-and-coherence} we showed that the more accurate Green's functions for $\overrightarrow{A}(s)$ and $\overrightarrow{B}(s)$ are $\mathcal{G}_A(s)$ and $\mathcal{G}_B(s)$ (see \cref{eqn:G-A,eqn:G-B-approx-G-B-0}). Should one obtain the equations of motion up to higher orders of perturbations, in RWA the poles would still be computed using $\mathbf{G}_{\txtA}(s)$ and $\mathbf{G}_{\txtB}(s)$. We learned from $\mathcal{G}_A(s)$ and $\mathcal{G}_B(s)$ that higher order corrections in poles receive contributions from $\mathbf{K}_{\text{AB}}(s)$ and $\mathbf{K}_{\text{BA}}(s)$, which will be ignored in RWA as off-resonance terms connecting part A and part B. Therefore, the poles calculated using RWA can only be exact up to the leading-order perturbation no matter how many higher order corrections we keep in the equations of motion and in $\mathbf{G}_{\txtA}(s)$ and $\mathbf{G}_{\txtB}(s)$.

\subsection{Connection to the Lindblad form}
\label{subsec:connection-to-the-Lindblad-form}

When studying the open quantum system, an alternative method commonly used is the QME of a Lindblad form, or, more generally, the GKSL equation named after Vittorio Gorini, Andrzej Kossakowski, George Sudarshan and Göran Lindblad.
\begin{equation}
    \dot{\hat{\rho}} = - \frac{i}{\hbar} \comm{H}{\rho} + \sum_{n,m} h_{nm} \pab{ A_n \rho A_m^\dagger - \frac{1}{2} \acomm{A_m^\dagger a_n}{\hat{\rho}} } \,.
    \label{eqn:GKSL}
\end{equation}
where $h_{nm}$ is a positive semidefinite Hermitian matrix and the operators $A_n$ are either given phenomenologically or derived using microscopic Hamiltonians. It is also the most general form of a Markovian QME \cite{10.1063/1.5115323,doi:10.1142/S1230161217400017}.

The integral in \cref{eqn:master-eqn-G-<>} clearly indicates the non-Markovian characteristic of the QME used throughout this article, which describes the memory effect due to the medium. It has been shown in many literatures, for example, \cite{Carmichael1993-qx,Alicki2007-yg}, how a GKSL equation is obtained from a non-Markovian QME after neglecting the memory effect. In what follows, we briefly show how this is done in \cref{eqn:master-eqn-G-<>} for the completeness of the discussion.

An alternative way to interpret a memoryless medium is that the correlations in the medium rapidly decay. Therefore, in the Markov approximation, we assume $G_{ab}^\lessgtr(x,x') \approx R_{ab}^{\lessgtr}(x,x') \delta(t-t')$, which transforms \cref{eqn:master-eqn-G-<>} into the following form.
\begin{multline}
    \dot{\hat{\rho}} = \sum_{a,b=1,2} \int \dn[3]{x} \dn[3]{y}
    \Big\{
        J_a(x) \hat{\rho}(t) J_b(y) R_{ab}^<(x,y) + J_b(y) \hat{\rho}(t) J_a(x) R_{ab}^>(x,y)
        \\
        - J_a(x) J_b(y) \hat{\rho}(t) R_{ab}^>(x,y) - \hat{\rho}(t) J_b(y) J_a(x) R_{ab}^<(x,y)
    \Big\}
    \,.
\end{multline}
where we carried out the time integral and defined $x=(\bvec{x},t)$ and $y=(\bvec{y},t)$. Definitions in \cref{eqn:G<>-definition} implies $G_{ab}^>(x,x') = G_{ba}^<(x',x)$, leading to $R_{ab}^>(x,x') = R_{ba}^<(x',x)$. We use this relation to write all ``$R$'' in the above equation as $R_{ab}^<(x,y)$. Relabeling $a\leftrightarrow b$ and $\bvec{x}\leftrightarrow\bvec{y}$ when necessary, we can rewrite the equation in the following form.
\begin{equation}
    \dot{\hat{\rho}} = \sum_{a,b=1,2} \int \dn[3]{x} \dn[3]{y} 2 R_{ab}^<(x,y) \pab{ J_a(x) \hat{\rho} J_b(y) - \acomm{J_b(y) J_a(x)}{\hat{\rho}(t)} }
    \,.
    \label{eqn:reduce-to-lindblad-form-interaction-picture}
\end{equation}
\cref{eqn:G<>-definition} also shows that $[G_{ab}^<(x,x')]^* = G_{ba}^<(x',x)$, leading to  $[R_{ab}^<(x,x')]^* = R_{ba}^<(x',x)$. This indicates that $R_{ab}^<(x,y)$ is a Hermitian matrix, with $\bvec{x}$ and $\bvec{y}$ also recognized as labels. We also need to show $R_{ab}^<(x,y)$, or equivalently $G_{ab}^<(x,y)$ is positive semidefinite. This requires us to show that for an arbitrary two-component complex function $\bvec{f} = [f_1(\bvec{x}), f_2(\bvec{x})]$, the following relation always holds.
\begin{equation}
    \sum_{a,b=1,2} \int \dn[3]{x} \dn[3]{y} f_{a}^*(\bvec{x}) G_{ab}^<(\bvec{x},t;\bvec{y},t) f_b(\bvec{y}) \geq 0 \,.
    \label{eqn:positive-R-ab}
\end{equation}
Notice that $G_{ab}^<(x,y) = \expval*{ \mathcal{O}_b(y) \mathcal{O}_a(x) }$ implies the following positive-semidefinite operator,
\begin{equation}
    \sum_{a,b=1,2} \int \dn[3]{x} \dn[3]{y} f_a^{*}(\bvec{x}) \mathcal{O}_b(\bvec{y},t) \mathcal{O}_a(\bvec{x},t) f_b(\bvec{y},t)
    = \abs{ \sum_{a=1} \int \dn[3]y f_a^*(\bvec{y}) \mathcal{O}_a(\bvec{y},t) }^2 \,.
\end{equation}
where we used the notation $\abs*{\hat{\mathcal{A}}}^2 = \hat{\mathcal{A}}^\dagger \hat{\mathcal{A}}$ for an operator $\hat{\mathcal{A}}$ and the assumption made in \cref{sec:effective-quantum-master-equation-of-field-mixing} that $\mathcal{O}_a(x)$ are Hermitian operators. Tracing this operator over $\hat{\rho}_\chi(0)$ proves \cref{eqn:positive-R-ab}. Therefore, \cref{eqn:reduce-to-lindblad-form-interaction-picture} is a GKSL equation. The result may be more straightforward if we write \cref{eqn:reduce-to-lindblad-form-interaction-picture} in Schr\"odinger picture by inverting \cref{eqn:interaction-picture}.
\begin{multline}
    \dot{\hat{\rho}}^{(\text{S})}
    = - i \comm{H_{\phi_1}^{(\text{S})} + H_{\phi_2}^{(\text{S})}}{\hat{\rho}^{(\text{S})}(t)}
    \\
    + \sum_{a,b=1,2} \int \dn[3]{x} \dn[3]{y} 2 R_{ab}^<(x,y) \pab{ J_a^{(\text{S})}(\bvec{x}) \hat{\rho}^{(\text{S})}(t) J_b^{(\text{S})}(\bvec{y}) - \acomm{J_b^{(\text{S})}(\bvec{y}) J_a^{(\text{S})}(\bvec{x})}{\hat{\rho}^{(\text{S})}(t)} }
    \,.
\end{multline}
In many cases the density matrix of the medium $\hat{\rho}_{\chi}(0)$ commute with the Hamiltonian $H_0$ that defines the interaction picture. For example, $\hat{\rho}_{\chi}$ is assumed to be thermal in this article. This property implies that $G_{ab}^<(x,y)$ is time-independent and so is $R_{ab}^<(x,y)$ as shown below, where we used $x=(\bvec{x},t), y=(\bvec{y},t)$ and the cyclic symmetry of the trace.
\begin{equation}
    G_{ab}^<(x,y) 
    = \Tr\Bab{ e^{i H_0 t} \mathcal{O}_b^{(\text{S})}(\bvec{y}) \mathcal{O}_a^{(\text{S})}(\bvec{x}) e^{-iH_0 t}  \hat{\rho}_\chi(0) }
    = \Tr\Bab{ \mathcal{O}_b^{(\text{S})}(\bvec{y}) \mathcal{O}_a^{(\text{S})}(\bvec{x}) \hat{\rho}_\chi(0)}
    \,.
\end{equation}

\subsection{Connection to the Weisskopf-Wigner theory}

The Weisskopf-Wigner theory is a well-established method for the study of meson-antimeson mixing in colliders. It assumes that the state of the meson $M$ and antimeson $\overline{M}$ is $\varphi_1(t) \ket{M} + \varphi_2(t) \ket{\overline{M}}$ and a non-Hermitian Hamiltonian $\mathcal{H}$ dictates the evolution in a vacuum background. Translating this assumption into the framework proposed in this article, we let $\phi_1$ be the meson field and $\phi_2$ the antimeson field with $\omega_1 = \omega_2 = \omegabar$. We also have to assume that the bath's temperature is zero and the density matrix is
\begin{equation}
\hat{\rho}(t) =
\begin{pNiceMatrix}[first-row,first-col]
        &  \bra{\text{vaccum}} & \bra{1_{\phi_1,\bvec{k}}; 0} & \bra{0 ; 1_{\phi_2, \bvec{k}}} \\
    \ket{\text{vaccum}} & 1 - \sum_{i=1,2} \abs{\varphi_i}^2 & 0 & 0  \\
    \ket{1_{\phi_1,\bvec{k}}; 0} & 0 & \abs{\varphi_1}^2 & \varphi_1 \varphi_2^*  \\
    \ket{0 ; 1_{\phi_2, \bvec{k}}} & 0 & \varphi_2 \varphi_1^* & \abs{\varphi_2}^2  
\end{pNiceMatrix} \,.
\label{eqn:rho-Weisskopf-Wigner}
\end{equation}
where the state labels for the entries are shown and all other entries in the infinitely-dimensional $\hat{\rho}(t)$ are zero. \cref{eqn:rho-Weisskopf-Wigner} means that there is only one particle at momentum $\bvec{k}$ and all coherence and entanglement are ignored except $\varphi_1^* \varphi_2$ and $\varphi_2^* \varphi_1$. A consequence of \cref{eqn:rho-Weisskopf-Wigner} is
\begin{equation}
    \pab{\varphi_1^* , \varphi_2^*} \otimes \pab{ \varphi_1, \varphi_2 } = \expval[big]{ \pab*{a_{1,\bvec{k}}^\dagger, a_{2,\bvec{k}}^\dagger} \otimes \pab{a_{1,\bvec{k}}, a_{2,\bvec{k}}} } \,.
    \label{eqn:varphi-a^dagger-a}
\end{equation}
where $\expval*{\pab*{ \hat{\mathcal{A}}, \hat{\mathcal{B}}, \dots }}$ is a shorthand for $\pab*{ \expval*{\hat{\mathcal{A}}}, \expval*{\hat{\mathcal{B}}}, \dots}$. Following the discussion in \cref{subsec:dynamics-of-the-vacuum-background-in-the-nearly-degenerate-case}, \cref{eqn:varphi-a^dagger-a} implies
\begin{equation}
    \boldsymbol{\varphi}^*(t) \otimes \boldsymbol{\varphi}(t)
    = \expval[big]{ \boldsymbol{a}_\bvec{k}^\dagger \otimes \boldsymbol{a}_\bvec{k} }
    = \frac{1}{2} \overrightarrow{A}_\bvec{k}^{\text{(exc)}}(t)
    = \bab[big]{ \mathbf{G}_{\boldsymbol{a}_{\bvec{k}}^\dagger}(t) \otimes \mathbf{G}_{\boldsymbol{a}_{\bvec{k}}}(t) } \cdot \bab[big]{ \boldsymbol{\varphi}^*(0) \otimes \boldsymbol{\varphi}(0) } \,.
\end{equation}
where $\boldsymbol{\varphi} \coloneqq \pab{ \varphi_1, \varphi_2 }^\T$ and $\boldsymbol{a}_\bvec{k}^\dagger, \boldsymbol{a}_\bvec{k}$ are defined below \cref{eqn:phi-pi-a-relation}. At the last ``='' \cref{eqn:G-A-k-relations} is used. The bath doesn't contribute to $\overrightarrow{A}_\bvec{k}^{\text{(exc)}}(t)$ because it is at zero temperature. It follows immediately from the conjugate symmetry in \cref{eqn:conjugate-symmetry-dimensionless-one-point-correlation-functions} that
\begin{equation}
    \boldsymbol{\varphi}(t) = \mathbf{G}_{\boldsymbol{a}_\bvec{k}}(t) \cdot \boldsymbol{\varphi}(0) \,,
    \qquad
    \boldsymbol{\varphi}(s) = \sum_{i=1}^2 \mathbb{G}_{a_{1,\bvec{k}}} \frac{1}{s - s_{a_{1,\bvec{k}}}} \cdot \boldsymbol{\varphi}(0) \,.
    \label{eqn:WW-result}
\end{equation}
where \cref{eqn:quasi-normal-modes-of-a-k} is used and we took the Laplace transform for a comparison later. \cref{eqn:WW-result} is exactly the result of the Weisskopf-Wigner theory. The non-Hermitian Hamiltonian can be obtained by reverse-engineering \cref{eqn:WW-result}, where we first solve the following effective equation in the $s$-domain.
\begin{equation}
    \diff**{t}{\boldsymbol{\varphi}} = \mathcal{H} \cdot \boldsymbol{\varphi}(t) 
    \quad\Longrightarrow\quad
    \boldsymbol{\varphi}(s) = \frac{1}{s - \mathcal{H}} \cdot \boldsymbol{\varphi}(0) \,.
    \label{eqn:non-Hermitian-equation}
\end{equation}
A comparison between \cref{eqn:WW-result,eqn:non-Hermitian-equation} shows that
\begin{equation}
    \mathcal{H} = s - \bab{ \sum_{i=1}^2 \mathbb{G}_{a_{1,\bvec{k}}} \frac{1}{s - s_{a_{1,\bvec{k}}}} }^{-1}
    = 
    \begin{pmatrix}
        -i\omega_1 - \frac{i \Sigma_{11}(-i\omegabar)}{2\omegabar} & - \frac{i \Sigma_{12}(-i\omegabar)}{2\omegabar} \\[0.5em]
        - \frac{i \Sigma_{21}(-i\omegabar)}{2\omegabar} & -i \omega_2 - \frac{i \Sigma_{22}(-i\omegabar)}{2\omegabar}
    \end{pmatrix} \,.
    \label{eqn:non-Hermitian-Hamiltonian}
\end{equation}
where we inserted the nearly-degenerate results in \cref{subapp:Green's-functions-of-dimensionless-one-point-correlation-functions} after noticing $\omega_1=\omega_2=\omegabar$ in this subsection.

In summary, we reproduced the result of the Weisskopf-Wigner theory in this subsection and learned that it ignores a lot of coherence in the subsystem. The effective Hamiltonian $\mathcal{H}$ only describes the effects from poles and doesn't account for the effects due to the multi-particle cut that we mentioned in \cref{subsec:power-counting-and-secular-contributions-of-solutions}. Furthermore, the Weisskopf-Wigner theory doesn't include the effects of the noise-kernel. Hence, it cannot show the dynamics of the subsystem's vacuum background (see \cref{subsec:dynamics-of-the-vacuum-background-in-the-nearly-degenerate-case}) and extending it for a finite-temperature bath is not easy. Besides, its extension to the non-degenerate case is not straightforward, either. For instance, if $\omega_1-\omega_2 \sim \order{1}$, what frequency should we put in $\Sigma_{12}(\cdot)$ and $\Sigma_{21}(\cdot)$ of \cref{eqn:non-Hermitian-Hamiltonian}?

\section{Summary of results and conclusions}
\label{sec:summary-of-results-and-conclusions}

We proposed a framework using an effective non-Markovian quantum master equation (QME) to study the indirect mixing of two generic (pseudo-)scalar fields induced by their couplings to their common decay channels in a medium. The QME yields the equations of motion for populations and coherence more readily and exhibits decoherence and thermalization more clearly, complementing as an alternative approach the study of the correlation functions in field mixing using the Keldysh–Schwinger formalism \cite{PhysRevD.108.025012,PhysRevD.109.036038}. We also showed how our method is related to the Markovian Lindblad form that is widely-used in quantum information and the Weisskopf-Wigner theory that is the cornerstone of the analysis of CP violation in flavored meson mixing in the vacuum.

The proposed framework extends the Weisskopf-Wigner theory in three ways:

\textbf{(i)} We used a multi-particle description, the density matrix and its QME, to describe the evolution of the quantum states of the two generic (pseudo-)scalar fields. This is in contrast to the single (or few) particle description in the Weisskopf-Wigner theory which uses a state vector $\varphi_1(t) \ket{M} + \varphi_2(t) \ket{\overline{M}}$ and an effective non-Hermtian Schr\"odinger equation. Using a state vector, or equivalently assuming a pure state for the excitation part, is a strong assumption that may not be valid in the evolution due to decoherence. Instead, the decoherence processes are readily described by the off-diagonal elements of a density matrix, for example, $\expval*{\phi_c}$ and $\expval*{a_{c,\bvec{k}} a_{d,-\bvec{k}}}$ calculated in \cref{sec:evolutions-of-field-amplitudes-and-their-mixing,sec:populations-and-coherence}. In addition, the density matrix allows us to obtain the evolution and correlation functions of a multi-particle state of the mixing fields, such as coherent states and statistical ensembles. Therefore, the framework we proposed provides useful guides to study the (pseudo-)scalar field mixing that starts with a generic initial state and is induced by the couplings with common intermediate states largely populated in a medium, which is a ubiquitous scenario in the universe.

\textbf{(ii)} We included the case where the mixing fields are not degenerate in the bare Lagrangian. Including a mass difference relaxes the assumption of charge conjugation, parity and time reversal (CPT) invariance and may provide a framework for the study of small violations of CPT symmetry. Besides, it extends the validity of our method beyond meson-antimeson mixing to a more general situation in particle physics, cosmology and condensed matter physics. Candidates of the mixing fields can be any (pseudo-)scalar particles in the SM and hypothetical dark sectors, and/or quasi-particles in condensed matter system. They are not necessarily related by the CPT symmetry but simply connected due to their couplings with common intermediate states.

\textbf{(iii)} We treated the evolution without resorting to Markov, secular or rotating wave approximations that typically neglect memory effects and off-resonance contributions. When deriving the effective QME, we only used Born approximation and assumed the initial state is a product state of the subsystem and the medium. As a consequence, the resulting QME is trace-preserving, Hermitian, non-Markovian in which both the self-energy $\Sigma_{ab}(x_1,x_2)$ and noise-kernel $\mathcal{N}_{ab}(x_1,x_2)$ are non-local and obey a generalized fluctuation-dissipation relation given a thermal medium. It yields the same equation for the field amplitudes as that of a quantum Brownian motion description of the field mixing \cite{PhysRevD.109.036038}. This treatment allows us to compare the solutions with the results of the rotating-wave approximation (RWA) and find that the RWA misses contributions from coherence terms in its solutions, which prevents it from extending beyond leading-order results.

We solved the evolution of the field amplitudes, populations and coherence up to the leading order of the couplings between the subsystem and the medium, in which the Green's functions are a sum of quasi-normal modes (QNM), i.e., exponential decays with/without oscillations. The Kronecker product of the field amplitudes' QNMs yield the QNMs of the populations and coherence. Therefore, their behaviours are consistent regardless of the magnitude of the mass degeneracy.

The non-diagonal QNMs of the field amplitudes imply a mixing of the two fields. If the amplitude of either field is initially non-zero, the amplitudes of both fields are non-zero at a later time , which means an initial condensate in either field may induce a condensate in the other field in the evolution. Both the populations and coherence of the two fields approach to nonvanishing values given by a subsystem in thermal equilibrium with the medium. It is the noise-kernel that causes the thermalization and determines the asymptotic values of the populations and coherence independent of the initial conditions. This suggests that the coherence in the subsystem is non-zero asymptotically even if it may not exist initially. We refer to this emerging long-lived coherence as medium-induced coherence.

The magnitude of the mass degeneracy of the two fields vastly affects the phenomena in the evolution. In the non-degenerate case, the poles of one field are only modified by the field's own coupling with the medium and are not affected by the other field. The mixing effects are weak in the field amplitudes. The quantum beats only appear in coherence and the asymptotic medium-induced coherence is of order $\order{g^2}$. The populations of the two fields thermalize in a way as if only $\phi_1$ (or $\phi_2$) existed. On the contrary, in the nearly-degenerate case, all poles of the two fields receive corrections from both fields' couplings with the medium. The asymptotic medium-induced coherence is enhanced to be of order $\order{1}$. The mixing effect in the field amplitudes is of order $\order{1}$ and strong quantum beats with $\order{1}$ magnitudes appear in the evolution of both populations and coherence. These strong interference effects may provide an observational signatures in the particles' evolution in the early universe and the search for cosmic axions using condensed matter systems. Furthermore, we found that the nearly-degenerate masses lead to prominent dynamics of the subsystem's vacuum in which the initially unperturbed vacuum evolves into the new vacuum of the total Hamiltonian. It is not the transient dynamics accounted for by renormalization because it takes the same form and has the same time scales as that of the excitations of the subsystem. We introduced normalized populations to address the time-dependent definitions of particles due to the changing vacuum, using which we showed that the asymptotic populations are consistent with that of a thermal subsystem and proved that the populations are non-negative in the evolution.

In the mixing between the short-lived and long-lived particles, the coupling strength hierarchy breaks the strong mixing in the nearly-degenerate case and results in a dynamics akin to that of the non-degenerate case with important difference. The self-energy corrections in the poles of field $\phi_1$ and $\phi_2$ are of order $\order{g_1^2}$ and $\order{g_2^2}$ respectively but both fields' poles receive corrections from the other fields. The populations of the two fields thermalize in largely separated time scales without quantum beats. The asymptotic medium-induced coherence is small but its power counting is $\order{g_2/g_1}$ instead of being $\order{g_1 g_2}$ as in the non-degenerate case. We want to emphasize that the relaxation time scale of coherence is of the same order as the lifetime of the short-lived particle. This is important as it suggests the possibility to detect an extremely long-lived and weakly coupled particle within a practical period via the coherence in its mixing with short-lived quasi-particles in a condensed matter system, such as ALP-synthetic axion coupling.

The phenomena revealed in this article, such as induced condensate, medium-induced coherence and quantum beats, are qualitatively general and independent of particular coupling operators or degrees of freedom in the medium. The quantitative form of the QNMs and the magnitudes of various effects clearly rely on particular models and parameters.

It is not difficult to see that the method introduced in this study can be easily extended for the mixing of $N$ (more than two) (pseudo-)scalar fields. The difference is that field indices will go from $1$ to $N$ and the discussion of mass degeneracy will be more complicated. Other potential extensions may include generalizing this method for direct mixing and/or fermionic fields. This may pave a way to study neutrino mixing in the mass basis, where the weak interaction vertices feature flavor off-diagonal terms after diagonalizing a mass matrix in the free part of the Lagrangian. It may complement the study of neutrinos in a medium using kinetic or Boltzman equations \cite{Kainulainen2024-js,Kainulainen:2023khg} or quantum optical master equation \cite{PhysRevD.101.056004}. We expect many phenomena found in this study to be common in other field mixing scenarios, such as thermalization, emergence of long-lived medium-induced coherence, and quantum beats. Robust consequence of field mixing may yield novel phenomena and observational signatures worth of studying. Among other further questions to be studied in future are the issues of renormalizability. In particular, counter terms introduced to renormalize divergences in off-diagonal elements of the self-energy may lead to direct mixing terms in the bare Lagrangian. These aspects can only be studied model by model as the renormalization depends on the specific forms of coupling operators. Another interesting question may be what if we include other types of couplings in the bare Lagrangian, such as a cubic self-interaction $\phi^3$, which may result in different timescales for coherence and thermalization. Interesting as they are, we leave these questions for future research.

\appendix

\section{Lehmann representation of correlation functions}
\label{app:Lehmann-representation-of-correlation-functions}

The correlation functions $G_{ab}^\lessgtr(x_1,x_2)$ can be written in an exact Lehmann (spectral) representation that is useful for solving equations in \cref{sec:evolutions-of-field-amplitudes-and-their-mixing,sec:populations-and-coherence}. Inserting the assumption in \cref{eqn:thermal-chi} into the definitions in \cref{eqn:G<>-definition} yields
\begin{subequations}
\begin{align}
    G_{ab}^>(x_1,x_2)
    & = \frac{1}{Z_\chi} \Tr_\chi\Bab{ e^{-\beta H_\chi} \mathcal{O}_a(x_1) \mathcal{O}_b(x_2) } \,.
    \\
    G_{ab}^<(x_1,x_2)
    & = \frac{1}{Z_\chi} \Tr_\chi\Bab{ e^{-\beta H_\chi} \mathcal{O}_b(x_2) \mathcal{O}_a(x_1)} \,.
\end{align}
\label{eqn:G-ab-<>-thermal}%
\end{subequations}
To proceed, we need to assume that $H_\chi$ is time-independent and spatially translationally invariant such that $\comm*{H_\chi}{\bvec{P}} = 0$, where $\bvec{P}$ is the total momentum operator. Then, there exists a complete set of simultaneous eigenstates $\Bab{\ket{n}}$ of $H_\chi$ and $\bvec{P}$ so that $(H_\chi, \bvec{P}) \ket{n} = (E_n, \bvec{P}_n) \ket{n}$. We also need to assume that in the interaction picture, the coupling operator $\mathcal{O}_a[\chi]$ can be written as
\begin{equation}
    \mathcal{O}_a(\bvec{x},t) = e^{i H_\chi t} e^{-i\bvec{P}\cdot\bvec{x}} \mathcal{O}_a(0) e^{i\bvec{P}\cdot\bvec{x}} e^{-iH_\chi t} \,.
    \label{eqn:O-space-time-translation}
\end{equation}
This is not a particularly special form in field theory since it is satisfied automatically as long as $\mathcal{O}_a[\chi]$ is a composite of bosonic and/or fermionic fields. Write the traces in \cref{eqn:G-ab-<>-thermal} out explicitly using the simultaneous eigen-basis $\Bab{\ket{n}}$ and insert an identity $\sum_{m} \ketbra{m}{m} \equiv \mathbf{1}$ in between the two coupling operators.
\begin{subequations}
\begin{align}
    G_{ab}^>(x_1,x_2)
    & = \frac{1}{Z_\chi} \sum_{n,m} e^{-\beta E_n} e^{-i(P_m - P_n)\cdot(x_1-x_2)}  \braket[3]{n}{\mathcal{O}_a(0)}{m} \braket[3]{m}{\mathcal{O}_b(0)}{n} \,.
    \\
    G_{ab}^<(x_1,x_2)
    & = \frac{1}{Z_\chi} \sum_{n,m} e^{-\beta E_n} e^{-i(P_n-P_m)\cdot(x_1-x_2)} \braket[3]{n}{\mathcal{O}_b(0)}{m} \braket[3]{m}{\mathcal{O}_a(0)}{n} \,.
\end{align}
\end{subequations}
where $P_m = (E_m, \bvec{P}_m)$ and $x_i=(t_i, \bvec{x}_i), i=1,2$. It is now obvious that correlation functions $G_{ab}^\lessgtr$ only depend on the relative difference $x_1 - x_2$ instead of on $x_1$ and $x_2$ respectively, hence are space-time translationally invariant. Thus, we write their arguments as $(x_1-x_2)$ instead of $(x_1,x_2)$ from now on. The space-time translational symmetry also endows us with a Fourier Transform dependent on just one set of frequencies $k=(k_0,\bvec{k})$ rather than two sets for a general function $F(x_1,x_2)$. By inserting 4-dim delta function $\delta^{(4)}(k)$, we are able to write the correlation functions as
\begin{equation}
    G_{ab}^\lessgtr(x_1-x_2) = \int \frac{\dn[4]k}{(2\pi)^4} \rho_{ab}^\lessgtr(k_0,\bvec{k}) e^{-ik\cdot(x_1-x_2)} \,.
    \label{eqn:G-ab-<>-fourier-transform}
\end{equation}
where $\rho_{ab}^\lessgtr(k)$ are spectral densities defined below.
\begin{subequations}
\begin{align}
    \rho_{ab}^>(k) 
    & = \frac{(2\pi)^4}{Z_\chi} \sum_{n,m} e^{-\beta E_n} \braket[3]{n}{\mathcal{O}_a(0)}{m} \braket[3]{m}{\mathcal{O}_b(0)}{n} \, \delta^{(4)}(k - (P_m - P_n)) \,.
    \label{eqn:rho-ab->}
    \\
    \rho_{ab}^<(k)
    & = \frac{(2\pi)^4}{Z_\chi} \sum_{n,m} e^{-\beta E_n} \braket[3]{n}{\mathcal{O}_b(0)}{m} \braket[3]{m}{\mathcal{O}_a(0)}{n} \, \delta^{(4)}(k - (P_n-P_m)) \,.
    \label{eqn:rho-ab-<}
\end{align}
\end{subequations}
Note that $\rho_{ab}^\lessgtr(k)$ are Hermitian matrices since $\mathcal{O}_a$ are assumed to be Hermitian in \cref{sec:effective-quantum-master-equation-of-field-mixing}. Besides, they are connected via the following relation where $\delta(-x) = \delta(x)$ is used.
\begin{equation}
    \rho_{ba}^>(-k)
    = \frac{(2\pi)^4}{Z_\chi} \sum_{n,m} e^{-\beta E_n} \braket[3]{n}{\mathcal{O}_b(0) }{m} \braket[3]{m}{\mathcal{O}_a(0)}{n} \delta^{(4)}(k - (P_n-P_m)) \equiv \rho_{ab}^<(k) \,.
    \label{eqn:rho-ba-(-k)}
\end{equation}
We can also switch $m\leftrightarrow n$ in the above expression and obtain another relation after noticing that $\exp\Bab{-\beta E_m} = \exp\Bab{-\beta k_0} \exp\Bab{-\beta E_n}$ if multiplied by $\delta(k_0 - (E_m-E_n))$.
\begin{equation}
    \rho_{ba}^>(-k)
    = \frac{(2\pi)^4}{Z_\chi} \sum_{n,m} e^{-\beta E_m} \braket[3]{n}{\mathcal{O}_a(0)}{m} \braket[3]{m}{\mathcal{O}_b(0)}{n} \delta^{(4)}(k-(P_m-P_n))
      \equiv e^{-\beta k_0} \rho_{ab}^>(k) \,.
\end{equation}
The generalized Kubo-Martin-Schwinger condition \cite{doi:10.1143/JPSJ.12.570,PhysRev.115.1342} immediately follows the above two ways to rewrite $\rho_{ba}^>(-k)$.
\begin{equation}
    \rho_{ab}^<(k) = e^{-\beta k_0} \rho_{ab}^>(k) \,.
\end{equation}
Introduce the spectral density
\begin{equation}
    \rho_{ab}(k) = \rho_{ab}^>(k) - \rho_{ab}^<(k) = (e^{\beta k_0} - 1) \rho_{ab}^<(k) \,.
    \label{eqn:spectral-rho-definition}
\end{equation}
Obviously, $\rho_{ab}(k)$ inherits the hermicity of $\rho_{ab}^\lessgtr(k)$. It follows the definition that
\begin{equation}
    \rho_{ab}^> = \pab{1 + n(k_0)} \rho_{ab}(k) \,,
    \qquad
    \rho_{ab}^< = n(k_0) \rho_{ab}(k) \,,
    \label{eqn:rho-<>-rho}
\end{equation}
where $n(k_0) = \frac{1}{e^{\beta k_0} - 1}$ is the Bose-Einstein distribution.
The relations in \cref{eqn:rho-ba-(-k),eqn:spectral-rho-definition} lead to a space-time inversion antisymmetry in $\rho_{ab}(k)$.
\begin{equation}
    \rho_{ba}(-k)
    = (e^{-\beta k_0} - 1) \rho_{ba}^<(-k)
    = (e^{-\beta k_0} - 1) e^{\beta k_0} \rho_{ab}^<(k)
    = - \rho_{ab}(k) \,.
    \label{eqn:rho-space-time-inversion-symmetry}
\end{equation}
It is usually the case that $\mathcal{O}_a[\chi]$ is rotationally invariant. One example is $\mathcal{O}_a[\chi]$ being a composite of field operators. Combined with the rotational invariance of a thermal state, this property results in rotational invariant $G_{ab}^\lessgtr(x_1,x_2)$ in \cref{eqn:G-ab-<>-thermal}. Then, the rotational symmetry of the spectral densities $\rho_{ab}^\lessgtr(k)$ and $\rho_{ab}(k)$ follow the definitions in \cref{eqn:G-ab-<>-fourier-transform,eqn:spectral-rho-definition}, which means
\begin{equation}
    \rho_{ab}(\bvec{k},k_0) = \rho_{ab}(\abs*{\bvec{k}},k_0) \,.
    \label{eqn:rho-rotational-symmetry}
\end{equation}
Combining \cref{eqn:rho-space-time-inversion-symmetry} and \cref{eqn:rho-rotational-symmetry} yields
\begin{equation}
    \rho_{ba}(\bvec{k},-k_0) = \rho_{ab}(-\bvec{k}, -k_0) = -\rho_{ab}(\bvec{k},k_0) \,.
    \label{eqn:rho-time-inversion-symmetry}
\end{equation}
Note that $\rho_{ab}(k)$ is Hermitian. \cref{eqn:rho-time-inversion-symmetry} leads to a conjugate antisymmetry of $\rho_{ab}(k)$.
\begin{equation}
    \rho_{ab}^*(\bvec{k},-k_0) = - \rho_{ab}(\bvec{k},k_0) \,.
    \label{eqn:rho-ab-k-conjugate-antisymmetry}
\end{equation}

\section{Symmetries of self-energy}
\label{app:symmetries-of-self-energy}

We introduce the 4-dimensional Fourier transform of the self-energy as below.
\begin{equation}
    i \Sigma_{ab}(x_1 - x_2) = \int \frac{\dn[4]k}{(2\pi)^4} i \Sigma_{ab}(k) e^{-ik\cdot(x_1-x_2)} \,.
\end{equation}
Then, a relation between $\Sigma_{ab}(k)$ and the spectral density follows \cref{eqn:self-energy-definition,eqn:G-ab-<>-fourier-transform,eqn:rho-<>-rho}.
\begin{equation}
    i\Sigma_{ab}(k) = \rho_{ab}(k) \,.
    \label{eqn:i-Sigma-k-rho-k}
\end{equation}
Thus, $\Sigma_{ab}(k)$ is anti-Hermitian, rotationally invariant, i.e., $\Sigma_{ab}(\bvec{k},k_0) = \Sigma_{ab}(\abs*{\bvec{k}},k_0)$, and conjugate symmetric, i.e., $\Sigma_{ab}^*(\bvec{k},-k_0) = \Sigma_{ab}(\bvec{k},k_0)$.

In this article, we solve equations in the spatial momentum basis where the spatial Fourier transform of the self-energy is used. It is given as below.
\begin{equation}
    \Sigma_{ab}(\bvec{k},t) = \int \frac{\dn{k_0}}{2\pi} (-i)\rho_{ab}(\bvec{k},k_0) e^{-ik_0 t} \,.
    \label{eqn:i-Sigma-k-t}
\end{equation}
where we inserted \cref{eqn:i-Sigma-k-rho-k}. Thus, $\Sigma_{ab}(\bvec{k},t)$ is rotationally invariant following \cref{eqn:rho-rotational-symmetry}. Take the complex conjugate of the definition and relabel $k_0 \to -k_0$. The conjugate antisymmetry of $\rho_{ab}(k)$ (see \cref{eqn:rho-ab-k-conjugate-antisymmetry}) implies that $\Sigma_{ab}(\bvec{k},t)$ is a real function.

The Laplace transform of the self-energy is used in this article, which is defined as
\begin{equation}
    \Sigma_{ab}(\bvec{k},s)
    = \int_0^\infty \dn{\tau} e^{-s\tau} \int \frac{\dn{k_0}}{2\pi} (-i) \rho_{ab}(\bvec{k},k_0) e^{-ik_0 \tau}
    = - \int \frac{\dn{k_0}}{2\pi} \frac{\rho_{ab}(\bvec{k},k_0)}{-i s + k_0} \,.
    \label{eqn:tilde-Sigma-s-domain}
\end{equation}
The time integral needs to converge for the Laplace transform to exist, indicating $\Re\Bab{s} > 0$. Therefore, $\Sigma_{ab}(\bvec{k},s)$ is only well-defined on the right half part of the  $s$-plane.

In this article, we compute the inverse Laplace transform of the Green's functions or solutions by choosing the contour $s=i\omega + \epsilon, -\infty<\omega<\infty, \epsilon \to 0^+$. In this limit, the self-energy in the $s$-domain (see \cref{eqn:tilde-Sigma-s-domain}) becomes
\begin{equation}
    \Sigma_{ab}(\bvec{k},i\omega + \epsilon)
    = - \int \frac{\dn{k_0}}{2\pi} \frac{\rho_{ab}(\bvec{k},k_0)}{\omega - i\epsilon + k_0}
    \eqqcolon \Sigma_{R,ab}(\bvec{k},i\omega) + i \Sigma_{I,ab}(\bvec{k},i\omega) \,.
    \label{eqn:Sigma-epsilon->0}
\end{equation}
where we used the formula
\begin{equation}
    \lim_{\epsilon \to 0^+} \frac{1}{x + i \epsilon} = \pv \pab{ \frac{1}{x} } - i \pi \delta(x), \qquad x\in\mathbb{R} \,.
    \label{eqn:Dirac-formula}
\end{equation}
to decompose $\Sigma_{ab}(\bvec{k},i\omega + \epsilon)$ into its Hermitian and anti-Hermitian part, which are
\begin{equation}
    \Sigma_{\txtR,ab}(\bvec{k},i\omega)
    = - \pv \int \frac{\dn{k_0}}{2\pi} \frac{\rho_{ab}(\bvec{k},k_0)}{\omega + k_0} \,,
    \qquad
    \Sigma_{\txtI,ab}(\bvec{k},i\omega) 
    = - \frac{1}{2} \rho_{ab}(\bvec{k},-\omega) \,.
    \label{eqn:Sigma-R-I-definitions}
\end{equation}
Following \cref{eqn:rho-time-inversion-symmetry}, they satisfy
\begin{equation}
    \Sigma_{\txtR,ab}(\bvec{k},-i\omega) = \Sigma_{R,ba}(\bvec{k},i\omega) \,,
    \qquad
    \Sigma_{\txtI,ab}(\bvec{k},-i\omega) = - \Sigma_{I,ba}(\bvec{k},i\omega) \,.
    \label{eqn:Sigma-R-I-symmetry}
\end{equation}
The matrices $\Sigma_{\txtR,ab}(\bvec{k},\omega)$ and $\Sigma_{\txtI,ab}(\bvec{k},\omega)$ inherit the hermicity from $\rho_{ab}(k)$. They are the matrix version of the real and imaginary part of the self-energy of one field, which explains the subscript ``$\txtR$'' and ``$\txtI$''. The conjugate antisymmetry of $\rho_{ab}(\bvec{k},k_0)$ (see \cref{eqn:rho-ab-k-conjugate-antisymmetry}) leads to the conjugate symmetry of $\Sigma_{ab}(\bvec{k},i\omega)$ via the definitions of $\Sigma_{\txtR,ab}(\bvec{k},i\omega)$ and $\Sigma_{\txtI,ab}(\bvec{k},i\omega)$ in \cref{eqn:Sigma-R-I-definitions}.
\begin{equation}
    \Sigma_{ab}(\bvec{k},-i\omega) = \bab{ \Sigma_{ab}(\bvec{k},i\omega) }^* \,.
    \label{eqn:Sigma-ab-omega-conjugate-symmetry}
\end{equation}
where $\pm i\omega$ should be understood as $\pm i \omega + 0^+$. Given the definition of $\widetilde{\Sigma}_{ab}(\bvec{k},t)$ in \cref{eqn:tilde-Sigma-N}, it has the same properties as $\Sigma_{ab}(\bvec{k},t)$.

\section{Symmetries of noise kernel}
\label{app:symmetries-of-noise-kernel}

The analysis of the noise kernel runs in parallel with that of the self-energy. We introduce the 4-dimensional Fourier transform of the noise-kernel as below.
\begin{equation}
    \mathcal{N}_{ab}(x_1 - x_2) = \int \frac{\dn[4]{k}}{(2\pi)^4} \mathcal{N}_{ab}(k) e^{-ik\cdot(x_1-x_2)} \,.
\end{equation}
Then, a relation between $\mathcal{N}_{ab}(k)$ and the spectral density follows \cref{eqn:noise-kernel-definition,eqn:G-ab-<>-fourier-transform,eqn:rho-<>-rho}.
\begin{equation}
    \mathcal{N}_{ab}(k) = \frac{1}{2} \coth\pab{\frac{\beta k_0}{2}} \rho_{ab}(k) \,.
    \label{eqn:N-ab-k-rho-ab-k}
\end{equation}
Thus, $\mathcal{N}_{ab}(k)$ is Hermitian, rotationally invariant, i.e., $\mathcal{N}_{ab}(\bvec{k},k_0) = \mathcal{N}_{ab}(\abs*{\bvec{k}},k_0)$, and conjugate symmetric, i.e., $\mathcal{N}_{ab}^*(\bvec{k},-k_0) = \mathcal{N}_{ab}(\bvec{k},k_0)$.

The spatial Fourier transform of the noise-kernel is defined as
\begin{equation}
    \mathcal{N}_{ab}(\bvec{k},t) = \int \frac{\dn{k_0}}{2\pi} \mathcal{N}_{ab}(\bvec{k},k_0) e^{-i k_0 t} \,.
\end{equation}
Hence, $\mathcal{N}_{ab}(\bvec{k},t)$ is rotationally invariant, too. Take the complex conjugate of the definition and relabel $k_0 \to -k_0$. The conjugate symmetry of $\mathcal{N}_{ab}(k)$ implies that $\mathcal{N}_{ab}(\bvec{k},t)$ is a real function.

The Laplace transform of the $\mathcal{N}_{ab}(\bvec{k},t)$ is defined as
\begin{equation}
    \mathcal{N}_{ab}(\bvec{k},s)
    = \int_0^\infty \dn{\tau} e^{-s\tau} \int \frac{\dn{k_0}}{2\pi} \mathcal{N}_{ab}(\bvec{k},k_0) e^{-ik_0\tau}
    = \int \frac{\dn{k_0}}{2\pi} \frac{\mathcal{N}_{ab}(\bvec{k},k_0)}{s+ik_0}
    \,.
    \label{eqn:noise-kernel-s-domain}
\end{equation}
Same as the Laplace transform of the self-energy, the region of convergence of $\mathcal{N}_{ab}(\bvec{k},s)$ is $\Re\Bab*{s}>0$. In this article, the inverse Laplace transform is computed using the contour $s=i\omega + \epsilon, -\infty<\omega<\infty, \epsilon \to 0^+$. We use \cref{eqn:Dirac-formula} to decompose $\mathcal{N}_{ab}(\bvec{k},i\omega + \epsilon)$ into a Hermitian and an anti-Hermitian part in this limit.
\begin{equation}
    \mathcal{N}_{ab}(\bvec{k},i\omega+\epsilon)
    = \int \frac{\dn{k_0}}{2\pi} \frac{\mathcal{N}_{ab}(\bvec{k},k_0)}{i\omega + \epsilon + i k_0}
    \eqqcolon \mathcal{N}_{\txtR,ab}(\bvec{k},i\omega) + i \mathcal{N}_{\txtI,ab}(\bvec{k},i\omega)
    \,.
    \label{eqn:noise-kernel-epsilon->0}
\end{equation}
where 
\begin{equation}
    \mathcal{N}_{\txtR,ab}(\bvec{k},i\omega)
    \coloneqq \frac{1}{2} \mathcal{N}_{ab}(\bvec{k},-\omega) \,,
    \qquad
    \mathcal{N}_{\txtI,ab}(\bvec{k},i\omega)
    \coloneq - \pv \int \frac{\dn{k_0}}{2\pi} \frac{\mathcal{N}_{ab}(\bvec{k},k_0)}{\omega + k_0} \,.
    \label{eqn:N-R-I-definitions}
\end{equation}
The matrices $\mathcal{N}_{\txtR,ab}(k,i\omega)$ and $\mathcal{N}_{\txtI,ab}(k,i\omega)$ inherit the hermicity of the Fourier transform $\mathcal{N}_{ab}(k)$. Following \cref{eqn:rho-time-inversion-symmetry,eqn:N-ab-k-rho-ab-k}, they satisfy
\begin{equation}
    \mathcal{N}_{\txtR,ab}(\bvec{k},i\omega)
    = \mathcal{N}_{\txtR,ba}(\bvec{k},-i\omega) \,.
    \qquad
    \mathcal{N}_{\txtI,ab}(\bvec{k},-i\omega)
    = - \mathcal{N}_{\txtI,ba}(\bvec{k},i\omega) \,.
    \label{eqn:noise-kernel-R-I-symmetry}
\end{equation}
The conjugate symmetry of the Fourier transform $\mathcal{N}_{ab}(k)$ results in the conjugate symmetry of $\mathcal{N}_{ab}(\bvec{k},i\omega)$ via the definitions  in \cref{eqn:N-R-I-definitions}.
\begin{equation}
    \mathcal{N}_{ab}(-i\omega) = \bab{\mathcal{N}_{ab}(i\omega)}^* \,.
\end{equation}
where $\pm i\omega$ should be understood as $\pm i\omega + 0^+$. Given the definition of $\widetilde{N}_{ab}(\bvec{k},t)$ in \cref{eqn:tilde-Sigma-N}, it has the same properties as $\mathcal{N}_{ab}(\bvec{k},t)$.

\section{Generalized fluctation-dissipation relation}
\label{app:generalized-fluctuation-dissipation-relation}

We have shown in this article that the self-energy yields frictions in the evolution and the noise-kernel describes fluctuations due to the medium. Both of them come from interactions between the subsystem and the medium, and are determined by couplings operators and the state of the medium. Therefore, they are closely connected. The generalized fluctuation-dissipation relation in the momentum-frequency domain follows a comparison between the Fourier transforms in \cref{eqn:i-Sigma-k-rho-k,eqn:N-ab-k-rho-ab-k}.
\begin{equation}
    i \Sigma_{ab}(\bvec{k},\omega) \coth\pab{\frac{\beta \omega}{2}} = 2 \mathcal{N}_{ab}(\bvec{k},\omega) \,.
    \label{eqn:fluctuation-dissipation-relation}
\end{equation}
Comparing \cref{eqn:Sigma-R-I-definitions,eqn:N-R-I-definitions} yields the generalized fluctuation-dissipation relation in the $s$-domain with the aid of \cref{eqn:N-ab-k-rho-ab-k}.
\begin{equation}
    \Sigma_{\txtI,ab}(\bvec{k},i\omega) \coth\pab{\frac{\beta \omega}{2}} = 2 \mathcal{N}_{\txtR,ab}(\bvec{k},i\omega) \,.
    \label{eqn:fluctuatio-dissipation-relation-s-domain}
\end{equation}
Note that in \cref{sec:evolutions-of-field-amplitudes-and-their-mixing,sec:populations-and-coherence}, $\Sigma_{\txtI,ab}(\bvec{k},i\omega)$ yield the decay rates, i.e., dissipation and $\mathcal{N}_{\txtR,ab}(\bvec{k},i\omega)$ yield the secular contributions, i.e., thermal fluctuations.

\section{Single species}
\label{app:single-species}

\subsection{Equations of motion and Green's functions}
\label{subapp:equations-of-motion-and-Green's-functions}

In this subsection, we compute the Green's functions of the equal-time two-point correlation functions in a subsystem of only one particle. Using such a smaller system is sufficient to reproduce and clarify the arguments in \cref{eqn:residue-G-A-at-sA-sB}. But smaller matrices in this calculation allows us to write everything out explicitly, in which the power counting is obvious. Many results in the single-species subsystem can be obtained by removing the indices of the results in other sections. We will use them directly in the following calculations.

Removing the indices in \cref{eqn:A-cd-k-B-cd-k-definitions} yields the definitions for the equal-time two-point correlation functions of the single-species subsystem, whose equations are obtained by removing the indices of the results in \cref{subapp:equations-of-motion-two-point-correlations}.
\begin{align*}
    \diff{A_\bvec{k}}{t} 
    & = -i \int_0^t \dn{t'} \bab{ - \pab{B_k(t') + A_k(t')} e^{-i\omega (t-t')} + \pab{A_k(t') + B_k^*(t')} e^{i\omega(t-t')} } \widetilde{\Sigma}(t-t')
    + \mathcal{I}_{A_\bvec{k}}(t) \,,
    \\
    \diff{B_\bvec{k}}{t} 
    & = - 2i\omega B_k
      -i \int_0^t \dn{t'} \bab{ \pab{B_k(t') + A_{-k}(t')} e^{-i\omega(t-t')} + \pab{B_k(t') + A_k(t')} e^{-i\omega(t-t')} } \widetilde{\Sigma}(t-t') + \mathcal{I}_{B_\bvec{k}}(t) \,,
    \\
    \diff{A_{-\bvec{k}}}{t} 
    & = -i \int_0^t \dn{t'} \bab{ -\pab{B_k(t') + A_{-k}(t')} e^{-i\omega (t-t')} + \pab{A_{-k}(t') + B_k^*(t')} e^{i\omega(t-t')} } \widetilde{\Sigma}(t-t') + \mathcal{I}_{A_{-\bvec{k}}}(t)
     \,,
    \\
    \diff{B_\bvec{k}^*}{t} 
    & = 2i\omega B_k^* 
      + i \int_0^t \dn{t'} \bab{ \pab{B_k^*(t') + A_{-k}(t')} e^{i\omega (t-t')} + \pab{ B_k^*(t') + A_k(t') } e^{i\omega(t-t')} } \widetilde{\Sigma}(t-t') + \mathcal{I}_{B_\bvec{k}^*}(t) \,.
\end{align*}
where we dropped ``$\bvec{k}$'' in $\omega_\bvec{k}$ and $\widetilde{\Sigma}(\bvec{k},t)$ for convenience. The inhomogeneous terms are
\begin{align*}
    \mathcal{I}_{A_\bvec{k}}(t) 
    & = + \int_0^t \dn{t'} 2 \bab{ e^{-i\omega t'} + e^{i\omega t'} } \widetilde{\mathcal{N}}(\bvec{k},t') \,,
    &
    \mathcal{I}_{B_\bvec{k}}(t)
    & = - \int_0^t \dn{t'} 4 e^{-i\omega t} \widetilde{\mathcal{N}}(\bvec{k},t') \,,
    \\
    \mathcal{I}_{A_{-\bvec{k}}}(t)
    & = + \int_0^t \dn{t'} 2 \bab{e^{-i\omega t'} + e^{i\omega t'}} \widetilde{\mathcal{N}}(\bvec{k},t') \,,
    &
    \mathcal{I}_{B_\bvec{k}^*}(t)
    & = - \int_0^t \dn{t'} 4 e^{i\omega t} \widetilde{\mathcal{N}}(\bvec{k},t') \,.
\end{align*}
Following the strategy in \cref{sec:populations-and-coherence}, we rewrite the above four equations as below after taking the Laplace transform.
\begin{equation}
    \mathbf{G}_{\mathcal{D}}^{-1}(s) \cdot \overrightarrow{\mathcal{D}}(s) = \overrightarrow{\mathcal{D}}(0) + \overrightarrow{\mathcal{I}}(s) \,.
\end{equation}
where we defined $\overrightarrow{\mathcal{D}} = \pab*{\overrightarrow{A}^\T, \overrightarrow{B}^\T}^\T, \overrightarrow{A}=\pab*{A_{\bvec{k}}, A_{-\bvec{k}}}^\T, \overrightarrow{B}=\pab*{B_{\bvec{k}}, B_{\bvec{k}}^*}^\T$, and  $\overrightarrow{\mathcal{I}}, \overrightarrow{\mathcal{I}}_A, \overrightarrow{\mathcal{I}}_B$ are correspondingly defined. Following similar notations in \cref{subapp:Greens-function-in-s-domain}, we read from the Laplace transform of the above four equations that
\begin{equation}
    \mathbf{G}_\mathcal{D}^{-1} =
    \begin{pmatrix}
        \mathbf{G}_{\txtA}^{-1} & \mathbf{K}_{\text{AB}} \\
        \mathbf{K}_{\text{BA}} & \mathbf{G}_{\txtB}^{-1}
    \end{pmatrix}
    \,,
    \qquad
    \mathbf{G}_{\txtA}^{-1} =
    \begin{pmatrix}
        \mathbf{G}_{A_{\bvec{k}}}^{-1} & 0 \\
        0 & \mathbf{G}_{A_{-\bvec{k}}}^{-1}
    \end{pmatrix}
    \,,
    \qquad
    \mathbf{G}_{\txtB}^{-1} =
    \begin{pmatrix}
        \mathbf{G}_{B_\bvec{k}}^{-1} & 0 \\
        0 & \mathbf{G}_{B_\bvec{k}^*}^{-1}
    \end{pmatrix}
    \,.
    \label{eqn:G-D-inverse-single-species}
\end{equation}
where we defined
\begin{align*}
    \mathbf{G}_{A_{\bvec{k}}}^{-1}(s)  
    & = s + i \widetilde{\Sigma}(s-i\omega) - i \widetilde{\Sigma}(s+i\omega) \,,
    &
    \mathbf{G}_{A_{-\bvec{k}}}^{-1}(s)  
    & = s + i \widetilde{\Sigma}(s-i\omega) - i \widetilde{\Sigma}(s+i\omega) \,,
    \\
    \mathbf{G}_{B_\bvec{k}}^{-1}(s) 
    & = s + 2i\omega + 2 i \widetilde{\Sigma}(s+i\omega) \,, 
    &
    \mathbf{G}_{B_\bvec{k}^*}^{-1}(s) 
    & = s - 2i\omega -2i\widetilde{\Sigma}(s-i\omega) \,,
    \\
    \mathbf{K}_{\text{AB}}(s) 
    & = \pab[bigg]{ \begin{matrix} 1 \\ 1 \end{matrix} } \cdot
        \begin{pmatrix}
            -i \widetilde{\Sigma}(s+i\omega) \; & i \widetilde{\Sigma}(s-i\omega)
        \end{pmatrix} \,,
    &
    \mathbf{K}_{\text{BA}}(s) 
    & =
        \begin{pmatrix}
            i \widetilde{\Sigma}(s+i\omega)  \\
            -i \widetilde{\Sigma}(s-i\omega) 
        \end{pmatrix}
        \cdot \pab[big]{1 \; 1} \,.
\end{align*}
Note that similar to \cref{eqn:G-A-0-G-B-0-block-diagonalized}, $\mathbf{G}_{A_\pm\bvec{k}}^{-1}(s)$ are the same in the case of single species as a consequence of the rotationally invariant Hamiltonian and medium, and the $\mathbf{G}_{\mathcal{D}}^{-1}(s)$ in \cref{eqn:G-D-inverse-single-species} can be decomposed using the same pattern as that in \cref{subsec:quasi-normal-modes-of-populations-and-coherence}.

We first solve $\det\Bab{\mathbf{G}_\mathcal{D}^{-1}(s)}=0$ directly for the poles of $\mathbf{G}_{\mathcal{D}}(s)$ for a comparison with the results from Gauss elimination. There is no need to discuss non-degenerate and nearly-degenerate cases as in \cref{subsec:quasi-normal-modes-of-populations-and-coherence} since we only have one species. The results up to $\order{g^2}$ are
\begin{equation}
\begin{aligned}
    s_{\mathcal{D},1} & 
        = s_{\mathcal{D},2} = -i \pab[big]{\widetilde{\Sigma}(-i\omega) - \widetilde{\Sigma}(i\omega)}
        = - 2 \widetilde{\Sigma}_{\txtI}(i\omega)
        \eqqcolon - \Gamma \,,
    \\
    s_{\mathcal{D},3}^* & 
        = s_{\mathcal{D},4} = 2i\omega + 2i \widetilde{\Sigma}(i\omega)
        = 2i \pab[big]{ \omega + \widetilde{\Sigma}_\txtR(i\omega) } - 2 \widetilde{\Sigma}_\txtI(i\omega)
        \eqqcolon 2i \Omega - \Gamma \,.
\end{aligned}
\label{eqn:G-D-poles-single-species}
\end{equation}
where we used the decomposition of $\widetilde{\Sigma}(\pm i\omega)$ (see \cref{eqn:Sigma-epsilon->0}) and defined the decay rate $\Gamma$ and renormalized frequency $\Omega$. We also used the conjugate symmetry of $\widetilde{\Sigma}(i\omega)$ (see \cref{eqn:Sigma-ab-omega-conjugate-symmetry}) to identity $s_{\mathcal{D},3}^* = s_{\mathcal{D},4}$. Following the same Gauss elimination as that in \cref{subsec:quasi-normal-modes-of-populations-and-coherence}, the single-species version of $\mathcal{G}_\txtA(s)$ is
\begin{equation}
    \mathcal{G}_\txtA(s) = \pab{ 1 - \mathbf{G}_\txtA(s) \cdot \mathbf{K}_{\txtA\txtB}(s) \cdot \mathbf{G}_\txtB(s) \cdot \mathbf{K}_{\txtB\txtA}(s) }^{-1} \cdot \mathbf{G}_\txtA(s) \,.
\end{equation}
$\mathcal{G}_\txtA(s)$ is a $2\times2$ matrix in the case of single species. We are able to calculate its elements directly. After writing all four elements of $\mathcal{G}_\txtA(s)$ as single rational functions, we find that they share the same denominator.
\begin{equation*}
    \pab[big]{s+i \pab[big]{\widetilde{\Sigma }(s-i \omega )-\widetilde{\Sigma }(s+i \omega ) } } \pab[big]{i (s-2 i \omega )^2 \widetilde{\Sigma }(s+i \omega )-i (s+2 i \omega )^2 \widetilde{\Sigma }(s-i \omega )+s^3+4 s \omega ^2 } \,.
\end{equation*}
The poles of $\mathcal{G}_\txtA(s)$ are given by the zeros of this expression. Up to $\order{g^2}$, they are
\begin{equation}
    s_{\txtA,1}  = s_{\txtA,2} = -i \pab[big]{ \widetilde{\Sigma}(-i\omega) - \widetilde{\Sigma}(i\omega) }\,, 
    \qquad
    s_{\txtA,3}^* = s_{\txtA,4} = 2i\omega + 2i\widetilde{\Sigma}(i\omega) \,.
\label{eqn:mathcal-G-A-poles-single-species}
\end{equation}
Note that the poles of $\mathcal{G}_\txtA(s)$ in \cref{eqn:mathcal-G-A-poles-single-species} are the same as the poles of $\mathbf{G}_\mathcal{D}(s)$ in \cref{eqn:G-D-poles-single-species}, which is exactly what we argued in the paragraph below \cref{eqn:G-A}. The residues of $\mathcal{G}_\txtA(s)$ at the four poles are
\begin{equation}
\begin{aligned}
    \residue\pab{\mathcal{G}_\txtA(s), s_{\txtA,1}} & = 
    \pab[bigg]{\begin{matrix}
        1 & 0 \\
        0 & 0
    \end{matrix}} \,,
    &\quad
    \residue\pab{\mathcal{G}_\txtA(s), s_{\txtA,3}} & = - \frac{\bab*{\widetilde{\Sigma}(-i\omega)}^2}{4\omega^2}
    \pab[bigg]{\begin{matrix}
        1 & 1 \\
        1 & 1
    \end{matrix}} \,,
    \\
    \residue\pab{\mathcal{G}_\txtA(s), s_{\txtA,2}} & = 
    \pab[bigg]{\begin{matrix}
        0 & 0 \\
        0 & 1
    \end{matrix}} \,,
    &\quad
    \residue\pab{\mathcal{G}_\txtA(s), s_{\txtA,4}} & = - \frac{\bab*{\widetilde{\Sigma}(i\omega)}^2}{4\omega^2}
    \pab[bigg]{\begin{matrix}
        1 & 1 \\
        1 & 1
    \end{matrix}} \,.
\end{aligned}
\label{eqn:residues-G-A-single-species}
\end{equation}
where the residues $s_{\txtA,1}$ and $s_{\txtA,2}$ are exact up to $\order{g^2}$ and we kept the leading contributions in the residues at $s_{\txtA,3}$ and $s_{\txtA,4}$. It is obvious that
\begin{equation}
    \residue\pab{\mathcal{G}_\txtA(s), s_{\txtA,3}} \sim \residue\pab{\mathcal{G}_\txtA(s), s_{\txtA,4}} \sim \order{g^4} \,.
\end{equation}
Thus, they do not contribute to $\mathcal{G}_\txtA(t)$ at the leading order of perturbations, which is exactly what we argued around \cref{eqn:residue-G-A-at-sA-sB}. Therefore, $\mathcal{G}_\txtA(s) \approx \mathbf{G}_\txtA(s)$ up to $\order{g^2}$.

We want to point out that the results in \cref{eqn:residues-G-A-single-species} don't include all $\order{g^4}$ in the residues at $s_{\txtA,3},s_{\txtA,4}$ because the equations of motion are only exact up to $\order{g^2}$. We present the residues at $s_{\txtA,3},s_{\txtA,4}$ only to show the orders of perturbations in their leading contributions. Besides, the poles $s_{\txtA,1}, s_{\txtA,2}$ of $\mathcal{G}_\txtA(s)$ are the same as a consequence of the rotationally invariant Hamiltonian and bath. Mathematically, they are the same one pole and only yields one residue $\operatorname{diag}(1,1)$. We separate the residue into $\operatorname{diag}(1,0)$ and $\operatorname{diag}(0,1)$ so that they become the unperturbed Green's functions of $A_{\pm\bvec{k}}$ respectively when the coupling is switched off. This is the same argument as that in \cref{subapp:residues-at-poles-of-equal-time-two-point-correlation-functions} for $\mathbf{G}_{B_{\bvec{k}}}$ and $\mathbf{G}_{B_{\bvec{k}}^*}$.

Applying similar steps to the B-part of the subsystem yields
\begin{equation}
    \mathcal{G}_\txtB(s)
    = \pab{ 1 - \mathbf{G}_\txtB(s) \cdot \mathbf{K}_{\txtB\txtA}(s) \cdot \mathbf{G}_\txtA(s) \cdot \mathbf{K}_{\txtA\txtB}(s) }^{-1} \cdot \mathbf{G}_\txtB(s) \,.
\end{equation}
The poles of $\mathcal{G}_\txtB(s)$ up to $\order{g^2}$ are the same as the poles of $\mathbf{G}_\mathcal{D}(s)$ in \cref{eqn:G-D-poles-single-species}.
\begin{equation}
    s_{\txtB,1} = s_{\txtB,2} = -i \pab[big]{ \widetilde{\Sigma}(-i\omega) - \widetilde{\Sigma}(i\omega) } \,, 
    \qquad
    s_{\txtB,3}^* = s_{\txtB,4} = 2i\omega + 2i\widetilde{\Sigma}(i\omega) \,.
\end{equation}
The residues at these poles are
\begin{equation}
\begin{aligned}
    \residue\pab{\mathcal{G}_{\txtB}(s), s_{\txtB,1}}
    & = \residue\pab{\mathcal{G}_{\txtB}(s), s_{\txtB,2}} = - \frac{1}{2\omega^2}
    \begin{pmatrix}
        \bab*{\widetilde{\Sigma}(i\omega)}^2 & \widetilde{\Sigma}(-i\omega) \widetilde{\Sigma}(i\omega) \\[0.5em]
        \widetilde{\Sigma}(-i\omega) \widetilde{\Sigma}(i\omega) & \bab*{\widetilde{\Sigma}(i\omega)}^2
    \end{pmatrix} \,,
    \\
    \residue\pab{\mathcal{G}_{\txtB}(s), s_{\txtB,3}} & =
    \pab[bigg]{\begin{matrix}
        1 & 0 \\
        0 & 0
    \end{matrix}} \,,
    \qquad
    \residue\pab{\mathcal{G}_{\txtB}(s), s_{\txtB,4}} =
    \pab[bigg]{\begin{matrix}
        0 & 0 \\
        0 & 1
    \end{matrix}} \,.
\end{aligned}
\end{equation}
With the same arguments, we conclude that up to $\order{g^2}$, $\mathcal{G}_\txtB(s) \approx \mathbf{G}_\txtB(s)$. 

This approximation is exactly what we argued in \cref{subsec:quasi-normal-modes-of-populations-and-coherence}.

\subsection{Thermalization and first order corrections}

We have shown that up to $\order{g^2}$, $\mathcal{G}_\txtA(s) \approx \mathbf{G}_\txtA(s)$ and $\mathcal{G}_\txtB(s) \approx \mathbf{G}_\txtB(s)$. Following the steps leading to \cref{eqn:A-s-elimination}, we reach a similar result for the single species case.
\begin{equation}
    \overrightarrow{A}(s) = \mathbf{G}_\txtA(s) \cdot \pab[big]{ \overrightarrow{A}(0^-) + \overrightarrow{\mathcal{I}}_\txtA(s) } - \mathbf{G}_\txtA(s) \cdot \mathbf{K}_{\txtA\txtB}(s) \cdot \mathbf{G}_\txtB(s) \cdot \pab[big]{ \overrightarrow{B}(0^-) + \overrightarrow{\mathcal{I}}_\txtB(s) }  \,.
\end{equation}
Inserting $\mathbf{G}_\txtA(s), \mathbf{G}_\txtB(s), \mathbf{K}_{\text{AB}}(s)$ defined in \cref{eqn:G-D-inverse-single-species} into the above equation yields
\begin{multline*}
    \begin{pmatrix}
        A_{\bvec{k}}(s) \\
        A_{-\bvec{k}}(s)
    \end{pmatrix}
    =
    \begin{pmatrix}
        \mathbf{G}_{A_\bvec{k}}(s) \bab{ A_\bvec{k}(0^-) + \mathcal{I}_{A_\bvec{k}}(s) } \\
        \mathbf{G}_{A_{-\bvec{k}}}(s)  \bab{ A_{-\bvec{k}}(0^-) + \mathcal{I}_{A_{-\bvec{k}}}(s) }
    \end{pmatrix}
    \\
    - \begin{pmatrix}
        \mathbf{G}_{A_\bvec{k}}(s) \\
        \mathbf{G}_{A_{-\bvec{k}}}(s)
    \end{pmatrix}
    \bab{
        \begin{pmatrix}
            -i \widetilde{\Sigma}(s+i\omega) \; & i \widetilde{\Sigma}(s-i\omega)
        \end{pmatrix}
        \begin{pmatrix}
            \mathbf{G}_{B_\bvec{k}}(s) & \\
            & \mathbf{G}_{B_\bvec{k}^*}(s)
        \end{pmatrix}
        \begin{pmatrix}
            B_\bvec{k}(0^-) + \mathcal{I}_{B_\bvec{k}}(s) \\
            B_{\bvec{k}}^*(0^-) + \mathcal{I}_{B_{\bvec{k}}^*}(s)
        \end{pmatrix}
    } \,.
\end{multline*}
The solutions of $A_\bvec{k}(s)$ and $A_{-\bvec{k}}(s)$ don't mix. To find the evolution of part A of the system, we only need to concentrate on
\begin{multline}
    A_\bvec{k}(s) = 
    \mathbf{G}_{A_\bvec{k}}(s) \bab{ A_\bvec{k}(0^-) + \mathcal{I}_{A_\bvec{k}}(s) }
    + \mathbf{G}_{A_\bvec{k}}(s) i \widetilde{\Sigma}(s+i\omega) \mathbf{G}_{B_\bvec{k}}(s) \bab{ B_\bvec{k}(0^-) + \mathcal{I}_{B_\bvec{k}}(s) }
    \\
    - \mathbf{G}_{A_\bvec{k}}(s) i \widetilde{\Sigma}(s-i\omega)  \mathbf{G}_{B_\bvec{k}^*}(s) \bab[big]{ B_\bvec{k}^*(0^-) + \mathcal{I}_{B_\bvec{k}^*}(s) } \,.
    \label{eqn:A_k-s-single-species}
\end{multline}
The same strategy as in \cref{subsec:power-counting-and-secular-contributions-of-solutions} is used to find $A_\bvec{k}(t)$. The results in \cref{subapp:equations-of-motion-and-Green's-functions} suggests that
\begin{equation}
    \mathbf{G}_{A_\bvec{k}}(s) 
    = \frac{1}{s - s_{\txtA,1}} \,,
    \qquad
    \mathbf{G}_{B_\bvec{k}}(s)
    = \frac{1}{s - s_{\txtB,3}} \,,
    \qquad
    \mathbf{G}_{B_\bvec{k}^*}(s) 
    = \frac{1}{s - s_{\txtB,4}} \,.
\end{equation}
Note that $\mathcal{L}^{-1}\bab*{ 1 / (s - s_{i}) } = \exp\Bab*{ s t} |_{s=s_i}$, which we can directly apply to $\mathbf{G}_{A_\bvec{k}}(s) A_\bvec{k}(0^-)$. For the terms including $B_\bvec{k}(0^-)$ or $B_\bvec{k}^*(0^-)$, we make use of the following decomposition.
\begin{equation}
    \frac{1}{\pab{s - s_{\txtA,1}} \pab{s - s_{\txtB,i}}} = \frac{1}{s_{\txtA,1} - s_{\txtB,i}} \pab{ \frac{1}{s - s_{\txtA,1}} - \frac{1}{s - s_{\txtB,i}} } \,, \qquad i = 3,4 \,.
\end{equation}
As for the inhomogeneous terms, we first let $\mathcal{I}_i(s) = \frac{1}{s} N_i(s), i=A_{\pm\bvec{k}}, B_\bvec{k}, B_\bvec{k}^*$ as in \cref{subapp:inhomogeneous-terms-in-s-domain}. Then, in $\mathbf{G}_{A_\bvec{k}}(s) \mathcal{I}_{A_\bvec{k}}(s)$, we use
\begin{equation}
    \frac{1}{s \pab{s - s_{\txtA,1}}} = - \frac{1}{s_{\txtA,1}} \pab{\frac{1}{s} - \frac{1}{s - s_{\txtA,1}}} \,.
\end{equation}
In the other two inhomogeneous terms associated with $\mathcal{I}_{B_\bvec{k}}$ and $\mathcal{I}_{B_\bvec{k}^*}$, we use
\begin{equation}
    \frac{1}{s \pab{s - s_{\txtA,1}} \pab{s - s_{\txtB,i}}} 
    = \frac{1}{s_{\txtA,1} s_{\txtB,i}} \frac{1}{s}
      + \frac{1}{\pab{ s_{\txtA,1} - s_{\txtB,i} } s_{\txtA,1} } \frac{1}{\pab{s - s_{\txtA,1}}}
      + \frac{1}{\pab{ s_{\txtB,i} - s_{\txtA,1} } s_{\txtB,i} } \frac{1}{\pab{s - s_{\txtB,i}}} \,.
    \label{eqn:G-A-G-B-I-B-decomposition}
\end{equation}
where $i=3,4$. In the two terms associated with $\mathcal{I}_{B_\bvec{k}}$ and $\mathcal{I}_{B_\bvec{k}^*}$, the numerators are of the form $\widetilde{\Sigma}(\cdot) \cdot \widetilde{\mathcal{N}}(\cdot) \sim \order{g^4}$. This means to keep contributions only up to $\order{g^2}$, we need to discard terms with $\order{1}$ factors in \cref{eqn:G-A-G-B-I-B-decomposition}. Given $s_{\txtA,1}\sim\order{g^2}$ and $s_{\txtB,3},s_{\txtB,4}\sim\order{1}$, the terms with pole $s_{\txtB,i}$, i.e., the last term in \cref{eqn:G-A-G-B-I-B-decomposition} should be ignored. $A_\bvec{k}(t)$ is obtained after we insert the above preparations into \cref{eqn:A_k-s-single-species}.
\begin{multline}
    A_\bvec{k}(t) = \bab{ A_\bvec{k}(0) + \frac{1}{\omega} \Re\Bab{ \pab{ \widetilde{\Sigma}(i\omega) - \widetilde{\Sigma}(-i\omega) e^{-2i\Omega t} } B_\bvec{k}(0^-) } } e^{-\Gamma t}
    \\
    + \bab{ \pab{1 - \frac{1}{\omega} \widetilde{\Sigma}_\txtR(i\omega)} \pab{2n(\omega) + 1} + \frac{2}{\omega} \widetilde{\mathcal{N}}_\txtI(i\omega) } \pab{1 - e^{-\Gamma t}}
    \label{eqn:A-k-t-single-species}
\end{multline}
where $\Omega$ and $\Gamma$ were defined in \cref{eqn:G-D-poles-single-species}. In this result we truncate $\widetilde{\Sigma}(\pm i\omega + \order{g^2})$ and $\widetilde{\mathcal{N}}(\pm i \omega + \order{g^2})$ to $\widetilde{\Sigma}(\pm i\omega)$ and $\mathcal{N}(\pm i \omega)$ with caveats discussed in \cref{eqn:Bose-delta} so as to be consistent with the order of perturbations of the equations of motion. The procedure to take the inverse Laplace transform is the same as that in \cref{sec:evolutions-of-field-amplitudes-and-their-mixing,sec:populations-and-coherence,app:symmetries-of-self-energy,app:symmetries-of-noise-kernel}. Thus, $\widetilde{\Sigma}(\pm i \omega)$ and $\widetilde{\mathcal{N}}(\pm i\omega)$ should be understood as $\widetilde{\Sigma}(\pm i \omega + \epsilon)$ and $\widetilde{\mathcal{N}}(\pm i\omega + \epsilon)$. Their decompositions (see \cref{eqn:Sigma-epsilon->0,eqn:noise-kernel-epsilon->0}) are used in the second line of \cref{eqn:A-k-t-single-species}. We used \cref{eqn:fluctuatio-dissipation-relation-s-domain} in the second line and recognized that $\coth\pab{\beta \omega / 2} = 2n(\omega) + 1$. In the first line of \cref{eqn:A-k-t-single-species}, we used the conjugate symmetry of $\widetilde{\Sigma}(\pm i\omega)$ (see \cref{eqn:Sigma-ab-omega-conjugate-symmetry}).

Following \cref{subsec:dynamics-of-the-vacuum-background-in-the-nearly-degenerate-case}, we write \cref{eqn:A-k-t-single-species} as $A_\bvec{k}(t) = A_\bvec{k}^{\text{(exc)}}(t) + A_\bvec{k}^{\text{(vac)}}(t)$, where
\begin{equation}
    A_\bvec{k}^{\text{(vac)}}(t) 
    = 1 + \frac{1}{\omega} \bab{ - \widetilde{\Sigma}_\txtR(i\omega) + 2 \widetilde{\mathcal{N}}_\txtI(i\omega) } \pab{1 - e^{-\Gamma t}}
    \label{eqn:A-k-t-vacuum-single-species}
\end{equation}
As discussed in \cref{subsec:dynamics-of-the-vacuum-background-in-the-nearly-degenerate-case}, the changing vacuum part implies a time-dependent vacuum background due to the interaction with the medium, and a time-dependent defintion of particles in the subsystem. Note the time-dependent term of the vacuum part is of order $\order{g^2}$ in \cref{eqn:A-k-t-vacuum-single-species} as opposed to of order $\order{1}$ in the nearly-degenerate case of a two-species subsystem. The single-species subsystem is more similar to the non-degenerate case of the two-species subsystem where the leading contribution of the vacuum part is constant. Similar to \cref{eqn:normalized-number-density}, the normalized population is
\begin{multline}
    \expval[big]{\widehat{\widetilde{N}}_\bvec{k}}(t) 
    = \frac{A_\bvec{k}^{\text{(exc)}}(t)}{2 A_\bvec{k}^{\text{(vac)}}(t)}
    = \bab[Big]{ 1 - \frac{1}{\omega} \widetilde{\Sigma}_\txtR(i\omega) e^{-\Gamma t} - \frac{2}{\omega} \widetilde{\mathcal{N}}_\txtI(i\omega) \pab{1 - e^{-\Gamma t}} } n(\omega) \pab{1 - e^{-\Gamma t}} +
    \\
    \bab[Big]{ \expval[big]{\widehat{N}_\bvec{k}}(0) \pab[Big]{ 1 + \frac{1}{\omega} \bab[Big]{ \widetilde{\Sigma}_\txtR(i\omega) - 2 \widetilde{\mathcal{N}}_\txtI(i\omega) } \pab{1 - e^{-\Gamma t}} } + \frac{1}{2\omega} \Re\Bab{ \pab{ \widetilde{\Sigma}(i\omega) - \widetilde{\Sigma}(-i\omega) e^{-2i\Omega t} } B_\bvec{k}(0^-) } } e^{-\Gamma t}
    \label{eqn:number-density-single-species}
\end{multline}
In this solution, it is clear that the initial value of $\expval*{\widehat{\widetilde{N}}_\bvec{k}}(t)$ decays and the population approaches to a thermal distribution in the long time limit.

Applying a similar procedure to the B-part yields the following solution.
\begin{multline}
    B_\bvec{k}(t) =
    \pab{ B_\bvec{k}(0^-) e^{-2i\Omega t} + \frac{1}{2\omega} \pab{A_{\bvec{k}}(0^-)+ A_{-\bvec{k}}(0^-)} \pab{e^{-2i\Omega t } \widetilde{\Sigma }(-i \omega)-\widetilde{\Sigma }(i \omega )} } e^{-\Gamma t}
    \\
    - \frac{2\widetilde{N}_\txtI(i\omega)}{\omega} \pab{1 + e^{-(2i\Omega + \Gamma)t}}
    + \frac{2n(\omega) + 1}{\omega}
    \pab{ i e^{-\Gamma t} \pab{1 - e^{-2i\Omega t}} \widetilde{\Sigma}_\txtI(i\omega) - \pab{1 - e^{-\Gamma t}} \widetilde{\Sigma}_\txtR(i\omega) } \,.
    \label{eqn:B-k-t-single-species}
\end{multline}
We only give $B_\bvec{k}(t)$ since $\bab*{B_\bvec{k}(t)}^* = B^*_\bvec{k}(t)$. We would like to emphasize that as discussed in \cref{subsec:power-counting-and-secular-contributions-of-solutions}, the $\order{g^2}$ contributions in \cref{eqn:A-k-t-single-species,eqn:A-k-t-vacuum-single-species,eqn:number-density-single-species,eqn:B-k-t-single-species} are not complete.

\section{Explicit results for one-point correlation functions}
\label{app:explicit-results-for-one-point-correlation-functions}

\subsection{Poles and residues in the Green's function of field amplitudes}
\label{subapp:pole-and-residues-in-the-Green's-function-of-field-amplitudes}

\cref{eqn:det-G-A^-1-s=0} can be solved formally by treating it as a bi-quadratic equation. The residues in $\mathbf{G}_\phi(t)$ can also be formally found after inserting the solutions into \cref{eqn:G-A-s-adj-det}. However, these formal results still depend on the unknown poles through the self-energy in them. Note that we only need the leading order results to be consistent with \cref{eqn:phi-k-dt2}. Using the zeroth order poles (see \cref{eqn:zero-order-poles-A}) as the input for the self-energy truncates the formal results to leading order, which are listed below.
\begin{equation}
\begin{aligned}
    s_{a_1} & = \eval{ - i \frac{\sqrt{ \Omega^2(s) + D(s)}}{\sqrt{2}} }_{s=s_{a_1}^{(0)}} \,,
    &\qquad&
    s_{a_1^\dagger} && = \eval{ i \frac{\sqrt{ \Omega^2(s) + D(s)}}{\sqrt{2}} }_{s=s_{a_1^\dagger}^{(0)}} \,,
    \\
    s_{a_2} & = \eval{ - i \frac{\sqrt{\Omega^2(s) - D(s)}}{\sqrt{2}} }_{s=s_{a_2}^{(0)}} \,,
    &\qquad&
    s_{a_2^\dagger} && = \eval{ i \frac{\sqrt{\Omega^2(s) - D(s)}}{\sqrt{2}} }_{s=s_{a_2^\dagger}^{(0)}} \,.
\end{aligned}
\label{eqn:A-poles-s}
\end{equation}
where we kept the subscripts in consistent with those in \cref{eqn:zero-order-poles-A} and defined
\begin{subequations}
\begin{align}
    \Omega^2(s) & \coloneq \omega_1^2 + \omega_2^2 + \Sigma_{11}(s) + \Sigma_{22}(s) \,,
    \\
    D(s) & \coloneq \sqrt{ \bab[big]{\Delta^2(s)}^2 + 4 \Sigma_{12}(s) \Sigma_{21}(s) } \,,
    \\
    \Delta^2(s) & \coloneq \omega_1^2 - \omega_2^2 + \Sigma_{11}(s) - \Sigma_{22}(s) \,.
\end{align}
\label{eqn:Omega^2-D-Delta^2-definitions}%
\end{subequations}
In \cref{eqn:A-poles-s} we singled out the imaginary unit "$i$" to ensure the consistency with the branch cut choice of the square root so that \cref{eqn:A-poles-s} reproduces \cref{eqn:zero-order-poles-A} when the couplings are switched off.

The residues corresponding to the poles are
\begin{equation}
\begin{aligned}
    \mathbb{G}_{\phi,a_1} & =
    \eval{
    \frac{1}{2 s_{a_1}}
    \begin{pmatrix}
         \frac{1}{2} + \frac{\Delta^2(s)}{2 D(s) }
        &
          \frac{\Sigma_{12}(s)}{ D(s) }
        \\[1em]
          \frac{\Sigma_{21}(s)}{ D(s) }
        &
         \frac{1}{2} - \frac{\Delta^2(s)}{2 D(s) }
    \end{pmatrix} }_{s=s_{a_1}^{(0)}}
    \,,
    \quad
    \mathbb{G}_{\phi,a_1^\dagger} & =
    \eval{
    \frac{1}{2 s_{a_1^\dagger}}
    \begin{pmatrix}
        \frac{1}{2} + \frac{\Delta^2(s)}{2 D(s) }
        &
        \frac{\Sigma_{12}(s)}{ D(s) }
        \\[1em]
        \frac{\Sigma_{21}(s)}{ D(s) }
        &
        \frac{1}{2} - \frac{\Delta^2(s)}{2 D(s) }
    \end{pmatrix} }_{s=s_{a_1^\dagger}^{(0)}}
    \,,
    \\
    \mathbb{G}_{\phi,a_2} & =
    \eval{
    \frac{1}{2 s_{a_2}}
    \begin{pmatrix}
        \frac{1}{2} - \frac{\Delta^2(s)}{2 D(s) }
        &
        \frac{ - \Sigma_{12}(s)}{ D(s) }
        \\[1em]
        \frac{-\Sigma_{21}(s)}{ D(s) }
        &
        \frac{1}{2} + \frac{  \Delta^2(s)}{2 D(s) }
    \end{pmatrix} }_{s=s_{a_2}^{(0)}}
    \,,
    \quad
    \mathbb{G}_{\phi,a_2^\dagger} & =
    \eval{
    \frac{1}{2 s_{a_2^\dagger}}
    \begin{pmatrix}
        \frac{1}{2} - \frac{\Delta^2(s)}{2 D(s) }
        &
        \frac{-\Sigma_{12}(s)}{ D(s) }
        \\[1em]
        \frac{-\Sigma_{21}(s)}{ D(s) }
        &
        \frac{1}{2} + \frac{\Delta^2(s)}{2 D(s) }
    \end{pmatrix} }_{s=s_{a_2^\dagger}^{(0)}}
    \,.
\end{aligned}
\label{eqn:residue-G-phi}%
\end{equation}

\subsection{Green's functions of dimensionless one-point correlation functions}
\label{subapp:Green's-functions-of-dimensionless-one-point-correlation-functions}

It has been shown in \cref{subsec:quasi-normal-modes-of-populations-and-coherence} that the evolutions of $\expval*{\boldsymbol{a}_{\bvec{k}}} = \bab{ \expval{a_{1,\bvec{k}}}, \expval{a_{2,\bvec{k}}}}^\T$ and $\expval*{\boldsymbol{a}_{\bvec{k}}^\dagger} = \bab*{ \expval*{a_{1,\bvec{k}}^\dagger}, \expval*{a_{2,\bvec{k}}}^\dagger}^\T$ are given by
\begin{subequations}
\begin{align}
    \expval{\boldsymbol{a}_\bvec{k}}(t)
    & = F_{a,a}[\mathbf{G}_\phi(t)] \expval{\boldsymbol{a}_{\bvec{k}}}(0^-)
      + F_{a,a^\dagger}[\mathbf{G}_\phi(t)] \expval*{\boldsymbol{a}_{\bvec{k}}^\dagger}(0^-) \,,
    \\
    \expval*{\boldsymbol{a}_\bvec{k}^\dagger}(t)
    & = F_{a^\dagger,a^\dagger}[\mathbf{G}_\phi(t)] \expval*{\boldsymbol{a}_{\bvec{k}}^\dagger}(0^-)
      + F_{a^\dagger,a}[\mathbf{G}_\phi(t)] \expval*{\boldsymbol{a}_{\bvec{k}}}(0^-) \,.
\end{align}
\end{subequations}
where
\begin{equation}
    U_\phi = 
    \begin{pmatrix}
        \frac{1}{\sqrt{2\omega_1}} & 0 \\
        0 & \frac{1}{\sqrt{2\omega_2}}
    \end{pmatrix}
    \,,
    \qquad
    U_\pi = 
    \begin{pmatrix}
        -i \sqrt{\frac{\omega_1}{2}} & 0 \\
        0 & -i \sqrt{\frac{\omega_2}{2}}
    \end{pmatrix}
    \,.
\end{equation}
\begin{subequations}
\begin{align}
    F_{a,a}[\mathbf{G}_\phi(t)]
    & = \frac{1}{2} \bab{ U_\phi^{-1} \dot{\mathbf{G}}_\phi(t) U_\phi + U_\pi^{-1} \dot{\mathbf{G}}_\phi(t) U_\pi + U_\phi^{-1} \mathbf{G}_\phi(t) U_\pi + U_\pi^{-1} \ddot{\mathbf{G}}_\phi(t) U_\phi} \,,
    \\
    F_{a,a^\dagger}[\mathbf{G}_\phi(t)]
    & = \frac{1}{2} \bab{ U_\phi^{-1} \dot{\mathbf{G}}_\phi(t) U_\phi - U_\pi^{-1} \dot{\mathbf{G}}_\phi U_\pi - U_\phi^{-1} \mathbf{G}_\phi(t) U_\pi + U_\pi^{-1} \ddot{\mathbf{G}}_\phi(t) U_\phi } \,,
    \\
    F_{a^\dagger,a^\dagger}[\mathbf{G}_\phi(t)]
    & =\frac{1}{2} \bab{ U_\phi^{-1} \dot{\mathbf{G}}_\phi(t) U_\phi + U_\pi^{-1} \dot{\mathbf{G}}_\phi(t) U_\pi - U_\phi^{-1} \mathbf{G}_\phi(t) U_\pi - U_\pi^{-1} \ddot{\mathbf{G}}_\phi(t) U_\phi} \,,
    \\
    F_{a^\dagger,a}[\mathbf{G}_\phi(t)]
    & =\frac{1}{2} \bab{ U_\phi^{-1} \dot{\mathbf{G}}_\phi(t) U_\phi - U_\pi^{-1} \dot{\mathbf{G}}_\phi U_\pi + U_\phi^{-1} \mathbf{G}_\phi(t) U_\pi - U_\pi^{-1} \ddot{\mathbf{G}}_\phi(t) U_\phi } \,.
\end{align}
\end{subequations}
Notice that $\mathbf{G}_\phi(t)$ is a sum of expotential functions. We can convert time derivatives to multiplications with poles and recognize quasinormal modes in the four ``$F$'' above by rewriting them in the following way.
\begin{equation}
    F_{m,n}[\mathbf{G}_\phi(t)] = \sum_{{i=a_1,a_1^\dagger,a_2,a_2^\dagger}} \mathbb{F}_{m,n}[\mathbb{G}_{\phi,i}] e^{s_i t} \,,
    \qquad\text{where}\; m,n = a, a^\dagger \,.
    \label{eqn:F-m-n-G-phi-t}
\end{equation}
where $\mathbb{G}_{\phi,i}$ are given in \cref{eqn:residue-G-phi}, $s_i$ are given in \cref{eqn:A-poles-s}, and the four ``$\mathbb{F}$'' are
\begin{subequations}
\begin{align}
    \mathbb{F}_{a,a}[\mathbb{G}_{\phi,i}]
    & = \frac{1}{2} \bab{ 
        U_\phi^{-1} \, s_i \mathbb{G}_{\phi,i} \, U_\phi 
        + U_\pi^{-1} \, s_i \mathbb{G}_{\phi,i} \, U_\pi 
        + U_\phi^{-1} \, \mathbb{G}_{\phi,i} \, U_\pi 
        + U_\pi^{-1} \, s_i^2 \mathbb{G}_{\phi,i} \, U_\phi
    } \,.
    \label{eqn:mathbb-F-aa}
    \\
        \mathbb{F}_{a,a^\dagger}[\mathbb{G}_{\phi,i}]
    & = \frac{1}{2} \bab{ 
        U_\phi^{-1} \, s_i \mathbb{G}_{\phi,i} \, U_\phi 
        - U_\pi^{-1} \, s_i \mathbb{G}_{\phi,i} \, U_\pi 
        - U_\phi^{-1} \, \mathbb{G}_{\phi,i} \, U_\pi 
        + U_\pi^{-1} \, s_i^2 \mathbb{G}_{\phi,i} \, U_\phi
    } \,,
    \label{eqn:mathbb-F-a-a^dagger}
    \\
    \mathbb{F}_{a^\dagger,a^\dagger}[\mathbb{G}_{\phi,i}]
    & = \frac{1}{2} \bab{ 
        U_\phi^{-1} \, s_i \mathbb{G}_{\phi,i} \, U_\phi 
        + U_\pi^{-1} \, s_i \mathbb{G}_{\phi,i} \, U_\pi 
        - U_\phi^{-1} \, \mathbb{G}_{\phi,i} \, U_\pi 
        - U_\pi^{-1} \, s_i^2 \mathbb{G}_{\phi,i} \, U_\phi
    } \,,
    \label{eqn:mathbb-F-a^dagger-a^dagger}
    \\
    \mathbb{F}_{a^\dagger,a}[\mathbb{G}_{\phi,i}]
    & = \frac{1}{2} \bab{ 
        U_\phi^{-1} \, s_i \mathbb{G}_{\phi,i} \, U_\phi 
        - U_\pi^{-1} \, s_i \mathbb{G}_{\phi,i} \, U_\pi 
        + U_\phi^{-1} \, \mathbb{G}_{\phi,i} \, U_\pi 
        - U_\pi^{-1} \, s_i^2 \mathbb{G}_{\phi,i} \, U_\phi
    } \,.
    \label{eqn:mathbb-F-a^dagger-a}
\end{align}
\label{eqn:mathbb-F}%
\end{subequations}
As shown in \cref{eqn:F-m-n-G-phi-t}, $\expval*{\boldsymbol{a}_\bvec{k}}, \expval*{\boldsymbol{a}_{\bvec{k}}^\dagger}$ have the same poles as the field amplitudes. We are particular interested in the approximations of \cref{eqn:mathbb-F} in both the non-degenerate and nearly-degenerate cases so as to compare the QNMs in \cref{eqn:F-m-n-G-phi-t} against the QNMs of the Green's functions of the equal-time two-point correlation functions, as discussed in \cref{subsec:quasi-normal-modes-of-populations-and-coherence}. The explicit results and their orders of perturbations are given in \cref{table:residues-of-qnm-of-a-a^dagger} where to keep notations simple, we define $M_{c,d}^{\pm\pm}(s)$ for results in the nearly-degenerate case.
\begin{equation}
    M_{cd}^{\pm\pm}(s) \coloneqq D(s) \pm \Delta^2(s) \pm 2 \Sigma_{cd}(s) \,.
\end{equation}
$M_{cd}^{\pm\pm}(s)$ is of order $\order{g^2}$ given $\omega_1^2 - \omega_2^2 \sim \Delta^{2}(s) \sim D(s) \sim \Sigma_{cd}(s) \sim \order{g^2}$.

The $\order{g^4}$ results in \cref{table:residues-of-qnm-of-a-a^dagger} are presented merely to show the order of perturbations of the leading contributions in these results. They may not contain all $\order{g^4}$ terms because our equation of motion is only derived up to the leading order perturbation. In our calculation, they should be ignored completely.

\afterpage{
\begingroup
\renewcommand{\arraystretch}{1.3}

\begin{landscape}
\begin{longtblr}[
    caption = {Results of $\mathbb{F}_{m,n}[\mathbb{G}_{\phi,l}],\; m,n = a,a^\dagger, \; l=a_1,a_1^\dagger,a_2,a_2^\dagger$},
    label = {table:residues-of-qnm-of-a-a^dagger},
]{
    colspec = {X[4,c,m]|X[15,c,m]|X[21.5,c,m]|X[3.5,c,m]},
    width = \linewidth,
    rowhead = 1, rowfoot = 0,
    column{1} = {mode=dmath},
    column{2} = {mode=dmath},
    column{3} = {mode=dmath},
    column{4} = {mode=dmath},
}
\hline
    \text{Residue} & \text{Non-degenerate Results} & \text{Nearly-degenerate Results} &  \text{Estimation}   \\
\hline
\mathbb{F}_{a,a}[\mathbb{G}_{\phi,a_1}]
    & 
    \left(
    \begin{array}{cc}
     1 & \frac{\Sigma _{12}\left(-i \omega _1\right)}{2 \left(\omega _1-\omega _2\right) \sqrt{\omega _1 \omega _2}} \\
     \frac{\Sigma _{21}\left(-i \omega _1\right)}{2 \left(\omega _1-\omega _2\right) \sqrt{\omega _1 \omega _2}} & 0 \\
    \end{array}
    \right)
    &
    \left(
    \begin{array}{cc}
     \frac{1}{2}+\frac{\Delta ^2\left(-i \omegabar\right)}{2 D\left(-i \omegabar\right)} & \frac{\Sigma _{12}\left(-i \omegabar\right)}{D\left(-i \omegabar\right)} \\
     \frac{\Sigma _{21}\left(-i \omegabar\right)}{D\left(-i \omegabar\right)} & \frac{1}{2}-\frac{\Delta ^2\left(-i \omegabar\right)}{2 D\left(-i \omegabar\right)} \\
    \end{array}
    \right)
    &
    \order{1}
\\
\mathbb{F}_{a,a}[\mathbb{G}_{\phi,a_1^\dagger}]
    & 
    \left(
    \begin{array}{cc}
     -\frac{\Sigma _{11}\left(i \omega _1\right){}^2}{16 \omega _1^4} & -\frac{\Sigma _{11}\left(i \omega _1\right) \Sigma _{12}\left(i \omega _1\right)}{8 \omega _1^{5/2} \sqrt{\omega _2} \left(\omega _1+\omega _2\right)} \\
     -\frac{\Sigma _{11}\left(i \omega _1\right) \Sigma _{21}\left(i \omega _1\right)}{8 \omega _1^{5/2} \sqrt{\omega _2} \left(\omega _1+\omega _2\right)} & -\frac{\Sigma _{12}\left(i \omega _1\right) \Sigma _{21}\left(i \omega _1\right)}{4 \omega _1 \omega _2 \left(\omega _1+\omega _2\right){}^2} \\
    \end{array}
    \right)
    &
    \left(
    \begin{array}{cc}
     -\left(\frac{1}{2}+\frac{\Delta ^2\left(i \omegabar\right)}{2 D\left(i \omegabar\right)}\right) \frac{M_{11}^{-+}\left(i \omegabar\right){}^2}{64 \omegabar^4} & - \frac{\Sigma _{12}\left(i \omegabar\right)}{D\left(i \omegabar\right)} \frac{M_{11}^{-+}\left(i \omegabar\right) M_{22}^{++}\left(i \omegabar\right)}{64 \omegabar^4} \\
     -\frac{\Sigma _{21}\left(i \omegabar\right)}{D\left(i \omegabar\right)} \frac{M_{11}^{-+}\left(i \omegabar\right) M_{22}^{++}\left(i \omegabar\right)}{64 \omegabar^4} & -\left(\frac{1}{2}-\frac{\Delta ^2\left(i \omegabar\right)}{2 D\left(i \omegabar\right)}\right) \frac{M_{22}^{++}\left(i \omegabar\right){}^2}{64 \omegabar^4} \\
    \end{array}
    \right)
    &
    \order{g^4}
\\
\mathbb{F}_{a,a}[\mathbb{G}_{\phi,a_2}]
    & 
    \left(
    \begin{array}{cc}
     0 & -\frac{\Sigma _{12}\left(-i \omega _2\right)}{2 \left(\omega _1-\omega _2\right) \sqrt{\omega _1 \omega _2}} \\
     -\frac{\Sigma _{21}\left(-i \omega _2\right)}{2 \left(\omega _1-\omega _2\right) \sqrt{\omega _1 \omega _2}} & 1 \\
    \end{array}
    \right)
    &
    \left(
    \begin{array}{cc}
     \frac{1}{2}-\frac{\Delta ^2\left(-i \omegabar\right)}{2 D\left(-i \omegabar\right)} & -\frac{\Sigma _{12}\left(-i \omegabar\right)}{D\left(-i \omegabar\right)} \\
     -\frac{\Sigma _{21}\left(-i \omegabar\right)}{D\left(-i \omegabar\right)} & \frac{1}{2}+\frac{\Delta ^2\left(-i \omegabar\right)}{2 D\left(-i \omegabar\right)} \\
    \end{array}
    \right)
    &
    \order{1}
\\
\mathbb{F}_{a,a}[\mathbb{G}_{\phi,a_2^\dagger}]
    & 
    \left(
    \begin{array}{cc}
     -\frac{\Sigma _{12}\left(i \omega _2\right) \Sigma _{21}\left(i \omega _2\right)}{4 \omega _1 \omega _2 \left(\omega _1+\omega _2\right){}^2} & -\frac{\Sigma _{12}\left(i \omega _2\right) \Sigma _{22}\left(i \omega _2\right)}{8 \sqrt{\omega _1} \omega _2^{5/2} \left(\omega _1+\omega _2\right)} \\
     -\frac{\Sigma _{21}\left(i \omega _2\right) \Sigma _{22}\left(i \omega _2\right)}{8 \sqrt{\omega _1} \omega _2^{5/2} \left(\omega _1+\omega _2\right)} & -\frac{\Sigma _{22}\left(i \omega _2\right){}^2}{16 \omega _2^4} \\
    \end{array}
    \right)
    &
    \left(
    \begin{array}{cc}
    -\left(\frac{1}{2}-\frac{\Delta ^2\left(i \omegabar\right)}{2 D\left(i \omegabar\right)}\right) \frac{M_{11}^{+-}\left(i \omegabar\right){}^2}{64 \omegabar^4} & \frac{\Sigma _{12}\left(i \omegabar\right)}{D\left(i \omegabar\right)} \frac{M_{11}^{+-}\left(i \omegabar\right) M_{22}^{--}\left(i \omegabar\right)}{64 \omegabar^4} \\
    \frac{\Sigma _{21}\left(i \omegabar\right)}{D\left(i \omegabar\right)} \frac{M_{11}^{+-}\left(i \omegabar\right) M_{22}^{--}\left(i \omegabar\right)}{64 \omegabar^4} & -\left(\frac{1}{2}+\frac{\Delta ^2\left(i \omegabar\right)}{2 D\left(i \omegabar\right)}\right) \frac{M_{22}^{--}\left(i \omegabar\right){}^2}{64 \omegabar^4} \\
    \end{array}
    \right)
    &
    \order{g^4}
\\
\hline
\mathbb{F}_{a,a^\dagger}[\mathbb{G}_{\phi,a_1}]
    &
    \left(
    \begin{array}{cc}
     \frac{\Sigma _{11}\left(-i \omega _1\right)}{4 \omega _1^2} & \frac{\Sigma _{12}\left(-i \omega _1\right)}{2 \sqrt{\omega _1 \omega _2} \left(\omega _1+\omega _2\right)} \\
     0 & 0 \\
    \end{array}
    \right)
    &
    \left(
    \begin{array}{cc}
     \left(\frac{1}{2}+\frac{\Delta ^2\left(-i \omegabar\right)}{2 D\left(-i \omegabar\right)}\right) \frac{M_{11}^{-+}\left(-i \omegabar\right)}{8 \omegabar^2} & \frac{M_{22}^{++}\left(-i \omegabar\right)}{8 \omegabar^2} \frac{\Sigma _{12}\left(-i \omegabar\right)}{D\left(-i \omegabar\right)} \\
     \frac{M_{11}^{-+}\left(-i \omegabar\right)}{8 \omegabar^2} \frac{\Sigma _{21}\left(-i \omegabar\right)}{D\left(-i \omegabar\right)} & \left(\frac{1}{2}-\frac{\Delta ^2\left(-i \omegabar\right)}{2 D\left(-i \omegabar\right)}\right) \frac{M_{22}^{++}\left(-i \omegabar\right)}{8 \omegabar^2} \\
    \end{array}
    \right)
    &
    \order{g^2}
\\
\mathbb{F}_{a,a^\dagger}[\mathbb{G}_{\phi,a_1^\dagger}]
    &
    \left(
    \begin{array}{cc}
     -\frac{\Sigma _{11}\left(i \omega _1\right)}{4 \omega _1^2} & 0 \\
     -\frac{\Sigma _{21}\left(i \omega _1\right)}{2 \sqrt{\omega _1 \omega _2} \left(\omega _1+\omega _2\right)} & 0 \\
    \end{array}
    \right)
    &
    \left(
    \begin{array}{cc}
     -\left(\frac{1}{2}+\frac{\Delta ^2\left(i \omegabar\right)}{2 D\left(i \omegabar\right)}\right) \frac{M_{11}^{-+}\left(i \omegabar\right)}{8 \omegabar^2} & -\frac{M_{11}^{-+}\left(i \omegabar\right)}{8 \omegabar^2} \frac{\Sigma _{12}\left(i \omegabar\right)}{D\left(i \omegabar\right)} \\
     -\frac{M_{22}^{++}\left(i \omegabar\right)}{8 \omegabar^2} \frac{\Sigma _{21}\left(i \omegabar\right)}{D\left(i \omegabar\right)} & -\left(\frac{1}{2}-\frac{\Delta ^2\left(i \omegabar\right)}{2 D\left(i \omegabar\right)}\right) \frac{M_{22}^{++}\left(i \omegabar\right)}{8 \omegabar^2} \\
    \end{array}
    \right)
    &
    \order{g^2}
\\
\mathbb{F}_{a,a^\dagger}[\mathbb{G}_{\phi,a_2}]
    &
    \left(
    \begin{array}{cc}
     0 & 0 \\
     \frac{\Sigma _{21}\left(-i \omega _2\right)}{2 \sqrt{\omega _1 \omega _2} \left(\omega _1+\omega _2\right)} & \frac{\Sigma _{22}\left(-i \omega _2\right)}{4 \omega _2^2} \\
    \end{array}
    \right)
    &
    \left(
    \begin{array}{cc}
     -\left(\frac{1}{2}-\frac{\Delta ^2\left(-i \omegabar\right)}{2 D\left(-i \omegabar\right)}\right) \frac{M_{11}^{+-}\left(-i \omegabar\right)}{8 \omegabar^2} & \frac{M_{22}^{--}\left(-i \omegabar\right)}{8 \omegabar^2} \frac{\Sigma _{12}\left(-i \omegabar\right)}{D\left(-i \omegabar\right)} \\
     \frac{M_{11}^{+-}\left(-i \omegabar\right)}{8 \omegabar^2} \frac{\Sigma _{21}\left(-i \omegabar\right)}{D\left(-i \omegabar\right)} & -\left(\frac{1}{2}+\frac{\Delta ^2\left(-i \omegabar\right)}{2 D\left(-i \omegabar\right)}\right) \frac{M_{22}^{--}\left(-i \omegabar\right)}{8 \omegabar^2} \\
    \end{array}
    \right)
    &
    \order{g^2}
\\
\mathbb{F}_{a,a^\dagger}[\mathbb{G}_{\phi,a_2^\dagger}]
    &
    \left(
    \begin{array}{cc}
     0 & -\frac{\Sigma _{12}\left(i \omega _2\right)}{2 \sqrt{\omega _1 \omega _2} \left(\omega _1+\omega _2\right)} \\
     0 & -\frac{\Sigma _{22}\left(i \omega _2\right)}{4 \omega _2^2} \\
    \end{array}
    \right)
    &
    \left(
    \begin{array}{cc}
     \left(\frac{1}{2}-\frac{\Delta ^2\left(i \omegabar\right)}{2 D\left(i \omegabar\right)}\right) \frac{M_{11}^{+-}\left(i \omegabar\right)}{8 \omegabar^2} & -\frac{M_{11}^{+-}\left(i \omegabar\right)}{8 \omegabar^2} \frac{\Sigma _{12}\left(i \omegabar\right)}{D\left(i \omegabar\right)} \\
     -\frac{M_{22}^{--}\left(i \omegabar\right)}{8 \omegabar^2} \frac{\Sigma _{21}\left(i \omegabar\right)}{D\left(i \omegabar\right)} & \left(\frac{1}{2}+\frac{\Delta ^2\left(i \omegabar\right)}{2 D\left(i \omegabar\right)}\right) \frac{M_{22}^{--}\left(i \omegabar\right)}{8 \omegabar^2} \\
    \end{array}
    \right)
    &
    \order{g^2}
\\
\hline
\mathbb{F}_{a^\dagger,a^\dagger}[\mathbb{G}_{\phi,a_1}]
    &
    \left(
    \begin{array}{cc}
     -\frac{\Sigma _{11}\left(-i \omega _1\right){}^2}{16 \omega _1^4} & -\frac{\Sigma _{11}\left(-i \omega _1\right) \Sigma _{12}\left(-i \omega _1\right)}{8 \omega _1^{5/2} \sqrt{\omega _2} \left(\omega _1+\omega _2\right)} \\
     -\frac{\Sigma _{11}\left(-i \omega _1\right) \Sigma _{21}\left(-i \omega _1\right)}{8 \omega _1^{5/2} \sqrt{\omega _2} \left(\omega _1+\omega _2\right)} & -\frac{\Sigma _{12}\left(-i \omega _1\right) \Sigma _{21}\left(-i \omega _1\right)}{4 \omega _1 \omega _2 \left(\omega _1+\omega _2\right){}^2} \\
    \end{array}
    \right)
    &
    \left(
    \begin{array}{cc}
     -\left(\frac{1}{2}+\frac{\Delta ^2\left(-i \omegabar\right)}{2 D\left(-i \omegabar\right)}\right) \frac{M_{11}^{-+}\left(-i \omegabar\right){}^2}{64 \omegabar^4} & -\frac{\Sigma _{12}\left(-i \omegabar\right)}{D\left(-i \omegabar\right)} \frac{M_{11}^{-+}\left(-i \omegabar\right) M_{22}^{++}\left(-i \omegabar\right)}{64 \omegabar^4} \\
     -\frac{\Sigma _{21}\left(-i \omegabar\right)}{D\left(-i \omegabar\right)} \frac{M_{11}^{-+}\left(-i \omegabar\right) M_{22}^{++}\left(-i \omegabar\right)}{64 \omegabar^4} & -\left(\frac{1}{2}-\frac{\Delta ^2\left(-i \omegabar\right)}{2 D\left(-i \omegabar\right)}\right) \frac{M_{22}^{++}\left(-i \omegabar\right){}^2}{64 \omegabar^4} \\
    \end{array}
    \right)
    &
    \order{g^4}
\\
\mathbb{F}_{a^\dagger,a^\dagger}[\mathbb{G}_{\phi,a_1^\dagger}]
    &
    \left(
    \begin{array}{cc}
     1 & \frac{\Sigma _{12}\left(i \omega _1\right)}{2 \left(\omega _1-\omega _2\right) \sqrt{\omega _1 \omega _2}} \\
     \frac{\Sigma _{21}\left(i \omega _1\right)}{2 \left(\omega _1-\omega _2\right) \sqrt{\omega _1 \omega _2}} & 0 \\
    \end{array}
    \right)
    &
    \left(
    \begin{array}{cc}
     \frac{1}{2}+\frac{\Delta ^2\left(i \omegabar\right)}{2 D\left(i \omegabar\right)} & \frac{\Sigma _{12}\left(i \omegabar\right)}{D\left(i \omegabar\right)} \\
     \frac{\Sigma _{21}\left(i \omegabar\right)}{D\left(i \omegabar\right)} & \frac{1}{2}-\frac{\Delta ^2\left(i \omegabar\right)}{2 D\left(i \omegabar\right)} \\
    \end{array}
    \right)
    &
    \order{1}
\\
\mathbb{F}_{a^\dagger,a^\dagger}[\mathbb{G}_{\phi,a_2}]
    &
    \left(
    \begin{array}{cc}
     -\frac{\Sigma _{12}\left(-i \omega _2\right) \Sigma _{21}\left(-i \omega _2\right)}{4 \omega _1 \omega _2 \left(\omega _1+\omega _2\right){}^2} & -\frac{\Sigma _{12}\left(-i \omega _2\right) \Sigma _{22}\left(-i \omega _2\right)}{8 \sqrt{\omega _1} \omega _2^{5/2} \left(\omega _1+\omega _2\right)} \\
     -\frac{\Sigma _{21}\left(-i \omega _2\right) \Sigma _{22}\left(-i \omega _2\right)}{8 \sqrt{\omega _1} \omega _2^{5/2} \left(\omega _1+\omega _2\right)} & -\frac{\Sigma _{22}\left(-i \omega _2\right){}^2}{16 \omega _2^4} \\
    \end{array}
    \right)
    &
    \left(
    \begin{array}{cc}
     -\left(\frac{1}{2}-\frac{\Delta ^2\left(-i \omegabar\right)}{2 D\left(-i \omegabar\right)}\right) \frac{M_{11}^{+-}\left(-i \omegabar\right){}^2}{64 \omegabar^4} & \frac{\Sigma _{12}\left(-i \omegabar\right)}{D\left(-i \omegabar\right)} \frac{M_{11}^{+-}\left(-i \omegabar\right) M_{22}^{--}\left(-i \omegabar\right)}{64 \omegabar^4} \\
     \frac{\Sigma _{21}\left(-i \omegabar\right)}{D\left(-i \omegabar\right)} \frac{M_{11}^{+-}\left(-i \omegabar\right) M_{22}^{--}\left(-i \omegabar\right)}{64 \omegabar^4} & -\left(\frac{1}{2}+\frac{\Delta ^2\left(-i \omegabar\right)}{2 D\left(-i \omegabar\right)}\right) \frac{M_{22}^{--}\left(-i \omegabar\right){}^2}{64 \omegabar^4} \\
    \end{array}
    \right)
    &
    \order{g^4}
\\
\mathbb{F}_{a^\dagger,a^\dagger}[\mathbb{G}_{\phi,a_2^\dagger}]
    &
    \left(
    \begin{array}{cc}
     0 & -\frac{\Sigma _{12}\left(i \omega _2\right)}{2 \left(\omega _1-\omega _2\right) \sqrt{\omega _1 \omega _2}} \\
     -\frac{\Sigma _{21}\left(i \omega _2\right)}{2 \left(\omega _1-\omega _2\right) \sqrt{\omega _1 \omega _2}} & 1 \\
    \end{array}
    \right)
    &
    \left(
    \begin{array}{cc}
     \frac{1}{2}-\frac{\Delta ^2\left(i \omegabar\right)}{2 D\left(i \omegabar\right)} & -\frac{\Sigma _{12}\left(i \omegabar\right)}{D\left(i \omegabar\right)} \\
     -\frac{\Sigma _{21}\left(i \omegabar\right)}{D\left(i \omegabar\right)} & \frac{1}{2}+\frac{\Delta ^2\left(i \omegabar\right)}{2 D\left(i \omegabar\right)} \\
    \end{array}
    \right)
    &
    \order{1}
\\
\hline
\mathbb{F}_{a^\dagger,a}[\mathbb{G}_{\phi,a_1}]
    &
    \left(
    \begin{array}{cc}
     -\frac{\Sigma _{11}\left(-i \omega _1\right)}{4 \omega _1^2} & 0 \\
     -\frac{\Sigma _{21}\left(-i \omega _1\right)}{2 \sqrt{\omega _1 \omega _2} \left(\omega _1+\omega _2\right)} & 0 \\
    \end{array}
    \right)
    &
    \left(
    \begin{array}{cc}
     -\left(\frac{1}{2}+\frac{\Delta ^2\left(-i \omegabar\right)}{2 D\left(-i \omegabar\right)}\right) \frac{M_{11}^{-+}\left(-i \omegabar\right)}{8 \omegabar^2} & -\frac{M_{11}^{-+}\left(-i \omegabar\right)}{8 \omegabar^2} \frac{\Sigma _{12}\left(-i \omegabar\right)}{D\left(-i \omegabar\right)} \\
     -\frac{M_{22}^{++}\left(-i \omegabar\right)}{8 \omegabar^2} \frac{\Sigma _{21}\left(-i \omegabar\right)}{D\left(-i \omegabar\right)} & -\left(\frac{1}{2}-\frac{\Delta ^2\left(-i \omegabar\right)}{2 D\left(-i \omegabar\right)}\right) \frac{M_{22}^{++}\left(-i \omegabar\right)}{8 \omegabar^2} \\
    \end{array}
    \right)
    &
    \order{g^2}
\\
\mathbb{F}_{a^\dagger,a}[\mathbb{G}_{\phi,a_1^\dagger}]
    &
    \left(
    \begin{array}{cc}
     \frac{\Sigma _{11}\left(i \omega _1\right)}{4 \omega _1^2} & \frac{\Sigma _{12}\left(i \omega _1\right)}{2 \sqrt{\omega _1 \omega _2} \left(\omega _1+\omega _2\right)} \\
     0 & 0 \\
    \end{array}
    \right)
    &
    \left(
    \begin{array}{cc}
     \left(\frac{1}{2}+\frac{\Delta ^2\left(i \omegabar\right)}{2 D\left(i \omegabar\right)}\right) \frac{M_{11}^{-+}\left(i \omegabar\right)}{8 \omegabar^2} & \frac{M_{22}^{++}\left(i \omegabar\right)}{8 \omegabar^2} \frac{\Sigma _{12}\left(i \omegabar\right)}{D\left(i \omegabar\right)} \\
     \frac{M_{11}^{-+}\left(i \omegabar\right)}{8 \omegabar^2} \frac{\Sigma _{21}\left(i \omegabar\right)}{D\left(i \omegabar\right)} & \left(\frac{1}{2}-\frac{\Delta ^2\left(i \omegabar\right)}{2 D\left(i \omegabar\right)}\right) \frac{M_{22}^{++}\left(i \omegabar\right)}{8 \omegabar^2} \\
    \end{array}
    \right)
    &
    \order{g^2}
\\
\mathbb{F}_{a^\dagger,a}[\mathbb{G}_{\phi,a_2}]
    &
    \left(
    \begin{array}{cc}
     0 & -\frac{\Sigma _{12}\left(-i \omega _2\right)}{2 \sqrt{\omega _1 \omega _2} \left(\omega _1+\omega _2\right)} \\
     0 & -\frac{\Sigma _{22}\left(-i \omega _2\right)}{4 \omega _2^2} \\
    \end{array}
    \right)
    &
    \left(
    \begin{array}{cc}
     \left(\frac{1}{2}-\frac{\Delta ^2\left(-i \omegabar\right)}{2 D\left(-i \omegabar\right)}\right) \frac{M_{11}^{+-}\left(-i \omegabar\right)}{8 \omegabar^2} & -\frac{M_{11}^{+-}\left(-i \omegabar\right)}{8 \omegabar^2} \frac{\Sigma _{12}\left(-i \omegabar\right)}{D\left(-i \omegabar\right)} \\
     -\frac{M_{22}^{--}\left(-i \omegabar\right)}{8 \omegabar^2} \frac{\Sigma _{21}\left(-i \omegabar\right)}{D\left(-i \omegabar\right)} & \left(\frac{1}{2}+\frac{\Delta ^2\left(-i \omegabar\right)}{2 D\left(-i \omegabar\right)}\right) \frac{M_{22}^{--}\left(-i \omegabar\right)}{8 \omegabar^2} \\
    \end{array}
    \right)
    &
    \order{g^2}
\\
\mathbb{F}_{a^\dagger,a}[\mathbb{G}_{\phi,a_2^\dagger}]
    &
    \left(
    \begin{array}{cc}
     0 & 0 \\
     \frac{\Sigma _{21}\left(i \omega _2\right)}{2 \sqrt{\omega _1 \omega _2} \left(\omega _1+\omega _2\right)} & \frac{\Sigma _{22}\left(i \omega _2\right)}{4 \omega _2^2} \\
    \end{array}
    \right)
    &
    \left(
    \begin{array}{cc}
     -\left(\frac{1}{2}-\frac{\Delta ^2\left(i \omegabar\right)}{2 D\left(i \omegabar\right)}\right) \frac{M_{11}^{+-}\left(i \omegabar\right)}{8 \omegabar^2} & \frac{M_{22}^{--}\left(i \omegabar\right)}{8 \omegabar^2} \frac{\Sigma _{12}\left(i \omegabar\right)}{D\left(i \omegabar\right)} \\
     \frac{M_{11}^{+-}\left(i \omegabar\right)}{8 \omegabar^2} \frac{\Sigma _{21}\left(i \omegabar\right)}{D\left(i \omegabar\right)} & -\left(\frac{1}{2}+\frac{\Delta ^2\left(i \omegabar\right)}{2 D\left(i \omegabar\right)}\right) \frac{M_{22}^{--}\left(i \omegabar\right)}{8 \omegabar^2} \\
    \end{array}
    \right)
    &
    \order{g^2}
\\
\hline
\end{longtblr}
\end{landscape}

\endgroup
}

\section{Explicit results of populations and coherence}
\label{app:explicit-results-in-evolutions-of-number-densities-and-coherence}

\subsection{Equations of motion for equal-time two-point correlations functions}
\label{subapp:equations-of-motion-two-point-correlations}

The equations of $A_{cd,\bvec{k}}$ and $B_{cd,\bvec{k}}$ are obtained by replacing $A_{db,\bvec{k}}^*$ in \cref{eqn:A-cd-k-dt} and $A_{db,-\bvec{k}}^*$ and $A_{cb,\bvec{k}}^*$ in \cref{eqn:B-cd-k-dt} using the relations in \cref{eqn:A-k-B-k-symmetries}. 
\begin{align}
    \diff**{t}{A_{cd,\bvec{k}}}(t) =
    & i \pab{\omega_{c,\bvec{k}} - \omega_{d,\bvec{k}}} A_{cd,\bvec{k}}(t)
    - i \sum_{a,b=1,2} \int_0^t \Big[ -\delta_{ca} \pab{ B_{db,\bvec{k}}(t') + A_{bd,\bvec{k}}(t') } e^{-i\omega_{d,\bvec{k}}(t-t')} 
    \nonumber \\*
    & + \delta_{da} \pab{ A_{cb,\bvec{k}}(t') + B_{cb,\bvec{k}}^*(t') } e^{i\omega_{c,\bvec{k}}(t-t')} \Big] \widetilde{\Sigma}_{ab}(\abs{\bvec{k}},t-t') \, \dn{t'}
    + \mathcal{I}_{A_\bvec{k},cd}(t) \,.
\label{eqn:A-k-equation}
    \\
    \diff**{t}{B_{cd,\bvec{k}}(t)} =
    & -i(\omega_{c,\bvec{k}} + \omega_{d,\bvec{k}}) B_{cd,\bvec{k}}(t)
    -i\sum_{a,b=1,2} \int_0^t \Big[ \delta_{ca} \pab{B_{bd,\bvec{k}}(t') + A_{bd,-\bvec{k}}(t')} e^{-i\omega_d(t-t')}
    \nonumber \\*
    & + \delta_{da} \pab{ B_{cb,\bvec{k}}(t') + A_{bc,\bvec{k}}(t') } e^{-i\omega_c(t-t')} \Big] \widetilde{\Sigma}_{ab}(\abs{\bvec{k}},t-t') \, \dn{t'}
    + \mathcal{I}_{B_\bvec{k},cd}(t) \,.
\label{eqn:B-k-equation}
\end{align}
Relabelling $\bvec{k} \to -\bvec{k}$ in \cref{eqn:A-k-equation} and using \cref{eqn:A-k-B-k-symmetries} yield the equation of $A_{cd,-\bvec{k}}$.
\begin{multline}
    \diff**{t}{A_{cd,-\bvec{k}}}(t) = 
    i(\omega_{c,\bvec{k}} - \omega_{d,\bvec{k}}) A_{cd,-\bvec{k}}(t)
    - i \sum_{a,b=1,2} \int_0^t \Big[ -\delta_{ca} \pab{ B_{bd,\bvec{k}}(t') + A_{bd,-\bvec{k}}(t') } e^{-i\omega_{d,\bvec{k}}(t-t')}  
    \\*
    + \delta_{da} \pab{ A_{cb,-\bvec{k}}(t') + B_{bc,\bvec{k}}^*(t') } e^{i\omega_{c,\bvec{k}}(t-t')} \Big] \widetilde{\Sigma}_{ab}(\abs{\bvec{k}},t-t') \, \dn{t'}
    + \mathcal{I}_{A_{-\bvec{k}},cd}(t)
    \,.
\label{eqn:A-(-k)-equation}
\end{multline}
Taking the complex conjugate of \cref{eqn:B-k-equation} and using \cref{eqn:A-k-B-k-symmetries} yield the equation of $B_{cd,\bvec{k}}^*$.
\begin{multline}
    \diff**{t}{B_{cd,\bvec{k}}^*}(t) = 
    i (\omega_{c,\bvec{k}} + \omega_{d,\bvec{k}}) B_{cd,\bvec{k}}^*(t)
    + i \sum_{a,b=1,2} \int_0^t \Big[ \delta_{ca} \pab{ B_{bd,\bvec{k}}^*(t') + A_{db,-\bvec{k}}(t') } e^{i\omega_{d,\bvec{k}}(t-t')} 
    \\*
    + \delta_{da} \pab{ B_{cb,\bvec{k}}^*(t') + A_{cb,\bvec{k}}(t') } e^{i\omega_{c,\bvec{k}}(t-t')} \Big] \widetilde{\Sigma}_{ab}(\abs{\bvec{k}},t-t') \, \dn{t'}
    + \mathcal{I}_{B_\bvec{k}^*,cd}(t)
    \,.
\label{eqn:B-k^*-equation}
\end{multline}
The inhomogeneous terms in the above four equationa are
\begin{subequations}
\begin{align}
    \mathcal{I}_{A_\bvec{k},cd}(t) 
    & = + \sum_{a,b=1,2} \int_0^t \dn{t'} 2 \bab{ \delta_{ca} \delta_{db} e^{-i\omega_{b,\bvec{k}} t'} + \delta_{da} \delta_{cb} e^{i\omega_{b,\bvec{k}} t'} } \widetilde{\mathcal{N}}_{ab}(\abs{\bvec{k}},t') \,,
    \label{eqn:I-A-k-cd}
    \\
    \mathcal{I}_{B_\bvec{k},cd}(t) 
    & = - \sum_{a,b=1,2} \int_0^t \dn{t'} 2 \bab{\delta_{da} \delta_{cb} + \delta_{ca} \delta_{db}} e^{-i\omega_b t'} \widetilde{\mathcal{N}}_{ab}(\abs{\bvec{k}},t') \,,
    \label{eqn:I-B-k-cd}
    \\
    \mathcal{I}_{A_{-\bvec{k}},cd}(t)
    & = + \sum_{a,b=1,2} \int_0^t \dn{t'} 2 \bab{ \delta_{ca} \delta_{db} e^{-i\omega_{b,\bvec{k}}t'} + \delta_{da} \delta_{cb} e^{i\omega_{b,\bvec{k}}t'} } \widetilde{\mathcal{N}}_{ab}(\abs{\bvec{k}},t') \,,
    \label{eqn:I-A-(-k)-cd}
    \\
    \mathcal{I}_{B_\bvec{k}^*,cd}(t) 
    & = - \sum_{a,b=1,2} \int_0^t \dn{t'} 2 \bab{ \delta_{da} \delta_{cb} + \delta_{ca} \delta_{db} } e^{i\omega_{b,\bvec{k}}t'} \widetilde{\mathcal{N}}_{ab}(\abs{\bvec{k}},t') \,.
    \label{eqn:I-B-k^*-cd}
\end{align}
\label{eqn:inhomogeneous-terms-in-time}%
\end{subequations}
In \cref{eqn:B-k^*-equation,eqn:I-B-k^*-cd}, we used that $\Sigma_{ab}(\bvec{k},t)$ and $\mathcal{N}_{ab}(\bvec{k},t)$ are real (see \cref{app:symmetries-of-self-energy,app:symmetries-of-noise-kernel}). It is easy to check that only $A_{cd,\bvec{k}}, A_{cd,-\bvec{k}}, B_{cd,\bvec{k}}, B_{cd,\bvec{k}}^*$ appear on the right hand side of the above equations. Therefore, the above four matrix equations are closed.

\subsection{Green's functions in s-domain}
\label{subapp:Greens-function-in-s-domain}

The order we used to list the equal-time two-point correlation functions in $\overrightarrow{\mathcal{D}}(t)$ (see \cref{eqn:A-B-separation}) also fixes the positions of nonzero elements in $\mathbf{\Omega}$ and $\mathbf{K}$ of \cref{eqn:16-equation-s-domain}, which can be obtained by comparing \cref{eqn:16-equation-s-domain,eqn:A-B-separation} and the four matrix equations in \cref{subapp:equations-of-motion-two-point-correlations}. In what follows, we drop the ``$\bvec{k}$'' in $\omega_{c,\bvec{k}}$ and $\widetilde{\Sigma}_{cd}(\bvec{k},s)$ to keep notations simple, since they only depend on the magnitude of $\bvec{k}$ and only one $\abs{\bvec{k}}$ is involved in the following results.

The matrix $\mathbf{\Omega}$ is diagonal and includes the bare frequencies of the correlation functions.
\begin{equation}
    \mathbf{\Omega} = \operatorname{diag}\pab[big]{ \mathbf{\Omega}_{A_\bvec{k}},\; \mathbf{\Omega}_{A_{-\bvec{k}}},\; \mathbf{\Omega}_{B_\bvec{k}},\; \mathbf{\Omega}_{B_\bvec{k}^*} } \,.
\end{equation}
where
\begin{align*}
    \mathbf{\Omega}_{A_\bvec{k}} 
    & = \operatorname{diag}\pab{0,\; \omega _1-\omega _2,\; \omega _2-\omega _1,\; 0}\,,
    &
    \mathbf{\Omega}_{B_\bvec{k}}
    & = \operatorname{diag}\pab{-2 \omega _1,\; -\omega _1-\omega _2,\; -\omega _1-\omega _2,\; -2 \omega _2} \,,
    \\
    \mathbf{\Omega}_{A_{-\bvec{k}}}
    & = \operatorname{diag}\pab{0,\; \omega _1-\omega _2,\; \omega _2-\omega _1,\; 0} \,,
    &
    \mathbf{\Omega}_{B_\bvec{k}^*}
    & = \operatorname{diag}\pab{2 \omega _1,\; \omega _1+\omega _2,\; \omega _1+\omega _2,\; 2 \omega _2} \,.
\end{align*}

Following the notations in \cref{subsec:quasi-normal-modes-of-populations-and-coherence}, we write the $16\times16$ matrix $\mathbf{K}$ block-wisely.
\begin{equation}
    \mathbf{K} = 
    \begin{pmatrix}
        \mathbf{K}_{\text{AA}} & \mathbf{K}_{\text{AB}} \\
        \mathbf{K}_{\text{BA}} & \mathbf{K}_{\text{BB}}
    \end{pmatrix}
    \,,\quad\text{where}\quad
    \mathbf{K}_{\text{AA}} = 
    \begin{pmatrix}
        \mathbf{K}_{A_{\bvec{k}}} & 0 \\
        0 & \mathbf{K}_{A_{-\bvec{k}}}
    \end{pmatrix}
    \,,\quad
    \mathbf{K}_{\text{BB}} = 
    \begin{pmatrix}
        \mathbf{K}_{B_\bvec{k}} & 0 \\
        0 & \mathbf{K}_{B_\bvec{k}^*}
    \end{pmatrix}
    \,.
    \label{eqn:K-blocks}
\end{equation}
$\mathbf{K}_{\text{AB}}$ and $\mathbf{K}_{\text{BA}}$ are $8\times8$ matrices and $\mathbf{K}_{A_{\bvec{k}}},\mathbf{K}_{A_{-\bvec{k}}},\mathbf{K}_{B_\bvec{k}},\mathbf{K}_{B_\bvec{k}^*}$ are $4\times4$ matrices. Their explicit expressions are given in \cref{table:blocks-of-G-D-inverse}. All of them are of order $\order{g^2}$ ($\sim \Sigma_{cd}(k)$). Given $\mathbf{G}_\mathcal{D}^{-1} = s - i \mathbf{\Omega} + \mathbf{K}(s)$ (see \cref{eqn:formal-solution-s-domain}), we can write $\mathbf{G}_\mathcal{D}^{-1}$ correspondingly in a block-wise way. 
\begin{equation}
    \mathbf{G}_\mathcal{D}^{-1} =
    \begin{pmatrix}
        \mathbf{G}_{\txtA}^{-1} & \mathbf{K}_{\text{AB}} \\
        \mathbf{K}_{\text{BA}} & \mathbf{G}_{\txtB}^{-1}
    \end{pmatrix}
    \,,
    \quad\text{where}\quad
    \mathbf{G}_{\txtA}^{-1} =
    \begin{pmatrix}
        \mathbf{G}_{A_{\bvec{k}}}^{-1} & 0 \\
        0 & \mathbf{G}_{A_{-\bvec{k}}}^{-1}
    \end{pmatrix}
    \,,
    \quad
    \mathbf{G}_{\txtB}^{-1} =
    \begin{pmatrix}
        \mathbf{G}_{B_\bvec{k}}^{-1} & 0 \\
        0 & \mathbf{G}_{B_\bvec{k}^*}^{-1}
    \end{pmatrix}
    \,.
\end{equation}
where we defined
\begin{equation}
\begin{aligned}
    \mathbf{G}_{A_\bvec{k}}^{-1} & \coloneq s - i \mathbf{\Omega}_{A_\bvec{k}} + \mathbf{K}_{A_\bvec{k}} \,,
    &\quad&&
    \mathbf{G}_{A_{-\bvec{k}}}^{-1} & \coloneqq s - i \mathbf{\Omega}_{A_{-\bvec{k}}} + \mathbf{K}_{A_{-\bvec{k}}} \,,
    \\
    \mathbf{G}_{B_\bvec{k}}^{-1} & \coloneqq s - i \mathbf{\Omega}_{B_\bvec{k}} + \mathbf{K}_{B_\bvec{k}} \,,
    &\quad&&
    \mathbf{G}_{B_\bvec{k}^*}^{-1} & \coloneqq s - i \mathbf{\Omega}_{B_\bvec{k}^*} + \mathbf{K}_{B_\bvec{k}^*} \,.
\end{aligned}
\end{equation}

\afterpage{
\begin{landscape}
\begin{longtblr}[
    caption = {Blocks of $\mathbf{K}$ in \cref{eqn:K-blocks}},
    label = {table:blocks-of-G-D-inverse},
]{
    colspec = {Q[c,m] X[c,m]},
    rowhead = 1, rowfoot = 0,
    column{1} = {mode=dmath,colsep=0pt},
    column{2} = {mode=dmath},
    rowsep = 10pt,
    row{1} = {rowsep=2pt},
}
\hline
    \text{Block} & \text{Expression}   \\
\hline
    \mathbf{K}_{A_\bvec{k}}
    &
    { 
    \left(
        \begin{array}{cccc}
         i \widetilde{\Sigma }_{11}\left(s-i \omega _1\right)-i \widetilde{\Sigma }_{11}\left(s+i \omega _1\right) & i \widetilde{\Sigma }_{12}\left(s-i \omega _1\right) & -i \widetilde{\Sigma }_{12}\left(s+i \omega _1\right) & 0 \\
         i \widetilde{\Sigma }_{21}\left(s-i \omega _1\right) & i \widetilde{\Sigma }_{22}\left(s-i \omega _1\right)-i \widetilde{\Sigma }_{11}\left(s+i \omega _2\right) & 0 & -i \widetilde{\Sigma }_{12}\left(s+i \omega _2\right) \\
         -i \widetilde{\Sigma }_{21}\left(s+i \omega _1\right) & 0 & i \widetilde{\Sigma }_{11}\left(s-i \omega _2\right)-i \widetilde{\Sigma }_{22}\left(s+i \omega _1\right) & i \widetilde{\Sigma }_{12}\left(s-i \omega _2\right) \\
         0 & -i \widetilde{\Sigma }_{21}\left(s+i \omega _2\right) & i \widetilde{\Sigma }_{21}\left(s-i \omega _2\right) & i \widetilde{\Sigma }_{22}\left(s-i \omega _2\right)-i \widetilde{\Sigma }_{22}\left(s+i \omega _2\right) \\
        \end{array}
    \right)
    }
\\
\hline
    \mathbf{K}_{A_{-\bvec{k}}}
    &
    {
    \left(
        \begin{array}{cccc}
         i \widetilde{\Sigma }_{11}\left(s-i \omega _1\right)-i \widetilde{\Sigma }_{11}\left(s+i \omega _1\right) & i \widetilde{\Sigma }_{12}\left(s-i \omega _1\right) & -i \widetilde{\Sigma }_{12}\left(s+i \omega _1\right) & 0 \\
         i \widetilde{\Sigma }_{21}\left(s-i \omega _1\right) & i \widetilde{\Sigma }_{22}\left(s-i \omega _1\right)-i \widetilde{\Sigma }_{11}\left(s+i \omega _2\right) & 0 & -i \widetilde{\Sigma }_{12}\left(s+i \omega _2\right) \\
         -i \widetilde{\Sigma }_{21}\left(s+i \omega _1\right) & 0 & i \widetilde{\Sigma }_{11}\left(s-i \omega _2\right)-i \widetilde{\Sigma }_{22}\left(s+i \omega _1\right) & i \widetilde{\Sigma }_{12}\left(s-i \omega _2\right) \\
         0 & -i \widetilde{\Sigma }_{21}\left(s+i \omega _2\right) & i \widetilde{\Sigma }_{21}\left(s-i \omega _2\right) & i \widetilde{\Sigma }_{22}\left(s-i \omega _2\right)-i \widetilde{\Sigma }_{22}\left(s+i \omega _2\right) \\
        \end{array}
    \right)
    }
\\
\hline
    \mathbf{K}_{B_\bvec{k}}
    &
    {
    \left(
        \begin{array}{cccc}
         2 i \widetilde{\Sigma }_{11}\left(s+i \omega _1\right) & i \widetilde{\Sigma }_{12}\left(s+i \omega _1\right) & i \widetilde{\Sigma }_{12}\left(s+i \omega _1\right) & 0 \\
         i \widetilde{\Sigma }_{21}\left(s+i \omega _1\right) & i \widetilde{\Sigma }_{11}\left(s+i \omega _2\right)+i \widetilde{\Sigma }_{22}\left(s+i \omega _1\right) & 0 & i \widetilde{\Sigma }_{12}\left(s+i \omega _2\right) \\
         i \widetilde{\Sigma }_{21}\left(s+i \omega _1\right) & 0 & i \widetilde{\Sigma }_{11}\left(s+i \omega _2\right)+i \widetilde{\Sigma }_{22}\left(s+i \omega _1\right) & i \widetilde{\Sigma }_{12}\left(s+i \omega _2\right) \\
         0 & i \widetilde{\Sigma }_{21}\left(s+i \omega _2\right) & i \widetilde{\Sigma }_{21}\left(s+i \omega _2\right) & 2 i \widetilde{\Sigma }_{22}\left(s+i \omega _2\right) \\
        \end{array}
    \right)
    }
\\
\hline
    \mathbf{K}_{B_\bvec{k}^*}
    &
    {
    \left(
    \begin{array}{cccc}
     -2 i \widetilde{\Sigma }_{11}\left(s-i \omega _1\right) & -i \widetilde{\Sigma }_{12}\left(s-i \omega _1\right) & -i \widetilde{\Sigma }_{12}\left(s-i \omega _1\right) & 0 \\
     -i \widetilde{\Sigma }_{21}\left(s-i \omega _1\right) & -i \widetilde{\Sigma }_{11}\left(s-i \omega _2\right)-i \widetilde{\Sigma }_{22}\left(s-i \omega _1\right) & 0 & -i \widetilde{\Sigma }_{12}\left(s-i \omega _2\right) \\
     -i \widetilde{\Sigma }_{21}\left(s-i \omega _1\right) & 0 & -i \widetilde{\Sigma }_{11}\left(s-i \omega _2\right)-i \widetilde{\Sigma }_{22}\left(s-i \omega _1\right) & -i \widetilde{\Sigma }_{12}\left(s-i \omega _2\right) \\
     0 & -i \widetilde{\Sigma }_{21}\left(s-i \omega _2\right) & -i \widetilde{\Sigma }_{21}\left(s-i \omega _2\right) & -2 i \widetilde{\Sigma }_{22}\left(s-i \omega _2\right) \\
    \end{array}
    \right)
    }
\\
\hline
    \mathbf{K}_{\text{AB}}
    &
    {
    \left(
        \begin{array}{cccccccc}
         -i \widetilde{\Sigma }_{11}\left(s+i \omega _1\right) & -i \widetilde{\Sigma }_{12}\left(s+i \omega _1\right) & 0 & 0 & i \widetilde{\Sigma }_{11}\left(s-i \omega _1\right) & i \widetilde{\Sigma }_{12}\left(s-i \omega _1\right) & 0 & 0 \\
         0 & 0 & -i \widetilde{\Sigma }_{11}\left(s+i \omega _2\right) & -i \widetilde{\Sigma }_{12}\left(s+i \omega _2\right) & i \widetilde{\Sigma }_{21}\left(s-i \omega _1\right) & i \widetilde{\Sigma }_{22}\left(s-i \omega _1\right) & 0 & 0 \\
         -i \widetilde{\Sigma }_{21}\left(s+i \omega _1\right) & -i \widetilde{\Sigma }_{22}\left(s+i \omega _1\right) & 0 & 0 & 0 & 0 & i \widetilde{\Sigma }_{11}\left(s-i \omega _2\right) & i \widetilde{\Sigma }_{12}\left(s-i \omega _2\right) \\
         0 & 0 & -i \widetilde{\Sigma }_{21}\left(s+i \omega _2\right) & -i \widetilde{\Sigma }_{22}\left(s+i \omega _2\right) & 0 & 0 & i \widetilde{\Sigma }_{21}\left(s-i \omega _2\right) & i \widetilde{\Sigma }_{22}\left(s-i \omega _2\right) \\
         -i \widetilde{\Sigma }_{11}\left(s+i \omega _1\right) & 0 & -i \widetilde{\Sigma }_{12}\left(s+i \omega _1\right) & 0 & i \widetilde{\Sigma }_{11}\left(s-i \omega _1\right) & 0 & i \widetilde{\Sigma }_{12}\left(s-i \omega _1\right) & 0 \\
         0 & -i \widetilde{\Sigma }_{11}\left(s+i \omega _2\right) & 0 & -i \widetilde{\Sigma }_{12}\left(s+i \omega _2\right) & i \widetilde{\Sigma }_{21}\left(s-i \omega _1\right) & 0 & i \widetilde{\Sigma }_{22}\left(s-i \omega _1\right) & 0 \\
         -i \widetilde{\Sigma }_{21}\left(s+i \omega _1\right) & 0 & -i \widetilde{\Sigma }_{22}\left(s+i \omega _1\right) & 0 & 0 & i \widetilde{\Sigma }_{11}\left(s-i \omega _2\right) & 0 & i \widetilde{\Sigma }_{12}\left(s-i \omega _2\right) \\
         0 & -i \widetilde{\Sigma }_{21}\left(s+i \omega _2\right) & 0 & -i \widetilde{\Sigma }_{22}\left(s+i \omega _2\right) & 0 & i \widetilde{\Sigma }_{21}\left(s-i \omega _2\right) & 0 & i \widetilde{\Sigma }_{22}\left(s-i \omega _2\right) \\
        \end{array}
    \right)
    }
\\
\hline
    \mathbf{K}_{\text{BA}}
    &
    {
    \left(
        \begin{array}{cccccccc}
         i \widetilde{\Sigma }_{11}\left(s+i \omega _1\right) & 0 & i \widetilde{\Sigma }_{12}\left(s+i \omega _1\right) & 0 & i \widetilde{\Sigma }_{11}\left(s+i \omega _1\right) & 0 & i \widetilde{\Sigma }_{12}\left(s+i \omega _1\right) & 0 \\
         i \widetilde{\Sigma }_{21}\left(s+i \omega _1\right) & 0 & i \widetilde{\Sigma }_{22}\left(s+i \omega _1\right) & 0 & 0 & i \widetilde{\Sigma }_{11}\left(s+i \omega _2\right) & 0 & i \widetilde{\Sigma }_{12}\left(s+i \omega _2\right) \\
         0 & i \widetilde{\Sigma }_{11}\left(s+i \omega _2\right) & 0 & i \widetilde{\Sigma }_{12}\left(s+i \omega _2\right) & i \widetilde{\Sigma }_{21}\left(s+i \omega _1\right) & 0 & i \widetilde{\Sigma }_{22}\left(s+i \omega _1\right) & 0 \\
         0 & i \widetilde{\Sigma }_{21}\left(s+i \omega _2\right) & 0 & i \widetilde{\Sigma }_{22}\left(s+i \omega _2\right) & 0 & i \widetilde{\Sigma }_{21}\left(s+i \omega _2\right) & 0 & i \widetilde{\Sigma }_{22}\left(s+i \omega _2\right) \\
         -i \widetilde{\Sigma }_{11}\left(s-i \omega _1\right) & -i \widetilde{\Sigma }_{12}\left(s-i \omega _1\right) & 0 & 0 & -i \widetilde{\Sigma }_{11}\left(s-i \omega _1\right) & -i \widetilde{\Sigma }_{12}\left(s-i \omega _1\right) & 0 & 0 \\
         -i \widetilde{\Sigma }_{21}\left(s-i \omega _1\right) & -i \widetilde{\Sigma }_{22}\left(s-i \omega _1\right) & 0 & 0 & 0 & 0 & -i \widetilde{\Sigma }_{11}\left(s-i \omega _2\right) & -i \widetilde{\Sigma }_{12}\left(s-i \omega _2\right) \\
         0 & 0 & -i \widetilde{\Sigma }_{11}\left(s-i \omega _2\right) & -i \widetilde{\Sigma }_{12}\left(s-i \omega _2\right) & -i \widetilde{\Sigma }_{21}\left(s-i \omega _1\right) & -i \widetilde{\Sigma }_{22}\left(s-i \omega _1\right) & 0 & 0 \\
         0 & 0 & -i \widetilde{\Sigma }_{21}\left(s-i \omega _2\right) & -i \widetilde{\Sigma }_{22}\left(s-i \omega _2\right) & 0 & 0 & -i \widetilde{\Sigma }_{21}\left(s-i \omega _2\right) & -i \widetilde{\Sigma }_{22}\left(s-i \omega _2\right) \\
        \end{array}
    \right)
    }
\\
\hline
\end{longtblr}
\end{landscape}
}

\subsection{Inhomogeneous terms in the s-domain}
\label{subapp:inhomogeneous-terms-in-s-domain}

The inhomogemeous terms are all in the form of an integral over time (see \cref{eqn:inhomogeneous-terms-in-time}). Hence, their Laplace transforms all take the following form.
\begin{equation}
    \mathcal{I}_{\text{label},cd} = \frac{1}{s} N_{\text{label},cd} \,,
    \qquad
    \text{label:}\;\; A_{\pm\bvec{k}}, B_\bvec{k}, B_\bvec{k}^* \,.
    \label{eqn:inhomogeneous-terms-in-s-domain}
\end{equation}
where
\begin{align*}
    N_{A_\bvec{k},cd}(s) & = 2 \bab{ \widetilde{\mathcal{N}}_{cd}(s+i\omega_d) + \widetilde{\mathcal{N}}_{dc}(s-i\omega_c) } \,,
    &
    N_{B_\bvec{k},cd}(s) & = -2 \bab{ \widetilde{N}_{cd}(s+i\omega_d) + \widetilde{N}_{dc}(s+i\omega_c) } \,,
    \\
    N_{A_{-\bvec{k}},cd}(s) & = 2 \bab{ \widetilde{N}_{cd}(s+i\omega_d) + \widetilde{N}_{dc}(s-i\omega_c) } \,,
    &
    N_{B_{\bvec{k}}^*,cd}(s) & = -2 \bab{ \widetilde{N}_{cd}(s-i\omega_d) + \widetilde{N}_{dc}(s-i\omega_c) } \,.
\end{align*}
We dropped the ``$\abs{\bvec{k}}$'' dependence of the noise-kernel for notational convenience. Following the order given in \cref{eqn:A-B-separation,eqn:decomposed-G-D-equation}, we define $\overrightarrow{\mathcal{I}} = \bab*{\overrightarrow{\mathcal{I}}_A, \overrightarrow{\mathcal{I}}_B}^\T$ with $\overrightarrow{\mathcal{I}}_A = \bab*{\overrightarrow{\mathcal{I}}_{A_{\bvec{k}}}, \overrightarrow{\mathcal{I}}_{A_{-\bvec{k}}}}^\T$ and $\overrightarrow{\mathcal{I}}_B = \bab*{ \overrightarrow{\mathcal{I}}_{B_\bvec{k}}, \overrightarrow{\mathcal{I}}_{B_\bvec{k}^*} }^\T$, where
\begin{equation}
    \overrightarrow{\mathcal{I}}_{i} = \bab{ \mathcal{I}_{i,11},\; \mathcal{I}_{i,12},\; \mathcal{I}_{i,21},\; \mathcal{I}_{i,22} } \,,
    \qquad
    i= A_{\pm\bvec{k}}, B_\bvec{k}, B_\bvec{k}^* \,.
\end{equation}
Correspondingly, we have $\overrightarrow{N} = \bab*{\overrightarrow{N}_{\txtA}, \overrightarrow{N}_{\txtB}}^\T$ with $\overrightarrow{N}_\txtA = \bab*{\overrightarrow{N}_{A_\bvec{k}}, \overrightarrow{N}_{A_{-\bvec{k}}}}^\T$ and $\overrightarrow{N}_\txtB = \bab*{\overrightarrow{N}_{B_\bvec{k}}, \overrightarrow{N}_{B_\bvec{k}^*}}^\T$ that are induced by \cref{eqn:inhomogeneous-terms-in-s-domain}. Note that $\overrightarrow{\mathcal{I}} \sim \overrightarrow{N} \sim \mathcal{N}_{cd}(k) \sim \order{g^2}$.

\subsection{Poles of Green's functions}
\label{subapp:poles-of-Green's-functions}

As discussed in \cref{subsec:quasi-normal-modes-of-populations-and-coherence}, the poles of $\mathbf{G}_{A_{\bvec{k}}}, \mathbf{G}_{A_{-\bvec{k}}}, \mathbf{G}_{B_\bvec{k}}$ and $\mathbf{G}_{B_\bvec{k}^*}$ are found by solving the following equations up to the leading order perturbation.
\begin{equation}
    \det\Bab{\mathbf{G}_{A_\bvec{k}}^{-1}} = 0 \,,
    \quad
    \det\Bab{\mathbf{G}_{A_{-\bvec{k}}}^{-1}} = 0 \,,
    \quad
    \det\Bab{\mathbf{G}_{B_\bvec{k}}^{-1}} = 0 \,,
    \quad
    \det\Bab{\mathbf{G}_{B_\bvec{k}^*}^{-1}} = 0 \,.
    \quad
\end{equation}
Since $\mathbf{G}_{A_\bvec{k}}^{-1}(s) = \mathbf{G}_{A_{-\bvec{k}}}^{-1}(s)$, we only need to solve three equations. The solutions are different in the non-degenerate case ($m_1^2 - m_2^2 \sim \order{1}$) and the nearly-degenerate case ($m_1^2 - m_2^2 \sim \order{g^2}$). The results are listed in \cref{table:poles-of-two-point-correlation-functions} where for simpler notations, we dropped $\bvec{k}$ when a quantity only depends on the magnitude $\abs{\bvec{k}}$ and defined
\begin{equation}
    W(-i\omegabar) = 
    \Sigma _{11}\left(-i \omegabar\right)-\Sigma _{11}\left(i \omegabar\right)+\Sigma _{22}\left(-i \omegabar\right)-\Sigma _{22}\left(i \omegabar\right) \,.
\end{equation}
Using properties shown in \cref{app:symmetries-of-noise-kernel}, $W(-i\omegabar)$ can also be simplified as $W(-i\omegabar) = -2 i \left(\Sigma _{\text{I},11}\left(i \omegabar\right)+\Sigma _{\text{I},22}\left(i \omegabar\right)\right)$. The $D(s)$ used in \cref{table:poles-of-two-point-correlation-functions} has been defined in \cref{eqn:Omega^2-D-Delta^2-definitions}. The labels of these pole are chosen such that they return to the bare frequencies of the operators indicated in the subscripts when the couplings are switched off.

\afterpage{
\begin{landscape}
    
\begin{table}[htbp!]
\centering
\begin{tblr}{
    colspec={Q[c,m]|Q[c,m]|X[c,m]|X[c,m]},
    column{2} = {mode=dmath},
    column{3} = {mode=dmath},
    column{4} = {mode=dmath}
}
\hline
{Green's \\ function} & \text{Pole} & \text{Non-degenerate} & \text{Nearly-degenerate} \\
\hline
\SetCell[r=4]{m} {$A_{\pm\bvec{p}}$ \\ $\bvec{p}=\pm\bvec{k}$}
    & s_{\acomm*{a_{1,\bvec{p}}^\dagger}{a_{1,\bvec{p}}}}
    & -\frac{i \left(\Sigma _{11}\left(-i \omega _1\right)-\Sigma _{11}\left(i \omega _1\right)\right)}{2 \omega _1}
    & -\frac{i \left(D\left(-i \omegabar\right)-D\left(i \omegabar\right)+W\left(-i \omegabar\right)\right)}{4 \omegabar}
    \\
    \hline
    & s_{\acomm*{a_{1,\bvec{p}}^\dagger}{a_{2,\bvec{p}}}}
    & +i \left(\omega _1-\omega _2\right) + i \left(\frac{\Sigma _{11}\left(i \omega _1\right)}{2\omega _1}-\frac{\Sigma _{22}\left(-i \omega _2\right)}{2\omega _2}\right)
    & \frac{i \left(D\left(-i \omegabar\right)+D\left(i \omegabar\right)-W\left(-i \omegabar\right)\right)}{4 \omegabar}
    \\
    \hline
    & s_{\acomm*{a_{2,\bvec{p}}^\dagger}{a_{1,\bvec{p}}}}
    & -i \left(\omega _1-\omega _2\right) - i \left(\frac{\Sigma _{11}\left(-i \omega _1\right)}{2\omega _1}-\frac{\Sigma _{22}\left(i \omega _2\right)}{2\omega _2}\right)
    & -\frac{i \left(D\left(-i \omegabar\right)+D\left(i \omegabar\right)+W\left(-i \omegabar\right)\right)}{4 \omegabar}
    \\
    \hline
    & s_{\acomm*{a_{2,\bvec{p}}^\dagger}{a_{2,\bvec{p}}}}
    & -\frac{i \left(\Sigma _{22}\left(-i \omega _2\right)-\Sigma _{22}\left(i \omega _2\right)\right)}{2 \omega _2}
    & \frac{i \left(D\left(-i \omegabar\right)-D\left(i \omegabar\right)-W\left(-i \omegabar\right)\right)}{4 \omegabar}
    \\
\hline
\SetCell[r=4]{m} $B_{\bvec{k}}$
    & s_{\acomm*{a_{1,\bvec{k}}}{a_{1,-\bvec{k}}}}
    & -2 i \omega _1 -\frac{i \Sigma _{11}\left(-i \omega _1\right)}{\omega _1}
    & -2 i \omegabar -\frac{i \left(D\left(-i \omegabar\right)+\Sigma _{11}\left(-i \omegabar\right)+\Sigma _{22}\left(-i \omegabar\right)\right)}{2 \omegabar}
    \\
    \hline
    & s_{\acomm*{a_{1,\bvec{k}}}{a_{2,-\bvec{k}}}}
    & - i \left(\omega _1+\omega _2\right) - i \left(\frac{\Sigma _{11}\left(-i \omega _1\right)}{2\omega _1}+\frac{\Sigma _{22}\left(-i \omega _2\right)}{2\omega _2}\right)
    & -2 i \omegabar -\frac{i \left(\Sigma _{11}\left(-i \omegabar\right)+\Sigma _{22}\left(-i \omegabar\right)\right)}{2 \omegabar}
    \\
    \hline
    & s_{\acomm*{a_{2,\bvec{k}}}{a_{1,-\bvec{k}}}}
    & -i \left(\omega _1+\omega _2\right) - i \left(\frac{\Sigma _{11}\left(-i \omega _1\right)}{2\omega _1}+\frac{\Sigma _{22}\left(-i \omega _2\right)}{2\omega _2}\right)
    & -2 i \omegabar -\frac{i \left(\Sigma _{11}\left(-i \omegabar\right)+\Sigma _{22}\left(-i \omegabar\right)\right)}{2 \omegabar}
    \\
    \hline
    & s_{\acomm*{a_{2,\bvec{k}}}{a_{2,-\bvec{k}}}}
    & -2 i \omega _2 -\frac{i \Sigma _{22}\left(-i \omega _2\right)}{\omega _2}
    & -2 i \omegabar -\frac{i \left(-D\left(-i \omegabar\right)+\Sigma _{11}\left(-i \omegabar\right)+\Sigma _{22}\left(-i \omegabar\right)\right)}{2 \omegabar}
    \\
\hline
\SetCell[r=4]{m} $B_{\bvec{k}}^*$
    & s_{\acomm*{a_{1,\bvec{k}}^\dagger}{a_{1,-\bvec{k}}^\dagger}}
    & +2 i \omega _1 + \frac{i \Sigma _{11}\left(i \omega _1\right)}{\omega _1}
    & +2 i \omegabar + \frac{i \left(D\left(i \omegabar\right)+\Sigma _{11}\left(i \omegabar\right)+\Sigma _{22}\left(i \omegabar\right)\right)}{2 \omegabar}
    \\
    \hline
    & s_{\acomm*{a_{1,\bvec{k}}^\dagger}{a_{2,-\bvec{k}}^\dagger}}
    & +i \left(\omega _1+\omega _2\right) + i \left(\frac{\Sigma _{11}\left(i \omega _1\right)}{2\omega _1}+\frac{\Sigma _{22}\left(i \omega _2\right)}{2\omega _2}\right)
    & +2 i \omegabar + \frac{i \left(\Sigma _{11}\left(i \omegabar\right)+\Sigma _{22}\left(i \omegabar\right)\right)}{2 \omegabar}
    \\
    \hline
    & s_{\acomm*{a_{2,\bvec{k}}^\dagger}{a_{1,-\bvec{k}}^\dagger}}
    & +i \left(\omega _1+\omega _2\right) + i \left(\frac{\Sigma _{11}\left(i \omega _1\right)}{2\omega _1}+\frac{\Sigma _{22}\left(i \omega _2\right)}{2\omega _2}\right)
    & +2 i \omegabar + \frac{i \left(\Sigma _{11}\left(i \omegabar\right)+\Sigma _{22}\left(i \omegabar\right)\right)}{2 \omegabar}
    \\
    \hline
    & s_{\acomm*{a_{2,\bvec{k}}^\dagger}{a_{2,-\bvec{k}}^\dagger}}
    & +2 i \omega _2 + \frac{i \Sigma _{22}\left(i \omega _2\right)}{\omega _2}
    & +2 i \omegabar + \frac{i \left(-D\left(i \omegabar\right)+\Sigma _{11}\left(i \omegabar\right)+\Sigma _{22}\left(i \omegabar\right)\right)}{2 \omegabar}
    \\
\hline
\end{tblr}
\caption{Poles of two-point correlation functions}
\label{table:poles-of-two-point-correlation-functions}
\end{table}

\end{landscape}
}

\subsection{Residues at poles of equal-time two-point correlation functions}
\label{subapp:residues-at-poles-of-equal-time-two-point-correlation-functions}

In this subsection, we write out all residues explicitly in both non-degenerate case and nearly-degenerate case. The residues are related to poles in \cref{table:poles-of-two-point-correlation-functions} through their subscripts. Similar to the poles, when couplings are switched off, these residues will return to the residues of free-field Green's functions of the operators given in the subscripts, in which the only non-zero element is a ``1'' on the diagonal. Because $\mathbf{G}_{A_\bvec{k}}^{-1} = \mathbf{G}_{A_{-\bvec{k}}}^{-1}$, the residues of $\mathbf{G}_{A_\bvec{k}}$ and $\mathbf{G}_{A_{-\bvec{k}}}$ are the same which is clearly shown in the following results. We will use $\delta = \omega_1 - \omega_2$ to keep notations simple.

Notice that the poles $s_{\acomm{a_{1,\bvec{k}}}{a_{2,-\bvec{k}}}}$ and $s_{\acomm{a_{2,\bvec{k}}}{a_{1,-\bvec{k}}}}$ of $\mathbf{G}_{B_\bvec{k}}$ are the same. This means they are degenerate poles in physics but represent the same pole in math. Mathematically, there is only one residue associated with them. Below is an example in the non-degenerate case.
\begin{equation*}
\left(
\begin{array}{cccc}
 0 & -\frac{\Sigma _{12}\left(-i \omega _2\right)}{2 \delta  \sqrt{\omega _1 \omega _2}} & -\frac{\Sigma _{12}\left(-i \omega _2\right)}{2 \delta  \sqrt{\omega _1 \omega _2}} & 0 \\
 -\frac{\Sigma _{21}\left(-i \omega _2\right)}{2 \delta  \sqrt{\omega _1 \omega _2}} & 1 & 0 & \frac{\Sigma _{12}\left(-i \omega _1\right)}{2 \delta  \sqrt{\omega _1 \omega _2}} \\
 -\frac{\Sigma _{21}\left(-i \omega _2\right)}{2 \delta  \sqrt{\omega _1 \omega _2}} & 0 & 1 & \frac{\Sigma _{12}\left(-i \omega _1\right)}{2 \delta  \sqrt{\omega _1 \omega _2}} \\
 0 & \frac{\Sigma _{21}\left(-i \omega _1\right)}{2 \delta  \sqrt{\omega _1 \omega _2}} & \frac{\Sigma _{21}\left(-i \omega _1\right)}{2 \delta  \sqrt{\omega _1 \omega _2}} & 0 \\
\end{array}
\right) \,.
\end{equation*}
However, we wish to relate one residue to each correlation function in $B_{cd,\bvec{k}}$ so as to interpret it as one mixing mode. This is also motivated by the ``1''s in the diagonal of the residue. They imply that this residue will become $\operatorname{diag}(0,1,0,0)$ plus $\operatorname{diag}(0,0,1,0)$ after switching off the couplings, which are exactly the residues for $B_{12,\bvec{k}}(t)$ and $B_{21,\bvec{k}}(t)$ respectively in free evolutions. Thus, in what follows, we decompose the above matrix into $\mathbb{G}_{\acomm{a_{1,\bvec{k}}}{a_{2,-\bvec{k}}}}$ and $\mathbb{G}_{\acomm{a_{2,\bvec{k}}}{a_{1,-\bvec{k}}}}$ such that they sum to the above matrix and satisfy the relation in \cref{eqn:modes-relations}. In the non-degenerate case, it turns out that this decomposition yields $\mathbb{G}_{\acomm{a_{1,\bvec{k}}}{a_{2,-\bvec{k}}}}$ and $\mathbb{G}_{\acomm{a_{2,\bvec{k}}}{a_{1,-\bvec{k}}}}$ in a form similar to other residues, where non-zero terms only appear in the row and column containing the ``1''. For the same reason, we define $\mathbb{G}_{\acomm*{a_{1,\bvec{k}}^\dagger}{a_{2,-\bvec{k}}^\dagger}}$ and $\mathbb{G}_{\acomm*{a_{2,\bvec{k}}^\dagger}{a_{1,-\bvec{k}}^\dagger}}$ for the residues of $\mathbf{G}_{B_\bvec{k}^*}$ whose poles are degenerate. The same arguments and manipulations will be applied in the nearly-degenerate case, too.

\begingroup
\renewcommand{\arraystretch}{1.3}

\noindent\textbf{Non-degenerate case}

\noindent
Below are the four residues of $\mathbf{G}_{A_\bvec{k}}$ (or $\mathbf{G}_{A_{-\bvec{k}}}$).
\begin{align*}
    & 
    \mathbb{G}_{\acomm{a_{1,\bvec{k}}^\dagger}{a_{1,\bvec{k}}}}
    = \mathbb{G}_{\acomm{a_{1,-\bvec{k}}^\dagger}{a_{1,-\bvec{k}}}}
    = 
    &&
    \mathbb{G}_{\acomm{a_{1,\bvec{k}}^\dagger}{a_{2,\bvec{k}}}}
    = \mathbb{G}_{\acomm{a_{1,-\bvec{k}}^\dagger}{a_{2,-\bvec{k}}}}
    =
    \\*    
    & \qquad 
    \left(
    \begin{array}{cccc}
     1 & \frac{\Sigma _{12}\left(-i \omega _1\right)}{2 \delta  \sqrt{\omega _1 \omega _2}} & \frac{\Sigma _{12}\left(i \omega _1\right)}{2 \delta  \sqrt{\omega _1 \omega _2}} & 0 \\
     \frac{\Sigma _{21}\left(-i \omega _1\right)}{2 \delta  \sqrt{\omega _1 \omega _2}} & 0 & 0 & 0 \\
     \frac{\Sigma _{21}\left(i \omega _1\right)}{2 \delta  \sqrt{\omega _1 \omega _2}} & 0 & 0 & 0 \\
     0 & 0 & 0 & 0 \\
    \end{array}
    \right) \,,
    && \qquad
    \left(
    \begin{array}{cccc}
     0 & -\frac{\Sigma _{12}\left(-i \omega _2\right)}{2 \delta  \sqrt{\omega _1 \omega _2}} & 0 & 0 \\
     -\frac{\Sigma _{21}\left(-i \omega _2\right)}{2 \delta  \sqrt{\omega _1 \omega _2}} & 1 & 0 & \frac{\Sigma _{12}\left(i \omega _1\right)}{2 \delta  \sqrt{\omega _1 \omega _2}} \\
     0 & 0 & 0 & 0 \\
     0 & \frac{\Sigma _{21}\left(i \omega _1\right)}{2 \delta  \sqrt{\omega _1 \omega _2}} & 0 & 0 \\
    \end{array}
    \right) \,,
    \\
    &
    \mathbb{G}_{\acomm{a_{2,\bvec{k}}^\dagger}{a_{1,\bvec{k}}}}
    = \mathbb{G}_{\acomm{a_{2,-\bvec{k}}^\dagger}{a_{1,-\bvec{k}}}}
    =
    &&
    \mathbb{G}_{\acomm{a_{2,\bvec{k}}^\dagger}{a_{2,\bvec{k}}}}
    = \mathbb{G}_{\acomm{a_{2,-\bvec{k}}^\dagger}{a_{2,-\bvec{k}}}}
    =
    \\*    
    & \qquad
    \left(
    \begin{array}{cccc}
     0 & 0 & -\frac{\Sigma _{12}\left(i \omega _2\right)}{2 \delta  \sqrt{\omega _1 \omega _2}} & 0 \\
     0 & 0 & 0 & 0 \\
     -\frac{\Sigma _{21}\left(i \omega _2\right)}{2 \delta  \sqrt{\omega _1 \omega _2}} & 0 & 1 & \frac{\Sigma _{12}\left(-i \omega _1\right)}{2 \delta  \sqrt{\omega _1 \omega _2}} \\
     0 & 0 & \frac{\Sigma _{21}\left(-i \omega _1\right)}{2 \delta  \sqrt{\omega _1 \omega _2}} & 0 \\
    \end{array}
    \right) \,,
    && \qquad
    \left(
    \begin{array}{cccc}
     0 & 0 & 0 & 0 \\
     0 & 0 & 0 & -\frac{\Sigma _{12}\left(i \omega _2\right)}{2 \delta  \sqrt{\omega _1 \omega _2}} \\
     0 & 0 & 0 & -\frac{\Sigma _{12}\left(-i \omega _2\right)}{2 \delta  \sqrt{\omega _1 \omega _2}} \\
     0 & -\frac{\Sigma _{21}\left(i \omega _2\right)}{2 \delta  \sqrt{\omega _1 \omega _2}} & -\frac{\Sigma _{21}\left(-i \omega _2\right)}{2 \delta  \sqrt{\omega _1 \omega _2}} & 1 \\
    \end{array}
    \right) \,.
\end{align*}

\noindent
Below are the four residues of $\mathbf{G}_{B_\bvec{k}}$.
\begin{align*}
    &
    \mathbb{G}_{\acomm{a_{1,\bvec{k}}}{a_{1,-\bvec{k}}}} =
    &&
    \mathbb{G}_{\acomm{a_{1,\bvec{k}}}{a_{2,-\bvec{k}}}} =
    \\*    
    & \qquad
    \left(
    \begin{array}{cccc}
     1 & \frac{\Sigma _{1,2}\left(-i \omega _1\right)}{2 \delta  \sqrt{\omega _1 \omega _2}} & \frac{\Sigma _{1,2}\left(-i \omega _1\right)}{2 \delta  \sqrt{\omega _1 \omega _2}} & 0 \\
     \frac{\Sigma _{2,1}\left(-i \omega _1\right)}{2 \delta  \sqrt{\omega _1 \omega _2}} & 0 & 0 & 0 \\
     \frac{\Sigma _{2,1}\left(-i \omega _1\right)}{2 \delta  \sqrt{\omega _1 \omega _2}} & 0 & 0 & 0 \\
     0 & 0 & 0 & 0 \\
    \end{array}
    \right) \,,
    && \qquad
    \left(
    \begin{array}{cccc}
     0 & -\frac{\Sigma _{1,2}\left(-i \omega _2\right)}{2 \delta  \sqrt{\omega _1 \omega _2}} & 0 & 0 \\
     -\frac{\Sigma _{2,1}\left(-i \omega _2\right)}{2 \delta  \sqrt{\omega _1 \omega _2}} & 1 & 0 & \frac{\Sigma _{1,2}\left(-i \omega _1\right)}{2 \delta  \sqrt{\omega _1 \omega _2}} \\
     0 & 0 & 0 & 0 \\
     0 & \frac{\Sigma _{2,1}\left(-i \omega _1\right)}{2 \delta  \sqrt{\omega _1 \omega _2}} & 0 & 0 \\
    \end{array}
    \right) \,,
    \\
    &
    \mathbb{G}_{\acomm{a_{2,\bvec{k}}}{a_{1,-\bvec{k}}}} =
    &&
    \mathbb{G}_{\acomm{a_{2,\bvec{k}}}{a_{2,-\bvec{k}}}} =
    \\*    
    & \qquad
    \left(
    \begin{array}{cccc}
     0 & 0 & -\frac{\Sigma _{1,2}\left(-i \omega _2\right)}{2 \delta  \sqrt{\omega _1 \omega _2}} & 0 \\
     0 & 0 & 0 & 0 \\
     -\frac{\Sigma _{2,1}\left(-i \omega _2\right)}{2 \delta  \sqrt{\omega _1 \omega _2}} & 0 & 1 & \frac{\Sigma _{1,2}\left(-i \omega _1\right)}{2 \delta  \sqrt{\omega _1 \omega _2}} \\
     0 & 0 & \frac{\Sigma _{2,1}\left(-i \omega _1\right)}{2 \delta  \sqrt{\omega _1 \omega _2}} & 0 \\
    \end{array}
    \right) \,,
    && \qquad
    \left(
    \begin{array}{cccc}
     0 & 0 & 0 & 0 \\
     0 & 0 & 0 & -\frac{\Sigma _{1,2}\left(-i \omega _2\right)}{2 \delta  \sqrt{\omega _1 \omega _2}} \\
     0 & 0 & 0 & -\frac{\Sigma _{1,2}\left(-i \omega _2\right)}{2 \delta  \sqrt{\omega _1 \omega _2}} \\
     0 & -\frac{\Sigma _{2,1}\left(-i \omega _2\right)}{2 \delta  \sqrt{\omega _1 \omega _2}} & -\frac{\Sigma _{2,1}\left(-i \omega _2\right)}{2 \delta  \sqrt{\omega _1 \omega _2}} & 1 \\
    \end{array}
    \right) \,.
\end{align*}

\noindent
Below are the four residues of $\mathbf{G}_{B_\bvec{k}^*}$.
\begin{align*}
    &
    \mathbb{G}_{\acomm{a_{1,\bvec{k}}^\dagger}{a_{1,-\bvec{k}}^\dagger}} =
    &&
    \mathbb{G}_{\acomm{a_{1,\bvec{k}}^\dagger}{a_{2,-\bvec{k}}^\dagger}} =
    \\*    
    & \qquad
    \left(
    \begin{array}{cccc}
     1 & \frac{\Sigma _{1,2}\left(i \omega _1\right)}{2 \delta  \sqrt{\omega _1 \omega _2}} & \frac{\Sigma _{1,2}\left(i \omega _1\right)}{2 \delta  \sqrt{\omega _1 \omega _2}} & 0 \\
     \frac{\Sigma _{2,1}\left(i \omega _1\right)}{2 \delta  \sqrt{\omega _1 \omega _2}} & 0 & 0 & 0 \\
     \frac{\Sigma _{2,1}\left(i \omega _1\right)}{2 \delta  \sqrt{\omega _1 \omega _2}} & 0 & 0 & 0 \\
     0 & 0 & 0 & 0 \\
    \end{array}
    \right) \,,
    && \qquad
    \left(
    \begin{array}{cccc}
     0 & -\frac{\Sigma _{1,2}\left(i \omega _2\right)}{2 \delta  \sqrt{\omega _1 \omega _2}} & 0 & 0 \\
     -\frac{\Sigma _{2,1}\left(i \omega _2\right)}{2 \delta  \sqrt{\omega _1 \omega _2}} & 1 & 0 & \frac{\Sigma _{1,2}\left(i \omega _1\right)}{2 \delta  \sqrt{\omega _1 \omega _2}} \\
     0 & 0 & 0 & 0 \\
     0 & \frac{\Sigma _{2,1}\left(i \omega _1\right)}{2 \delta  \sqrt{\omega _1 \omega _2}} & 0 & 0 \\
    \end{array}
    \right) \,,
    \\
    &
    \mathbb{G}_{\acomm{a_{2,\bvec{k}}^\dagger}{a_{1,-\bvec{k}}^\dagger}} =
    &&
    \mathbb{G}_{\acomm{a_{2,\bvec{k}}^\dagger}{a_{1,-\bvec{k}}^\dagger}} =
    \\*    
    & \qquad
    \left(
    \begin{array}{cccc}
     0 & 0 & -\frac{\Sigma _{1,2}\left(i \omega _2\right)}{2 \delta  \sqrt{\omega _1 \omega _2}} & 0 \\
     0 & 0 & 0 & 0 \\
     -\frac{\Sigma _{2,1}\left(i \omega _2\right)}{2 \delta  \sqrt{\omega _1 \omega _2}} & 0 & 1 & \frac{\Sigma _{1,2}\left(i \omega _1\right)}{2 \delta  \sqrt{\omega _1 \omega _2}} \\
     0 & 0 & \frac{\Sigma _{2,1}\left(i \omega _1\right)}{2 \delta  \sqrt{\omega _1 \omega _2}} & 0 \\
    \end{array}
    \right) \,,
    && \qquad
    \left(
    \begin{array}{cccc}
     0 & 0 & 0 & 0 \\
     0 & 0 & 0 & -\frac{\Sigma _{1,2}\left(i \omega _2\right)}{2 \delta  \sqrt{\omega _1 \omega _2}} \\
     0 & 0 & 0 & -\frac{\Sigma _{1,2}\left(i \omega _2\right)}{2 \delta  \sqrt{\omega _1 \omega _2}} \\
     0 & -\frac{\Sigma _{2,1}\left(i \omega _2\right)}{2 \delta  \sqrt{\omega _1 \omega _2}} & -\frac{\Sigma _{2,1}\left(i \omega _2\right)}{2 \delta  \sqrt{\omega _1 \omega _2}} & 1 \\
    \end{array}
    \right) \,.
\end{align*}
\endgroup 

\begin{landscape}
\begingroup
\renewcommand{\arraystretch}{1.5}
\addtolength{\jot}{0.5em}

\noindent\textbf{Nearly-degenerate case}

\noindent
Below are the four residues of $\mathbf{G}_{A_\bvec{k}}$ (or $\mathbf{G}_{A_{-\bvec{k}}}$). The residues of $\mathbf{G}_{A_\bvec{k}}$ and $\mathbf{G}_{A_{-\bvec{k}}}$ are the same.
\begin{align*}
    \left.\begin{aligned}
        & \mathbb{G}_{\acomm{a_{1,\bvec{k}}^\dagger}{a_{1,\bvec{k}}}}
        \\
        & \mathbb{G}_{\acomm{a_{1,-\bvec{k}}^\dagger}{a_{1,-\bvec{k}}}}
    \end{aligned}\right\}
    & = 
    \left(
\begin{array}{cccc}
 \left(\frac{1}{2}+\frac{\Delta ^2\left(-i \omegabar\right)}{2 D\left(-i \omegabar\right)}\right) \left(\frac{1}{2}+\frac{\Delta ^2\left(i \omegabar\right)}{2 D\left(i \omegabar\right)}\right) & \left(\frac{1}{2}+\frac{\Delta ^2\left(i \omegabar\right)}{2 D\left(i \omegabar\right)}\right) \frac{\Sigma _{12}\left(-i \omegabar\right)}{D\left(-i \omegabar\right)} & \left(\frac{1}{2}+\frac{\Delta ^2\left(-i \omegabar\right)}{2 D\left(-i \omegabar\right)}\right) \frac{\Sigma _{12}\left(i \omegabar\right)}{D\left(i \omegabar\right)} & \frac{\Sigma _{12}\left(-i \omegabar\right)}{D\left(-i \omegabar\right)} \frac{\Sigma _{12}\left(i \omegabar\right)}{D\left(i \omegabar\right)} \\
 \left(\frac{1}{2}+\frac{\Delta ^2\left(i \omegabar\right)}{2 D\left(i \omegabar\right)}\right) \frac{\Sigma _{21}\left(-i \omegabar\right)}{D\left(-i \omegabar\right)} & \left(\frac{1}{2}-\frac{\Delta ^2\left(-i \omegabar\right)}{2 D\left(-i \omegabar\right)}\right) \left(\frac{1}{2}+\frac{\Delta ^2\left(i \omegabar\right)}{2 D\left(i \omegabar\right)}\right) & \frac{\Sigma _{12}\left(i \omegabar\right)}{D\left(i \omegabar\right)} \frac{\Sigma _{21}\left(-i \omegabar\right)}{D\left(-i \omegabar\right)} & \left(\frac{1}{2}-\frac{\Delta ^2\left(-i \omegabar\right)}{2 D\left(-i \omegabar\right)}\right) \frac{\Sigma _{12}\left(i \omegabar\right)}{D\left(i \omegabar\right)} \\
 \left(\frac{1}{2}+\frac{\Delta ^2\left(-i \omegabar\right)}{2 D\left(-i \omegabar\right)}\right) \frac{\Sigma _{21}\left(i \omegabar\right)}{D\left(i \omegabar\right)} & \frac{\Sigma _{12}\left(-i \omegabar\right)}{D\left(-i \omegabar\right)} \frac{\Sigma _{21}\left(i \omegabar\right)}{D\left(i \omegabar\right)} & \left(\frac{1}{2}+\frac{\Delta ^2\left(-i \omegabar\right)}{2 D\left(-i \omegabar\right)}\right) \left(\frac{1}{2}-\frac{\Delta ^2\left(i \omegabar\right)}{2 D\left(i \omegabar\right)}\right) & \left(\frac{1}{2}-\frac{\Delta ^2\left(i \omegabar\right)}{2 D\left(i \omegabar\right)}\right) \frac{\Sigma _{12}\left(-i \omegabar\right)}{D\left(-i \omegabar\right)} \\
 \frac{\Sigma _{21}\left(-i \omegabar\right)}{D\left(-i \omegabar\right)} \frac{\Sigma _{21}\left(i \omegabar\right)}{D\left(i \omegabar\right)} & \left(\frac{1}{2}-\frac{\Delta ^2\left(-i \omegabar\right)}{2 D\left(-i \omegabar\right)}\right) \frac{\Sigma _{21}\left(i \omegabar\right)}{D\left(i \omegabar\right)} & \left(\frac{1}{2}-\frac{\Delta ^2\left(i \omegabar\right)}{2 D\left(i \omegabar\right)}\right) \frac{\Sigma _{21}\left(-i \omegabar\right)}{D\left(-i \omegabar\right)} & \left(\frac{1}{2}-\frac{\Delta ^2\left(-i \omegabar\right)}{2 D\left(-i \omegabar\right)}\right) \left(\frac{1}{2}-\frac{\Delta ^2\left(i \omegabar\right)}{2 D\left(i \omegabar\right)}\right) \\
\end{array}
\right) \,,
    \\
    \left.\begin{aligned}
        & \mathbb{G}_{\acomm{a_{1,\bvec{k}}^\dagger}{a_{2,\bvec{k}}}}
        \\
        & \mathbb{G}_{\acomm{a_{1,-\bvec{k}}^\dagger}{a_{2,-\bvec{k}}}}
    \end{aligned}\right\}
    & =
\left(
\begin{array}{cccc}
 \left(\frac{1}{2}-\frac{\Delta ^2\left(-i \omegabar\right)}{2 D\left(-i \omegabar\right)}\right) \left(\frac{1}{2}+\frac{\Delta ^2\left(i \omegabar\right)}{2 D\left(i \omegabar\right)}\right) & -\left(\frac{1}{2}+\frac{\Delta ^2\left(i \omegabar\right)}{2 D\left(i \omegabar\right)}\right) \frac{\Sigma _{12}\left(-i \omegabar\right)}{D\left(-i \omegabar\right)} & \left(\frac{1}{2}-\frac{\Delta ^2\left(-i \omegabar\right)}{2 D\left(-i \omegabar\right)}\right) \frac{\Sigma _{12}\left(i \omegabar\right)}{D\left(i \omegabar\right)} & -\frac{\Sigma _{12}\left(-i \omegabar\right)}{D\left(-i \omegabar\right)} \frac{\Sigma _{12}\left(i \omegabar\right)}{D\left(i \omegabar\right)} \\
 -\left(\frac{1}{2}+\frac{\Delta ^2\left(i \omegabar\right)}{2 D\left(i \omegabar\right)}\right) \frac{\Sigma _{21}\left(-i \omegabar\right)}{D\left(-i \omegabar\right)} & \left(\frac{1}{2}+\frac{\Delta ^2\left(-i \omegabar\right)}{2 D\left(-i \omegabar\right)}\right) \left(\frac{1}{2}+\frac{\Delta ^2\left(i \omegabar\right)}{2 D\left(i \omegabar\right)}\right) & -\frac{\Sigma _{12}\left(i \omegabar\right)}{D\left(i \omegabar\right)} \frac{\Sigma _{21}\left(-i \omegabar\right)}{D\left(-i \omegabar\right)} & \left(\frac{1}{2}+\frac{\Delta ^2\left(-i \omegabar\right)}{2 D\left(-i \omegabar\right)}\right) \frac{\Sigma _{12}\left(i \omegabar\right)}{D\left(i \omegabar\right)} \\
 \left(\frac{1}{2}-\frac{\Delta ^2\left(-i \omegabar\right)}{2 D\left(-i \omegabar\right)}\right) \frac{\Sigma _{21}\left(i \omegabar\right)}{D\left(i \omegabar\right)} & -\frac{\Sigma _{12}\left(-i \omegabar\right)}{D\left(-i \omegabar\right)} \frac{\Sigma _{21}\left(i \omegabar\right)}{D\left(i \omegabar\right)} & \left(\frac{1}{2}-\frac{\Delta ^2\left(-i \omegabar\right)}{2 D\left(-i \omegabar\right)}\right) \left(\frac{1}{2}-\frac{\Delta ^2\left(i \omegabar\right)}{2 D\left(i \omegabar\right)}\right) & -\left(\frac{1}{2}-\frac{\Delta ^2\left(i \omegabar\right)}{2 D\left(i \omegabar\right)}\right) \frac{\Sigma _{12}\left(-i \omegabar\right)}{D\left(-i \omegabar\right)} \\
 -\frac{\Sigma _{21}\left(-i \omegabar\right)}{D\left(-i \omegabar\right)} \frac{\Sigma _{21}\left(i \omegabar\right)}{D\left(i \omegabar\right)} & \left(\frac{1}{2}+\frac{\Delta ^2\left(-i \omegabar\right)}{2 D\left(-i \omegabar\right)}\right) \frac{\Sigma _{21}\left(i \omegabar\right)}{D\left(i \omegabar\right)} & -\left(\frac{1}{2}-\frac{\Delta ^2\left(i \omegabar\right)}{2 D\left(i \omegabar\right)}\right) \frac{\Sigma _{21}\left(-i \omegabar\right)}{D\left(-i \omegabar\right)} & \left(\frac{1}{2}+\frac{\Delta ^2\left(-i \omegabar\right)}{2 D\left(-i \omegabar\right)}\right) \left(\frac{1}{2}-\frac{\Delta ^2\left(i \omegabar\right)}{2 D\left(i \omegabar\right)}\right) \\
\end{array}
\right) \,,
    \\
    \left.\begin{aligned}
        & \mathbb{G}_{\acomm{a_{2,\bvec{k}}^\dagger}{a_{1,\bvec{k}}}}
        \\
        & \mathbb{G}_{\acomm{a_{2,-\bvec{k}}^\dagger}{a_{1,-\bvec{k}}}}
    \end{aligned}\right\}
    & =
\left(
\begin{array}{cccc}
 \left(\frac{1}{2}+\frac{\Delta ^2\left(-i \omegabar\right)}{2 D\left(-i \omegabar\right)}\right) \left(\frac{1}{2}-\frac{\Delta ^2\left(i \omegabar\right)}{2 D\left(i \omegabar\right)}\right) & \left(\frac{1}{2}-\frac{\Delta ^2\left(i \omegabar\right)}{2 D\left(i \omegabar\right)}\right) \frac{\Sigma _{12}\left(-i \omegabar\right)}{D\left(-i \omegabar\right)} & -\left(\frac{1}{2}+\frac{\Delta ^2\left(-i \omegabar\right)}{2 D\left(-i \omegabar\right)}\right) \frac{\Sigma _{12}\left(i \omegabar\right)}{D\left(i \omegabar\right)} & -\frac{\Sigma _{12}\left(-i \omegabar\right)}{D\left(-i \omegabar\right)} \frac{\Sigma _{12}\left(i \omegabar\right)}{D\left(i \omegabar\right)} \\
 \left(\frac{1}{2}-\frac{\Delta ^2\left(i \omegabar\right)}{2 D\left(i \omegabar\right)}\right) \frac{\Sigma _{21}\left(-i \omegabar\right)}{D\left(-i \omegabar\right)} & \left(\frac{1}{2}-\frac{\Delta ^2\left(-i \omegabar\right)}{2 D\left(-i \omegabar\right)}\right) \left(\frac{1}{2}-\frac{\Delta ^2\left(i \omegabar\right)}{2 D\left(i \omegabar\right)}\right) & -\frac{\Sigma _{12}\left(i \omegabar\right)}{D\left(i \omegabar\right)} \frac{\Sigma _{21}\left(-i \omegabar\right)}{D\left(-i \omegabar\right)} & -\left(\frac{1}{2}-\frac{\Delta ^2\left(-i \omegabar\right)}{2 D\left(-i \omegabar\right)}\right) \frac{\Sigma _{12}\left(i \omegabar\right)}{D\left(i \omegabar\right)} \\
 -\left(\frac{1}{2}+\frac{\Delta ^2\left(-i \omegabar\right)}{2 D\left(-i \omegabar\right)}\right) \frac{\Sigma _{21}\left(i \omegabar\right)}{D\left(i \omegabar\right)} & -\frac{\Sigma _{12}\left(-i \omegabar\right)}{D\left(-i \omegabar\right)} \frac{\Sigma _{21}\left(i \omegabar\right)}{D\left(i \omegabar\right)} & \left(\frac{1}{2}+\frac{\Delta ^2\left(-i \omegabar\right)}{2 D\left(-i \omegabar\right)}\right) \left(\frac{1}{2}+\frac{\Delta ^2\left(i \omegabar\right)}{2 D\left(i \omegabar\right)}\right) & \left(\frac{1}{2}+\frac{\Delta ^2\left(i \omegabar\right)}{2 D\left(i \omegabar\right)}\right) \frac{\Sigma _{12}\left(-i \omegabar\right)}{D\left(-i \omegabar\right)} \\
 -\frac{\Sigma _{21}\left(-i \omegabar\right)}{D\left(-i \omegabar\right)} \frac{\Sigma _{21}\left(i \omegabar\right)}{D\left(i \omegabar\right)} & -\left(\frac{1}{2}-\frac{\Delta ^2\left(-i \omegabar\right)}{2 D\left(-i \omegabar\right)}\right) \frac{\Sigma _{21}\left(i \omegabar\right)}{D\left(i \omegabar\right)} & \left(\frac{1}{2}+\frac{\Delta ^2\left(i \omegabar\right)}{2 D\left(i \omegabar\right)}\right) \frac{\Sigma _{21}\left(-i \omegabar\right)}{D\left(-i \omegabar\right)} & \left(\frac{1}{2}-\frac{\Delta ^2\left(-i \omegabar\right)}{2 D\left(-i \omegabar\right)}\right) \left(\frac{1}{2}+\frac{\Delta ^2\left(i \omegabar\right)}{2 D\left(i \omegabar\right)}\right) \\
\end{array}
\right) \,,
    \\
    \left.\begin{aligned}
        & \mathbb{G}_{\acomm{a_{2,\bvec{k}}^\dagger}{a_{2,\bvec{k}}}}
        \\
        & \mathbb{G}_{\acomm{a_{2,-\bvec{k}}^\dagger}{a_{2,-\bvec{k}}}}
    \end{aligned}\right\}
    & =
\left(
\begin{array}{cccc}
 \left(\frac{1}{2}-\frac{\Delta ^2\left(-i \omegabar\right)}{2 D\left(-i \omegabar\right)}\right) \left(\frac{1}{2}-\frac{\Delta ^2\left(i \omegabar\right)}{2 D\left(i \omegabar\right)}\right) & -\left(\frac{1}{2}-\frac{\Delta ^2\left(i \omegabar\right)}{2 D\left(i \omegabar\right)}\right) \frac{\Sigma _{12}\left(-i \omegabar\right)}{D\left(-i \omegabar\right)} & -\left(\frac{1}{2}-\frac{\Delta ^2\left(-i \omegabar\right)}{2 D\left(-i \omegabar\right)}\right) \frac{\Sigma _{12}\left(i \omegabar\right)}{D\left(i \omegabar\right)} & \frac{\Sigma _{12}\left(-i \omegabar\right)}{D\left(-i \omegabar\right)} \frac{\Sigma _{12}\left(i \omegabar\right)}{D\left(i \omegabar\right)} \\
 -\left(\frac{1}{2}-\frac{\Delta ^2\left(i \omegabar\right)}{2 D\left(i \omegabar\right)}\right) \frac{\Sigma _{21}\left(-i \omegabar\right)}{D\left(-i \omegabar\right)} & \left(\frac{1}{2}+\frac{\Delta ^2\left(-i \omegabar\right)}{2 D\left(-i \omegabar\right)}\right) \left(\frac{1}{2}-\frac{\Delta ^2\left(i \omegabar\right)}{2 D\left(i \omegabar\right)}\right) & \frac{\Sigma _{12}\left(i \omegabar\right)}{D\left(i \omegabar\right)} \frac{\Sigma _{21}\left(-i \omegabar\right)}{D\left(-i \omegabar\right)} & -\left(\frac{1}{2}+\frac{\Delta ^2\left(-i \omegabar\right)}{2 D\left(-i \omegabar\right)}\right) \frac{\Sigma _{12}\left(i \omegabar\right)}{D\left(i \omegabar\right)} \\
 -\left(\frac{1}{2}-\frac{\Delta ^2\left(-i \omegabar\right)}{2 D\left(-i \omegabar\right)}\right) \frac{\Sigma _{21}\left(i \omegabar\right)}{D\left(i \omegabar\right)} & \frac{\Sigma _{12}\left(-i \omegabar\right)}{D\left(-i \omegabar\right)} \frac{\Sigma _{21}\left(i \omegabar\right)}{D\left(i \omegabar\right)} & \left(\frac{1}{2}-\frac{\Delta ^2\left(-i \omegabar\right)}{2 D\left(-i \omegabar\right)}\right) \left(\frac{1}{2}+\frac{\Delta ^2\left(i \omegabar\right)}{2 D\left(i \omegabar\right)}\right) & -\left(\frac{1}{2}+\frac{\Delta ^2\left(i \omegabar\right)}{2 D\left(i \omegabar\right)}\right) \frac{\Sigma _{12}\left(-i \omegabar\right)}{D\left(-i \omegabar\right)} \\
 \frac{\Sigma _{21}\left(-i \omegabar\right)}{D\left(-i \omegabar\right)} \frac{\Sigma _{21}\left(i \omegabar\right)}{D\left(i \omegabar\right)} & -\left(\frac{1}{2}+\frac{\Delta ^2\left(-i \omegabar\right)}{2 D\left(-i \omegabar\right)}\right) \frac{\Sigma _{21}\left(i \omegabar\right)}{D\left(i \omegabar\right)} & -\left(\frac{1}{2}+\frac{\Delta ^2\left(i \omegabar\right)}{2 D\left(i \omegabar\right)}\right) \frac{\Sigma _{21}\left(-i \omegabar\right)}{D\left(-i \omegabar\right)} & \left(\frac{1}{2}+\frac{\Delta ^2\left(-i \omegabar\right)}{2 D\left(-i \omegabar\right)}\right) \left(\frac{1}{2}+\frac{\Delta ^2\left(i \omegabar\right)}{2 D\left(i \omegabar\right)}\right) \\
\end{array}
\right) \,.
\end{align*}

\noindent
Below are the four residues of $\mathbf{G}_{B_\bvec{k}}$.
\begin{align*}
    \mathbb{G}_{\acomm{a_{1,\bvec{k}}}{a_{1,-\bvec{k}}}}
    & =
\left(
\begin{array}{cccc}
 \left(\frac{1}{2}+\frac{\Delta ^2\left(-i \omegabar\right)}{2 D\left(-i \omegabar\right)}\right)^2 & \left(\frac{1}{2}+\frac{\Delta ^2\left(-i \omegabar\right)}{2 D\left(-i \omegabar\right)}\right) \frac{\Sigma _{12}\left(-i \omegabar\right)}{D\left(-i \omegabar\right)} & \left(\frac{1}{2}+\frac{\Delta ^2\left(-i \omegabar\right)}{2 D\left(-i \omegabar\right)}\right) \frac{\Sigma _{12}\left(-i \omegabar\right)}{D\left(-i \omegabar\right)} & \frac{\Sigma _{12}\left(-i \omegabar\right){}^2}{D\left(-i \omegabar\right)^2} \\
 \left(\frac{1}{2}+\frac{\Delta ^2\left(-i \omegabar\right)}{2 D\left(-i \omegabar\right)}\right) \frac{\Sigma _{21}\left(-i \omegabar\right)}{D\left(-i \omegabar\right)} & \left(\frac{1}{2}-\frac{\Delta ^2\left(-i \omegabar\right)}{2 D\left(-i \omegabar\right)}\right) \left(\frac{1}{2}+\frac{\Delta ^2\left(-i \omegabar\right)}{2 D\left(-i \omegabar\right)}\right) & \frac{\Sigma _{12}\left(-i \omegabar\right) \Sigma _{21}\left(-i \omegabar\right)}{D\left(-i \omegabar\right)^2} & \left(\frac{1}{2}-\frac{\Delta ^2\left(-i \omegabar\right)}{2 D\left(-i \omegabar\right)}\right) \frac{\Sigma _{12}\left(-i \omegabar\right)}{D\left(-i \omegabar\right)} \\
 \left(\frac{1}{2}+\frac{\Delta ^2\left(-i \omegabar\right)}{2 D\left(-i \omegabar\right)}\right) \frac{\Sigma _{21}\left(-i \omegabar\right)}{D\left(-i \omegabar\right)} & \frac{\Sigma _{12}\left(-i \omegabar\right) \Sigma _{21}\left(-i \omegabar\right)}{D\left(-i \omegabar\right)^2} & \left(\frac{1}{2}-\frac{\Delta ^2\left(-i \omegabar\right)}{2 D\left(-i \omegabar\right)}\right) \left(\frac{1}{2}+\frac{\Delta ^2\left(-i \omegabar\right)}{2 D\left(-i \omegabar\right)}\right) & \left(\frac{1}{2}-\frac{\Delta ^2\left(-i \omegabar\right)}{2 D\left(-i \omegabar\right)}\right) \frac{\Sigma _{12}\left(-i \omegabar\right)}{D\left(-i \omegabar\right)} \\
 \frac{\Sigma _{21}\left(-i \omegabar\right){}^2}{D\left(-i \omegabar\right)^2} & \left(\frac{1}{2}-\frac{\Delta ^2\left(-i \omegabar\right)}{2 D\left(-i \omegabar\right)}\right) \frac{\Sigma _{21}\left(-i \omegabar\right)}{D\left(-i \omegabar\right)} & \left(\frac{1}{2}-\frac{\Delta ^2\left(-i \omegabar\right)}{2 D\left(-i \omegabar\right)}\right) \frac{\Sigma _{21}\left(-i \omegabar\right)}{D\left(-i \omegabar\right)} & \left(\frac{1}{2}-\frac{\Delta ^2\left(-i \omegabar\right)}{2 D\left(-i \omegabar\right)}\right)^2 \\
\end{array}
\right) \,,
    \\
    \mathbb{G}_{\acomm{a_{1,\bvec{k}}}{a_{2,-\bvec{k}}}}
    & = 
\left(
\begin{array}{cccc}
 \left(\frac{1}{2}-\frac{\Delta ^2\left(-i \omegabar\right)}{2 D\left(-i \omegabar\right)}\right) \left(\frac{1}{2}+\frac{\Delta ^2\left(-i \omegabar\right)}{2 D\left(-i \omegabar\right)}\right) & -\left(\frac{1}{2}+\frac{\Delta ^2\left(-i \omegabar\right)}{2 D\left(-i \omegabar\right)}\right) \frac{\Sigma _{12}\left(-i \omegabar\right)}{D\left(-i \omegabar\right)} & \left(\frac{1}{2}-\frac{\Delta ^2\left(-i \omegabar\right)}{2 D\left(-i \omegabar\right)}\right) \frac{\Sigma _{12}\left(-i \omegabar\right)}{D\left(-i \omegabar\right)} & -\frac{\Sigma _{12}\left(-i \omegabar\right){}^2}{D\left(-i \omegabar\right)^2} \\
 -\left(\frac{1}{2}+\frac{\Delta ^2\left(-i \omegabar\right)}{2 D\left(-i \omegabar\right)}\right) \frac{\Sigma _{21}\left(-i \omegabar\right)}{D\left(-i \omegabar\right)} & \left(\frac{1}{2}+\frac{\Delta ^2\left(-i \omegabar\right)}{2 D\left(-i \omegabar\right)}\right)^2 & -\frac{\Sigma _{12}\left(-i \omegabar\right) \Sigma _{21}\left(-i \omegabar\right)}{D\left(-i \omegabar\right)^2} & \left(\frac{1}{2}+\frac{\Delta ^2\left(-i \omegabar\right)}{2 D\left(-i \omegabar\right)}\right) \frac{\Sigma _{12}\left(-i \omegabar\right)}{D\left(-i \omegabar\right)} \\
 \left(\frac{1}{2}-\frac{\Delta ^2\left(-i \omegabar\right)}{2 D\left(-i \omegabar\right)}\right) \frac{\Sigma _{21}\left(-i \omegabar\right)}{D\left(-i \omegabar\right)} & -\frac{\Sigma _{12}\left(-i \omegabar\right) \Sigma _{21}\left(-i \omegabar\right)}{D\left(-i \omegabar\right)^2} & \left(\frac{1}{2}-\frac{\Delta ^2\left(-i \omegabar\right)}{2 D\left(-i \omegabar\right)}\right)^2 & -\left(\frac{1}{2}-\frac{\Delta ^2\left(-i \omegabar\right)}{2 D\left(-i \omegabar\right)}\right) \frac{\Sigma _{12}\left(-i \omegabar\right)}{D\left(-i \omegabar\right)} \\
 -\frac{\Sigma _{21}\left(-i \omegabar\right){}^2}{D\left(-i \omegabar\right)^2} & \left(\frac{1}{2}+\frac{\Delta ^2\left(-i \omegabar\right)}{2 D\left(-i \omegabar\right)}\right) \frac{\Sigma _{21}\left(-i \omegabar\right)}{D\left(-i \omegabar\right)} & -\left(\frac{1}{2}-\frac{\Delta ^2\left(-i \omegabar\right)}{2 D\left(-i \omegabar\right)}\right) \frac{\Sigma _{21}\left(-i \omegabar\right)}{D\left(-i \omegabar\right)} & \left(\frac{1}{2}-\frac{\Delta ^2\left(-i \omegabar\right)}{2 D\left(-i \omegabar\right)}\right) \left(\frac{1}{2}+\frac{\Delta ^2\left(-i \omegabar\right)}{2 D\left(-i \omegabar\right)}\right) \\
\end{array}
\right) \,,
    \\
    \mathbb{G}_{\acomm{a_{2,\bvec{k}}}{a_{1,-\bvec{k}}}}
    & =
\left(
\begin{array}{cccc}
 \left(\frac{1}{2}-\frac{\Delta ^2\left(-i \omegabar\right)}{2 D\left(-i \omegabar\right)}\right) \left(\frac{1}{2}+\frac{\Delta ^2\left(-i \omegabar\right)}{2 D\left(-i \omegabar\right)}\right) & \left(\frac{1}{2}-\frac{\Delta ^2\left(-i \omegabar\right)}{2 D\left(-i \omegabar\right)}\right) \frac{\Sigma _{12}\left(-i \omegabar\right)}{D\left(-i \omegabar\right)} & -\left(\frac{1}{2}+\frac{\Delta ^2\left(-i \omegabar\right)}{2 D\left(-i \omegabar\right)}\right) \frac{\Sigma _{12}\left(-i \omegabar\right)}{D\left(-i \omegabar\right)} & -\frac{\Sigma _{12}\left(-i \omegabar\right){}^2}{D\left(-i \omegabar\right)^2} \\
 \left(\frac{1}{2}-\frac{\Delta ^2\left(-i \omegabar\right)}{2 D\left(-i \omegabar\right)}\right) \frac{\Sigma _{21}\left(-i \omegabar\right)}{D\left(-i \omegabar\right)} & \left(\frac{1}{2}-\frac{\Delta ^2\left(-i \omegabar\right)}{2 D\left(-i \omegabar\right)}\right)^2 & -\frac{\Sigma _{12}\left(-i \omegabar\right) \Sigma _{21}\left(-i \omegabar\right)}{D\left(-i \omegabar\right)^2} & -\left(\frac{1}{2}-\frac{\Delta ^2\left(-i \omegabar\right)}{2 D\left(-i \omegabar\right)}\right) \frac{\Sigma _{12}\left(-i \omegabar\right)}{D\left(-i \omegabar\right)} \\
 -\left(\frac{1}{2}+\frac{\Delta ^2\left(-i \omegabar\right)}{2 D\left(-i \omegabar\right)}\right) \frac{\Sigma _{21}\left(-i \omegabar\right)}{D\left(-i \omegabar\right)} & -\frac{\Sigma _{12}\left(-i \omegabar\right) \Sigma _{21}\left(-i \omegabar\right)}{D\left(-i \omegabar\right)^2} & \left(\frac{1}{2}+\frac{\Delta ^2\left(-i \omegabar\right)}{2 D\left(-i \omegabar\right)}\right)^2 & \left(\frac{1}{2}+\frac{\Delta ^2\left(-i \omegabar\right)}{2 D\left(-i \omegabar\right)}\right) \frac{\Sigma _{12}\left(-i \omegabar\right)}{D\left(-i \omegabar\right)} \\
 -\frac{\Sigma _{21}\left(-i \omegabar\right){}^2}{D\left(-i \omegabar\right)^2} & -\left(\frac{1}{2}-\frac{\Delta ^2\left(-i \omegabar\right)}{2 D\left(-i \omegabar\right)}\right) \frac{\Sigma _{21}\left(-i \omegabar\right)}{D\left(-i \omegabar\right)} & \left(\frac{1}{2}+\frac{\Delta ^2\left(-i \omegabar\right)}{2 D\left(-i \omegabar\right)}\right) \frac{\Sigma _{21}\left(-i \omegabar\right)}{D\left(-i \omegabar\right)} & \left(\frac{1}{2}-\frac{\Delta ^2\left(-i \omegabar\right)}{2 D\left(-i \omegabar\right)}\right) \left(\frac{1}{2}+\frac{\Delta ^2\left(-i \omegabar\right)}{2 D\left(-i \omegabar\right)}\right) \\
\end{array}
\right) \,,
    \\
    \mathbb{G}_{\acomm{a_{2,\bvec{k}}}{a_{2,-\bvec{k}}}}
    & = 
\left(
\begin{array}{cccc}
 \left(\frac{1}{2}-\frac{\Delta ^2\left(-i \omegabar\right)}{2 D\left(-i \omegabar\right)}\right)^2 & -\left(\frac{1}{2}-\frac{\Delta ^2\left(-i \omegabar\right)}{2 D\left(-i \omegabar\right)}\right) \frac{\Sigma _{12}\left(-i \omegabar\right)}{D\left(-i \omegabar\right)} & -\left(\frac{1}{2}-\frac{\Delta ^2\left(-i \omegabar\right)}{2 D\left(-i \omegabar\right)}\right) \frac{\Sigma _{12}\left(-i \omegabar\right)}{D\left(-i \omegabar\right)} & \frac{\Sigma _{12}\left(-i \omegabar\right){}^2}{D\left(-i \omegabar\right)^2} \\
 -\left(\frac{1}{2}-\frac{\Delta ^2\left(-i \omegabar\right)}{2 D\left(-i \omegabar\right)}\right) \frac{\Sigma _{21}\left(-i \omegabar\right)}{D\left(-i \omegabar\right)} & \left(\frac{1}{2}-\frac{\Delta ^2\left(-i \omegabar\right)}{2 D\left(-i \omegabar\right)}\right) \left(\frac{1}{2}+\frac{\Delta ^2\left(-i \omegabar\right)}{2 D\left(-i \omegabar\right)}\right) & \frac{\Sigma _{12}\left(-i \omegabar\right) \Sigma _{21}\left(-i \omegabar\right)}{D\left(-i \omegabar\right)^2} & -\left(\frac{1}{2}+\frac{\Delta ^2\left(-i \omegabar\right)}{2 D\left(-i \omegabar\right)}\right) \frac{\Sigma _{12}\left(-i \omegabar\right)}{D\left(-i \omegabar\right)} \\
 -\left(\frac{1}{2}-\frac{\Delta ^2\left(-i \omegabar\right)}{2 D\left(-i \omegabar\right)}\right) \frac{\Sigma _{21}\left(-i \omegabar\right)}{D\left(-i \omegabar\right)} & \frac{\Sigma _{12}\left(-i \omegabar\right) \Sigma _{21}\left(-i \omegabar\right)}{D\left(-i \omegabar\right)^2} & \left(\frac{1}{2}-\frac{\Delta ^2\left(-i \omegabar\right)}{2 D\left(-i \omegabar\right)}\right) \left(\frac{1}{2}+\frac{\Delta ^2\left(-i \omegabar\right)}{2 D\left(-i \omegabar\right)}\right) & -\left(\frac{1}{2}+\frac{\Delta ^2\left(-i \omegabar\right)}{2 D\left(-i \omegabar\right)}\right) \frac{\Sigma _{12}\left(-i \omegabar\right)}{D\left(-i \omegabar\right)} \\
 \frac{\Sigma _{21}\left(-i \omegabar\right){}^2}{D\left(-i \omegabar\right)^2} & -\left(\frac{1}{2}+\frac{\Delta ^2\left(-i \omegabar\right)}{2 D\left(-i \omegabar\right)}\right) \frac{\Sigma _{21}\left(-i \omegabar\right)}{D\left(-i \omegabar\right)} & -\left(\frac{1}{2}+\frac{\Delta ^2\left(-i \omegabar\right)}{2 D\left(-i \omegabar\right)}\right) \frac{\Sigma _{21}\left(-i \omegabar\right)}{D\left(-i \omegabar\right)} & \left(\frac{1}{2}+\frac{\Delta ^2\left(-i \omegabar\right)}{2 D\left(-i \omegabar\right)}\right)^2 \\
\end{array}
\right) \,.
\end{align*}

\noindent
Below are the four residues of $\mathbf{G}_{B_\bvec{k}^*}$.
\begin{align*}
    \mathbb{G}_{\acomm{a_{1,\bvec{k}}^\dagger}{a_{1,-\bvec{k}}^\dagger}}
    & =
\left(
\begin{array}{cccc}
 \left(\frac{1}{2}+\frac{\Delta ^2\left(i \omegabar\right)}{2 D\left(i \omegabar\right)}\right)^2 & \left(\frac{1}{2}+\frac{\Delta ^2\left(i \omegabar\right)}{2 D\left(i \omegabar\right)}\right) \frac{\Sigma _{12}\left(i \omegabar\right)}{D\left(i \omegabar\right)} & \left(\frac{1}{2}+\frac{\Delta ^2\left(i \omegabar\right)}{2 D\left(i \omegabar\right)}\right) \frac{\Sigma _{12}\left(i \omegabar\right)}{D\left(i \omegabar\right)} & \frac{\Sigma _{12}\left(i \omegabar\right){}^2}{D\left(i \omegabar\right)^2} \\
 \left(\frac{1}{2}+\frac{\Delta ^2\left(i \omegabar\right)}{2 D\left(i \omegabar\right)}\right) \frac{\Sigma _{21}\left(i \omegabar\right)}{D\left(i \omegabar\right)} & \left(\frac{1}{2}-\frac{\Delta ^2\left(i \omegabar\right)}{2 D\left(i \omegabar\right)}\right) \left(\frac{1}{2}+\frac{\Delta ^2\left(i \omegabar\right)}{2 D\left(i \omegabar\right)}\right) & \frac{\Sigma _{12}\left(i \omegabar\right) \Sigma _{21}\left(i \omegabar\right)}{D\left(i \omegabar\right)^2} & \left(\frac{1}{2}-\frac{\Delta ^2\left(i \omegabar\right)}{2 D\left(i \omegabar\right)}\right) \frac{\Sigma _{12}\left(i \omegabar\right)}{D\left(i \omegabar\right)} \\
 \left(\frac{1}{2}+\frac{\Delta ^2\left(i \omegabar\right)}{2 D\left(i \omegabar\right)}\right) \frac{\Sigma _{21}\left(i \omegabar\right)}{D\left(i \omegabar\right)} & \frac{\Sigma _{12}\left(i \omegabar\right) \Sigma _{21}\left(i \omegabar\right)}{D\left(i \omegabar\right)^2} & \left(\frac{1}{2}-\frac{\Delta ^2\left(i \omegabar\right)}{2 D\left(i \omegabar\right)}\right) \left(\frac{1}{2}+\frac{\Delta ^2\left(i \omegabar\right)}{2 D\left(i \omegabar\right)}\right) & \left(\frac{1}{2}-\frac{\Delta ^2\left(i \omegabar\right)}{2 D\left(i \omegabar\right)}\right) \frac{\Sigma _{12}\left(i \omegabar\right)}{D\left(i \omegabar\right)} \\
 \frac{\Sigma _{21}\left(i \omegabar\right){}^2}{D\left(i \omegabar\right)^2} & \left(\frac{1}{2}-\frac{\Delta ^2\left(i \omegabar\right)}{2 D\left(i \omegabar\right)}\right) \frac{\Sigma _{21}\left(i \omegabar\right)}{D\left(i \omegabar\right)} & \left(\frac{1}{2}-\frac{\Delta ^2\left(i \omegabar\right)}{2 D\left(i \omegabar\right)}\right) \frac{\Sigma _{21}\left(i \omegabar\right)}{D\left(i \omegabar\right)} & \left(\frac{1}{2}-\frac{\Delta ^2\left(i \omegabar\right)}{2 D\left(i \omegabar\right)}\right)^2 \\
\end{array}
\right) \,,
    \\
    \mathbb{G}_{\acomm{a_{1,\bvec{k}}^\dagger}{a_{2,-\bvec{k}}^\dagger}}
    & = 
\left(
\begin{array}{cccc}
 \left(\frac{1}{2}-\frac{\Delta ^2\left(i \omegabar\right)}{2 D\left(i \omegabar\right)}\right) \left(\frac{1}{2}+\frac{\Delta ^2\left(i \omegabar\right)}{2 D\left(i \omegabar\right)}\right) & -\left(\frac{1}{2}+\frac{\Delta ^2\left(i \omegabar\right)}{2 D\left(i \omegabar\right)}\right) \frac{\Sigma _{12}\left(i \omegabar\right)}{D\left(i \omegabar\right)} & \left(\frac{1}{2}-\frac{\Delta ^2\left(i \omegabar\right)}{2 D\left(i \omegabar\right)}\right) \frac{\Sigma _{12}\left(i \omegabar\right)}{D\left(i \omegabar\right)} & -\frac{\Sigma _{12}\left(i \omegabar\right){}^2}{D\left(i \omegabar\right)^2} \\
 -\left(\frac{1}{2}+\frac{\Delta ^2\left(i \omegabar\right)}{2 D\left(i \omegabar\right)}\right) \frac{\Sigma _{21}\left(i \omegabar\right)}{D\left(i \omegabar\right)} & \left(\frac{1}{2}+\frac{\Delta ^2\left(i \omegabar\right)}{2 D\left(i \omegabar\right)}\right)^2 & -\frac{\Sigma _{12}\left(i \omegabar\right) \Sigma _{21}\left(i \omegabar\right)}{D\left(i \omegabar\right)^2} & \left(\frac{1}{2}+\frac{\Delta ^2\left(i \omegabar\right)}{2 D\left(i \omegabar\right)}\right) \frac{\Sigma _{12}\left(i \omegabar\right)}{D\left(i \omegabar\right)} \\
 \left(\frac{1}{2}-\frac{\Delta ^2\left(i \omegabar\right)}{2 D\left(i \omegabar\right)}\right) \frac{\Sigma _{21}\left(i \omegabar\right)}{D\left(i \omegabar\right)} & -\frac{\Sigma _{12}\left(i \omegabar\right) \Sigma _{21}\left(i \omegabar\right)}{D\left(i \omegabar\right)^2} & \left(\frac{1}{2}-\frac{\Delta ^2\left(i \omegabar\right)}{2 D\left(i \omegabar\right)}\right)^2 & -\left(\frac{1}{2}-\frac{\Delta ^2\left(i \omegabar\right)}{2 D\left(i \omegabar\right)}\right) \frac{\Sigma _{12}\left(i \omegabar\right)}{D\left(i \omegabar\right)} \\
 -\frac{\Sigma _{21}\left(i \omegabar\right){}^2}{D\left(i \omegabar\right)^2} & \left(\frac{1}{2}+\frac{\Delta ^2\left(i \omegabar\right)}{2 D\left(i \omegabar\right)}\right) \frac{\Sigma _{21}\left(i \omegabar\right)}{D\left(i \omegabar\right)} & -\left(\frac{1}{2}-\frac{\Delta ^2\left(i \omegabar\right)}{2 D\left(i \omegabar\right)}\right) \frac{\Sigma _{21}\left(i \omegabar\right)}{D\left(i \omegabar\right)} & \left(\frac{1}{2}-\frac{\Delta ^2\left(i \omegabar\right)}{2 D\left(i \omegabar\right)}\right) \left(\frac{1}{2}+\frac{\Delta ^2\left(i \omegabar\right)}{2 D\left(i \omegabar\right)}\right) \\
\end{array}
\right) \,,
    \\
    \mathbb{G}_{\acomm{a_{2,\bvec{k}}^\dagger}{a_{1,-\bvec{k}}^\dagger}}
    & = 
\left(
\begin{array}{cccc}
 \left(\frac{1}{2}-\frac{\Delta ^2\left(i \omegabar\right)}{2 D\left(i \omegabar\right)}\right) \left(\frac{1}{2}+\frac{\Delta ^2\left(i \omegabar\right)}{2 D\left(i \omegabar\right)}\right) & \left(\frac{1}{2}-\frac{\Delta ^2\left(i \omegabar\right)}{2 D\left(i \omegabar\right)}\right) \frac{\Sigma _{12}\left(i \omegabar\right)}{D\left(i \omegabar\right)} & -\left(\frac{1}{2}+\frac{\Delta ^2\left(i \omegabar\right)}{2 D\left(i \omegabar\right)}\right) \frac{\Sigma _{12}\left(i \omegabar\right)}{D\left(i \omegabar\right)} & -\frac{\Sigma _{12}\left(i \omegabar\right){}^2}{D\left(i \omegabar\right)^2} \\
 \left(\frac{1}{2}-\frac{\Delta ^2\left(i \omegabar\right)}{2 D\left(i \omegabar\right)}\right) \frac{\Sigma _{21}\left(i \omegabar\right)}{D\left(i \omegabar\right)} & \left(\frac{1}{2}-\frac{\Delta ^2\left(i \omegabar\right)}{2 D\left(i \omegabar\right)}\right)^2 & -\frac{\Sigma _{12}\left(i \omegabar\right) \Sigma _{21}\left(i \omegabar\right)}{D\left(i \omegabar\right)^2} & -\left(\frac{1}{2}-\frac{\Delta ^2\left(i \omegabar\right)}{2 D\left(i \omegabar\right)}\right) \frac{\Sigma _{12}\left(i \omegabar\right)}{D\left(i \omegabar\right)} \\
 -\left(\frac{1}{2}+\frac{\Delta ^2\left(i \omegabar\right)}{2 D\left(i \omegabar\right)}\right) \frac{\Sigma _{21}\left(i \omegabar\right)}{D\left(i \omegabar\right)} & -\frac{\Sigma _{12}\left(i \omegabar\right) \Sigma _{21}\left(i \omegabar\right)}{D\left(i \omegabar\right)^2} & \left(\frac{1}{2}+\frac{\Delta ^2\left(i \omegabar\right)}{2 D\left(i \omegabar\right)}\right)^2 & \left(\frac{1}{2}+\frac{\Delta ^2\left(i \omegabar\right)}{2 D\left(i \omegabar\right)}\right) \frac{\Sigma _{12}\left(i \omegabar\right)}{D\left(i \omegabar\right)} \\
 -\frac{\Sigma _{21}\left(i \omegabar\right){}^2}{D\left(i \omegabar\right)^2} & -\left(\frac{1}{2}-\frac{\Delta ^2\left(i \omegabar\right)}{2 D\left(i \omegabar\right)}\right) \frac{\Sigma _{21}\left(i \omegabar\right)}{D\left(i \omegabar\right)} & \left(\frac{1}{2}+\frac{\Delta ^2\left(i \omegabar\right)}{2 D\left(i \omegabar\right)}\right) \frac{\Sigma _{21}\left(i \omegabar\right)}{D\left(i \omegabar\right)} & \left(\frac{1}{2}-\frac{\Delta ^2\left(i \omegabar\right)}{2 D\left(i \omegabar\right)}\right) \left(\frac{1}{2}+\frac{\Delta ^2\left(i \omegabar\right)}{2 D\left(i \omegabar\right)}\right) \\
\end{array}
\right) \,,
    \\
    \mathbb{G}_{\acomm{a_{2,\bvec{k}}^\dagger}{a_{2,-\bvec{k}}^\dagger}}
    & =
\left(
\begin{array}{cccc}
 \left(\frac{1}{2}-\frac{\Delta ^2\left(i \omegabar\right)}{2 D\left(i \omegabar\right)}\right)^2 & -\left(\frac{1}{2}-\frac{\Delta ^2\left(i \omegabar\right)}{2 D\left(i \omegabar\right)}\right) \frac{\Sigma _{12}\left(i \omegabar\right)}{D\left(i \omegabar\right)} & -\left(\frac{1}{2}-\frac{\Delta ^2\left(i \omegabar\right)}{2 D\left(i \omegabar\right)}\right) \frac{\Sigma _{12}\left(i \omegabar\right)}{D\left(i \omegabar\right)} & \frac{\Sigma _{12}\left(i \omegabar\right){}^2}{D\left(i \omegabar\right)^2} \\
 -\left(\frac{1}{2}-\frac{\Delta ^2\left(i \omegabar\right)}{2 D\left(i \omegabar\right)}\right) \frac{\Sigma _{21}\left(i \omegabar\right)}{D\left(i \omegabar\right)} & \left(\frac{1}{2}-\frac{\Delta ^2\left(i \omegabar\right)}{2 D\left(i \omegabar\right)}\right) \left(\frac{1}{2}+\frac{\Delta ^2\left(i \omegabar\right)}{2 D\left(i \omegabar\right)}\right) & \frac{\Sigma _{12}\left(i \omegabar\right) \Sigma _{21}\left(i \omegabar\right)}{D\left(i \omegabar\right)^2} & -\left(\frac{1}{2}+\frac{\Delta ^2\left(i \omegabar\right)}{2 D\left(i \omegabar\right)}\right) \frac{\Sigma _{12}\left(i \omegabar\right)}{D\left(i \omegabar\right)} \\
 -\left(\frac{1}{2}-\frac{\Delta ^2\left(i \omegabar\right)}{2 D\left(i \omegabar\right)}\right) \frac{\Sigma _{21}\left(i \omegabar\right)}{D\left(i \omegabar\right)} & \frac{\Sigma _{12}\left(i \omegabar\right) \Sigma _{21}\left(i \omegabar\right)}{D\left(i \omegabar\right)^2} & \left(\frac{1}{2}-\frac{\Delta ^2\left(i \omegabar\right)}{2 D\left(i \omegabar\right)}\right) \left(\frac{1}{2}+\frac{\Delta ^2\left(i \omegabar\right)}{2 D\left(i \omegabar\right)}\right) & -\left(\frac{1}{2}+\frac{\Delta ^2\left(i \omegabar\right)}{2 D\left(i \omegabar\right)}\right) \frac{\Sigma _{12}\left(i \omegabar\right)}{D\left(i \omegabar\right)} \\
 \frac{\Sigma _{21}\left(i \omegabar\right){}^2}{D\left(i \omegabar\right)^2} & -\left(\frac{1}{2}+\frac{\Delta ^2\left(i \omegabar\right)}{2 D\left(i \omegabar\right)}\right) \frac{\Sigma _{21}\left(i \omegabar\right)}{D\left(i \omegabar\right)} & -\left(\frac{1}{2}+\frac{\Delta ^2\left(i \omegabar\right)}{2 D\left(i \omegabar\right)}\right) \frac{\Sigma _{21}\left(i \omegabar\right)}{D\left(i \omegabar\right)} & \left(\frac{1}{2}+\frac{\Delta ^2\left(i \omegabar\right)}{2 D\left(i \omegabar\right)}\right)^2 \\
\end{array}
\right) \,.
\end{align*}

\endgroup 
\end{landscape}

\acknowledgments

The author gratefully acknowledges support from the U.S. National Science Foundation through grant NSF 2111743 and NSF 2412374.



\bibliographystyle{JHEP}
\bibliography{biblio.bib}






\end{document}